%% file: thesis.tex
\documentclass[]{report}

\usepackage{amssymb, amsmath}
\usepackage{graphicx}
\usepackage{psfrag,float}
\usepackage{mathrsfs}
\usepackage{fancyhdr}

\fancypagestyle{plain}{\fancyhead{}}
\DeclareSymbolFontAlphabet{\mathrsfs}{rsfs}
\DeclareMathAlphabet{\mathcal}{OMS}{cmsy}{m}{n}
\newcommand{\scri}{\mathrsfs{I}}

\providecommand{\abs}[1]{\lvert#1\rvert}
\providecommand{\norm}[1]{\lVert#1\rVert}

\newcommand{\be}{\begin{equation}}
\newcommand{\ee}{\end{equation}}

\newcommand{\tit}{\mbox{$\tilde{t}$}}
\newcommand{\tir}{\mbox{$\tilde{r}$}}

\begin{document}
\pagestyle{empty}
\include{title}
\include{acknowledge}\thispagestyle{headings}
\include{prelude}\thispagestyle{headings}
\include{intro}
\include{null}

\include{i0}

\include{conclusion}

\include{appendix}

\bibliography{references}\bibliographystyle{plain}
\end{document}

%% file: title.tex
\thispagestyle{empty}
\begin{titlepage}
\begin{center}
{\Huge\bf A conformal approach \\ \vspace{3mm}
          to numerical calculations\\ \vspace{3mm}
          \mbox{of asymptotically flat spacetimes}\\
}
\vspace{2cm}
{\huge Dissertation}
\vspace{7mm}\\
{\large von}\\
\vspace{7mm}
{\Large\bf An\i l  \c{C}olpan Zengino\u{g}lu} \\
\vspace{2cm}
{\large eingereicht bei der} \\
\vspace{3mm}
{\Large{Mathematisch-Naturwissenschaftlichen Fakult\"at \\
\vspace{1mm}
der Universit\"at Potsdam}}  \\
\vspace{1cm}
{\large durchgef\"uhrt in Golm am} \\
\vspace{3mm}
{\Large{Max Planck Institut f\"ur Gravitationsphysik\\
\vspace{1mm}
Albert Einstein Institut}}\\
\vspace{1.3cm}
{\large unter der Betreuung von}  \\
\vspace{5mm}
{\Large{Prof.~Dr.~Helmut Friedrich}}\\
\vspace{2cm}
{\Large Potsdam, Juni 2007} \\
\end{center}
\end{titlepage}

\pagestyle{empty}

%% file: acknowledge.tex
\chapter*{Acknowledgements}

I thank my supervisor Helmut Friedrich for suggesting the field of
research and for forming my speculations into arguments with his supportive 
remarks. His deep understanding ranging 
from the broad picture to the minute details of technical questions showed me
the way in many problems. I am especially grateful
for his support and trust in difficult times during my thesis.

I thank Sascha Husa for guiding me in my first steps in numerical relativity
and for his motivation, encouragement and support. 

There are many people that I would like to thank who, directly or indirectly, 
helped me to develop some of the ideas presented in this thesis.
I thank especially to Robert Beig, Christiane Lechner, 
B\'{e}la Szil\'{a}gyi and Jeffrey Winicour.

For their help with computational infrastructure that I used in my numerical 
experiments I would like to thank Thomas Radke, Christian Reisswig, 
Erik Schnetter and Jonathan Thornburg.

I thank my office mates Florian Beyer, Roger Bieli and Carsten 
Schneemann for the pleasant office atmosphere we shared. I thank Andres 
Acena, Markus Ansorg, Badri Krishnan, Mark Heinzle, Jennifer Seiler
and Tilman Vogel for fruitful discussions.

The Max Planck Institute for Gravitational Physics 
has been an efficient working environment for me with all the facilities 
that a researcher can wish for. In that respect I would like to thank the 
people working on backstage: Our cleaning lady Frau Pappa for a warm 
"G\"unaydin!" each morning, our librarians Elisabeth Schlenk and Anja 
Lehmann, our secretary Anne Lampe and the computer support.

\vspace{3cm}

\hfill \emph{To Oya \c{C}olpan}

\vspace{3mm}

\hfill \emph{and Talha Zengino\u{g}lu}

\setcounter{page}{-1}
\tableofcontents
\thispagestyle{headings}

%% file: prelude.tex
\chapter*{Preface}
The main question that this thesis tries to answer is the following:
\textit{How can we calculate numerically the gravitational field of isolated 
systems in an efficient way based upon a clear geometric framework?}

Isolated systems in general relativity are models of self-gravitating
astrophysical sources. Depending on the scales in consideration, 
these sources can be planets, stars, black holes and even 
galaxy clusters. It is not the source that makes a system isolated, 
but the behavior of its gravitational field far away from the source. 
The systems that we might consider isolated share the property that as we move 
away from the source, the gravitational field becomes weak. 
We model such systems by suitably attaching to the far-field zone
an asymptotic region in which the spacetime becomes flat in a certain sense. 
These models are commonly referred to as asymptotically flat spacetimes.

In this idealization procedure, the details of the asymptotic behavior 
need to be decided upon in accordance with physical intuition and 
the equations describing the model, as we can not yet observe 
the precise behavior of asymptotic gravitational fields
to test and modify our assumptions.

An elegant and powerful characterization of the asymptotic behavior of 
spacetimes can be given in terms of the conformal structure. Conformal
techniques make use of an important interplay between physics, conformal 
geometry and the theory of partial differential equations in the study of
Einstein equations. The conformal approach has been applied successfully 
in mathematical relativity to various problems concerned with 
the large scale structure of spacetimes.

To relate observations of gravitational radiation 
to properties of astrophysical sources, we need to understand spacetimes 
representing isolated systems arising from a large class of initial data that 
correspond to astrophysical configurations.
As the Einstein equations can not be solved explicitly for interesting 
dynamical systems emitting gravitational radiation, we need to apply 
approximation schemes to make quantitative predictions that can be compared 
with observations. 
The post-Newtonian and perturbative schemes cover the weak field 
and low velocity limits. To make reliable predictions on highly dynamical 
strong fields, we need numerical simulations.

There are many open problems in numerical calculations of spacetimes. 
At least two of these problems, namely the treatment of the outer grid boundary
and the interpretation of numerically generated spacetimes, are related to the 
asymptotic region. Applying conformal techniques to deal with these problems
can lead to accurate and efficient codes that allow us to gain control over
numerical approximations.

An important goal of this thesis is the development of a numerical
code that would allow us to calculate entire, asymptotically flat,
radiative solutions to the Einstein equations. With such a code we
should be able to follow the maximal development of asymptotically
flat initial data starting from a Cauchy surface reaching spatial
infinity up to the region close to timelike infinity. The achievement
of this goal seems to require different methods adapted to different
asymptotic regions, i.e.~spatial infinity, null infinity and timelike
infinity. With this goal in mind, we will discuss two novel
applications of conformal techniques to numerical simulations.  These
new methods share the important advantage that the representation of
the conformal factor is known a priori in terms of grid coordinates.
I will argue that this feature is very convenient, because it is
crucial to have numerical access to and some control over the
asymptotic region for accurate and efficient calculations of
asymptotically flat spacetimes.

\subsection*{Organization}
In the introduction we describe shortly the development of some basic
notions in conformal geometry. We mention the controversy on the
nature of gravitational radiation and then present the idea of
conformal infinity introduced by Penrose that is widely used in
mathematical relativity today.  We discuss conditions for the
feasibility of the conformal approach investigated by Friedrich.  An
overview of current methods in numerical relativity with respect to
the treatment of the far-field region of asymptotically flat
spacetimes is given to motivate the numerical application of conformal
techniques further.

Chapter \ref{chapter:null} concentrates on infinity in null
directions, called null infinity.  In \ref{sec:hyp_sur} and
\ref{sec:scr_fix} we investigate spacelike slices reaching null
infinity, so-called hyperboloidal slices, and conformal compactification
of the Minkowski and the extended Schwarzschild spacetimes.  We see
that one can prescribe the representation of a conformal factor in
terms of a suitably chosen compactifying radial coordinate on
spherically symmetric hyperboloidal slices such that null infinity is
at a fixed spatial coordinate location.

In \ref{sec:hyp_evol} and \ref{sec:gauge} we construct a method to
numerically treat the hyperboloidal initial value problem including
null infinity in the computational domain without making any symmetry
assumptions.  This method allows us to find solutions to the Einstein
equations in coordinates in which null infinity is fixed to a spatial
coordinate location.  It is based on the general wave gauge that is
commonly used in numerical relativity. It introduces a suitable
coupling of the conformal and the coordinate gauge to guarantee
certain geometric properties of null infinity with an appropriate
choice of the gauge source functions for the coordinates.  It requires
the numerical calculation of formally singular terms arising from the
conformal compactification.

In \ref{ss_num} we discuss numerical test results obtained with the suggested 
method in the special case of spherical symmetry on the example 
of the extended Schwarzschild spacetime.
We use a simple choice of evolution variables and numerical boundary 
treatment. While our numerical setup does not allow us to do long time 
evolutions of the extended Schwarzschild spacetime, one can see that the method
can be applied even with crude numerical techniques without the formally 
singular terms leading to an immediate blow-up of numerical errors at the outer
boundary. 

\thispagestyle{headings}

The chapter ends in \ref{sec:outlook} with an outlook on next steps to 
establish this new approach for simulating dynamical isolated systems, 
and a short discussion of its open problems and limitations.

In chapter \ref{chapter:i0} we include not just null infinity but also
spatial infinity in the numerical domain which gives us direct access
to the global structure of asymptotically flat spacetimes.  The
reduced general conformal field equations developed by Friedrich
provide the only available system for the treatment of spatial infinity.  The
system is based on the conformal Gauss gauge that we study using
numerical methods.  In \ref{cgg} we reproduce Friedrich's construction
of a conformal Gauss gauge in the Schwarzschild-Kruskal spacetime
numerically covering the entire solution in a smooth way. Going beyond
analytical studies we find out numerically that one can also cover the
Kerr solution using conformal geodesics including null infinity,
timelike infinity and the Cauchy horizon.

In \ref{sec:rgcfe} we discuss certain aspects of the reduced general
conformal field equations relevant for their numerical implementation.
In \ref{sec:ss} we solve numerically the Cauchy problem for the
equations in spherical symmetry with initial data from the
Schwarzschild-Kruskal spacetime which gives us the first numerical
calculation of an entire asymptotically flat black hole spacetime
including timelike, null and spacelike infinity and the region close
to the singularity.

In order to include dynamical gravitational fields into our discussion
of spacetimes in a neighborhood of spatial infinity, we calculate in
\ref{sec:point_data} asymptotically flat, axially symmetric initial
data based on studies by Friedrich. The calculated initial data has
the special property that its ADM-mass vanishes but its development
has a non-vanishing radiation field.  In \ref{sec:cart_rgcfe} we
develop this data on a three dimensional Cartesian grid using the
reduced general conformal field equations such that spatial infinity
is represented by the point at the origin of our Cartesian grid.  We
calculate a certain component of the Weyl tensor in a suitably adapted
Newman-Penrose tetrad representing the radiation field at null
infinity and show that it does not vanish along null infinity in
accordance with expectation.

The Cartesian code can not be used to study spacetimes with non-vanishing 
ADM-mass. For a generalizable code we implement the regular finite initial 
value problem near spatial infinity formulated by Friedrich based on the 
reduced general conformal field equations in a gauge that allows us to 
represent spatial infinity as a cylinder. The cylinder at infinity imposes a 
certain geometry that requires a numerical code which can handle a spherical 
grid topology in a frame formalism. In \ref{sec:gzps} we discuss a numerical 
implementation of a frame-based evolution system using spherical grid topology 
in three spatial dimensions with overlapping grids.

In \ref{sec:cylinder} we implement the regular finite initial value problem 
with the cylinder at spatial infinity.
A difficulty caused by the degeneracy of the equations 
at the set where the cylinder at spatial infinity touches null infinity 
is dealt with by freezing the evolution in the unphysical domain by 
exploiting the a priori knowledge of the conformal factor in terms of initial 
data and grid coordinates. The chapter ends with a discussion of possible 
improvements of the code.

In chapter \ref{sec:conclusion} I summarize the main results of the thesis 
and give an outlook for possible directions for future work. 
The chapter ends with remarks on the idea of conformal infinity.

In the appendix \ref{app:A} we discuss the generation of causal diagrams.

\thispagestyle{headings}

%% file: intro.tex
\chapter{Introduction}
\pagestyle{fancy}
This thesis is concerned with numerical studies of asymptotic fields of 
isolated systems. Conformal techniques play a fundamental role
in the discussion of the asymptotic structure of spacetimes. To develop an 
intuitive understanding of conformal rescalings, I will refer to
their application in cartography where their advantages and disadvantages 
can be addressed in a simple manner and describe the argumentation of Hermann 
Weyl to emphasize the role of the conformal structure of spacetimes in 
general relativity.

The historical controversy concerning the physical nature 
of gravitational radiation is especially demonstrative for being cautious 
about interpretations of coordinate dependent calculations in general 
relativity. I will indicate how an unambiguous formulation of gravitational
radiation that uses conformal techniques has been achieved.

The Penrose conformal compactification technique will be discussed and 
results on its feasibility due to Friedrich will be described. 
It is a goal of this thesis to combine a rigorous treatment of the far-field 
region of asymptotically flat spacetimes with advantages of widely used 
numerical methods. To motivate this goal I will point out certain properties 
of common numerical methods with special emphasis on their treatment 
of the asymptotic region. 

The introduction follows chronological lines because I believe that some of the 
current problems in numerical relativity have their counterparts 
in earlier problems that have been thoroughly discussed. We can
learn from these discussions to avoid similar confusions. Naturally, 
the presentation is highly subjective and makes no claim of completeness.
\section{Conformal geometry}\label{sec:con_geo}
A map that preserves local angles is called conformal.  The first
problem where conformal mappings played an important role, beside the
trivial rescaling where the conformal factor is just a positive
constant, was the problem of projecting the surface of a globe onto a
plane.  The stereographic projection, probably already known to the
Egyptians, has been applied in the 2nd century BC by
Hipparchus to represent the celestial sphere on an astrolabe. 
Ptolemy used the stereographic projection for his cartography, a field which
historically saw the most influential application of conformal
techniques \cite{Cartography1}.  The Mercator projection of the Earth
constructed in the 16th century is also a conformal projection. It was
devised for nautical charts with the property that all lines of
constant compass direction are represented by straight segments, but
its use was not only restricted to marine navigation.  The Mercator
projection was so influential that its wide use was subject 
to political controversies related to the distortion of scales in
conformal maps \cite{Crampton94}.

The development of methods in cartography has paved the way for the 
discussion on geometry of curved surfaces by Lambert, Bolyai, 
Lobachevsky, Gauss, Riemann and others. 
Indeed, some of the basic terminology in differential geometry has its 
origin in cartography. Therefore, it does not surprise that certain problems of
cartography are similar to certain problems of general relativity. 
For example, it is clear today that in cartography different map projections 
serve different purposes. 
There is no single map of the Earth that covers all requirements. 
Similarly, there is no best gauge choice for a spacetime in general relativity.
Emphasis on different aspects and parts of the solution 
require different gauge choices as will be demonstrated in this thesis
by studies of various representations of the Minkowski and the
Schwarzschild-Kruskal spacetimes.

A key feature leading to the wide use of the Mercator projection is
the fact that it is a conformal projection. As a consequence the
projection keeps relative local directions -in the sense of local
angles- invariant. Clearly, relative local directions are important
for marine navigation. In general relativity, the fundamental
important structure is the causal structure which is equivalent to the
null cone structure as well as to the conformal structure.  We will
see in the next sections how this fact can be used to get access to
the large scale structure of spacetimes and to describe gravitational
radiation rigorously, in an unambiguous way.

The importance of conformal geometry in general relativity has been
recognized already in 1918 by Hermann Weyl \cite{Weyl18a}. His
motivation to study conformal rescalings of the metric has partly been
philosophical \cite{Scholz2}. He was disturbed by what he called an
inconsistency due to which lengths of vectors at different points in
Riemannian geometry can be directly compared with each other. Weyl
observes in \cite{Weyl18a}: "The metric allows the magnitudes of two
vectors to be compared, not only at the same point, but at any two
arbitrarily separated points."  To remove this remnant of non-local
geometry ("Ferngeometrie"), he proposed to consider the conformal
class of a spacetime.

Let $\widetilde{\mathcal{M}}$ be a four dimensional, smooth manifold
and $\tilde{g}$ a Lorentzian metric. The \textit{conformal class}
$[\tilde{g}]$ of a spacetime $(\widetilde{\mathcal{M}},\tilde{g})$ is
given by metrics $g$ related to $\tilde{g}$ by a \textit{conformal
  rescaling} $g=\Omega^2 \tilde{g}$ with a positive, point dependent
conformal factor $\Omega > 0$.  A choice of the unit of measurement,
i.e.~the \textit{gauge}, corresponds to the choice of a metric in the
conformal class $[\tilde{g}]$ or equivalently to the choice of a point
dependent conformal factor $\Omega$.  As a consequence, in Weyl's
"pure infinitesimal geometry" \cite{Weyl18b}, the choice of the unit
of measurement is subject to local variations.  Weyl states in
\cite{Weyl18a} that for the explicit representation of the spacetime
in such a geometry we have to choose a coordinate system and at each
point determine the conformal factor.  Therefore each formula must
have a double-invariance, namely invariance with respect to arbitrary
smooth coordinate transformations and with respect to conformal
rescalings of the metric.  For applications of conformal techniques in
this thesis, especially in chapter \ref{chapter:null}, a suitable
coupling of the coordinate and the conformal gauge freedom will be
essential to gain control over the asymptotic region.

\pagebreak
The main step from Euclidean geometry to Riemannian geometry is the removal of 
the assumption of integrability of vectors by parallel transport. 
In Riemannian geometry, a vector parallel transported along a closed curve 
changes in general its direction due to curvature, 
while its norm stays the same. 
Weyl demanded ``the non-integrability of the transference of distances'' 
\cite{Weyl23}. He states in \cite{Weyl18a}:
"\textit{A true infinitesimal geometry should, however, recognize 
only a principle for transferring the magnitude of a vector to an 
infinitesimally close point} and then, on transfer to an arbitrarily distant 
point, the integrability of the magnitude of a vector is no more to be expected
than the integrability of its direction." To allow for comparison of lengths in
each small neighborhood, he introduced what we call today the Weyl connection.
A concise description of these ideas can be found for example in 
\cite{Adler,Pauli}. We use a modern representation following the notation 
of \cite{Friedrich04}.

A \textit{Weyl connection} $\hat{\nabla}$ is a torsion free connection
which is not necessarily the Levi-Civita connection of a metric in the
conformal class $[\tilde{g}]$ but preserves the conformal structure so
that the covariant derivative of any metric $\tilde{g}$ in
$[\tilde{g}]$ is proportional to itself, \be \label{weyl-conn}
\hat{\nabla}_{\lambda} \tilde{g}_{\mu\nu}=-2
f_{\lambda}\tilde{g}_{\mu\nu},\ee with a one-form $f$. Under conformal
transformations of the metric $g'=\omega^2 \tilde{g}$ with a function
$\omega>0$, the one-form $f$ transforms according to
$f'=f-\omega^{-1}\,d\omega$.  If $f$ is exact, it can be written as $f
= \Omega^{-1} d\Omega$ and the Weyl connection $\hat{\nabla}$ is the
Levi-Civita connection $\nabla$ of a metric $g$ in the conformal class
given by $g = \Omega^{2} \tilde{g}$.

The relation of the Weyl connection $\hat{\nabla}$ with a metric connection 
$\tilde{\nabla}$ of a metric $\tilde{g}$ in the conformal class is given by 
\be \label{intro:sf} \hat{\nabla} - \tilde{\nabla} = S(\tilde{f}), \qquad 
{\rm where} \qquad S(\tilde{f})^{\ \rho}_{\mu \ \nu} \equiv 
\delta^{\rho}_{\ \mu}\tilde{f}_{\nu} + \delta^{\rho}_{\ \nu}\tilde{f}_{\mu}-
\tilde{g}_{\mu\nu}\tilde{g}^{\rho\lambda} \tilde{f}_{\lambda}.\ee
We see that Weyl connections characterized by 1-forms $\tilde{f}$
allow us a more general discussion of the conformal structure 
than conformal rescalings of the metric.

The introduction of the Weyl connection was the first step 
in the study of generalized connections and 
it led to a rich geometry, which Weyl used in his attempt to unify 
the gravitational and electromagnetic fields. 
His attempt for a unified theory was unsuccessful due to physical observations 
\cite{Goenner}, however the main idea has been the basis 
during the construction of modern gauge theories 
\cite{Oraifeartaigh}. It is interesting to note that the concept of 
\textit{gauge invariance} in theoretical physics originates in Weyl's ideas 
on conformal invariance \cite{ORaif-Straumann}. 

A torsion free connection implies the notion of parallel transport and allows 
the construction of geodesics. Auto-parallel curves with respect to a Weyl 
connection are referred to as conformal geodesics \cite{Friedrich87}. 
Remarkably, conformal geodesics have been useful in investigations 
of the asymptotic behavior of solutions to Einstein equations carried out
by Helmut Friedrich \cite{Friedrich04}. Following his investigations, 
we will make use of conformal geodesics in chapter \ref{chapter:i0} 
for a numerical treatment of spatial infinity.

The application of conformal techniques in general relativity 
played an important role in the clarification of a heated controversy 
in mathematical relativity on the physical nature of gravitational radiation 
that we discuss in the next section.
\pagebreak
\section{Gravitational radiation and asymptotic flatness} \label{sec:gr}
The Newtonian theory of gravitation is governed by a linear elliptic
Poisson equation for the Newtonian potential and does not allow for
free gravitational degrees of freedom carrying energy.  The
action-at-a-distance property of Newtonian gravitation related to the
elliptic nature of the underlying Poisson equation was already
questionable to Newton and his contemporaries.  A finite speed of
gravity has been studied by Laplace who concluded from this assumption
to a damping force on the orbital motion of planetary systems. In
1908, Poincar\'e suggested an acceleration of planetary orbits due to
loss of energy by emission of gravitational waves.

The first relativistic description of gravitational waves is due to
Albert Einstein \cite{Einstein16,Einstein18}. He studied wave
phenomena in linearized gravity and introduced the quadrupole formula
which, in the weak field limit, relates the motion of a source to
generation of gravitational waves. In his linearized calculations he
made use of a coordinate dependent quantity which he called the pseudo
energy momentum tensor to calculate the energy flux of gravitational
waves.  Einstein's use of the coordinate dependent pseudo-tensor was
criticized by many scientists of his time \cite{Cattani}, among others
Eddington who suggested to put more emphasis on the curvature tensor,
showed that certain classes of wave-like solutions to the linearized
Einstein equations are not physical and pointed out limitations of
linearized calculations \cite{Eddington22}.

Until the 1960s the status of gravitational waves was quite uncertain. 
Even Einstein seems to have believed for some time that gravitational waves 
did not exist \cite{Kennefick}.
The use of linearized calculations whose qualitative predictions 
are not necessarily valid in the full theory, 
the application of coordinate-dependent methods and 
the non-local nature of gravitational field energy have been 
sources of confusion. The difficulty with a local energy concept for the 
gravitational field lies in the background independence of
general relativity. Due to the equivalence principle, 
the local gravitational field can be transformed away by passing to a 
freely falling frame of reference.  

With the work of Lichnerowicz on mathematical foundations of general
relativity \cite{Lichnerowicz55} it turned out that certain
calculations supporting the view that gravitational waves are
unphysical put too restrictive conditions on coordinates.  Based on
the development of mathematical foundations and a stronger emphasis on
curvature invariants, the question on the status of gravitational
waves was gradually answered by the works of Bondi, Newman, Penrose,
Petrov, Pirani, Robinson, Sachs and Trautman among others.

It is difficult to give a precise definition to radiation in general relativity
as the gravitational field acts as its own source so that there is no 
background we can rely on. However, the gravitational field becomes weak far 
away from sources. When we concentrate our studies on an astrophysical system
we assume that the interaction of this system with the rest of the universe to 
be negligible. In the study of such "isolated systems" the astrophysical system
is modeled by an asymptotically flat spacetime. 
This corresponds to an idealized situation in 
which massive sources are confined to spatially compact domains and the 
spacetime metric is required to approach a flat metric at infinity.
One might hope to find a precise notion of radiation for such systems 
in the asymptotic region using the flat limit metric 
as a kind of background structure.

For the discussion of gravitational radiation, it is natural to
require asymptotic flatness in null directions.  Historically,
however, first discussions of asymptotic flatness have been made with
respect to spatial infinity.  Note that asymptotic flatness is an
assumption that is "put in by hand".  Today there are several
mathematically inequivalent definitions of the concept and we will
refer to some of them in later sections. In essence, one imposes some
condition determining the approach of the metric to a flat metric at
(spatial or null) infinity. A formal definition based on the existence
of distinguished classes of coordinates, say near spatial infinity,
can be roughly given as follows.  We may call a spacetime
$(\widetilde{\mathcal{M}},\tilde{g})$ asymptotically flat at spatial
infinity when there exists a coordinate system $\{\tilde{x}^\mu\}$ in
a neighborhood of spatial infinity with respect to which
\be\label{eq_as_flat} \tilde{g}_{\mu\nu}= \tilde{\eta}_{\mu\nu} +
O_k(\tilde{r}^{-p}),\quad \textrm{as}\quad \tir\to\infty, \quad
\textrm{for some}\ k,p>0, \ee 
where $\tir^2=\delta_{\alpha\beta}\tilde{x}^\alpha \tilde{x}^\beta$ with
$\alpha,\beta=1,2,3$. We denote by $\tilde{\eta}$ the Minkowski metric,
$\delta_{\alpha\beta}$ is the Kronecker symbol and the terms
$O_k(\tir^{-p})$ behave like $O(\tir^{-p+j})$ under differentiations
of order $j\leq k$.  The strength of the fall-off behavior is
determined when we give a precise description of $O_k(\tir^{-p})$.  A
question is how strong this fall-off behavior needs to be so that
suitable physical notions of total energy, linear and angular momentum
can be defined. It is known, for example, that for $p>1/2$ and $k\geq
1$ a well-defined notion of total energy at spatial infinity can be
associated with an asymptotically flat spacetime \cite{Szabados}.

The linearized theory suggests that gravitational perturbations travel
along null directions. Therefore, to define gravitational radiation,
one takes limits to infinity in null directions. 
One may define a notion of asymptotic flatness in null 
directions by requiring (\ref{eq_as_flat}) in a neighborhood of null infinity
with respect to a suitable coordinate system.
We might intuitively imagine that the spacetime should become simply 
close to the Minkowski spacetime in whatever direction we approach infinity.
The geometric picture, however, is quite different and it is a delicate
question how the notions of asymptotic flatness at spatial and null infinity 
fit together. We will come back to this issue at the end of the section. 
Let us first follow the historical development of the concept of gravitational
radiation.

An important step in the clarification of questions regarding gravitational 
radiation was the demonstration that outgoing gravitational radiation carries 
positive energy away from an isolated system \cite{Bondi62,Sachs62}. 
This analysis used a special class of coordinate systems 
which is natural for studying radiation phenomena. The coordinates are 
generated by distinguished families of outgoing null hypersurfaces. 
On the hypersurfaces the luminosity distance $\tir$ has 
been defined. Today such coordinates are called Bondi coordinates.
The field equations have been studied along null directions and certain 
conditions at infinity have been put that are similar to the no incoming
radiation conditions suggested by Trautman \cite{Trautman58}. 
These conditions included the requirement that metric 
components have certain smoothness properties in $1/\tir$ in the limit 
$\tir\to\infty$ along null directions. 
An important open problem was whether and to what extend these
conditions excluded interesting solutions \cite{Sachs62}. 

Instead of putting conditions on metric components corresponding to
asymptotic flatness as above, it is more convenient to impose
restrictions on certain components of the Weyl tensor
$C^\mu_{\ \nu\lambda\rho}$ representing the free gravitational
field. Sachs analyzed the vacuum Bianchi identities $\nabla_\mu
C^\mu_{\ \nu\lambda\rho}=0$ and suggested that the Weyl tensor
satisfies a "peeling property" along outgoing null geodesics relating
the fall-off behavior of its components to the Petrov classification
\cite{Sachs61}. Shortly after this analysis, Newman and Penrose found
a way to formulate fall-off conditions in an elegant way and showed
that they imply the peeling behavior \cite{Newman62a}. They introduced
the spin frame formalism and analyzed expansions of gravitational
fields in $1/\tir$ where $\tir$ is an affine parameter along
outgoing null geodesics.  Imposing a certain uniform smoothness
condition, they found out that if a particular complex tetrad
component of the Weyl tensor with respect to a distinguished tetrad at
infinity has an asymptotic behavior $\psi_0 =O(\tir^{-5})$ and
$\partial_{\tilde{r}}\psi_0 =O(\tir^{-6})$ as $\tir\to\infty$, one can
derive the Sachs peeling behavior $\psi_k =O(\tir^{k-5})$ as
$\tir\to\infty$ for ${k=0,1,2,3,4}$. In analogy to electromagnetism,
the field with the weakest fall-off in the order $1/\tir$, namely
$\psi_4$, is interpreted as representing the outgoing radiation field.
Note that this interpretation of $\psi_4$ is strongly related to the
Sachs peeling behavior and requires a distinguished choice of a tetrad
field at infinity. In contrast to electromagnetism, where the
linearity of the Maxwell's equations allows a clean global separation
between near field terms and far field terms, in general relativity no
such separation can be expected in general as the peeling behavior is
valid only in the asymptotic region.  In the strong field regime one
would expect a mixture of the fall-off behavior.  Therefore it is
misleading to talk about $\psi_4$ calculated with respect to some
tetrad field in the interior as the radiation field.

In the studies just discussed, the close relationship between
gravitational radiation and asymptotic behavior of solutions to the
Einstein equations has been demonstrated. However, the studies still
relied on the use of special classes of coordinate systems and
included awkward limits to infinity. Such limits arise because a
rigorous meaning to gravitational radiation can only be given in the
asymptotic region.

To study physical properties of isolated systems, one would like to
have coordinate independent definitions. The coordinate dependence 
in the early studies of gravitational radiation blurred the global geometric 
structure of asymptotically flat spacetimes. In the early development of the
field, the global picture was by no means clear, not even to those
who contributed substantially to the development of the picture.
Sachs, for example, refers to null infinity as "the cylinder at spatial 
infinity" in \cite{Sachs62}.

To motivate a coordinate independent approach to the concept of gravitational 
radiation further and to point out a subtlety regarding the notion of 
asymptotic flatness in spacelike and null directions, we shall discuss 
an illustrative example taken from \cite{Penrose80}. 
The Schwarzschild metric with mass $m$ in standard Schwarzschild 
coordinates $(t,\tilde{r}_s,\vartheta,\varphi)$ reads
\be\label{eq:st_ss} 
\tilde{g}_s = -\left(1-\frac{2 m}{\tilde{r}_s}\right)\,dt^2 + 
\left(1-\frac{2m}{\tilde{r}_s}\right)^{-1}d\tilde{r}_s^2 + 
\tilde{r}_s^2 \,d\sigma^2, \qquad \tilde{r}_s>2m,\ee
where $d\sigma^2=d\vartheta^2+\sin^2\vartheta\,d\varphi^2$. 
This metric obviously approaches the flat Minkowski metric $\tilde{\eta}_s$ 
given by $\tilde{\eta}_s=-dt^2+d\tilde{r}_s^2+\tilde{r}_s^2 d\sigma^2$,
as $\tilde{r}_s\to\infty$, so that we may call the Schwarzschild spacetime 
asymptotically flat at spatial infinity. This, however, does not by itself 
imply that the Schwarzschild spacetime is asymptotically flat 
in null directions. With regard to null geodesics the Schwarzschild 
metric (\ref{eq:st_ss}) differs greatly from the Minkowski metric 
$\tilde{\eta}_s$ at large distances. Consider outgoing null geodesics given by
$u=\mathrm{const.}$, $\vartheta=\mathrm{const.}$, $\varphi=\mathrm{const.}$, 
where $u$ is the Schwarzschild retarded time 
$u=t-(\tilde{r}_s+2m\,\ln(\tilde{r}_s-2m))$.
We see that as $\tilde{r}_s\to\infty$ the value of the Minkowski retarded time 
$t-\tilde{r}_s$ is unbounded above along outgoing Schwarzschild null geodesics
(the argument applies also for ingoing null geodesics).
Schwarzschild null geodesics do not correspond at all to 
Minkowski null geodesics with respect to $\tilde{\eta}_s$ for large 
$\tilde{r}_s$. 

The solution to this apparent problem is to relate a Minkowski 
metric to the Schwarzschild metric in a different way. Define a coordinate 
$\tilde{r}_\ast = \tilde{r}_s + 2m \,\ln(\tilde{r}_s-2m)$. We can choose on the 
Schwarzschild manifold another Minkowski metric $\tilde{\eta}_\ast$ given by
$\tilde{\eta}_\ast=-dt^2+d\tilde{r}_\ast^2+\tilde{r}_\ast^2\,d\sigma^2$. 
Now the described difficulty does not arise because outgoing null geodesics 
given by $u=t-\tilde{r}_\ast=\mathrm{const.}$ have the same form for both 
metrics $\tilde{g}_s$ and $\tilde{\eta}_\ast$. While the condition 
for asymptotic flatness (\ref{eq_as_flat}) has the same form in spacelike and
null directions, its geometric interpretation, i.e.~the way we relate the 
Minkowski metric $\tilde{\eta}$ to the curved spacetime metric $\tilde{g}$, 
is different.

We mentioned that for studying gravitational radiation one might use
the flat limit metric as a kind of background structure. This example
makes clear that the relation of the flat limit metric to the curved
spacetime is determined by the null cone structure. The null cone
structure, however, is part of the unknown in the Einstein equations.
It is therefore a difficult question how the flat background structure
can be used conveniently in dynamical spacetimes to discuss their
asymptotic properties.

It turns out that the structure that we are looking for can be given in terms 
of the idea of conformal infinity presented in the next section. 
This construction also allows to replace asymptotic calculations by local 
differential geometry.

\section{Conformal infinity} 
\label{sec:con_comp}
Conformal techniques in general relativity played
an important role in the investigation of the solution space to the Einstein 
equations. The analysis of elliptic constraint equations implied by the 
Einstein equations on spacelike hypersurfaces has made intensive use of 
conformal methods \cite{Lichnerowicz44,York73}. 
In the following, we will concentrate on conformal rescalings of the full 
spacetime metric. We will also ignore the application of conformal techniques 
in cosmological models (see \cite{Friedrich98b} for an overview and 
\cite{Beyer-phd} for a recent work).

It was discovered by Penrose that a suitable notion of asymptotic
flatness can be formulated in an elegant way using the conformal
equivalence class of the metric. The conformal compactification
technique enabled a geometric formulation of the fall-off
behaviour. Penrose could also deduce the peeling behavior from a few
assumptions on the conformal structure of the metric not favoring any
coordinate systems \cite{Penrose63,Penrose65}.  Since then, conformal
techniques to study in detail the structure of gravitational fields in
the asymptotic region in a coordinate independent way have been very
fruitful in mathematical relativity.  Many physical concepts like
mass, momentum or gravitational radiation of isolated systems have
been unambiguously defined in the asymptotic region of an isolated
system using conformal techniques.

Two important observations suggest a conformal approach for discussing 
radiation. The first observation is the importance of the null cone structure. 
Conformal rescalings of the metric preserve the null cone structure,
which is equivalent to the characteristic and the causal structures.  
It turns out that essential features of free gravitational fields strongly
related to the null cone structure can be discussed with respect to the 
conformal equivalence class of the physical metric. 
The second observation is related to the non-local feature 
of gravitational radiation which requires limits to infinity 
as discussed in the previous section. Conformal rescalings can be used
to avoid such limits by compactifying the spacetime such that infinity 
corresponds to a finite hypersurface. Asymptotic calculations can then 
be replaced by local differential methods under certain conditions.
\pagebreak

The idea of conformal compactification is similar to the construction 
of the Riemann sphere in complex analysis by adjoining infinity to the complex 
plane. Due to the Lorentzian signature of the metric, 
infinity for a spacetime corresponds not to a point but has a richer 
structure, as Weyl already observed in \cite{Weyl23a}.
To demonstrate the technique we discuss the conformal compactification
of the Minkowski spacetime (see also \cite{Frauendiener04,Penrose67,Wald84}). 

The Minkowski metric $\tilde{\eta}$ in coordinates
adapted to spherical symmetry is given by  
\be\label{intro:eta}\tilde{\eta}=-d\tilde{t}^2+d\tilde{r}^2+\tilde{r}^2\,
d\sigma^2,\qquad \mathrm{on} \quad  \tilde{t}\in \mathbb{R}, \quad 
\tilde{r}\geq 0, \ee
where $d\sigma^2$ is the standard metric on the unit sphere $S^2$. 
Introducing null coordinates $\tilde{u}=\tilde{t}-\tilde{r}$ and 
$\tilde{v}=\tilde{t}+\tilde{r}$ for $\tilde{v} \geq \tilde{u}$, and 
compactifying them by $U= \arctan \tilde{u}$ and $V=\arctan \tilde{v}$, we get
\[\tilde{\eta}=\frac{1}{\cos^2 V \cos^2 U} \left(-dV\,dU+
\frac{1}{4}\sin^2(V-U)d\sigma^2\right), \quad \mathrm{on} \quad 
(-\pi/2 < U \leq V < \pi/2). \]
Points at infinity with respect to the original coordinates have finite values
with respect to the compactifying coordinates ($V=\pi/2$ or $U=-\pi/2$), 
however, the physical metric $\tilde{\eta}$ in compactifying coordinates 
is singular at these points. This singular behavior can be compensated 
by a conformal rescaling with the conformal factor 
$\Omega= \cos V \cos U$, so that the rescaled metric
\[ \eta = \Omega^2\tilde{\eta}= -dU\,dV+\frac{1}{4}\sin^2(V-U)d\sigma^2, \]
is well defined on the domain $(-\pi/2 \leq U \leq V \leq \pi/2)$ including 
points that are at infinity with respect to $\tilde{\eta}$. 
We say that $\tilde{\eta}$ can be extended beyond infinity. 

\begin{figure}[t]
  \centering
  \psfrag{i-}{$i^-$}
  \psfrag{i+}{$i^+$}
  \psfrag{scr+}{$\scri^+$}
  \psfrag{scr-}{$\scri^-$}
  \psfrag{i0}{$i^0$}
  \psfrag{t=const}{$\tit=\textrm{const.}$}
  \psfrag{r=const}{$\tir=\textrm{const.}$}
  \includegraphics[width=0.2\textwidth]{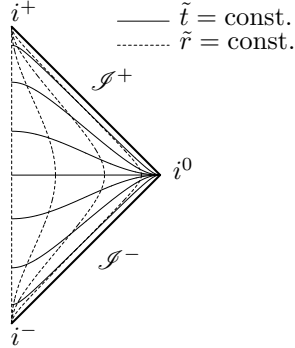}
  \caption{Penrose diagram of the Minkowski spacetime
    \label{fig:minkowski}}
\end{figure}

For Fig.~\ref{fig:minkowski} time and space coordinates
\mbox{$t=(V+U)/2$} and $r=(V-U)/2$ have been introduced. The resulting
metric $\eta=-dt^2+dr^2+\sin^2 r\,d\sigma^2$ is the standard metric on
the Einstein cosmos $\mathbb{R}\times S^3$.  The embedding
$\phi:\mathbb{R}^4\to\mathbb{R}\times S^3$ of the Minkowski spacetime
into the Einstein cosmos is given by
\[\tilde{t}\mapsto\frac{1}{2} \left(\tan\left(\frac{t+r}{2}\right) 
+ \tan\left(\frac{t-r}{2}\right)\right), \quad \tilde{r}\mapsto \frac{1}{2} 
\left(\tan\left(\frac{t+r}{2}\right)-\tan\left(\frac{t-r}{2}\right)\right),\]
from $\widetilde{\mathcal{M}}=\mathbb{R}^4=\{\tilde{t}\in(-\infty,\infty),
\tilde{r}\in[0,\infty)\}$ into $\mathcal{M}=\mathbb{R}\times S^3=
\{t\in(-\infty,\infty)$, \mbox{$r\in[0,\pi]\}$}, suppressing the
angular coordinates. The part of $\mathcal{M}$ that corresponds to the
Minkowski spacetime is given by $\{|t+r|<\pi,|t-r|<\pi\}$.

We see that the completion of $\widetilde{\mathcal{M}}$ is a manifold
with boundary $\{t=\pm(\pi-r)$, \mbox{$r\in[0,\pi]\}$} where the
boundary points correspond to points at infinity with respect to the
physical metric.  Asymptotic behavior of fields on
$\widetilde{\mathcal{M}}$ can be studied using local differential
geometry on this boundary where the conformal factor $\Omega=\cos
t+\cos r$ vanishes. The part of the boundary without the points at
$r=0,\pi$ is denoted by $\scri=\{t=\pm(\pi-r),r\in(0,\pi)\}$.  The
differential of the conformal factor does not vanish at $\scri$,
$d\Omega|_{\scri}\ne 0$, and $\scri$ consists of two parts
$\scri^\pm$, each of them with the topology $\mathbb{R}\times S^2$.

Fig.~\ref{fig:minkowski} shows curves that correspond to constant
values of the coordinates $\tilde{t}$ and $\tilde{r}$ in the new
representation.  Each point of the diagram represents a sphere except
the dashed vertical line segment connecting $i^+$ and $i^-$ which
corresponds to the origin $\{\tir=0\}$. Surfaces of
$\tilde{r}=\textrm{const.}$ approach timelike infinity
$i^{\pm}=\{t=\pm\pi,r=0\}$, surfaces of $\tilde{t}=\textrm{const.}$
approach spatial infinity $i^0=\{t=0,r=\pi\}$.  The figure is a
Penrose diagram which is especially useful for depicting causal
structures. In Penrose diagrams, radial light rays are straight line
segments with 45 degrees to the horizontal. Null rays
($U=\textrm{const.}$ and $V=\textrm{const.}$ surfaces) reach $\scri$
for an infinite value of the physical affine parameter along them,
hence $\scri$ is called null infinity.

The idea is now to take certain properties of the asymptotic structure of the 
Minkowski spacetime as being representative for the asymptotic structure 
of isolated systems. We make the following assumptions 
about an asymptotically flat spacetime $(\widetilde{\mathcal{M}},\tilde{g})$: 
There exists a regular spacetime $(\mathcal{M},g)$, 
a sufficiently differentiable (say $C^3$) function $\Omega$ on $\mathcal{M}$
and an embedding $\phi:\widetilde{\mathcal{M}}\to\mathcal{M}$, 
which is conformal with conformal factor $\Omega$ such that   
$\tilde{g} = (\Omega\circ\phi)^{-2} \phi^{\ast} g = \phi^{\ast}(\Omega^{-2} g).$
The conformal factor satisfies $\Omega>0$ on 
$\phi(\widetilde{\mathcal{M}})\subset\mathcal{M}$ and the completion of 
$\phi(\widetilde{\mathcal{M}})$ is a submanifold with non-empty boundary 
$\scri$, that is 
$\partial\overline{\phi(\widetilde{\mathcal{M}})}=\scri\ne\emptyset$, 
on which $\Omega|_{\scri}=0,\ d\Omega|_{\scri}\ne 0$. 
These assumptions are closely related to the notion of \textit{weak asymptotic 
simplicity} \cite{Penrose65}. In the following, we identify 
$\phi(\widetilde{\mathcal{M}})$ with $\widetilde{\mathcal{M}}$ so that we 
write $\widetilde{\mathcal{M}}\subset\mathcal{M}$ or $
g=\Omega^2\tilde{g}$. The triple $(\mathcal{M},g,\Omega)$ is called 
\textit{conformal extension of} $(\widetilde{\mathcal{M}},\tilde{g})$. 
Note that the conformal extension is not unique.

\begin{figure}[t]
  \centering
  \psfrag{I}{\bf{I}}
  \psfrag{II}{\bf{II}}
  \psfrag{i-}{$i^-$}
  \psfrag{i-}{$i^-$}
  \psfrag{i+}{$i^+$}
  \psfrag{scr+}{$\scri^+$}
  \psfrag{scr-}{$\scri^-$}
  \psfrag{i0}{$i^0$}
  \psfrag{r=0}{$\tilde{r}_s=0$}
  \psfrag{t=const}{$\tit=\textrm{const.}$}
  \psfrag{r=const}{$\tir_s=\textrm{const.}$}
  \includegraphics[width=0.75\textwidth]{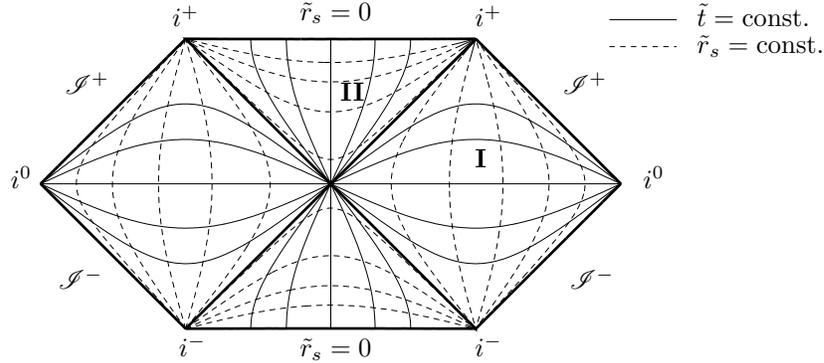}
  \caption{Penrose diagram of the Schwarzschild-Kruskal spacetime.
    \label{fig:schwarzschild}}
\end{figure}

The basic assumption here is that the rescaled metric $g$ is regular
across $\scri$. A similar construction where this assumption is
satisfied can be made for other explicit solutions that can be
regarded as asymptotically flat, for example the Schwarzschild-Kruskal
solution (see Fig.~\ref{fig:schwarzschild} for the resulting Penrose
diagram). However, we do not have a large class of explicit radiative
solutions where we can test our requirements on the conformal
extension. Therefore we do not know whether the above assumptions
cover sufficiently general spacetimes
so that we can take a conformal approach in the isolated system
idealization of interesting astrophysical configurations .

A strong support for the conformal compactification technique comes
from the studies described in the previous section.  Results achieved
by coordinate dependent methods and limits to infinity can be derived
very elegantly for spacetimes admitting a smooth conformal boundary
using the conformal technique \cite{Penrose63,Penrose65}.

To utilize the conformal compactification technique for calculating 
gravitational radiation we would like to answer: 
\textit{How general is the description proposed by Penrose?}  \cite{Penrose82}

The details of the conformal compactification technique have been motivated 
by explicit examples or well studied assumptions about the fall-off behavior 
of certain fields as above and not by a detailed study of the full non-linear 
Einstein equations. 
The assumption on smoothness of $\scri$ is geometrical and includes only 
requirements on the conformal class $[\tilde{g}]$. 
Smoothness of the rescaled metric $g$ along the conformal 
boundary $\scri$ corresponds to a fall-off behavior for the physical metric 
$\tilde{g}$, which is however a solution to the Einstein equations 
that impose their own asymptotic behavior. It might
be that the smoothness requirement on the conformal extension is too 
restrictive such that only a very special class of solutions satisfies it.
Whether the geometrical requirement of smoothness of $\scri$
is compatible with the fall-off behavior of the metric fields 
as a solution to the Einstein equations is a delicate question. 

We would like to know by some general argument whether we have a large class of
non-trivial, asymptotically flat, radiative spacetimes that admit a smooth 
conformal boundary such that we can apply the conformal compactification 
technique. To answer this question, the solution space to the Einstein 
equations needs to be studied with an emphasis on the asymptotic structure 
of gravitational fields.

The available explicit solutions are not general enough to study the solution 
space by direct means, therefore we are led to abstract analysis. 
By sufficient knowledge on properties of spacetimes, we can get some results 
in certain classes. We know, for example, that asymptotically flat, vacuum, 
stationary spacetimes admit an analytic compactification at null infinity 
\cite{Dain01d}. For more general results that include radiative spacetimes, 
the initial value problem needs to be studied in a general setting.

The analysis of the initial value problem for Einstein equations
in the compactified picture is not straightforward, as the Einstein equations 
are not conformally invariant and compactification leads to formally singular 
equations at infinity. However, the equations are conformally regular 
as Friedrich showed by constructing a system which is equivalent 
to the Einstein equations for $\Omega>0$ and is regular for all values 
of the conformal factor so that the equations for the rescaled metric $g$ 
can be analyzed on a conformal extension $\mathcal{M}$ including $\scri$ 
where $\Omega=0$ \cite{Friedrich81a}.

\begin{figure}[t]
  \centering
  \psfrag{i+}{$i^+$}
  \psfrag{scr+}{$\scri^+$}
  \psfrag{i0}{$i^0$}
  \psfrag{S}{$\mathcal{S}$}
  \psfrag{sigma}{$\Sigma=\mathcal{S}\cap\scri^+$}
  \psfrag{sig}{$\Sigma$}
  \includegraphics[width=0.35\textwidth]{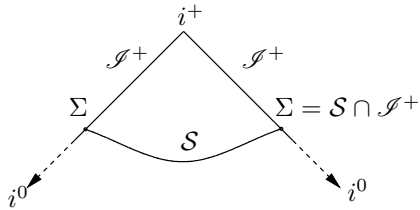}
  \caption{Penrose diagram for a hyperboloidal initial value problem.
    \label{fig:hivp}}
\end{figure}

Friedrich analyzed the initial value problem for the conformally regular field 
equations based on spacelike surfaces that extend smoothly through 
null infinity. These surfaces are called \textit{hyperboloidal} as their 
asymptotic behavior is similar to the standard hyperboloids in Minkowski 
spacetime \cite{Friedrich83a}. Let $\mathcal{S}$ be such a hyperboloidal 
surface that cuts $\scri^+$ in a spacelike two dimensional surface 
$\Sigma=\mathcal{S}\cap\scri^+$ and extends smoothly through $\scri^+$ as in 
Fig.~\ref{fig:hivp} (the discussion of $\scri^-$ follows by time reversal).
The analysis of the hyperboloidal initial value problem revealed that if 
regular data for the conformally regular field equations is given on
$\mathcal{S}$, the smoothness of $\scri^+$ will be preserved into the future 
of $\mathcal{S}$, at least for a while \cite{Friedrich83a}. While this result
is local in time, Friedrich also proved a semi-global stability result stating 
that, for small data, $\scri^+$ will admit future complete null generators 
and a regular timelike infinity $i^+$ \cite{Friedrich86}.
The existence of regular data for the conformally regular field equations 
has been shown in \cite{ACF92}. The analysis of \cite{ACF92} showed further 
that, in general, even with the strongest smoothness requirements on free data,
logarithmic singularities can arise at $\Sigma$ unless some mild regularity 
conditions on the geometry of $\Sigma$ are satisfied. 

These analytic results are important for practical purposes in numerical 
calculations. One should be aware that the requirement 
of the existence of a smooth $\scri^+$ implies a restriction on the data.
One may expect that this restriction does not exclude physically relevant 
solutions, but this is yet an unresolved question. Still, for a non-smooth 
$\scri^+$ some of the relevant structure may be recovered 
\cite{Chrusciel95,Winicour85}. 
It is an interesting open question whether one can deal with some mildly 
singular behavior numerically. We will study the hyperboloidal initial value 
problem in chapter \ref{chapter:null}.

The analytic results just described show that the decision on smoothness 
of null infinity is made at spatial infinity $i^0$.
Remarkably, our knowledge on the asymptotic behavior of solutions like 
Schwarzschild or Kerr can be applied to this question. 
By gluing techniques developed by Corvino and Schoen \cite{Corvino-Schoen},
one knows that there exists a large class of non-trivial initial data 
which is Schwarzschild or Kerr in a neighborhood of spatial infinity. 
As stationary spacetimes admit an analytic compactification at null infinity, 
we know that the development of such data leads to 
regular hyperboloidal initial data. Combined with Friedrich's results 
on the hyperboloidal initial value problem this leads to the existence 
of a large class of non-trivial, radiative, vacuum spacetimes 
with smooth null infinity \cite{Chrusciel-Delay-2002}.
These spacetimes, however, have a special asymptotic structure. 
The question still remains whether more general spacetimes exist. 
To answer it, the behavior of gravitational fields 
in a neighborhood of spatial infinity needs to be studied in detail.

A detailed study of solutions in a neighborhood of spatial infinity is
complicated by the fact that when the ADM-mass of the spacetime does
not vanish point compactification at spatial infinity leads to a
certain singular behavior. Physical fields do not admit smooth limits
at the point $i^0$.  The limits depend on the spacelike direction
along which one approaches $i^0$ \cite{Geroch77,Ashtekar80}.
Therefore, to study fields at spatial infinity, it might be better not
to represent it as a point. In \cite{Ashtekar92,Bobby82,Sommers}, a
representation of spatial infinity as a unit timelike hyperboloid has
been used.  This representation of spatial infinity, however, does not
allow a study of the full system of the Einstein equations in a
unified picture with spatial and null infinity.

A regular finite initial value problem near spatial infinity has been
formulated by Friedrich using the reduced general conformal field
equations that he introduced in \cite{Friedrich95}.  These equations
are written in a conformal Gauss gauge based on conformal geodesics
which are auto-parallel curves with respect to a Weyl connection
\cite{Friedrich03,Friedrich87}.  In the conformal Gauss gauge, spatial
infinity can be represented as a cylinder. The cylinder at infinity
allows a detailed discussion of gravitational fields at spatial and
null infinity in a single, regular setting that only depends on the
conformal structure of the spacetime \cite{Friedrich98, Friedrich04}.

Friedrich's analysis showed that in general, logarithmic singularities 
arise in a small neighborhood of spatial infinity. 
He obtained necessary regularity conditions on the Cotton tensor for these 
singularities to vanish. 
Later on, Valiente Kroon showed that these conditions 
are not sufficient and obtained further obstructions to smoothness of 
null infinity \cite{ValienteKroon04}. The question on necessary and sufficient
conditions on Cauchy data for smoothness of $\scri$ is still open and a field 
of active research. A rich structure can be expected in a neighborhood of 
spatial infinity, as suggested for example by the existence of initial data 
whose development may lead to different smoothness properties 
on past and future null infinity \cite{ValienteKroon06}.

For having numerical access to spatial and null infinity in a single
setting, the reduced general conformal field equations are the only
tool we have at present. A numerical implementation of this system
needs to deal with certain difficulties.
The regular finite initial value problem at spatial infinity distinguishes 
directions at spatial infinity. The regular representation of conformal data
on a Cauchy surface crucially relies on a conformal rescaling of a 
distinguished tetrad with respect to which the fields are calculated.
For numerical studies using this system, the geometry imposed by the cylinder 
needs to be implemented. We will discuss an implementation and tests
in chapter \ref{chapter:i0}.

In this section, I described conformal techniques and some analytic results
on their applicability to present the analytic foundation on which the 
numerical studies of this thesis are build. 
The studies are complementary to existing numerical methods 
but they are not an aim by themselves. 
A motivation to use conformal techniques in numerical calculations is 
given by the hope that interaction between mathematical and numerical 
methods will lead to fruitful studies of the solution space to the Einstein
equations. Especially interesting seems the question, whether a mildly singular
behavior of solutions at infinity can be treated by numerical methods 
so that conjectures on properties of solutions can be made which in turn 
might be studied by rigorous analyses.

Another basic motivation is achieving efficient numerical simulations 
of isolated systems using controlled approximations so that reliable 
predictions on gravitational radiation can be made. Current numerical methods 
to calculate gravitational radiation have certain disadvantages that we will
discuss in the next section to give an overview on the subject and to put
the conformal approach in numerical relativity into its broader context.
\section{The outer boundary in numerical relativity}\label{sec:nr}
To study the development of large classes of data and to make quantitative 
predictions on highly dynamical, strong gravitational fields, 
we need to use numerical methods. The Einstein equations 
written in arbitrary coordinates are of no known type as they only determine
the isometry class of the metric. To numerically calculate 
a solution metric with given initial data on some hypersurface, 
one typically reduces the Einstein equations to a hyperbolic system of partial 
differential equations. The reduced system is discretized and solved 
iteratively by some numerical algorithm. A reduction can be done in various 
ways which suggests a variety of methods in numerical relativity.

Special properties of general relativity among other theories of
physics, particularly the lack of a background and related
difficulties like the non-linearity of the equations or the non-local
feature of gravitational field energy, require a strong interplay
between mathematical and numerical methods not only on numerical
analysis or analysis of partial differential equations, but also on
differential geometry.  The importance of abstract methods that can
suggest well-defined techniques for numerical calculations has been
recognized only recently.  Today, the numerical stability problem for
some astrophysically interesting classes of data seems to be
solved. The numerical development of various binary black hole initial
data, which was a big challenge for many years, can now be followed
from the late inspiral through the coalescence phase
\cite{Baker05,Bruegmann06,Campanelli05,Koppitz07,Pretorius05b,Scheel06,
  Szilagyi06}.

There remain, however, many open problems that need to be studied.  In
this thesis we are mainly concerned with numerical studies of the
asymptotic behavior of isolated systems. We will therefore concentrate
on problems concerning the numerical treatment of the asymptotic
region. To illustrate the computational domain in various methods that
we discuss, the regions I and II in the Penrose diagram of the
Schwarzschild-Kruskal spacetime from Fig.~\ref{fig:schwarzschild} will
be plotted. We will also discuss the causal structure on the grid
depicted by ingoing and outgoing null surfaces. We will assume that
the interior of the event horizon is dealt with by the excision
method. This corresponds to the introduction of a spacelike boundary
inside the black hole so that no boundary data needs to be given.  The
requirement for the numerical treatment of such a spacelike inner
boundary is numerical stability.  There are other possibilities for
treating the singularity that we ignore.
\subsection{Timelike artificial  boundary} \label{sec:tab}
\begin{figure}
  \centering
  \begin{minipage}[t]{0.5\textwidth}
    \centering
    \psfrag{I}{\bf{I}}
    \psfrag{II}{\bf{II}}
    \psfrag{i+}{$i^+$}
    \psfrag{i-}{$i^-$}
    \psfrag{scr+}{$\scri^+$}
    \psfrag{i0}{$i^0$}
    \psfrag{r=0}{$\tilde{r}_s=0$}
    \psfrag{h}{$\mathcal{H}$}
    \includegraphics[width=\textwidth,height=0.8\textwidth]
{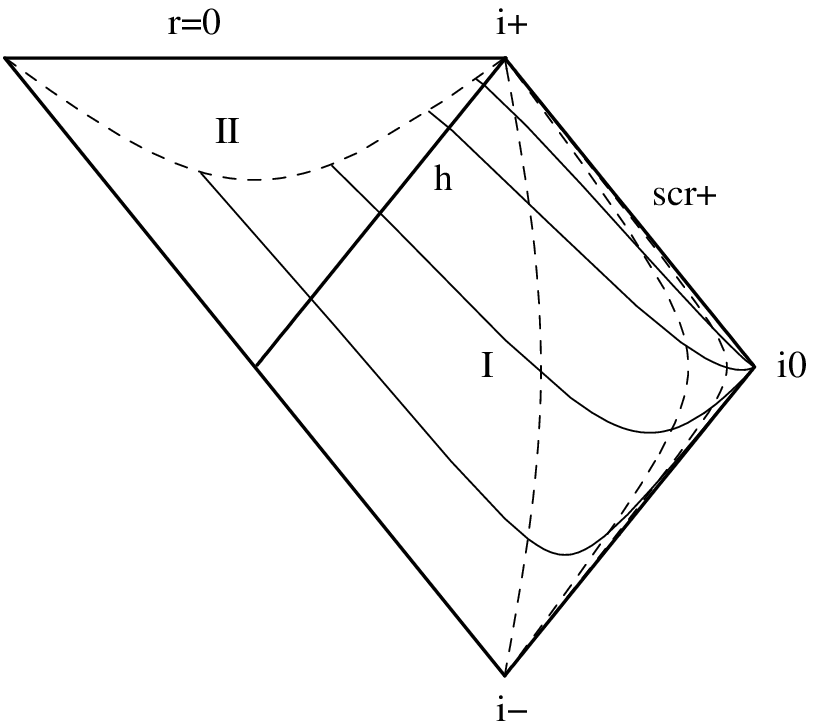}
    \caption{The Penrose diagram for a numerical evolution in ingoing 
      Eddington-Finkelstein coordinates. Dashed curves represent
      timelike surfaces at
      $\tilde{r}_s=\{\frac{3}{2}m,3m,5m,7m\}$. The event horizon $\mathcal{H}$ 
      is at $\tilde{r}_s=2m$.
      \label{fig:ief}}
  \end{minipage}%
  \hspace{0.03\linewidth}%
  \begin{minipage}[t]{0.45\textwidth}
    \centering
    \psfrag{t}{$\tilde{t}$}
    \psfrag{r}{$\tilde{r}_s/m$}
    \psfrag{3/2}{$\frac{3}{2}$}
    \psfrag{2m}{$2$}
    \psfrag{7m}{$7$}
    \psfrag{h}{$\mathcal{H}$}
    \psfrag{outgoing}{outgoing}
    \psfrag{ingoing}{ingoing}
    \psfrag{timelike}{$\tilde{r}_s=\textrm{const.}$}
    \includegraphics[width=0.95\textwidth]{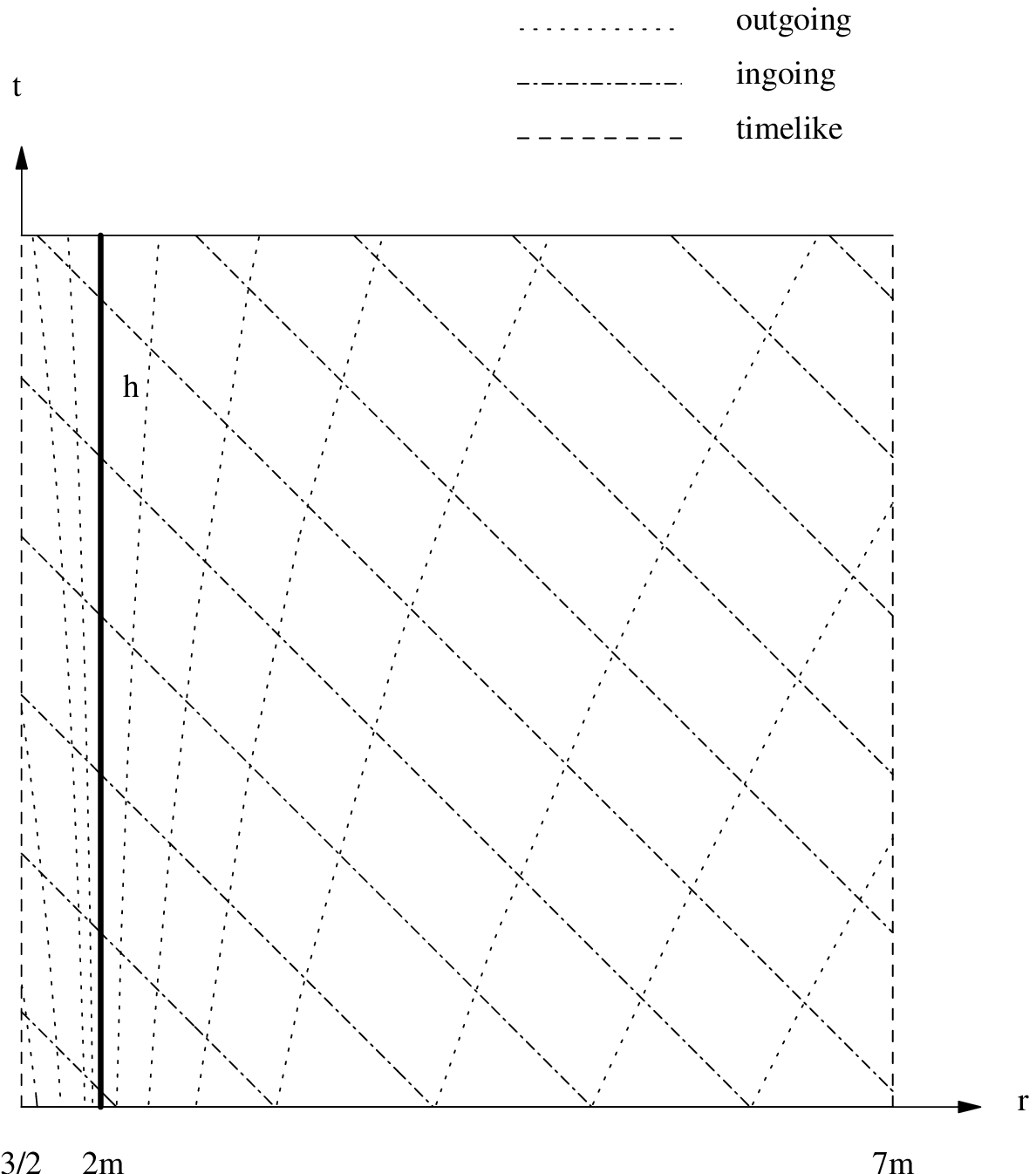}
    \caption{Causal structure on the grid bounded by the dashed surfaces 
      at $\tilde{r}_s=\frac{3}{2} m$ and $\tilde{r}_s=7 m$. Depicted are 
      ingoing and outgoing null surfaces.
      \label{fig:ief_grid}}
  \end{minipage}
\end{figure}

The asymptotic region of isolated systems extends to infinity. To simulate such
spacetimes numerically on a finite grid, the most common method truncates the
computational domain by introducing an artificial, timelike outer boundary
which introduces certain problems that we discuss in the
following. Fig.~\ref{fig:ief} shows a typical foliation of the
Schwarzschild-Kruskal spacetime along with timelike surfaces that may
act as artificial boundaries. The diagram has been calculated 
using the ingoing Eddington-Finkelstein coordinates described in Appendix 
\ref{app:A}.

The introduction of an artificial, timelike, outer boundary implies an 
initial boundary value problem for a hyperbolic reduction 
of the Einstein equations which should be well-posed. The requirement of 
well-posedness is not just a mathematical subtlety 
but is also physically motivated. It means that unique solutions are required 
to exist that depend continuously on initial data. This is strongly related 
to the requirement of predictability on a physical theory.

Further, the solutions to the reduced system should correspond to 
solutions to the full system in the sense that the vanishing of the constraints
on an initial hypersurface is preserved into the future of that surface. 
Therefore, the boundary conditions at the outer boundary should preserve the 
vanishing of the constraints.
Such a well-posed, constraint preserving initial boundary value problem 
for Einstein equations has been constructed by Friedrich and Nagy  
\cite{Friedrich99}. Their frame-based, first order formulation 
allows the adaptation of the evolution system to a timelike outer
boundary such that the constraints are propagated by maximally dissipative 
boundary conditions without the need to restrict the boundary data. 
However, their reduction has not yet been used in numerical simulations. 
Currently, there are many studies on the the initial
boundary value problem for the Einstein equations using methods that
are simpler to implement in numerical calculations but do not solve
the full problem
\cite{Babiuc06,Babiuc06a,Kidder04,Kreiss06,Rinne06,Rinne07,Sarbach04}.

Another difficulty concerns the choice of boundary data. We see in
Fig.~\ref{fig:ief_grid} that ingoing characteristics enter the
spacetime from the outer boundary. One might wish to prescribe data on
these characteristics that would correspond to a no incoming radiation
condition. However, the timelike outer boundary surface is
geometrically arbitrary, without invariant meaning. Apart from the
problem related to its gauge dependent choice, it seems difficult to
give a precise definition of gravitational radiation on such a
surface. This difficulty is related to the lack of a quasi-local
energy concept in general relativity. Besides, such a no incoming
radiation condition will not in general correspond to the transparency
of the artificial boundary due to backscattering of gravitational
radiation, which is related to the non-linearity of the theory. To
give an example for potential problems, we mention that it has been
proven that certain types of commonly used boundary conditions in
numerical relativity destroy the tail behavior
\cite{Dafermos04}. Nonetheless, there are detailed analyses on how to
deal with these problems in the linearized regime
\cite{Buchman06,Buchman07}. 

The problems described above can be treated numerically provided that
the outer boundary is sufficiently far away so that linearized
analysis is a good approximation. When the outer boundary is far
enough, a numerically stable treatment might be good enough as one can
trust the solution up to numerical accuracy on the domain of
dependence of the initial hypersurface.  As can be seen in
Fig.~\ref{fig:ief}, the domain of dependence grows when the outer
boundary is put farther away. A difficulty arises due to the slow
$1/\tilde{r}$ fall-off rate of the radiation field. To gain a factor
of $2$ in accuracy, the outer boundary needs to be pushed to twice the
distance which requires $2^3=8$ times the number of grid points on a
homogeneous grid in a simulation in three space dimensions.
Therefore, today's numerical codes use inhomogeneous grids.  By using
mesh refinement techniques one can put the outer boundary to as far as
$700m$ \cite{Baker:2006yw,Marronetti07}. Mesh boundaries, however,
cause numerical errors and there is still a problem of efficiency on
foliations that approach spatial infinity with increasing physical
distance.  Such foliations waste computational resources, as one needs
long time evolutions to calculate the emission of radiation in the
far-field zone.

This brings us to the radiation extraction problem. In the analysis of
numerically generated spacetimes, one calculates certain quantities
interpreted as gravitational radiation along a timelike surface
representing a family of observers far away from the source whereas
$\scri^+$ is the natural place to measure emission
\cite{Frauendiener98c}.  As there is no unambiguous definition of
gravitational radiation at a finite distance away from the source one
uses approximative methods. A common method, called the
Regge-Wheeler-Zerilli method assumes that the full metric in the
extraction region is a perturbation of the Schwarzschild spacetime in
a certain gauge and tries to extract radiation information in a
coordinate invariant way \cite{Regge57,Zerilli70a}.  Another method
relies on calculating Weyl scalars with respect to a special tetrad
class, called the quasi-Kinnersley frame, that can be constructed in
an invariant way in spacetimes of Petrov type D
\cite{Nerozzi04}. These methods rely on linearized approximations. As
shown in \cite{Pazos06}, different extraction methods can deliver
different results in an accurate code so that numerical error is
dominated by systematic error from the extraction method. Note that
this error can not be estimated by convergence tests. Even in the
continuum limit of infinite resolution, the calculation of
gravitational radiation will have the same systematic error.

To estimate the extraction error, one can put the extraction surface
farther away so that the systematic error decreases typically with
$1/\tir$.  However, there are limitations on the choice of the
observer location that also limit the accuracy of radiation
extraction. On the one hand, one should not put the observer too close
to the sources because the assumptions underlying the extraction
method are not valid in the strong field regime.  On the other hand,
one should not put the observer too far, because then the
contamination from the outer boundary hits the extraction surface
before the waves reach the observer. A further difficulty arising from
putting the extraction surface far away is due to numerical
dissipation which lowers the accuracy of the extraction as the waves
propagate slowly along the grid. Today, the observer location for wave
extraction is typically set to about $50m$ \cite{Berti,Marronetti07}.

Although recent years saw quite impressive advances in the problems
mentioned, one can say that there is still a lack of appreciation of
global or semi-global considerations. While the problems we alluded to
can be regarded as subtleties and not crucial for the detection of
gravitational waves, the approximation errors might turn out to be
relevant for gravitational wave astronomy where accurate waveforms are
needed that might in principle allow us to decide on different
equations of state for inspiralling neutron stars or to extract the
remnant gravitational radiation from the big bang. Such questions
require accurate predictions under controllable approximations. As
large parameter studies need to be done, the efficiency of the codes
will also play an important role.

Beside the requirement of an unambiguous radiation extraction we can
deduce the following requirements on a smooth foliation for a clean
boundary treatment just by considering the causal structure on the
grid as depicted in Fig.~\ref{fig:ief_grid}:
\begin{itemize}
\item No characteristics should enter the computational domain 
from the boundary.
\item Outgoing characteristics should leave the computational domain
through the boundary.
\end{itemize}
Note that both requirements are fulfilled for excision along a spacelike inner
boundary inside the black hole.

In the following we will review alternative methods that have been
suggested to deal with the problems of the artificial timelike outer
boundary approach.
\subsection{Coordinate compactification at spatial infinity}
\label{sec:coord_comp}

To avoid problems related to the treatment of the outer boundary
within the standard 3+1 approach, one can use a compactifying
coordinate system on the Cauchy hypersurfaces. Then the outer boundary
of the computational domain is at spatial infinity where natural
boundary conditions can be used \cite{Choptuik03,Pretorius05b}.

To elucidate the idea, take an asymptotically flat, two dimensional spacetime 
metric and a time function $t$ such that $t=\mathrm{const.}$
surfaces are Cauchy surfaces. Introduce on these surfaces a
coordinate $\tilde{r}$. We write the metric on the coordinate domain 
$\{t\in(-\infty,\infty),\tir\in[0,\infty)\}$ as
\[ \tilde{g} = \tilde{g}_{tt}\,dt^2+2 \tilde{g}_{t\tilde{r}}\,dt\,d\tilde{r} + 
\tilde{g}_{\tilde{r}\tilde{r}}\,d\tilde{r}^2.\] To avoid a timelike
outer boundary, we can introduce a compactifying coordinate $r$ by
setting $r(\tir)=\tir/(1+\tir)$. The coordinate transformed metric
takes the form \be\label{intro:trafo} \tilde{g} =
\tilde{g}_{tt}\,dt^2+ 2 \tilde{g}_{tr}\,dt\,dr + \tilde{g}_{rr}\,dr^2,
\quad \textrm{where}\quad
\tilde{g}_{tr}=\frac{\tilde{g}_{t\tilde{r}}}{(1-r)^2}, \quad
\tilde{g}_{rr}=\frac{\tilde{g}_{\tilde{r}\tilde{r}}}{(1-r)^4}.\ee The
coordinate value $\{r=1\}$ corresponds to spatial infinity $i^0$ as
the surfaces $t=\mathrm{const.}$ are Cauchy surfaces.  While the
transformed metric components $\tilde{g}_{tr}$ and $\tilde{g}_{rr}$
are singular at this point, the untransformed ones attain their
Minkowski values for an asymptotically flat spacetime.  A numerical
code can evolve the untransformed components by substituting the
relation (\ref{intro:trafo}) into the Einstein equations. The
equations become formally singular at $\{r=1\}$ where the following outer
boundary conditions can be applied
\[\tilde{g}_{tt}(t,1)=-1,\quad \tilde{g}_{t\tilde{r}}(t,1)=0,\quad 
\tilde{g}_{\tilde{r}\tilde{r}}(t,1)=1.\]

\begin{figure}
  \centering
  \begin{minipage}[t]{0.45\textwidth}
    \centering
    \psfrag{t}{$\tilde{t}/m$}
    \psfrag{r}{$r=\frac{\tilde{r}_s/m}{1+\tilde{r}_s/m}$}
    \psfrag{0.4}{$\frac{2}{5}$}
    \psfrag{3/2}{$\frac{3}{5}$}
    \psfrag{2}{$\frac{2}{3}$}
    \psfrag{infty}{$1$}
    \psfrag{i0}{$i^0$}
    \psfrag{h}{$\mathcal{H}$}
    \psfrag{outgoing}{outgoing}
    \psfrag{ingoing}{ingoing}
    \psfrag{timelike}{$r=\frac{3}{5}$}
    \includegraphics[width=0.95\textwidth]{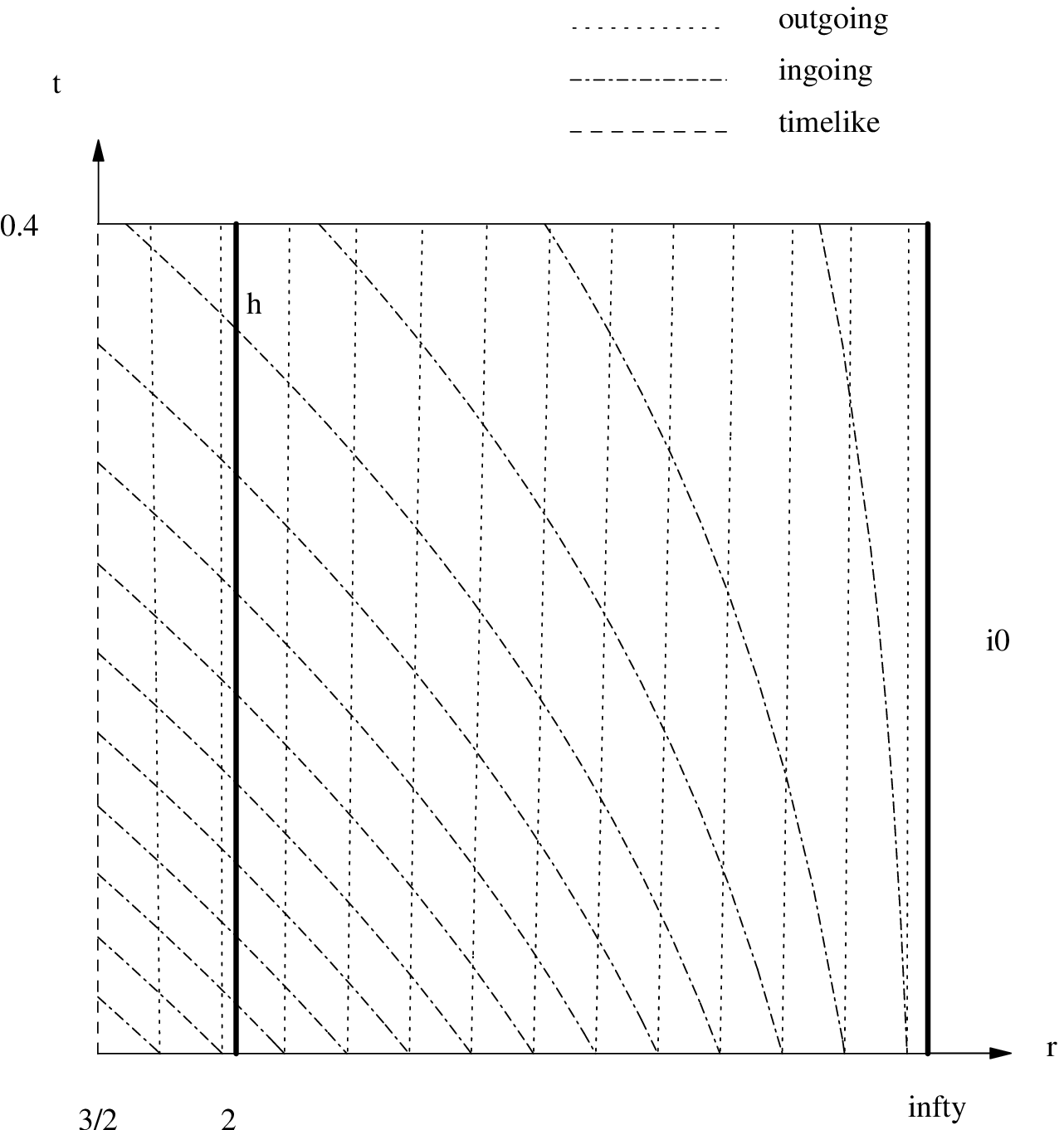}
    \caption{Causal structure on the grid bounded by spatial infinity and 
      a spacelike surface in the interior of the event horizon.
      \label{fig:i0_grid1}}
  \end{minipage}%
  \hspace{0.1\linewidth}%
  \begin{minipage}[t]{0.45\textwidth}
    \centering
    \psfrag{t}{$\tilde{t}/m$}
    \psfrag{r}{$\tilde{r}_s/m$}
    \psfrag{20}{$20$}
    \psfrag{3/2}{$\frac{3}{2}$}
    \psfrag{2}{$2$}
    \psfrag{infty}{$\infty$}
    \psfrag{i0}{$i^0$}
    \psfrag{h}{$\mathcal{H}$}
    \psfrag{outgoing}{outgoing}
    \psfrag{ingoing}{ingoing}
    \psfrag{timelike}{$\tilde{r}_s=\frac{3}{2}m$}
    \includegraphics[width=0.95\textwidth]{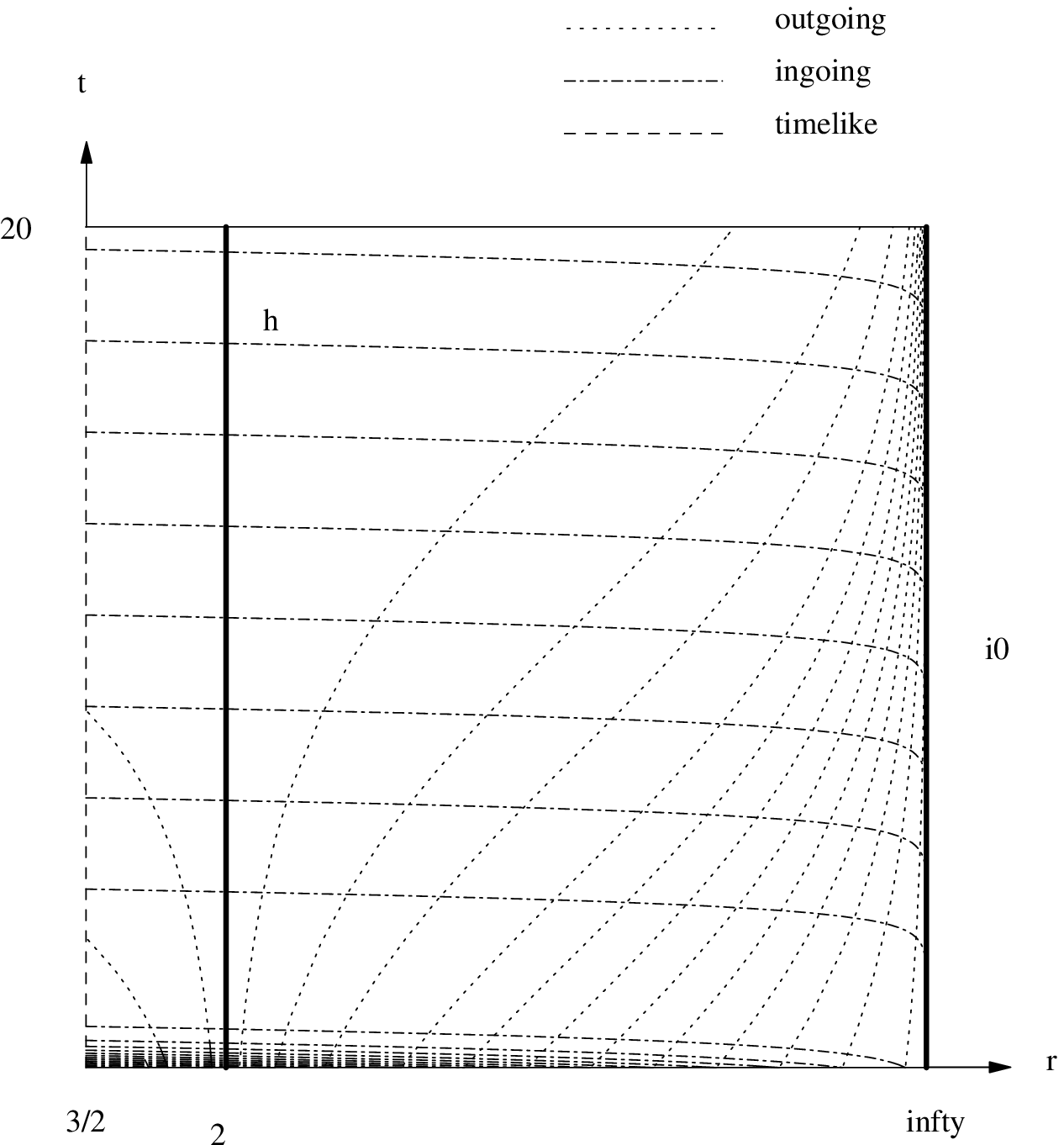}
    \caption{Causal structure on the same grid at a later time. 
      \label{fig:i0_grid2}}
  \end{minipage}
\end{figure}

The outlined method solves the Einstein equations in a domain that is not 
bounded by a timelike dashed curve in region I of Fig.~\ref{fig:ief}, 
but by the point $i^0$. Note that although $i^0$ is a point, in the coordinate
compactification at spatial infinity it is blown up in terms of the coordinates
as illustrated in Fig.~\ref{fig:i0_grid1} and Fig.~\ref{fig:i0_grid2}.
In \cite{Pretorius05b,Pretorius06} spatial infinity becomes a cylinder with 
cubical cuts due to the use of a Cartesian coordinate system.

While this method avoids the introduction of an artificial boundary, 
it does not solve the problem of radiation extraction. 
Using compactifying coordinates on Cauchy surfaces is very awkward for studying
radiation. The grid velocity of outgoing characteristics decreases to zero 
towards the outer grid boundary. 
In simulations, this is seen by a slowing down of outgoing waves.
Gravitational radiation, traveling along outgoing null rays in 
Fig.~\ref{fig:i0_grid1} and Fig.~\ref{fig:i0_grid2}, can not leave 
the numerical domain and piles up during the evolution near the boundary. 
This leads to instabilities as the waves can not be 
resolved after some time. Dissipation can be used to deal with such 
instabilities, but this leads to a loss of accuracy.
Radiation extraction is done at some timelike, gauge dependent surface at a 
finite distance away from the source as in the artificial timelike boundary 
case which implies similar problems. Extraction at outer domains is not 
accurate due to the decrease of wave resolution at large spatial distances (see
\cite{Pretorius05b,Rinne07} for detailed discussions). In addition, errors
generated in the vicinity of the outer boundary can travel along ingoing null 
rays depicted in Fig.~\ref{fig:i0_grid2} and contaminate the interior solution.

In this method, our first requirement that no incoming characteristics
should enter the computational domain is fulfilled, while the second
requirement is not. The use of hyperboloidal foliations instead of
Cauchy foliations might lead to a method that satisfies the second
requirement too. To my knowledge this idea has not yet been tried
out. It is an interesting open question whether the coordinate
compactification technique can be applied using hyperboloidal
foliations.
\subsection{Characteristic evolution and matching}
The characteristic approach is based on studies by Bondi, Sachs 
and others \cite{Bondi62,Sachs62}.
For the treatment of the outer boundary in numerical calculations with the 
characteristic approach, one uses compactifying coordinates 
along outgoing characteristic surfaces with respect to which 
$\scri^+$ is at a known coordinate location and rescales the metric with an 
appropriate conformal factor \cite{Tamburino66}. In a certain conformal gauge,
this procedure leads to equations which are formally singular but have a 
regular limit at $\scri^+$ that can be calculated by numerical methods. 

\begin{figure}
  \centering
  \begin{minipage}[t]{0.5\textwidth}
    \psfrag{i+}{$i^+$}
    \psfrag{i-}{$i^-$}
    \psfrag{scr+}{$\scri^+$}
    \psfrag{i0}{$i^0$}
    \psfrag{r=0}{$\tilde{r}_s=0$}
    \psfrag{h}{$\mathcal{H}$}
    \includegraphics[width=1\textwidth,height=0.8\textwidth]
    {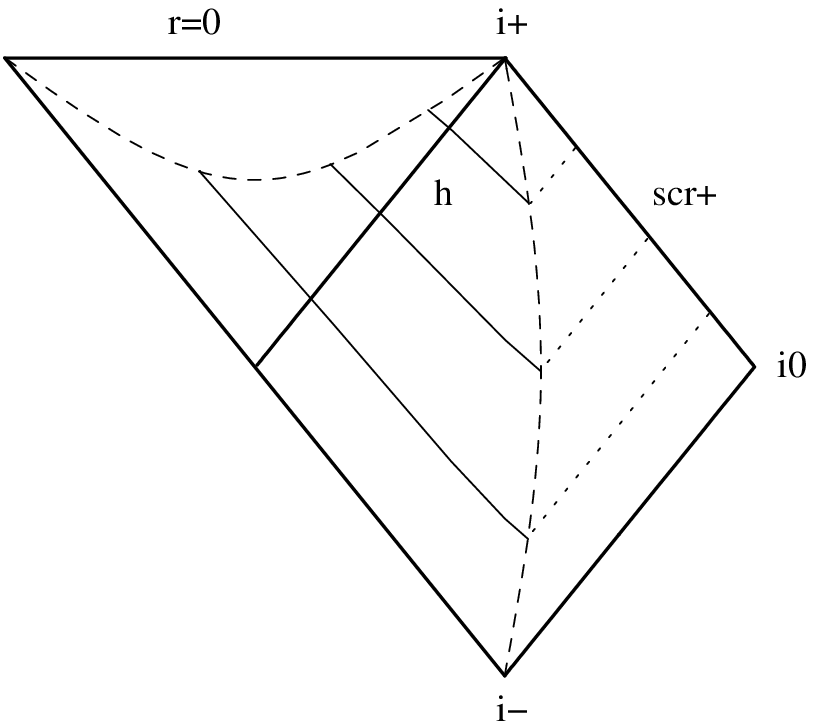}
    \caption{Penrose diagram of the CCM-approach with a matching boundary  
      at $\tilde{r}_s=3m$.
      \label{fig:ccm_causal}}
  \end{minipage}%
  \hspace{0.03\linewidth}%
  \begin{minipage}[t]{0.45\textwidth}
    \psfrag{t}{$\tilde{t}$}
    \psfrag{u}{$\tilde{u}$}
    \psfrag{r}{$\tilde{r}_s/m$}
    \psfrag{3/2}{$\frac{3}{2}$}
    \psfrag{2}{$2$}
    \psfrag{5}{$5$}
    \psfrag{h}{$\mathcal{H}$}
    \psfrag{scr+}{$\scri^+$}
    \psfrag{infty}{$\infty$}
    \psfrag{outgoing}{outgoing}
    \psfrag{ingoing}{ingoing}
    \psfrag{timelike}{$\tilde{r}_s=\textrm{const.}$}
    \includegraphics[height=0.31\textheight,width=\textwidth]
    {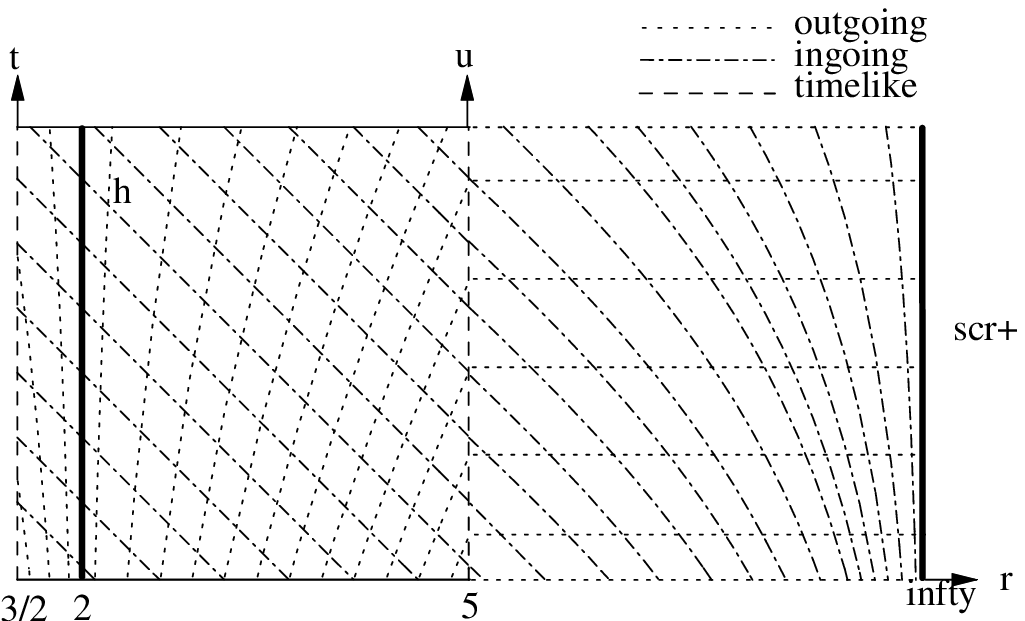}
    \caption{Causal structure on the grid for CCM with a matching boundary
      at $\tilde{r}_s=5m$. \label{fig:ccm_grid}}
  \end{minipage}
\end{figure}

The number of variables in a characteristic evolution scheme reduce
dramatically compared to the corresponding version of a Cauchy problem
because the equations do not involve second time derivatives. Also the
structure of the equations is simpler and there are no elliptic
constraints so that initial data are free \cite{Winicour05}. These
features make the characteristic approach very attractive. This
approach has been successful in cases where null foliations smoothly
cover the numerical domain, such as a single black hole or a
relativistic star. Unfortunately the coordinates are not flexible and
there is little gauge freedom one can use to avoid coordinate
singularities.  The main problem with this approach is the fact that
characteristic foliations are not well-behaved in regions of strong
dynamical gravitational fields due to formation of caustics in the
light rays generating the null hypersurfaces \cite{Friedrich83}.
Therefore, a modification has been suggested which matches a
calculation in the interior based on a spacelike foliation to a
characteristic calculation in the asymptotic region where the fields
become weak and a null foliation can be expected to smoothly cover the
simulation domain.  This approach is called Cauchy-characteristic
matching (CCM) \cite{Bishop93, Winicour05}.

The conformal diagram for a CCM evolution scheme has been plotted 
in Fig.~\ref{fig:ccm_causal}. Note that the solid curves represent pieces of
spacelike Cauchy surfaces as in Fig.~\ref{fig:ief}. Outgoing characteristics 
from the interior calculation are developed up to $\scri^+$ 
and ingoing characteristics to the interior are calculated 
by interpolating the solution in the characteristic region 
to a timelike curve. This timelike curve represents an outer boundary 
for the Cauchy evolution and an inner boundary for the characteristic 
evolution. The causal structure on the grid resulting from the matching has 
been plotted in Fig.~\ref{fig:ccm_grid}.

Both the inner and the outer boundaries fulfill our requirements for a good 
numerical boundary treatment, however, the foliation is not smooth.
The main difficulty with CCM is the matching along the timelike 
boundary between the interior and the asymptotic region, 
where the causal nature of the foliation changes. 
Stability problems caused by interpolation between a Cartesian code solving a
Cauchy problem and a spherical code solving a characteristic problem using 
different sets of variables impeded a further development of this approach.
One can hope that with recent improvements 
on the treatment of timelike boundaries, 
it will be possible to overcome the difficulties the CCM approach 
has been facing in the past. 

In the next subsection we will see that the 
required behavior at both boundaries can be achieved smoothly without changing 
the causal nature of the foliation or the variables of the reduction.

\subsection{Conformally regular field equations}
\begin{figure}[ht]
  \centering
  \begin{minipage}[h]{0.53\textwidth}
    \centering
    \psfrag{i+}{$i^+$}
    \psfrag{i-}{$i^-$}
    \psfrag{scr+}{$\scri^+$}
    \psfrag{scr-}{$\scri^-$}
    \psfrag{i0}{$i^0$}
    \psfrag{r=0}{$\tilde{r}_s=0$}
    \psfrag{h}{$\mathcal{H}$}
    \includegraphics[width=1\textwidth]{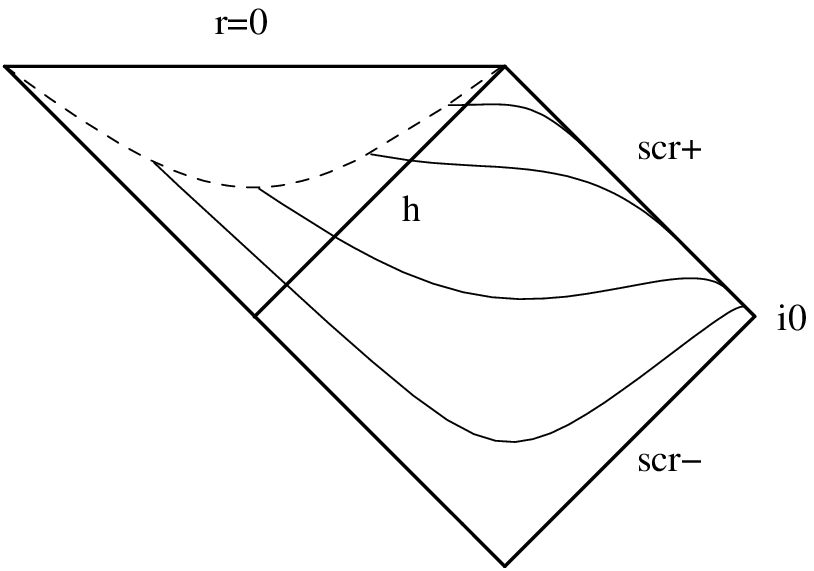}
    \caption{Penrose diagram of a hyperboloidal initial value problem
      on the extended Schwarzschild spacetime.
      \label{fig:cmc_causal}}
  \end{minipage}%
  \hspace{0.07\linewidth}%
  \begin{minipage}[h]{0.37\textwidth}
    \centering
    \psfrag{t}{$\tilde{t}$}
    \psfrag{r}{$\tilde{r}_s/m$}
    \psfrag{7m}{$\infty$}
    \psfrag{3/2}{$\frac{3}{2}$}
    \psfrag{2m}{$2$}
    \psfrag{h}{$\mathcal{H}$}
    \psfrag{scr+}{$\scri^+$}
    \psfrag{outgoing}{outgoing}
    \psfrag{ingoing}{ingoing}
    \psfrag{timelike}{$\tilde{r}_s=\frac{3}{2}m$}
    \includegraphics[width=1\textwidth]{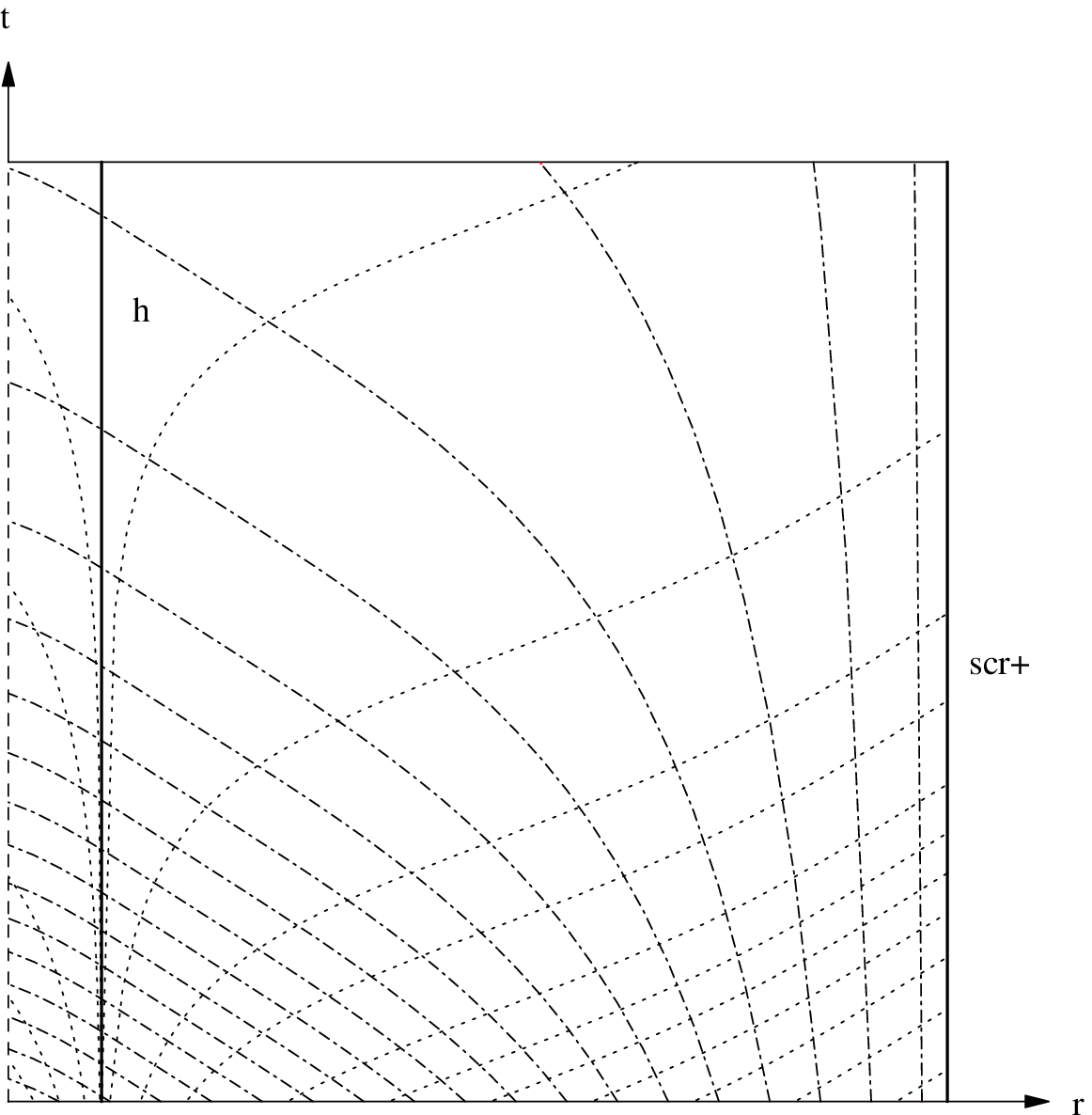}
    \caption{The corresponding causal structure on the grid.
      \label{fig:cmc_grid}}
  \end{minipage}
\end{figure}

The conformally regular approach in numerical relativity is based on analytic 
studies by Friedrich \cite{Friedrich81a,Friedrich83a,Friedrich86} and it
started with numerical studies in spherical symmetry by H\"ubner 
\cite{Huebner93}. In this approach one solves numerically a hyperboloidal 
initial value problem for the conformally regular field equations 
on a domain illustrated in Fig.~\ref{fig:cmc_causal} 
(see \cite{Frauendiener04,Husa01} for reviews). 
The use of compactified hyperboloidal foliations is promising 
because they combine advantages of Cauchy and characteristic 
foliations. On the one hand, instead of approaching spatial infinity 
as Cauchy surfaces do, these surfaces reach null infinity 
which makes them suitable for unambiguous radiation extraction, 
and on the other hand, in contrast to characteristic foliations,
hyperboloidal foliations are spacelike everywhere and therefore 
as flexible as Cauchy surfaces.

An exemplary causal structure on the grid is depicted in 
Fig.~\ref{fig:cmc_grid} for constant mean curvature slices 
in the Schwarzschild-Kruskal spacetime in $\scri^+$-fixing coordinates, 
i.e. in coordinates in which the spatial coordinate location of $\scri^+$ 
has been fixed as described in \ref{sec:cmc_ss}. 
Both requirements for a good boundary treatment are fulfilled. 
Outgoing characteristics leave the computational domain through null infinity 
where a rigorous definition of gravitational radiation allows us to construct
an unambiguous numerical radiation extraction method and there are no incoming 
characteristics so that no boundary conditions are needed. 
The foliation is smooth throughout the simulation domain and very flexible.

A difficulty with the conformally regular approach in numerical calculations
is that the equations include, among others, 
evolution equations for the conformal factor which results in a solution
dependent representation of $\scri^+=\{\Omega=0\}_{+}$.
Note that in Fig.~\ref{fig:cmc_grid}, $\scri^+$ has been fixed 
to a spatial coordinate location.
In the metric based conformally regular field equations, however, 
the location of $\scri^+$ on the grid is not known
a priori. The numerical boundary does not coincide with the conformal boundary
$\scri^+$ which leads to problems of efficiency and requires 
a numerical boundary treatment outside the physical spacetime 
$\widetilde{\mathcal{M}}$ \cite{Husa:2002kk}. 
A gauge condition to fix $\scri^+$ to a spatial coordinate 
location has been suggested by Frauendiener in \cite{Frauendiener98b} 
in the context of frame-based conformally regular field equations. 
To my knowledge, this gauge has not been implemented in a general scheme 
in which $\scri^+$ corresponds to the outer grid boundary 
due to technical complications with numerical 
implementation of a frame-based evolution system 
that requires spherical grid topology.

Another difficulty is that the conformally regular field equations are 
significantly larger than usual formulations of Einstein equations 
including many additional evolution variables subject to constraints.
Due to the large number of constraint equations, numerical errors require 
a stronger control on constraint propagation properties of the system. 
As there is not enough numerical experience with the equations, 
one can not use established methods easily to deal with the
encountered instabilities. While these problems are not of a principal nature, 
they have made progress in the conformally regular approach difficult. 

Instead of trying to overcome the described difficulties within the context 
of the outlined methods, we will discuss in the following chapters 
two novel conformal approaches for the numerical calculation of 
asymptotically flat spacetimes. 

The approach taken in chapter \ref{chapter:i0} falls into
the category of the conformally regular approach. It is based
on the system of reduced general conformal field equations which has certain
advantages over the systems that have been tried in the
conformally regular approach until now.  Using this system, we will
numerically study a neighborhood of spatial infinity including a
smooth piece of null infinity.

Once a full neighborhood of spatial infinity has been calculated,
including a piece of null infinity, one can evolve the system further by
solving a hyperboloidal initial value problem. For accurate
numerical calculations of gravitational radiation along null infinity,
however, the reduced general conformal field equations do not seem to
be the appropriate tool. The reason is that the underlying gauge based
on timelike conformal geodesics leads
to a loss of resolution in the physical domain.  
A further motivation to construct a different method for the numerical
solution of a hyperboloidal initial value problem is the wish to
employ the extensive experience in numerical relativity gathered over
the last decades.  In chapter \ref{chapter:null}, I will
present a method in which null infinity can be included in the
computational domain for a common reduction of the Einstein
equations. In this method the location of null infinity is
fixed on the numerical grid so that no resolution loss appears.


%% file: null.tex
\chapter{Null Infinity}\label{chapter:null}
The discussion on numerical outer boundary treatments in the
introduction suggests that one should include null infinity 
in the computational domain for a proper treatment of the outer boundary 
as well as for unambiguous radiation extraction.
In this chapter, we will see how $\scri^+$ can be 
included in the computational domain for a common reduction of the 
Einstein equations.

We start with a study of a certain class of spherically symmetric 
hyperboloidal surfaces in the Minkowski spacetime to develop an intuition 
about their behavior. We observe in spherical symmetry 
that a conformal factor $\Omega$ together with a radial spatial coordinate $r$ 
can be chosen such that $\scri^+$ is given by $\{r=1\}$ as plotted in
Fig.~\ref{fig:cmc_grid}. Then we address the full problem without symmetry 
assumptions. We see for a common reduction of the Einstein equations 
how $\scri^+$-fixing can be achieved by prescribing the representation of a
preferred conformal factor in terms of suitably specified coordinates 
in a hyperboloidal initial value problem. 
We present numerical tests of the method in spherical symmetry.
The chapter ends with a discussion.
\section{Hyperboloidal surfaces in Minkowski spacetime}
\label{sec:hyp_sur}
A hypersurface $\mathcal{S}$ in a spacetime $(\mathcal{M},g)$ can be
given as a level set of a scalar function $\Phi(p)=\textrm{const}.$
which satisfies $d\Phi\ne 0$ for all $p\in \mathcal{S}$. We assume
$(\mathcal{M},g)$ to be time-orientable, with metric signature
$(-,+,+,+)$ and coordinates $\{x^\mu\}$. We set the positive
$x^0$-direction to be the future direction. The unit normal to the
hypersurface is given by \be\label{null:unit_n}
n^\mu=g^{\mu\nu}\frac{\nabla_\nu\Phi}{\|\nabla\Phi\|_g}.\ee We require
the hypersurface to be space-like so that the unit normal is
time-like, $g_{\mu\nu}n^\mu n^\nu = -1$, and we choose $n^\mu$ to be
future-pointing, $n^0>0$.  If the level sets of $\Phi(x)$ define a
local foliation in $\mathcal{M}$, we can introduce a new time
coordinate $t(x)=\Phi(x)$ and construct a new coordinate system on a
certain domain by choosing space coordinates on one of the
hypersurfaces $\mathcal{S}_{\Phi}=\{\Phi=\textrm{const.}\}$.  We
define the lapse function $\alpha$ and the shift vector $\beta^\mu$ in
this coordinate system by the decomposition $\partial_t^\mu=\alpha\,
n^\mu+\beta^\mu,$ with $g(n,\beta)=0$.  The induced Riemannian metric
on $\mathcal{S}_{t}$ is given by $h_{\mu\nu}=g_{\mu\nu}+n_\mu n_\nu$.
The extrinsic curvature $K_{\mu\nu}$, given by
$K_{\mu\nu}=h_\mu^{\ \lambda}\nabla_\lambda n_\nu$, is a measure for
the variation of the unit normal along the hypersurfaces
$\mathcal{S}_t$.  The mean extrinsic curvature $K$ is the trace of the
extrinsic curvature,
\[ K = g^{\mu\nu}K_{\mu\nu} = \nabla_\mu n^\mu = 
\frac{1}{\sqrt{-g}}\partial_\mu(\sqrt{-g}\, n^\mu).\]
In our notation positive $K$ means expansion.
The mean curvature of a hyperboloidal surface $\mathcal{S}$ 
approaches a strictly positive value at $\scri^+$. We will discuss some 
examples of spherically symmetric hyperboloidal surfaces in the Minkowski 
spacetime $(\mathbb {R}^4, \tilde{\eta})$ with the metric $\tilde{\eta}$ given 
as in (\ref{intro:eta}). The mean curvature $\tilde{K}$ of such a 
hypersurface with unit normal $\tilde{n}^\mu(\tit,\tir)$ becomes
\be\label{null:trk} \tilde{K}=\partial_{\tilde{t}} \tilde{n}^{\tilde{t}} + 
\partial_{\tilde{r}}\tilde{n}^{\tilde{r}}+
\frac{2}{\tir}\,\tilde{n}^{\tilde{r}}. \ee
\subsection{Standard hyperboloids}
The standard hyperboloids in the Minkowski spacetime can be given as level sets
of the function
\be \label{null:sthyp} \Phi(\tit,\tir)=\tit^2-\tir^2. \ee
We investigate the surfaces $\mathcal{S}_{\Phi}=\{\Phi=\mathrm{const.}\}$ 
strictly inside the upper light cone, $\tit>\tir$, where they are spacelike and
$\Phi>0$. The unit future-directed normal (\ref{null:unit_n}) 
and the mean curvature (\ref{null:trk}) on $\mathcal{S}_\Phi$ read
\[ \tilde{n}=\frac{1}{\sqrt{\tit^2-\tir^2}}\,\left(\tit\,\partial_{\tilde{t}}+
\tir\,\partial_{\tilde{r}}\right),\qquad 
\tilde{K} = \frac{3}{\sqrt{\tit^2-\tir^2}} = \frac{3}{\sqrt{\Phi}}. \]
Standard hyperboloids given by different values of $\Phi$ have 
spatially constant mean curvature (CMC), 
however, if we introduce $\Phi$ as a foliation parameter we do not get a 
CMC foliation as the value of $\tilde{K}$ varies from slice to slice.

\begin{figure}[ht]
  \begin{minipage}[ht]{0.55\textwidth}
    \centering
    \psfrag{i+}{$i^+$}
    \psfrag{S_1}{$\mathcal{S}_1$}
    \psfrag{i0}{$i^0$}
    \psfrag{scr+}{$\scri^+$}
    \psfrag{v}{$V$}
    \psfrag{u}{$U$}
    \includegraphics[width=1\textwidth]{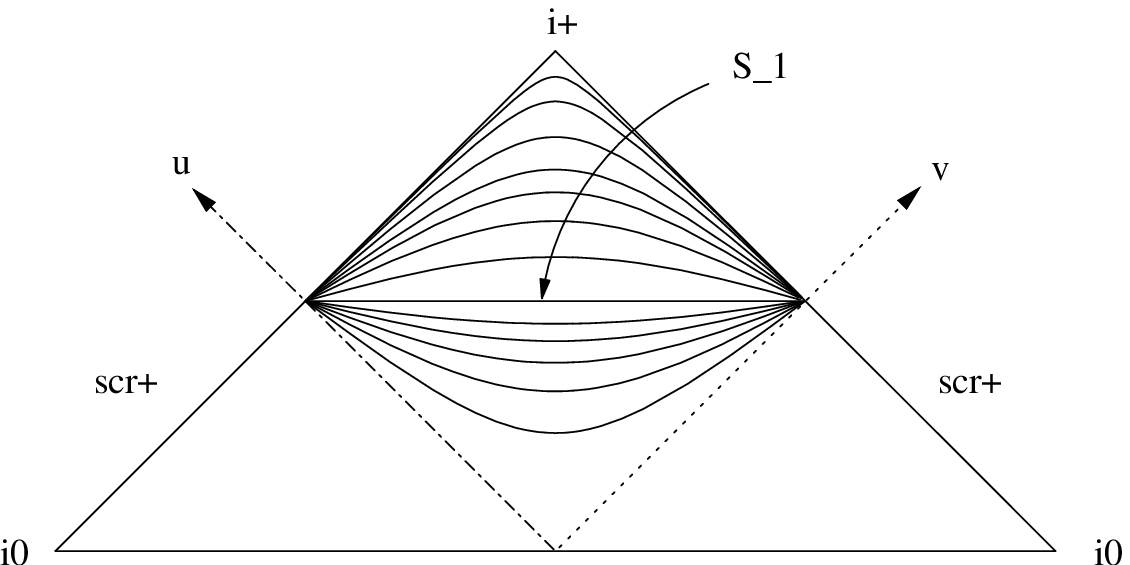}
    \caption{Standard hyperboloids. 
      \label{fig:sthyp}}
  \end{minipage}%
  \hspace{0.02\linewidth}%
  \begin{minipage}[ht]{0.4\textwidth}
    \centering
    \psfrag{S_1}{$\mathcal{S}_1$}
    \psfrag{alpha}{$\alpha=\frac{\pi}{4}$}
    \psfrag{pi/2}{$\frac{\pi}{2}$}
    \psfrag{scr+}{$\scri^+$}
    \psfrag{v}{$V$}
    \psfrag{u}{$U$}
    \includegraphics[width=0.65\textwidth]{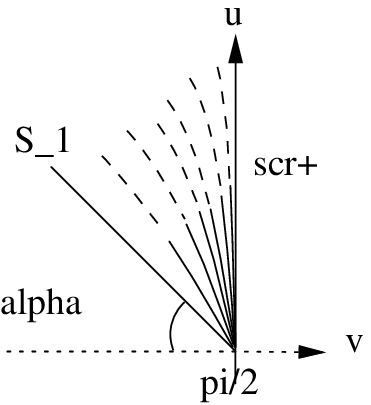}
    \caption{Cutting angle at $\scri^+$.
      \label{fig:angle}}
  \end{minipage}
\end{figure}

The embedding of the Minkowski spacetime into the Einstein universe
from \ref{sec:con_comp} allows us to study the asymptotic behavior of
standard hyperboloids by using local differential geometry. We get
\[\Phi\left(\tit(V,U),\tir(V,U)\right) = \tan V \tan U.\] 
The intersection $\Sigma_{\Phi}=\mathcal{S}_{\Phi}\cap\scri^+$ 
is independent of $\Phi$,
\[U|_{\Sigma_{\Phi}}=\arctan(\Phi \cot V)\big|_{V=\frac{\pi}{2}} = 0,\]
as seen in Fig \ref{fig:sthyp}. Therefore $\Phi$ is unsuitable as a foliation 
parameter for $\scri^+$.
It determines the angle $\alpha$ of the cut at $\Sigma=\{U=0,V=\pi/2\}$,
\[ d\Phi|_{\mathcal{S}_\Phi} =\frac{\tan U}{\cos^2 V} dV+
\frac{\tan V}{\cos^2 U} dU = 0 \
\Rightarrow \ \frac{dU}{dV}\Big|_{\mathcal{S}_{\Phi}} =
 -\frac{\tan U \cos^2 U}{\tan V \cos^2 V} = -\Phi\, \frac{\cos^2 U}{\sin^2 V},\]
\be\label{eq:angle1} \tan \alpha=-\frac{dU}{dV}\Big|_{\Sigma} = 
\Phi \quad \Rightarrow \quad 
\alpha = \arctan \Phi=\arctan\left(\frac{9}{\tilde{K}^2}\right). \ee
We can use this relationship to find the hyperboloid that intersects $\scri^+$
in a certain angle. The hyperboloid which is "parallel" to the $\tit=0$ 
surface in the above embedding has a cutting angle $\alpha=\pi/4$
which implies $\Phi=1$. The surface $\mathcal{S}_1=\{\Phi=1\}$ has been plotted 
in Fig.~\ref{fig:sthyp} and Fig.~\ref{fig:angle}. 

We see from (\ref{eq:angle1}) that for large values of $\tilde{K}$, the
angle approaches 0.  This suggests that $\tilde{K}$ can be used as an
intuitive measure for how close the surfaces are to null surfaces. In
subsequent sections we will see how this aspect plays a role in
numerical calculations.

\subsection{Simple hyperboloidal foliations}
\begin{figure}[t]
  \centering
  \psfrag{i+}{$i^+$} \psfrag{i0}{$i^0$}\psfrag{i-}{$i^-$}\psfrag{S-}{$\scri^-$}
  \psfrag{S+}{$\scri^+$}\psfrag{u}{$V$}\psfrag{v}{$U$}
  \includegraphics[width=0.9\textwidth]{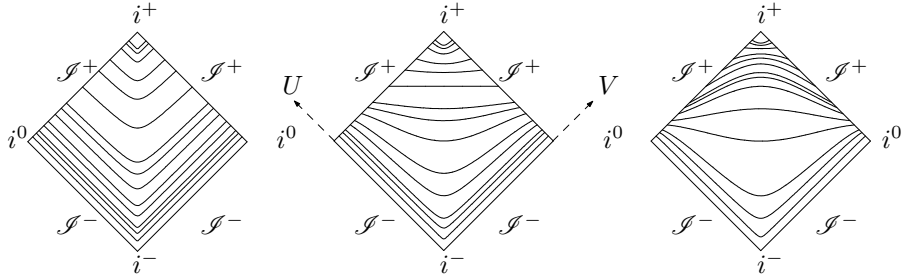}
  \caption{\label{fig:cmc_fol} CMC foliations of the Minkowski spacetime with 
    $\tilde{K}=\{15,3,1.5\}$.}
\end{figure}

A hyperboloidal foliation can be constructed by translating a standard 
hyperboloid, given by $a^2=\tit^2-\tir^2$ with $a\in\mathbb{R}$, along the 
timelike Killing vector field  $\partial_{\tilde{t}}$ of the Minkowski spacetime.
The translated surfaces satisfy $a^2=(\tit-\Phi)^2-\tir^2$. So, they are given 
by level sets of
\be\label{eq:st_hyp_fol} \Phi(\tit,\tir)=\tit-\sqrt{a^2+\tir^2}. \ee
The unit future-directed normal and the mean curvature read
\[ \tilde{n}=\frac{1}{a}\left(\sqrt{a^2+\tir^2}\,\partial_{\tilde{t}} + 
\tir\,\partial_{\tilde{r}}\right),\qquad \tilde{K} = \frac{3}{a}. \]
In this case, each surface not only has a spatially constant mean curvature, 
but also different surfaces of the foliation share the same $\tilde{K}$. 
The value of the mean extrinsic curvature does not depend on $\Phi$ but on a 
real parameter $a$ that can be fixed freely. We embed the surfaces 
into the Einstein universe as before to get
\[ \Phi(V,U) = \frac{1}{2} (\tan V + \tan U) - \sqrt{a^2 + \frac{1}{4}
(\tan V - \tan U)^2}. \]
To study their asymptotic behavior near $\scri^+$ where $V\to\pi/2$ and 
$\cot V\to 0$ we write
\[ \Phi(V,U)= \frac{2 \tan U-2\, a^2\cot V}
{ 1+\tan U \cot  V+\sqrt{ 4 a^2 \cot^2 V + 
\left( 1- \tan U \cot V \right)^2}}. \]
We get a Taylor expansion in $\cot V$ near $\scri^+$
\[ \Phi(V,U) = \tan U - a^2 \cot V - a^2 \tan U \cot^2 V + O(\cot^3 V) 
\quad \mathrm{for} \quad V\to \pi/2. \]
The cut at $\scri^+$ depends on $\Phi$ only
\[ \Phi|_{\scri^+} =\tan U\quad\Rightarrow\quad U\big|_{\scri^+}=\arctan \Phi.\]
The angle of the cut can be calculated by
\be \label{eq:angle} \tan\alpha =-\frac{dU}{dV}\Big|_{\scri^+}= 
a^2 \cos^2 U|_{\scri^+} \quad \Rightarrow \quad 
\alpha = \arctan\left(\frac{9}{\tilde{K}^2(1+\Phi^2)}\right). \ee
Fig.~\ref{fig:cmc_fol} shows three CMC foliations in the Penrose 
diagram of the Minkowski spacetime with the same set of values of $\Phi$, 
but different values of $\tilde{K}$. Each plotted surface is spacelike, 
so their angle is smaller then 45 degrees to the horizontal.
For small $\tilde{K}$ some of the surfaces look similar to Cauchy
surfaces while for large $\tilde{K}$ the surfaces rather look like null 
surfaces. The diagrams illustrate the sense in which we say that
hyperboloidal surfaces mediate between Cauchy and characteristic 
surfaces. This behavior affects also the dynamics of wave-like solutions. 
The propagation speed of waves on the grid in adapted coordinates depends 
among others on the choice of $\tilde{K}$ 
as can be seen in Fig.~\ref{fig:cmc_mink_causal}.
It is a nice feature that by choosing the mean curvature 
we can control the behavior of the surfaces 
in the interior without changing their cut at $\scri^+$.

For a given foliation with some $\tilde{K}$, we can find the surface 
which is "parallel" to the Cauchy surface $\tilde{t}=0$ 
with the slope $\alpha=\pi/4$ at $\scri^+$. From (\ref{eq:angle}) we get 
$\Phi=\pm\sqrt{a^2-1}=\pm\sqrt{9/\tilde{K}^2-1}$.
Such surfaces exist only for foliations with $\tilde{K}\leq 3$ 
which can also be seen in Fig.~\ref{fig:cmc_fol}.

Another interesting set of foliations can be constructed by requiring that each
surface of the foliation intersects $\scri^+$ with the same angle. 
Choosing $\alpha=\pi/4$ we substitute $a^2=1+\Phi^2$ into (\ref{eq:st_hyp_fol}) 
to get
\[\Phi(\tit,\tir)=\frac{1}{2 \tit}\left(\tit^2-\tir^2-1\right). \]
The mean curvature reads by construction 
$\tilde{K}=3/\sqrt{1+\Phi^2}$. As in the case of the standard hyperboloids, 
each surface has a spatially constant mean curvature 
but $\tilde{K}$ depends on the foliation parameter $\Phi$. 
In this case we get a foliation. The embedding results in 
\[ \Phi = -\cot(V+U) = -\cot(2t),\]
where we used the compactifying time coordinate $t=(V+U)/2$. As might have been
expected, we get the $t=\mathrm{const.}$ surfaces.
\section{$\scri^+$-fixing in spherical symmetry}\label{sec:scr_fix}
As seen in Penrose diagrams like Fig.~\ref{fig:cmc_fol}, 
$\scri^+$ is an ingoing null surface. This 
property can be made manifest in suitable coordinates such that the null 
generators of $\scri^+$ converge. Such coordinates can be useful in 
numerical studies for discussing global properties of spacetimes, 
for example with respect to the existence of a regular point
corresponding to timelike infinity $i^+$ \cite{Huebner01}.
These coordinates, however, do not seem to be convenient for calculating 
gravitational radiation accurately as they lead to a loss of resolution in the 
physical part of the conformal extension. Besides, when the outer boundary is a
timelike surface in the unphysical spacetime, our requirements for a boundary 
treatment from \ref{sec:nr} are not fulfilled.

It has been suggested
\cite{Andersson02a,Frauendiener98b,Husa05,Zeng06b} that
conformal compactifications in which $\scri^+$ is kept at a fixed
spatial coordinate location might be suitable for numerical
calculations.  This is also suggested by illustrations of the causal
structure as in Fig.~\ref{fig:ccm_grid} or Fig.~\ref{fig:cmc_grid}.
In this section, we discuss the general construction of such
$\scri^+$-fixing coordinates for spherically symmetric hyperboloidal
foliations.  The explicit examples concentrate on CMC foliations.

We introduce some conformal transformation rules. Assume a time coordinate $t$ 
and space coordinates $\{x^\gamma\}_{\gamma=1,2,3}$ have been chosen in 
$\mathcal{M}$. We write the conformal rescaling in terms of variables 
of a 3+1 decomposition in the given coordinate system
\begin{eqnarray*} 
g &=& \left(-\alpha^2 + h_{\gamma\delta}\beta^\gamma \beta^\delta\right)\,dt^2 + 
2 h_{\gamma\delta}\beta^\gamma \, dt\,dx^\delta + 
h_{\gamma\delta} \,dx^\gamma dx^\delta = \Omega^2 \tilde{g} = \\ &=&
\left(-(\Omega\tilde{\alpha})^2+\Omega^2\tilde{h}_{\gamma\delta}
\tilde{\beta}^\gamma \tilde{\beta}^\delta\right)\,dt^2 + 
2 \Omega^2\tilde{h}_{\gamma\delta}\tilde{\beta}^\gamma \, dt\,dx^\delta + 
\Omega^2\tilde{h}_{\gamma\delta} \,dx^\gamma dx^\delta, \end{eqnarray*}
and deduce the transformation rules
\[ \alpha = \Omega \tilde{\alpha},\quad \beta^\gamma=\tilde{\beta}^\gamma,\quad 
h_{\alpha\beta} = \Omega^2 \tilde{h}_{\alpha\beta}.\]
The unit normal $n^\mu=\frac{1}{\alpha}\left(\partial_t^\mu-\beta^\mu\right)$ 
transforms as $n^\mu = \frac{1}{\Omega} \tilde{n}^\mu$. The transformation of 
the covariant derivative under the conformal rescaling
$g=\Omega^2\tilde{g}$ reads \cite{Wald84}
\be\label{conf_trafo} \nabla = \tilde{\nabla} + S(\Omega^{-1}d\Omega), \qquad 
S^\lambda_{\mu\nu}(\Omega^{-1}d\Omega)=\frac{1}{\Omega}\left(\delta^\lambda_\mu
\Omega_\nu+\delta^\lambda_\nu\Omega_\mu-
g^{\lambda\rho}g_{\mu\nu} \Omega_{\lambda}\right),\ee
where $\Omega_\mu := (d\Omega)_\mu = \nabla_\mu \Omega $. By the above 
transformation rules and by the definition of the extrinsic curvature
$K_{\mu\nu}=h_\mu^{\ \lambda}\nabla_\lambda n_\nu$, we derive
$K_{\mu\nu}=\Omega (\tilde{K}_{\mu\nu} + \Omega_n \tilde{h}_{\mu\nu})$,
where $\Omega_n=n^\alpha \Omega_\alpha$. The mean curvature
transforms as $K=\Omega^{-1}(\tilde{K}+3\Omega_n)$. We note that a CMC foliation
with respect to the physical metric $\tilde{g}$ does not necessarily imply 
a CMC foliation with respect to the rescaled metric $g$.

The transformation rules above have been written in a given coordinate system.
For $\scri^+$-fixing, we need to construct the coordinate system together
with the choice of the conformal rescaling. This can be done explicitly in 
spherical symmetry. The physical line element in spherical symmetry 
in coordinates $\{t,\tilde{r},x^A\}$ can be written as
\be \label{sph_sym} \tilde{g} = \left(-\tilde{\alpha}^2 +
\tilde{h}_{\tilde{r}}^2(\tilde{\beta}^{\tilde{r}})^2\right)
\,dt^2 + 2 \tilde{h}_{\tilde{r}}^2\tilde{\beta}^{\tilde{r}}\, dt\,d\tilde{r} + 
\tilde{h}_{\tilde{r}}^2 \,d\tilde{r}^2 + \tilde{r}^2 \,d\sigma^2. \ee
The lapse $\tilde{\alpha}$, the shift component $\tilde{\beta}^{\tilde{r}}$, 
and the spatial metric function $\tilde{h}_{\tilde{r}}$ are functions 
of the coordinates $(t,\tilde{r})$ only. We assume that  the metric 
(\ref{sph_sym}) admits a smooth conformal boundary and that the time 
coordinate $t$ is such that $t=\textrm{const}.$ hypersurfaces are 
hyperboloidal. We do not compactify the time direction. 
The conformal compactification, $g=\Omega^2 \tilde{g}$, can be done such that
\be \label{conf_comp} \Omega^2(\tilde{h}_{\tilde{r}}^2d\tilde{r}^2+
\tilde{r}^2d\sigma^2) = h_r^2\,dr^2 + r^2 \,d\sigma^2, \ee
with respect to a compactifying radial coordinate $r$. 
Note that we have some freedom here. One can require for example $h_r=1$ 
which leads to $r$ being the proper distance, but then the conformal 
factor is determined by the above relation and the radial 
coordinate transformation can not, in general, be written in explicit form. 
By keeping $h_r$, we have the freedom to prescribe the representation of a
conformal factor in terms of a suitable compactifying radial coordinate $r$ 
and the coordinate transformation can be made explicit. 
The relation (\ref{conf_comp}) implies for a given representation 
$\Omega(r)$ a coordinate transformation $\tilde{r}=\Omega^{-1}\,r$ so that 
$d\tilde{r}=(\Omega - r\,\Omega')\Omega^{-2}\,dr$. 
Then the spatial metric function transforms as 
$h_r = \Omega\,\tilde{h}_r=(\Omega - r\,\Omega')\Omega^{-1}\tilde{h}_{\tilde{r}}$.
For regularity of this conformal compactification, 
$\tilde{h}_{\tilde{r}}(t,\tilde{r})$ needs to have a specific asymptotic 
fall-off behavior for $\tilde{r}\to\infty$ on the hyperboloidal surfaces 
of constant $t$. 

A convenient representation for the conformal factor is
$\Omega=(1-r)$. This is not a good choice at the origin, but we are
interested in the asymptotic region. A possible representation of a
conformal factor in the interior will be discussed in
\ref{sec:interior} in connection with a hyperboloidal initial value
problem.  Our choice fixes the compactifying coordinate $r$ via
\be \label{null:radial}\tir(r) = \frac{r}{1-r}=\frac{r}{\Omega}, \quad
d\tir = \frac{dr}{(1-r)^2}=\frac{dr}{\Omega^2},\qquad
r(\tir)=\frac{\tir}{1+\tir},\ee which implies \be\label{null:trafo}
\alpha =\Omega \tilde{\alpha},\qquad
\beta^r=\Omega^2\tilde{\beta}^{\tilde{r}}, \qquad h_r =
\Omega^{-1}\tilde{h}_{\tilde{r}}. \ee

In the following we will explicitly construct a $\scri^+$-fixing
conformal compactification for Minkowski and extended Schwarzschild
spacetimes using a CMC foliation. There are many other possibilities
for constructing simple hyperboloidal foliations, for example constant
scalar curvature surfaces \cite{Frauendiener06}.
\subsection{The Minkowski spacetime}
We use the CMC foliation (\ref{eq:st_hyp_fol}) to introduce a new time 
coordinate $t=\Phi(\tit,\tir)$ for the Minkowski spacetime 
$(\mathbb{R}^4,\tilde{\eta})$. We have $d\tit=dt+\tir/\sqrt{a^2+\tir^2}\,d\tir$,
so that the Minkowski metric (\ref{intro:eta}) becomes 
\[ \tilde{\eta} = -dt^2-\frac{2\tilde{r}}{\sqrt{a^2+
\tilde{r}^2}}\,dt d\tilde{r} + \frac{a^2}{a^2+\tilde{r}^2}\,d\tilde{r}^2 +
\tilde{r}^2\,d\sigma^2. \]
The variables of the 3+1 decomposition read
\[ \tilde{\alpha} = \frac{1}{a}\sqrt{a^2+\tir^2},\quad 
\tilde{\beta}^{\tilde{r}}=-\frac{\tir}{a}\tilde{\alpha},\quad
\tilde{h}_{\tilde{r}} = \frac{1}{\tilde{\alpha}}. \]
Conformal compactification, $\eta=\Omega^2\tilde{\eta}$, with $\Omega=1-r$
using (\ref{null:trafo}) results in 
\be \label{cmc_mink}\eta=-\Omega^2dt^2-
\frac{2r}{\sqrt{a^2\,\Omega^2+r^2}}\,dt dr 
+ \frac{a^2}{a^2\,\Omega^2+r^2}\,dr^2+r^2d\sigma^2. \ee
The variables of the 3+1 decomposition are
\[\alpha(r) = \frac{1}{a}\,\sqrt{a^2\,\Omega^2+r^2}=\Omega\,\tilde{\alpha},\quad
\beta^r = -\frac{r}{a}\,\alpha = \Omega^2\tilde{\beta}^{\tilde{r}}, \quad 
h_r = \frac{1}{\alpha} = \Omega^{-1} \tilde{h}_{\tilde{r}}. \]
With our choice of the conformal factor we get 
$\Omega_n=\beta^r/\alpha=-r/a = -\tilde{K} r/3$.
Using the transformation rule of the mean curvature under 
conformal rescalings we derive
$K=\Omega^{-1}(\tilde{K}+3\Omega_n)=\Omega^{-1}(\tilde{K}-\tilde{K}r)=\tilde{K}$.
In this case the mean extrinsic curvatures with respect to the physical and
the rescaled metric are equal so that a CMC foliation of the physical
Minkowski spacetime gives a CMC foliation of the conformal extension
that we constructed.

\begin{figure}[t]
  \begin{minipage}[t]{0.3\textwidth}
    \centering
    \psfrag{0}{$0$}
    \psfrag{1}{$1$}
    \psfrag{t}{$t$}
    \psfrag{r}{$r$}
    \psfrag{scr+}{$\scri^+$}
    \includegraphics[width=\textwidth]{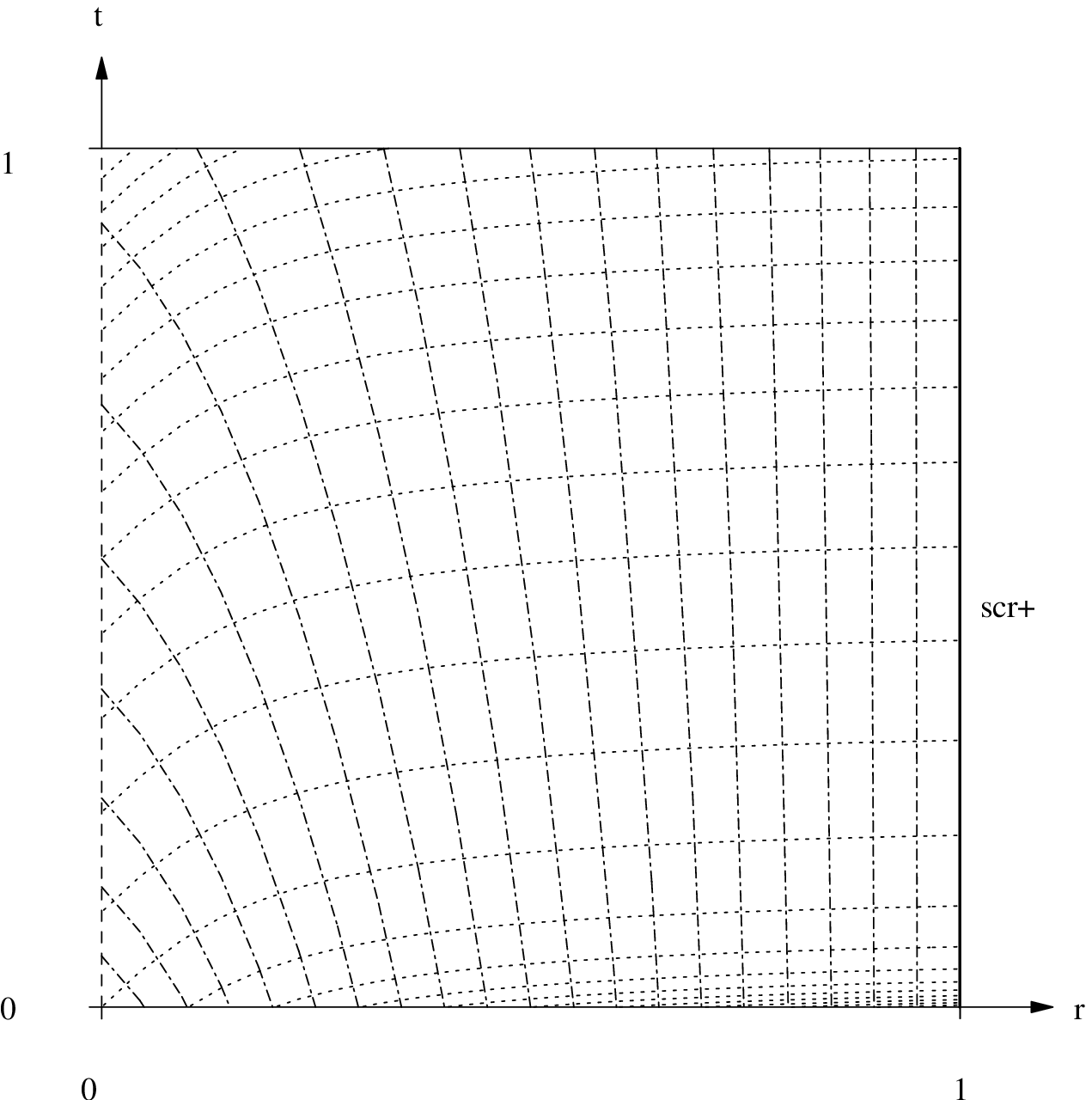}
  \end{minipage}%
  \hspace{0.04\linewidth}%
  \begin{minipage}[t]{0.3\textwidth}
    \centering
    \psfrag{0}{$0$}
    \psfrag{1}{$1$}
    \psfrag{t}{$t$}
    \psfrag{r}{$r$}
    \psfrag{scr+}{$\scri^+$}
    \psfrag{outgoing}{outgoing}
    \psfrag{ingoing}{ingoing}
    \psfrag{timelike}{$r=0$}
    \includegraphics[width=\textwidth]{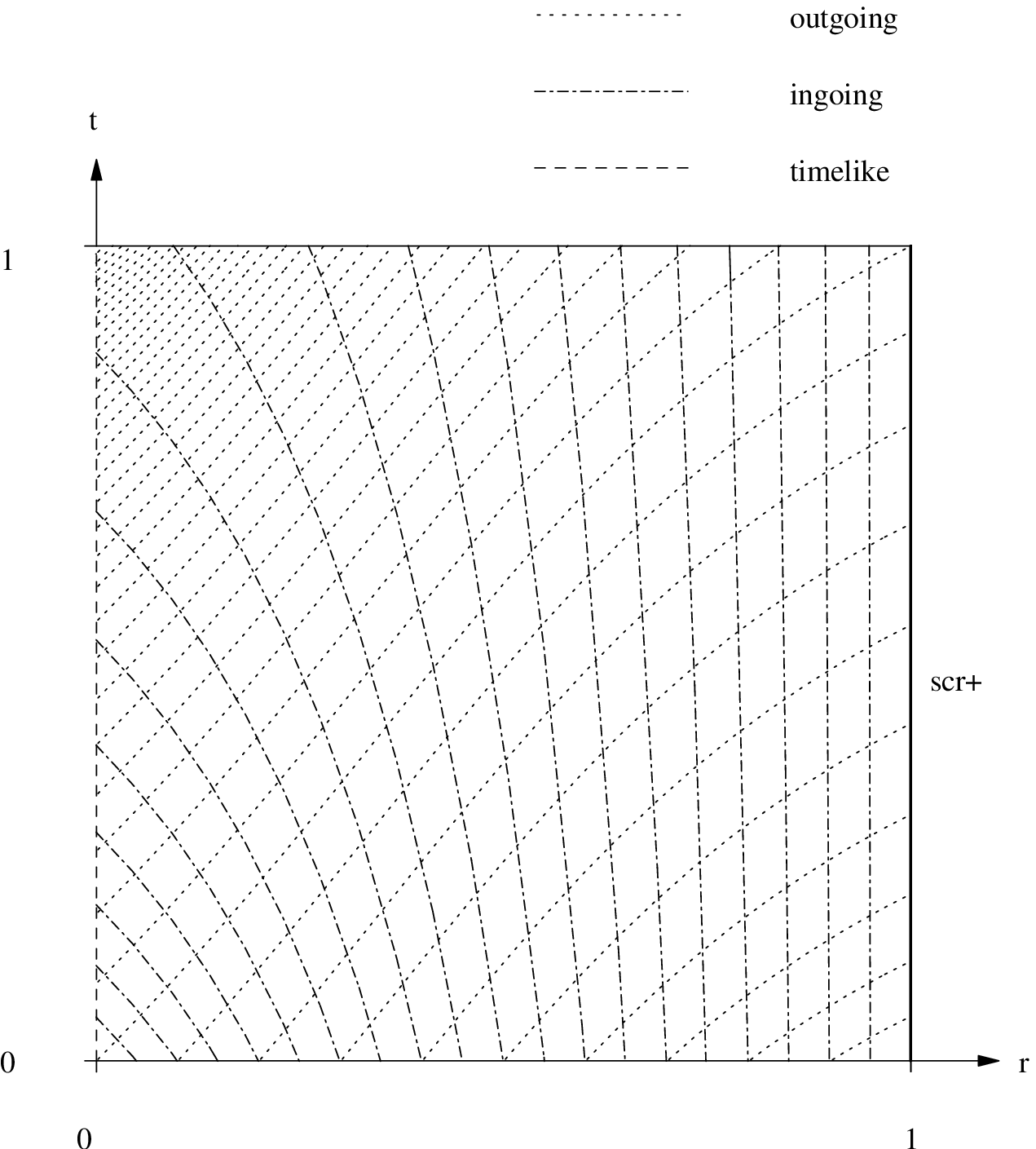}
  \end{minipage}
  \hspace{0.04\linewidth}%
  \begin{minipage}[t]{0.3\textwidth}
    \centering
    \psfrag{0}{$0$}
    \psfrag{1}{$1$}
    \psfrag{t}{$t$}
    \psfrag{r}{$r$}
    \psfrag{scr+}{$\scri^+$}
    \includegraphics[width=\textwidth]{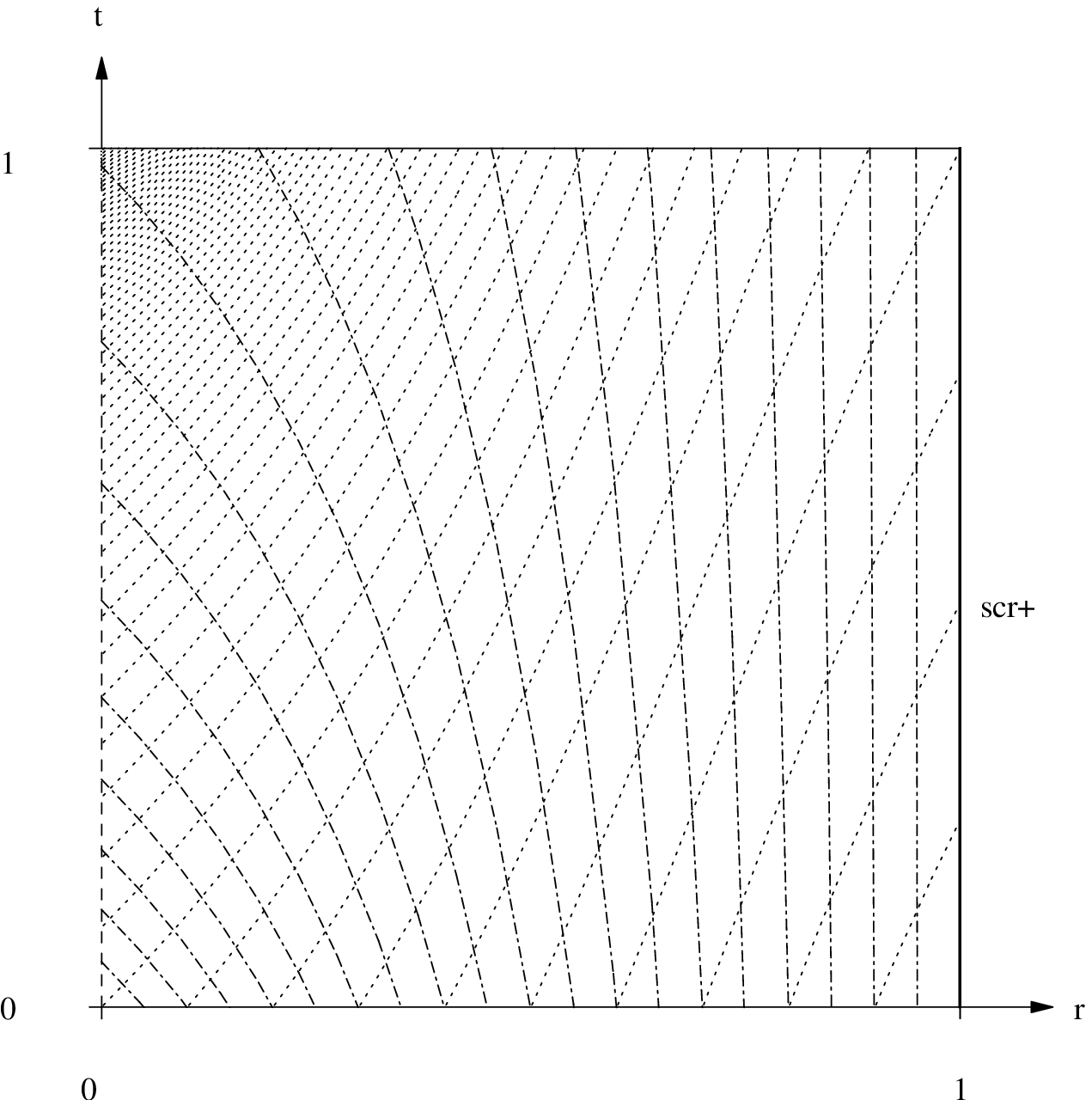}
  \end{minipage}
    \caption{Causal structures on the grid for CMC foliations of Minkowski 
      spacetime with ${\tilde{K}=\{15,3,1.5\}}$. \label{fig:cmc_mink_causal}}
\end{figure}

The causal structure for $\eta$ plotted in Fig.~\ref{fig:cmc_mink_causal} 
is equivalent to the causal structure for $\tilde{\eta}$. 
The radial compactification allows us to include $\scri^+$ in the grid domain 
and the rescaling of the metric results in a regular conformal extension. 
We see that the value of $\tilde{K}$ has a strong effect on the grid speed 
of outgoing characteristics.

\subsection{The extended Schwarzschild spacetime} 
\label{sec:cmc_ss}
To find a representation of the Schwarzschild spacetime 
based on hyperboloidal surfaces, we use the family of spherically 
symmetric CMC surfaces in the extended Schwarzschild spacetime constructed in 
\cite{MalecMurch03} (see Appendix \ref{app:cmc} for the definition of the 
surfaces and \cite{Gentle00} for a numerical study). 
In coordinates adapted to the CMC slicing, 
the standard Schwarzschild metric is obtained in the form
\[ \tilde{g}_s = -\left(1-\frac{2 m}{\tir}\right)\, dt^2 -
\frac{2(J(\tir)-C)}{\widetilde{P}(\tir)}\,dt d\tir+
\frac{\tir^4}{\widetilde{P}^2(\tir)}d\tir^2 + \tir^2\,d\sigma^2, \]
where $C$ is a constant parameter and we have defined
\[ J(\tir) := \frac{\tilde{K}}{3}\,\tir^3, \qquad \widetilde{P}(\tir) := 
\sqrt{(J(\tir)-C)^2+\left(1-\frac{2m}{\tir}\right)\tir^4}. \]
The constant mean curvature of the surface
$\tilde{\mathcal{S}}_t=\{t=\textrm{const}.\}$ is denoted by $\tilde{K}$.
The unit normal to the hypersurfaces $\tilde{\mathcal{S}}_t$ is
\[ n(\tir) = \frac{\tir^2}{\tilde{P}(\tir)}\,\partial_t +
\frac{(J(\tir)-C)}{\tir^2}\,\partial_{\tilde{r}}. \] 
The variables of the 3+1 decomposition read
\[ \tilde{\alpha}=\frac{\tilde{P}(\tir)}{\tir^2},\quad 
\tilde{\beta}^{\tilde{r}} = -\frac{J(\tir)-C}{\tir^2}\,\tilde{\alpha}, \quad 
\tilde{h}_{\tilde{r}} = \frac{1}{\tilde{\alpha}}.\]

Conformal compactification, $g_s=\Omega^2 \tilde{g}_s$, as in (\ref{conf_comp})
with $\Omega=1-r$ and (\ref{null:radial}) results in
\be \label{cmc_ss} g_s = -\left(1-\frac{2 m \,\Omega}{r}\right)\Omega^2 dt^2 - 
\frac{2\left(J(r)-C\,\Omega^3\right)}{P(r)}\,dt dr + \frac{r^4}{P^2(r)}\,dr^2 +
r^2 d\sigma^2, \ee
where $J(r)=\Omega^3 \,J(\tir(r))$ and $P(r):=\Omega^3\,\tilde{P}(\tir(r))$, or
\[J(r)=\frac{\tilde{K}}{3}\,r^3, \quad P(r):=\left(\left(J(r)-C\Omega^3\right)^2
+\left(1-\frac{2 m (1-r)}{r}\right)(1-r)^2 r^4\right)^{\frac{1}{2}}.\]
The variables of the 3+1 decomposition read
\be \label{eq:cmc_ss_3+1} \alpha=\frac{P(r)}{r^2},\quad 
\beta^r = -\frac{J(r)-C\,\Omega^3}{r^2}\,\alpha, \quad 
h_r = \frac{1}{\alpha}.\ee
The mean curvature reads
\[ K = \tilde{K}+\frac{3C\Omega^2}{r^2}.\]
The slicing of the rescaled metric (\ref{cmc_ss}) is not a CMC slicing for 
$C\ne 0$, but the values of $K$ and $\tilde{K}$ are equal at $\scri^+$.

\begin{figure}[t]
  \begin{minipage}[t]{0.34\textwidth}
    \centering
    \psfrag{t}{$t$}
    \psfrag{h}{$\mathcal{H}^{-}$}
    \psfrag{r}{$r$}
    \psfrag{scr+}{$\scri^+$}
    \includegraphics[width=\textwidth]{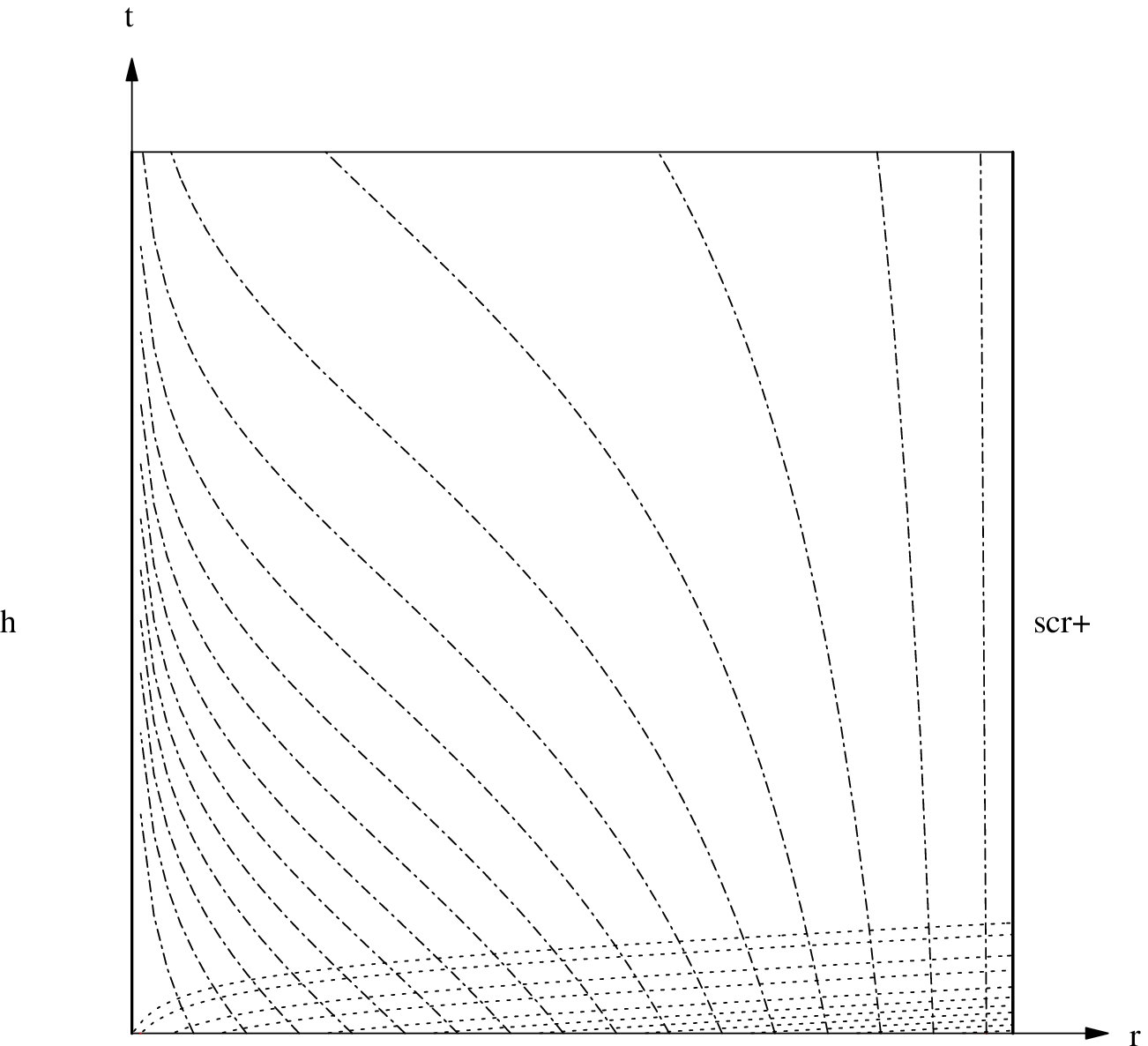}
  \end{minipage}%
  \hspace{0.02\linewidth}%
  \begin{minipage}[t]{0.3\textwidth}
    \centering
    \psfrag{t}{$t$}
    \psfrag{h}{$\mathcal{H}$}
    \psfrag{r}{$r$}
    \psfrag{scr+}{$\scri^+$}
    \includegraphics[width=\textwidth]{figures/causal0.3.eps}
  \end{minipage}
  \hspace{0.02\linewidth}%
  \begin{minipage}[t]{0.3\textwidth}
    \centering
    \psfrag{t}{$t$}
    \psfrag{h}{$\mathcal{H}$}
    \psfrag{r}{$r$}
    \psfrag{scr+}{$\scri^+$}
    \includegraphics[width=\textwidth]{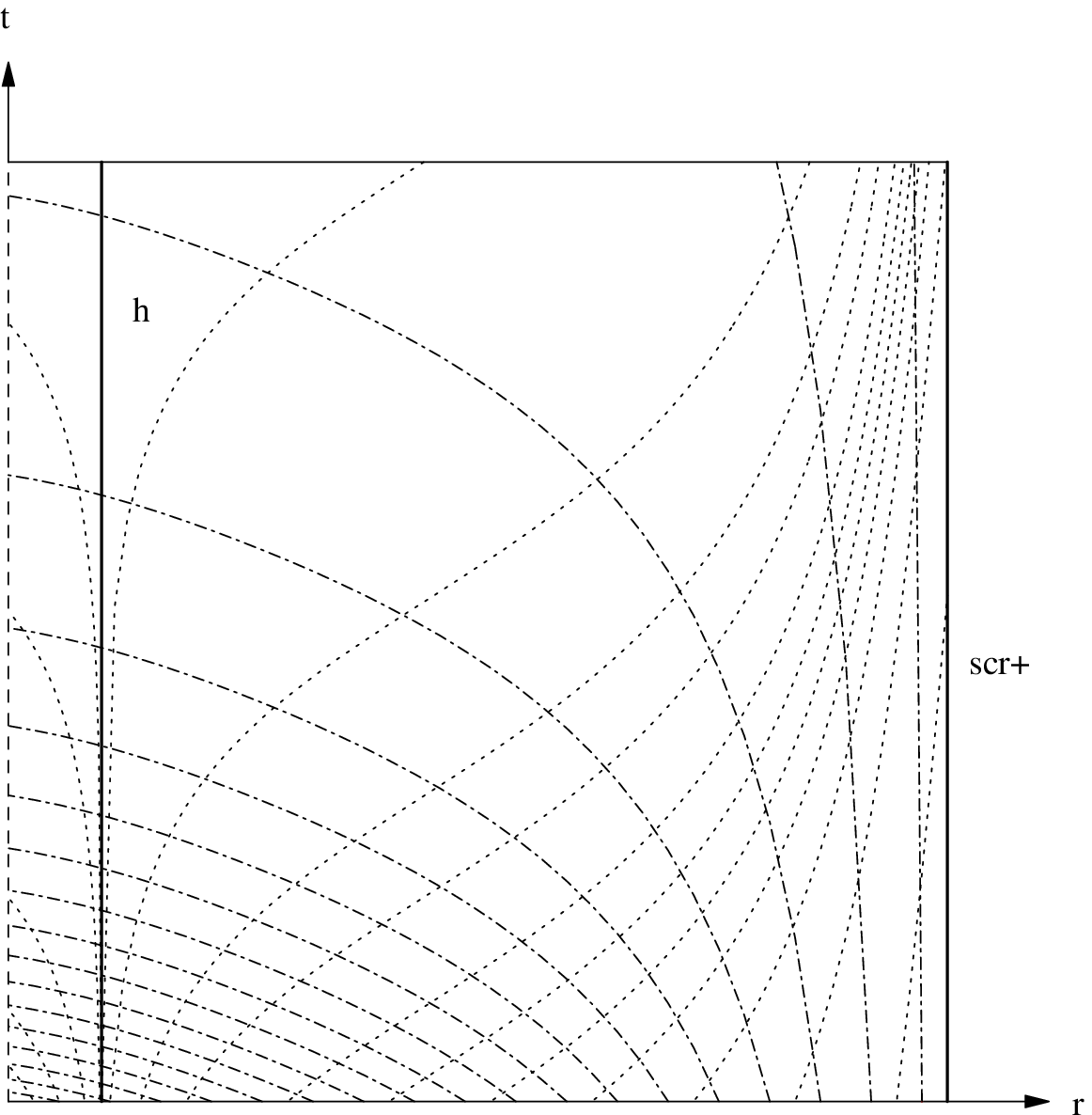}
  \end{minipage}
    \caption{Causal structures on the grid for CMC foliations of the
      Schwarzschild spacetime with $C=2$, $m=1$ and $\tilde{K}=\{1,0.3,0.07\}$.
      See Fig.~\ref{fig:cmc_ss_conformal} for the corresponding Penrose 
      diagrams. \label{fig:cmc_ss_causal}}
\end{figure}

Different choices of $\tilde{K}$ by constant $C$ have a similar effect
on the grid speed of outgoing characteristics as in the case of CMC
slicings of the Minkowski spacetime. Large values of $\tilde{K}$
correspond to high grid velocities of outgoing characteristics as seen
in Fig.~\ref{fig:cmc_ss_causal}.  For low values of $\tilde{K}$, the
causal structure on the grid resembles the causal structure on Cauchy
surfaces (Fig.~\ref{fig:i0_grid2}) in that the grid speed of outgoing
characteristics decreases. In contrast to a Cauchy foliation, this
speed stays larger than zero allowing outgoing characteristics to
leave the spacetime through the outer grid boundary.

\section{Hyperboloidal evolution with a prescribed $\Omega>0$}
\label{sec:hyp_evol}
We have seen in spherical symmetry that we can prescribe the
representation of a conformal factor in terms of a suitably chosen
radial coordinate which can be determined to achieve explicit
$\scri^+$-fixing.  The freedom to prescribe the representation of a
conformal factor might be a special feature of spherical symmetry.  We
will see in the following, however, that one can construct general
solutions to the Einstein equations where the representation of a
conformal factor can be prescribed freely in terms of suitably specified
coordinates.

Consider the transformation behavior of the Ricci tensor under conformal 
rescalings $g=\Omega^2\tilde{g}$ with $\Omega>0$ \cite{Wald84}
\[ 
R_{\mu\nu}[g]=R_{\mu\nu}[\tilde{g}] - \frac{1}{\Omega}\left(2\, \nabla_\mu 
\nabla_\nu \Omega + \Box \Omega\, g_{\mu\nu} \right) + \frac{3}{\Omega^2} 
(\nabla_\lambda \Omega)\, \nabla^\lambda \Omega \, g_{\mu\nu}. \]
The Einstein tensor, $G_{\mu\nu} = R_{\mu\nu} -\frac{1}{2}g_{\mu\nu} R$, transforms
as
\be \label{einst_trafo}
G_{\mu\nu}[g] =  G_{\mu\nu}[\tilde{g}] -\frac{2}{\Omega}\left(\nabla_\mu 
\nabla_\nu \Omega - \Box\Omega\, g_{\mu\nu}\right) - \frac{3}{\Omega^2}
(\nabla_\lambda \Omega) \nabla^\lambda \Omega \, g_{\mu\nu}. \ee
As seen by this relation, the Einstein vacuum field equations for the 
physical metric, $G_{\mu\nu}[\tilde{g}]=0$, are equivalent to a similar system
\be \label{comp_einst} G_{\mu\nu}[g] = T_{\mu\nu}[\Omega]:=
-\frac{2}{\Omega}\left(\nabla_\mu \nabla_\nu 
\Omega - \Box\Omega \, g_{\mu\nu}\right) - \frac{3}{\Omega^2} (\nabla_\lambda 
\Omega) \nabla^\lambda \Omega\, g_{\mu\nu}, \ee
for the conformally rescaled metric $g$. 
We want to formulate a well-posed initial value problem for this system.

As mentioned in \cite{Friedrich02} there are two difficulties when we
try to solve (\ref{comp_einst}) for a conformally compactified metric
$g$.  The first one is caused by the terms involving divisions by
factors of $\Omega$ which are formally singular at $\scri^+$. These
terms attain regular limits at $\scri^+$ if the spacetime admits a
smooth conformal boundary.  In the following we will assume that
hyperboloidal initial data has been chosen whose development admits a
smooth conformal boundary as indicated in \ref{sec:con_comp}. The
question how these limits can be calculated numerically will be
discussed in \ref{ss_num}.

The second difficulty is related to the question of how to determine $\Omega$.
The conformal extension is what we are solving the equations for and the 
conformal factor is related to the asymptotic structure of the solution metric,
therefore it must be determined jointly with the metric. Although it seems 
difficult to control the evolution of the conformal factor we will see that 
the conformal invariance of our evolution system (\ref{comp_einst}) 
combined with certain geometric properties of $\scri^+$ 
allow us to prescribe the conformal and the coordinate gauge 
in a way that $\scri^+$ is kept at a spatial grid coordinate, the representation
of a suitable conformal factor in terms of grid coordinates is known a priori 
and each of the formally singular terms in (\ref{comp_einst}) attains a regular
limit at $\scri^+$. 

In this section we concentrate on the case of a positive,
sufficiently differentiable conformal factor.
The system (\ref{comp_einst}) has the form of Einstein equations with source
terms. In general, such a system must be completed by additional equations 
derived from the Bianchi identity. The divergence freeness condition, 
$\nabla^\mu T_{\mu\nu}=0$, implies equations for the source functions. 
Some notable exceptions to this procedure are the Vaidya metric 
and the null dust \cite{Stephani03}.
In those cases, the divergence freeness condition is satisfied without 
implying additional equations for the source functions so that certain 
functions can be prescribed freely, up to physically reasonable energy 
conditions, in terms of coordinates adapted to the null cones. 
It is interesting to note that these free functions are prescribed in a way 
that keeps the null cone structure invariant which is equivalent to the 
conformal structure. We show now that also in our case 
there are no additional equations required for the conformal factor 
$\Omega$ to satisfy the divergence freeness condition. We calculate
\[ \nabla^\mu T_{\mu\nu}[\Omega] = -\frac{2}{\Omega^2}\, \nabla^\mu
\Omega \, \left(2 \nabla_\mu \nabla_\nu \Omega + \Box\Omega\,
g_{\mu\nu} \right) - \frac{2}{\Omega}\, \left(\Box\nabla_\nu \Omega -
\nabla_\nu \Box \Omega\right) + \]\be \label{contr_tmunu} +
\frac{6}{\Omega^3}\,(\nabla_\nu \Omega) (\nabla_\lambda \Omega)
\nabla^\lambda \Omega. \ee Contracting the commutation relation,
$\nabla_\lambda \nabla_\nu \, \nabla_\rho \Omega - \nabla_\nu
\nabla_\lambda \, \nabla_\rho \Omega = R_{\lambda\nu\rho}^{\ \quad
  \sigma} \nabla_\sigma \Omega$, with $g^{\lambda\rho}$ and exchanging
derivatives we get, $\Box\nabla_\nu \Omega - \nabla_\nu \Box \Omega =
R_\nu^{\ \sigma}\nabla_\sigma \Omega$. Using this, the definition of
the Einstein tensor and the conformal source tensor as given in
(\ref{comp_einst}), we get the identity \be \label{div_tmunu}
\nabla^\mu T_{\mu\nu}[\Omega] = -\frac{2}{\Omega}\,\nabla^\mu \Omega
\, \left(G_{\mu\nu}[g] - T_{\mu\nu}[\Omega]\right).\ee We see that the
divergence freeness condition, $\nabla^\mu T_{\mu\nu}[\Omega] = 0$, is
satisfied identically by the conformally transformed Einstein equations for a
non-vanishing $\Omega$ that is at least $C^3$.  Therefore, $\Omega$
can be regarded as a free function.  We can write some favorable
equation for its evolution consistent with the above calculation or
prescribe it directly in terms of arbitrary spacetime coordinates as
long as $\Omega\ne 0$.
\subsection{A hyperbolic reduction} \label{wp_red}
We showed that the Einstein vacuum field equations for the physical metric 
$\tilde{g}$ are equivalent to the conformally transformed Einstein equations 
(\ref{comp_einst}) for the rescaled metric $g=\Omega^2 \tilde{g}$ 
with a non-vanishing, sufficiently differentiable conformal factor 
$\Omega\ne 0$ written as a function of some yet unspecified coordinate system. 
However, this is not enough to work numerically with the system 
(\ref{comp_einst}). 
We also would like to see that we can {\it find} solutions to 
(\ref{comp_einst}). To show this we use the hyperbolic reduction technique 
\cite{Friedrich96}.

The source terms in (\ref{comp_einst}) involve second order
derivatives of the conformal factor which translate to first order
derivatives of the rescaled metric via Christoffel symbols,
$\nabla_\mu \nabla_\nu \Omega = \partial_\mu \partial_\nu \Omega -
\Gamma_{\mu\nu}^{\lambda} \partial_\lambda\, \Omega$.  In a fully
first order reduction of the Einstein equations, these Christoffel
symbols can change the principal part. In the general wave gauge
reduction we discuss below, however, the principal part does not
change and the argument for well-posedness from
\cite{Friedrich:2000qv} can be directly used with a minor
modification.

Regarding the Ricci tensor $R^{\mu\nu}$ as a differential operator acting 
on the physical metric $\tilde{g}$, we can write the Einstein vacuum field 
equations in a local coordinate system $\{\tilde{x}^\mu\}_{\mu=0,1,2,3}$ as
\be \label{einst_vac} R^{\mu\nu}[\tilde{g}]=\frac{1}{2}\tilde{g}^{\lambda\rho}
\partial_\lambda\partial_\rho \tilde{g}^{\mu\nu}+ 
\tilde{\nabla}^{(\mu}\tilde{\Gamma}^{\nu)} - 
\tilde{g}^{\lambda\rho}\tilde{g}^{\sigma\tau}\,
\tilde{\Gamma}^\mu_{\lambda\sigma}\tilde{\Gamma}^\nu_{\rho\tau}=0,\ee
where we have defined the contracted Christoffel symbols
$\tilde{\Gamma}^\mu:=\tilde{g}^{\sigma\tau}\tilde{\Gamma}_{\sigma\tau}^\mu =
-\tilde{\Box}_{\tilde{g}} \tilde{x}^\mu$ and set 
$\tilde{\nabla}^\mu\tilde{\Gamma}^\nu=
\tilde{g}^{\mu\rho}(\partial_\rho\tilde{\Gamma}^\nu+
\tilde{\Gamma}_{\rho\lambda}^\nu \tilde{\Gamma}^\lambda)$.
The principal part of the operator $R^{\mu\nu}$ is of no known type. It was
recognized by Choquet-Bruhat that one can always choose a wave gauge 
(historically referred to as harmonic gauge), 
at least locally, so that the contracted Christoffel symbols vanish, 
$\tilde{\Gamma}^\mu=-\tilde{\Box}_{\tilde{g}} \tilde{x}^\mu=0$, and the Einstein 
vacuum equations (\ref{einst_vac}), reduce to a quasi-linear system of wave 
equations. This reduction technique led to the first local existence result in 
general relativity \cite{Choquet52}.

The reduction based on the wave gauge was generalized to arbitrary coordinate 
systems by Friedrich with the introduction of gauge source functions 
\cite{Friedrich85}. In the general wave gauge, 
the coordinates are constructed as solutions to an initial value problem for 
the semi-linear system of wave equations $\tilde{\Box}_{\tilde{g}}\tilde{x}^\mu=
-\tilde{\Gamma}^\mu=-F^\mu$ with 
prescribed functions $F^\mu(\tilde{x}, \tilde{g})$ that can depend on the 
coordinates and the metric. These functions act as source functions for the 
coordinate gauge, hence the name gauge source functions. Note that the general
wave gauge, in contrast to the wave gauge described above, is not a specific 
choice of coordinates but a way to prescribe general coordinates in an initial 
value formulation. 

The reduced system to (\ref{comp_einst}) is obtained by replacing 
the contracted Christoffel symbols with the gauge source functions $F^\mu$. 
The resulting system becomes a quasi-linear system of wave equations 
for the metric components which can be written as
\be\label{reduced_einst} G^{\mu\nu}[g] = T^{\mu\nu}[\Omega] + 
\nabla^{(\mu} C^{\nu)} - \frac{1}{2} (\nabla_\lambda C^\lambda) g^{\mu\nu}, \ee
where $C^\mu=\Gamma^\mu-F^\mu$ are called the constraint fields. This
is a coupled system of quasi-linear wave equations for the unknown
$g^{\mu\nu}$ (\ref{eq:written}).
We need to study the Cauchy problem for this system.
I will just mention certain aspects that play a role in later considerations or
that are different from the detailed discussion in \cite{Friedrich:2000qv}. 

The Cauchy data on an initial hypersurface
$\mathcal{S}\equiv\{x^0=0\}$ consists of $g^{\mu\nu}|_{\mathcal{S}}$
and $\partial_0 g^{\mu\nu}|_{\mathcal{S}}$.  Assume we are given on
$\mathcal{S}$ a Riemannian metric $h_{\alpha\beta}$ and a symmetric
tensor $K_{\alpha\beta}$ as a solution to the Einstein constraint
equations and we have chosen gauge source functions
$F^\mu(x^\lambda)$.  We further choose four functions on $\mathcal{S}$
that correspond to initial data for the lapse function $\alpha>0$ and
the three components of the shift vector $\beta^\delta$. In the
interior, these functions should be chosen such that $\partial_0$ is
timelike which implies $\alpha^2-h_{\gamma\delta}\beta^\gamma
\beta^\delta>0$. We will later allow $\partial_0$ to become null at
the outer boundary (see the discussion leading to
(\ref{eq:lapse_shift})). We obtain the data
$g^{\mu\nu}|_{\mathcal{S}}$ via the decomposition \be\label{eq:deco} g
= g^{\mu\nu} \partial_\mu \partial_\nu =
-\frac{1}{\alpha^2}\partial_0^2 + \frac{2}{\alpha^2}\beta^\gamma
\partial_0 \partial_\gamma + \left(h^{\gamma\delta}-
\frac{\beta^\gamma
  \beta^\delta}{\alpha^2}\right)\partial_\gamma\partial_\delta, \quad
\gamma,\delta=1,2,3.\ee The data $\partial_0g^{\mu\nu}|_{\mathcal{S}}$ 
is determined so that $C^\mu|_{\mathcal{S}}=0$ and
$K_{\alpha\beta}$ is the second fundamental form on $\mathcal{S}$.
Standard theorems guarantee that we can find a unique solution to the
Cauchy problem for the reduced equations (\ref{reduced_einst}) that
depends continuously on the initial data.  Now we need to show that
this solution is also a solution to (\ref{comp_einst}). The solution
spaces of (\ref{reduced_einst}) and (\ref{comp_einst}) are equivalent
if the constraint fields vanish, $C^\mu = 0$. We can derive a system
of partial differential equations for the constraint fields by taking
the divergence of our reduced system (\ref{reduced_einst}).  The
Bianchi identity, $\nabla_\mu G^{\mu\nu}=0$, and (\ref{div_tmunu})
imply the following subsidiary system for the constraint fields
\be \label{sub_sys} \Box C^\mu + R^{\mu}_{\nu}C^\nu
-\frac{4}{\Omega}\nabla_\nu \Omega \left(\nabla^{(\mu} C^{\nu)} -
\frac{1}{2} (\nabla_\lambda C^\lambda) g^{\mu\nu}\right)=0.\ee This is
a homogeneous, semi-linear system of wave equations for $C^\mu$.  The
Cauchy problem for this system has unique solutions. The initial data
for the evolution equations has been constructed such that the $C^\mu$
vanish on the initial surface, $C^\mu|_{\mathcal{S}}=0$. From the
evolution equations evaluated on $\mathcal{S}$ it follows that also
the first derivatives of the constraint fields vanish on the initial
surface, $\partial_0 C^\mu|_{\mathcal{S}}=0$.  The uniqueness of
solutions to the Cauchy problem for the subsidiary system
(\ref{sub_sys}) then implies that the solution to the reduced system
(\ref{reduced_einst}) satisfies $C^\mu=0$ away from the initial
surface $\mathcal{S}$. Therefore, we can use (\ref{reduced_einst}) in
a numerical calculation to find solutions to (\ref{comp_einst}).

The above argument does not change when we add homogeneous 
combinations of the constraint fields to the reduced system 
(\ref{reduced_einst}). These "constraint adjustment" terms can 
be of the form $A^\mu_{\lambda}(x,g,\partial g)\,C^\lambda$ without changing the
principal part of the reduced system and the uniqueness of the vanishing 
solution to the subsidiary system. Such terms can be useful for changing the 
behavior of the subsidiary system to damp constraint violations 
when the constraints are only satisfied up to an error. 
It is known that non-linearities in the subsidiary
system can lead to catastrophic growth of constraint violations 
\cite{Friedrich:2005cqg}. There are suggestions for constraint adjustment terms
based on linearized studies \cite{Gundlach2005:constraint-damping} and 
numerical experiments with certain solutions \cite{Babiuc05}. A general 
procedure for their choice can hardly be expected.

Note the appearance of a division by the conformal factor in the 
subsidiary system (\ref{sub_sys}). With our current limited understanding of 
the subsidiary system away from the constraint surface, 
it can only be decided by numerical experiments whether this division causes 
difficulties for the propagation of constraint errors for a large class of 
dynamical solutions in regions where $\Omega$ is small.

We have seen that we can prescribe the representation 
of a conformal factor a priori in an initial value problem for the
Einstein equations in $\widetilde{\mathcal{M}}$, 
that is we can set $\Omega=f(x)$ in (\ref{reduced_einst}) 
where $x^\mu$ are coordinates in
$\widetilde{\mathcal{M}}$ and $f$ is a positive function $f(x)>0$, which
is at least three times continuously differentiable.
Such a prescription does not determine the conformal factor 
a priori as a function from the manifold to the 
positive real line. I shall point out how this prescription is to be understood 
to avoid confusion due to my sloppy but common notation. 

In an initial value problem we do not know the manifold a priori. The 
coordinates on the manifold are constructed during the solution process and 
they are determined by initial data as well as by the choice of the gauge 
source functions.
The prescription of a function for a conformal factor determines only 
the representation of a conformal factor in terms of coordinates 
which are yet to be constructed during the solution 
of (\ref{reduced_einst}). Properties of the resulting conformal factor 
will depend on initial data and the choice of the gauge source functions 
for the coordinates, however, we can choose the coordinate representation 
of $\Omega$ to be a convenient one for numerical calculations.

The essential property of (\ref{comp_einst}) that is responsible for this 
feature is its conformal invariance, 
in the sense that if $(\widetilde{\mathcal{M}},g,\Omega)$
is a solution to $G_{\mu\nu}[g] = T_{\mu\nu}[\Omega]$, then
$(\widetilde{\mathcal{M}},\omega^2 g,\omega \Omega)$ with a 
sufficiently differentiable positive function $\omega$ is a solution to 
$G_{\mu\nu}[\omega^2 g] = T_{\mu\nu}[\omega\Omega]$. 
The system (\ref{comp_einst}) determines the conformal class of the metric $g$ 
in contrast to the vacuum Einstein equations which determine the isometry class
of $\tilde{g}$. In the spirit of Weyl, an arbitrary choice of a point-dependent 
unit of measurement, i.e.~an arbitrary conformal gauge, is allowed. 
By prescribing  a coordinate representation for a conformal factor, we do not
fix the unit of measurement, but its dependence on coordinates. The conformal
factor is then constructed jointly with the metric, the manifold and the 
coordinates.

\section{Including null infinity}
\label{sec:gauge}
The freedom to prescribe a convenient representation for the conformal
factor in terms of arbitrary coordinates is a feature valid for
$\Omega\ne 0$ where no invariant requirements are to be fulfilled by
the conformal factor. However, our motivation to study conformal
rescalings was to include $\scri^+$ in the computational domain. To achieve
this, we need to allow \mbox{$\Omega=0$} at the outer boundary of the future
domain of dependence of an initial hyperboloidal hypersurface
$\mathcal{S}$ while keeping $\Omega> 0$ in the physical manifold
$\widetilde{\mathcal{M}}$. Of course, this can only be done, if at
all, when $(\widetilde{\mathcal{M}},\tilde{g})$ admits a sufficiently
regular conformal boundary.

When $\Omega$ is a prescribed function in the above sense, the
requirement that $\{\Omega=0\}$ shall correspond to $\scri^+$ implies
certain conditions on the coordinates.  The choice of the conformal gauge is
then coupled to the choice of the coordinate gauge. In this section,
we see how a suitable coupling can be achieved.
\subsection{The preferred conformal gauge at $\scri^+$}\label{sec:pref_gauge}
We assume that the Einstein vacuum field equations 
with vanishing cosmological constant are satisfied 
in a neighborhood of $\scri^+$. We restrict ourselves to the case 
where matter fields have spatially compact support, 
although a sufficiently strong fall-off behavior of matter fields 
can also be regarded as compatible with our discussion. 
As a consequence of the Einstein vacuum field equations, 
$\scri^+$ is a shear-free null surface as discussed below 
(cf.~\cite{Penrose65}). To find an appropriate conformal gauge for our 
calculations, we will use this property of $\scri^+$.

Consider the transformation behavior of the Einstein tensor under conformal 
rescalings. Multiplying (\ref{einst_trafo}) with $\Omega^2$ and 
evaluating it along $\scri^+$ where $\Omega=0$ we see that
$g^{\lambda\rho}\nabla_\lambda\Omega\nabla_\rho\Omega|_{\scri^+}=0$. This shows 
together with $d\Omega|_{\scri^+}\ne 0$
that $\scri^+$ is a null surface. 

We multiply (\ref{einst_trafo}) with $\Omega$ and take its trace-free part 
along $\scri^+$ where $\Omega=0$ to get
\be \label{shear_free} 
\left(\nabla_\mu \nabla_\nu \Omega-\frac{1}{4} g_{\mu\nu}\, 
\Box \Omega\right)\Big|_{\scri^+} = 0, \ee
The relation above is independent of the conformal gauge because we 
derived it from the transformation behavior of the Einstein 
tensor (\ref{einst_trafo}) under some conformal rescaling. Another way to see 
this independence is to consider the transformation behavior of 
(\ref{shear_free}) under a further conformal rescaling of the metric given by
\be\label{con_res} g'=\omega^2 g,\quad \Omega'=\omega \Omega,\quad \omega>0 \ 
\textrm{on}\ \mathcal{M}.\ee
Using (\ref{conf_trafo}) with $\nabla'=\nabla + S(\omega^{-1}d\omega)$, we 
derive the following transformation behavior for the second covariant 
derivative of the conformal factor along $\scri^+$
\[\nabla'_\mu \nabla'_\nu\Omega'|_{\scri^+} = \omega \nabla_\mu \nabla_\nu \Omega+
g_{\mu\nu}\, \Omega^\lambda \omega_\lambda,\]
where we have set $\Omega_\lambda=\nabla_\lambda \Omega$ and 
$\omega_\lambda=\nabla_\lambda \omega$. The trace of the above relation reads
\be\label{box_trafo}\Box' \Omega'|_{\scri^+}= \frac{1}{\omega^2}
\left(\omega\Box\Omega+ 4\, \Omega^\lambda \omega_\lambda \right). \ee
We get
\[\left(\nabla'_\mu \nabla'_\nu \Omega' - \frac{1}{4}g'_{\mu\nu}\,
\Box' \Omega'\right)\Big|_{\scri^+} =\omega\,\left(\nabla_\mu \nabla_\nu \Omega - 
\frac{1}{4}g_{\mu\nu}\,\Box \Omega\right)\Big|_{\scri^+}=0. \]

The relation (\ref{shear_free}) can be interpreted as implying shear-freeness 
of $\scri^+$. To see this, we introduce in a neighborhood of $\scri^+$ a null 
vector field $l^\mu$ that satisfies on $\scri^+$ the relation 
$l^\mu = \nabla^\mu \Omega$. We complete $l^\mu$ to a Newman-Penrose complex 
tetrad $(l,k,m,\bar{m})$ satisfying
\begin{eqnarray}\label{eq:np_tetrad}
g(l,l)=g(k,k)=g(m,m)=g(\bar{m},\bar{m})=0, \qquad g(l,k)=-g(m,\bar{m})=-1,
\nonumber\\ g(l,m)=g(k,m)=g(l,\bar{m})=g(k,\bar{m})= 0.\end{eqnarray}
In \cite{Newman62a}, Newman and Penrose introduce twelve complex functions 
called spin coefficients. We are interested in two of them defined by
$\sigma := m^\mu m^\nu \nabla_\mu l_\nu$ and 
$\rho := m^\mu \bar{m}^\nu \nabla_\mu l_\nu$. As discussed in \cite{Newman62a}, 
when $l^\mu$ is tangent to an affinely parametrized null geodesic, 
$\sigma$ can be interpreted as the complex shear of the null geodesic 
congruence given by $l^\mu$ and the expansion of the congruence is 
characterized by $\rho$. 
Evaluating $\sigma$ and $\rho$ at $\scri^+$ and using (\ref{shear_free}) we see
\[\sigma|_{\scri^+} = \frac{1}{4}\,m^\mu m^\nu g_{\mu\nu} \Box \Omega = 0,\qquad 
\rho|_{\scri^+} =\frac{1}{4}\,m^\mu \bar{m}^\nu g_{\mu\nu} \Box \Omega = 
\frac{1}{4}\Box \Omega. \]
In our case, the null generators of $\scri^+$ are not necessarily
geodesic, i.e.~do not satisfy in general 
$l^\lambda\nabla_\lambda l^\mu=0$ on $\scri^+$. However, under a
rescaling of $l^\mu$ given by $(l')^\mu= \theta l^\mu$ with a positive function 
$\theta$, the spin coefficient $\sigma$ transforms as $\sigma'=\theta \sigma$, 
so the vanishing of $\sigma$ is invariant under a rescaling of $l^\mu$ which we
can use to make $l^\mu$ geodesic. We therefore conclude that $\scri^+$ is a 
shear free surface.

While the vanishing of $\sigma$ and the shear freeness of $\scri^+$ 
is valid in any conformal gauge, the vanishing of the expansion of $\scri^+$ 
characterized by $\Box\Omega|_{\scri^+}$ depends 
on the conformal rescaling as seen from the transformation behavior 
(\ref{box_trafo}).
The relation (\ref{box_trafo}) implies that given $\Omega$ whose 
expansion along $\scri^+$ does not vanish, we can always 
find a conformal rescaling (\ref{con_res}) such that $\Box'\Omega'|_{\scri^+}=0$.
To construct this conformal gauge in a given conformal extension, 
we need to solve an ordinary differential equation for the rescaling $\omega$ 
along the null generators of $\scri^+$ which reads
\[ \Omega^\lambda (\nabla_\lambda \ln\omega)|_{\scri^+}=
-\frac{1}{4}\Box\Omega|_{\scri^+}.\] This equation can always be
solved in a given regular conformal extension and so, one can always
construct the conformal gauge in which $\scri^+$ is manifestly
expansion-free.

We call the conformal gauge in which the expansion of $\scri^+$
vanishes, $\Box\Omega|_{\scri^+}=0$, a \textit{preferred conformal
  gauge}. This gauge is useful for analytic studies because of its
special properties \cite{Penrose84,Stewart}. It is also used in
numerical calculations in the characteristic approach
\cite{Tamburino66}.  A direct consequence of (\ref{shear_free}) in a
preferred conformal gauge is that $\nabla_\mu \nabla_\nu \Omega
|_{\scri^+}=0$. By multiplying (\ref{comp_einst}) with $\Omega$ we see
that in this gauge
\[\lim_{\Omega\to 0}\,\frac{1}{\Omega}\,g^{\lambda\beta}\nabla_\lambda\Omega
\nabla_\beta\Omega=0.\]
So a useful property of this gauge is that each term in the 
conformally compactified Einstein equations (\ref{comp_einst}) attains a 
regular limit at $\scri^+$.

Summarizing, a preferred conformal gauge in which $\scri^+$ is expansion free
satisfies the following relations along $\scri^+$
\be \label{preferred}
\Box\Omega|_{\scri^+} = 0 \quad \Rightarrow \quad 
\nabla_\mu\nabla_\nu\Omega |_{\scri^+}=0 \quad \textrm{and} \quad 
\lim_{\Omega\to 0}\,\frac{1}{\Omega}\,g^{\lambda\beta}\nabla_\lambda\Omega
\nabla_\beta\Omega=0.\ee
\subsection{Coupling the conformal and the coordinate gauge at $\scri^+$}
In a preferred conformal gauge, we adapt the coordinates to $\scri^+$ 
so that $\scri^+$ is at a fixed spatial coordinate location. This
secures that no resolution loss in the physical part of the conformal
extension appears.

Assume a conformal extension $(\mathcal{M},g, \Omega)$ with a regular $\scri^+$
has been given in a preferred conformal gauge. As $d\Omega|_{\scri^+}\ne 0$, 
one can use the conformal factor as a coordinate in a neighborhood of $\scri^+$,
which has the topology $\mathbb{R}\times S^2$.
We can then introduce another coordinate system $\{x^\mu\}=\{t,r,x^A\}$ 
with respect to which $\Omega=1-r$ in a 
neighborhood of $\scri^+$.
Here, $t$ is some time coordinate, 
$r$ is a radial coordinate and $x^A$ are angular coordinates. 
This essentially corresponds to using the
conformal factor as a coordinate in a neighborhood of $\scri^+$ which now 
corresponds to the coordinate surface $\{r=1\}$. 

We use a specific representation of the conformal factor 
in terms of coordinates, $\Omega=1-r$, which couples the conformal and the
coordinate gauge so that geometric properties of $\scri^+$ translate into 
coordinate conditions. The relation
$\Omega=1-r$ satisfies $d\Omega|_{\scri^+}\ne 0$. The requirement that 
$\Omega$ is a preferred conformal factor such that the expansion of 
$\scri^+$ vanishes manifestly in $\Omega$ implies
\be \label{expansion}
\Box \Omega|_{\scri^+} = \Gamma^r|_{r=1}=0. \ee
By (\ref{preferred}), we further have as a consequence of (\ref{expansion})
\begin{eqnarray} \label{gauge} 
g^{\mu\nu}\nabla_\mu\Omega\nabla_\nu \Omega|_{\scri^+} =g^{rr}|_{r=1} = 0,\quad
\nabla_\mu \nabla_\nu \Omega|_{\scri^+} = \Gamma^r_{\mu\nu}|_{r=1} = 0, \nonumber \\
0=\lim_{\Omega\to 0}\,\frac{1}{\Omega}g^{\mu\nu}\nabla_\mu\Omega\nabla_\nu\Omega= 
\lim_{r\to1}\,\frac{g^{rr}}{1-r}=-\partial_r g^{rr}|_{r=1}.
\end{eqnarray}
In this representation of the conformal factor, the source terms 
(\ref{comp_einst}) for the conformally transformed Einstein equations 
(\ref{comp_einst}) read
\be \label{pres_tmunu} 
T_{\mu\nu} = - \frac{2}{1-r}(\Gamma_{\mu\nu}^r - g_{\mu\nu}\Gamma^r)
- \frac{3}{(1-r)^{2}}g_{\mu\nu}g^{rr}. \ee
By our assumptions on the spacetime and our choice of coordinates, each term
in $T_{\mu\nu}$ attains a regular limit along $\{r=1\}$.

We can specify $\partial_t$ further such that it is equal to 
$\Omega^\mu=g^{\mu\nu}\nabla_{\nu}\Omega$ along $\scri^+$. Then we get
\be \label{null} \Omega^\mu|_{\scri^+}= -g^{\mu r}|_{r=1} = 
\partial_t^\mu = \delta^\mu_t, \ee
The relation (\ref{null}) implies along $\scri^+$:
$0=g^{rr}=g^{r\mu}g^{r\nu}g_{\mu\nu}=  g_{tt}$ 
which is compatible with the fact that $t$ is a parameter along a null
surface. The vanishing of $\Gamma^r_{\mu\nu}|_{r=1}$ implies then in combination 
with (\ref{null}) the following 10 relations along $\scri^+$
\begin{eqnarray*} 
\partial_t g_{tt} = \partial_r g_{tt}=\partial_A g_{tt}=0, \quad 
\partial_r g_{tr}-\frac{1}{2}\partial_t g_{rr}=0, \\ 
\partial_r g_{At}+\partial_A g_{tr} - \partial_t g_{rA}=0, \quad
\partial_A g_{Bt}+\partial_B g_{At} - \partial_t g_{AB}=0  
\end{eqnarray*}
The coordinate conditions we get are similar but more flexible than those 
in the characteristic approach \cite{Tamburino66}. This is because the 
hyperboloidal foliations allow for more gauge freedom.

\subsection{Constructing preferred coordinates}
We have studied implications (\ref{expansion}, \ref{gauge}) of the specific 
prescription of a preferred conformal factor by $\Omega=1-r$ in terms of a 
radial coordinate $r$. In a numerical calculation of a previously unknown 
spacetime, we would like to know how to \textit{construct} preferred 
coordinates such
that the coordinate surface $\{r=1\}$ has the desired geometric properties of 
$\scri^+$. In general, it is not clear how to control geometric properties of 
given coordinate surfaces during time evolution because the spacetime 
which determines the geometry is constructed together with the coordinates, 
however, in the general wave gauge the condition (\ref{expansion})
can be controlled in a hyperboloidal initial value problem directly by an 
appropriate choice of the radial component $F^r$ of the gauge source functions.

Assume hyperboloidal initial data is given on a spacelike hypersurface 
$\mathcal{S}$ with a two dimensional boundary $\Sigma$ such that its evolution 
admits a regular $\scri^+$ in accordance with the theorems 
of \cite{ACF92,Friedrich83a}.
We can always do a coordinate transformation on $\mathcal{S}$ such that the 
conformal factor has the form $\Omega|_{\mathcal{S}}=1-r$ in a neighborhood of 
$\Sigma$. 

In the hyperboloidal initial value problem for the conformally compactified 
Einstein equations using the general wave gauge reduction 
(\ref{reduced_einst}), the coordinates are constructed during 
time evolution implicitly as solutions to the wave equations
with source terms $\Box_g x^\mu = -\Gamma^\mu = -F^\mu$. 
As seen by this relation and $\Omega=1-r$, the radial gauge source function 
$F^r$ at $\{r=1\}$ determines directly the expansion of $\scri^+$. 
We can make sure that (\ref{expansion}) is satisfied by simply setting 
$F^r|_{r=1}=0$ initially and during time evolution. 
The initial data for lapse and shift should be chosen according to (\ref{null}).
By (\ref{eq:deco}) we set
\be\label{eq:lapse_shift} \alpha|_{\Sigma} =\sqrt{h^{rr}},\qquad 
\beta^{\gamma}|_{\Sigma}=-h^{\gamma r}.\ee
By our choice of initial data and gauge source functions we have a 
preferred conformal gauge and therefore also (\ref{gauge}). 
Note that while it is not clear whether (\ref{null}) can be satisfied by the 
choice of gauge source functions, (\ref{gauge}) is sufficient for the 
conformal source terms given by (\ref{pres_tmunu}) to attain regular limits at 
$\scri^+$ that can in principal be calculated numerically. 
A possibility to deal with these formally singular terms is discussed 
in section \ref{ss_num}.

We note again that, even in a given manifold, a prescription such as 
$\Omega=1-r$ will result in different functions on the manifold
depending on the choice of the coordinate $r$.
It is a remarkable feature of the conformally compactified Einstein equations 
that they allow us to prescribe the conformal and the coordinate gauge in a way
that geometric properties of $\scri^+$ are respected by the coordinates 
in an initial value formulation and the representation of a preferred conformal
factor in terms of coordinates is known a priori.

\subsection{Choice of a conformal factor in the interior}\label{sec:interior}
The notion of an isolated system implies the existence of an interior and 
a far-field region. The suggested method to include null 
infinity in the numerical domain is tailored for the treatment 
of the asymptotic region. The numerical evolution scheme in the interior 
does not need to be changed in this approach as we are free to prescribe 
the representation of a conformal factor in $\widetilde{\mathcal{M}}$.

The interior domain on a spacelike surface will be confined to the
interior of a ball of radius $r_i$ which includes the spatially
compact support of matter sources. Assume that a radial coordinate $r$
has its origin in the center of this ball and choose a large
$r_a>r_i$.  The domain $r_i<r<r_a$ can be regarded as the transition
domain between the interior and the asymptotic region.  The conformal
factor in the interior can be set to unity, $\Omega=1$ for $r\leq
r_i$, so that the conformal source terms in (\ref{comp_einst})
vanish. In the asymptotic region $r\geq r_a$ we can choose
$\Omega=R-r$ with any large constant $R>r_a$ without changing the
analysis in the previous sections made for a neighborhood of $\scri^+$
which is now given by $\{r=R\}$. On $r_i<r<r_a$, any transition
between the different domains can be used that is at least $C^3$.  A
smooth transition is given for example by \[\Omega =
e^{-\frac{(r-r_i)^2}{(r-r_a)^2}}+(R-r)\left(1-e^{-\frac{(r-r_i)^2}{(r-r_a)^2}}\right).\]
There are of course many other possibilities.  The important point is
that we are allowed to prescribe a non-vanishing, sufficiently
differentiable function for the representation of a conformal factor
to solve the Einstein equations in the interior which allows us in
principle to attach a compactified asymptotic region to standard
numerical relativistic calculations. The width of the transition
region or the form of the transition function should be decided upon
by empirical studies.
\section{Numerical tests in spherical symmetry}
\label{ss_num}
A question that we did not discuss so far is how to set up a numerical code
such that the formally singular source terms $T_{\mu\nu}$ 
as given in (\ref{pres_tmunu}) can be dealt with. In this 
section, we study a simple possibility and discuss numerical test 
results on the example of the extended Schwarzschild spacetime as given in
(\ref{cmc_ss}).

The system (\ref{reduced_einst}) is not suitable for analytic studies 
on questions of smoothness properties of $\scri^+$, 
as the equations are formally singular at the set 
whose properties we are interested in. In this section, however, the main 
interest does not lie in questions of differentiability or existence but 
in the numerical construction of spacetimes which admit a sufficiently smooth 
conformal boundary.
To achieve this in the case of spherical symmetry, we numerically
solve the conformally transformed Einstein equations in the general 
wave gauge discussed in previous sections for the extended 
Schwarzschild spacetime. The example we study is clearly very special.

The general wave gauge has been used in various numerical studies
\cite{Babiuc06a,Babiuc05,Garfinkle02,Lindblom05,Pazos06,Pretorius06,Rinne06,
  Rinne07,Scheel06,Szilagyi06,Szilagyi02b,Szilagyi02a}.  There are
many aspects to numerical evolutions that a successful calculation
needs to deal with. The numerical results presented below are not optimal
with respect to the numerical grid boundary treatment or the
constraint propagation. The aim is only to see whether the suggested
method is robust enough to deliver results with a simple choice of
evolution variables and a straightforward treatment of the outer grid
boundary.

\pagebreak
\subsection{The evolution system}
The evolution system can be written as
\be \label{ss_eq} R_{\mu\nu}[g] - \nabla_{(\mu} C_{\nu)} + A_{\mu\nu}=
T_{\mu\nu}[\Omega]-\frac{1}{2}\,g_{\mu\nu} T[\Omega], \ee
with $C_\mu=\Gamma_\mu-F_\mu$. Here, $A_{\mu\nu}$ are constraint adjustment terms 
of the form
\[ A_{\mu\nu} =  C_\lambda \, A^\lambda_{\mu\nu}(x, g,\partial g). \]
These terms vanish when the constraints are satisfied. They change
the propagation properties of constraint errors without effecting the 
principal part of the system. In terms of metric components we get for 
(\ref{ss_eq}) 
\[ \frac{1}{2}g^{\lambda\rho}\partial_\lambda\partial_\rho g_{\mu\nu} -
\nabla_{(\mu} F_{\nu)} - \Gamma^\eta_{\lambda\mu}\Gamma^\sigma_{\rho\nu}
g^{\lambda\rho}g_{\eta\sigma} - 2 \Gamma^\lambda_{\sigma\eta}g^{\sigma\rho}
g_{\lambda(\mu}\Gamma^\eta_{\nu)\rho} - A_{\mu\nu}= \]
\be\label{eq:written}
= \frac{1}{1-r}\left(2\, \Gamma^r_{\mu\nu} + F^r\, g_{\mu\nu} \right) 
- \frac{3}{(1-r)^2} g^{rr} \, g_{\mu\nu}. \ee
For the right hand side, $\Omega=1-r$ has been set explicitly and the 
contracted Christoffel symbol $\Gamma^r$ has been replaced by the gauge source 
function $F^r$. This replacement modifies the subsidiary system (\ref{sub_sys}) 
by a single term without changing the results we discussed in \ref{wp_red}.

The calculations are done for a spherically symmetric metric in adapted 
coordinates $(t,r,\vartheta,\varphi)$. The line element can be written as
\be \label{ss_metric} g = g_{tt}\,dt^2+2 g_{tr}\,dt\,dr + g_{rr}\,dr^2 + 
g_{\vartheta\vartheta}\,d\sigma^2, \ee
where $d\sigma^2=d\vartheta^2+\sin^2\vartheta\,d\varphi^2$. All metric 
components are functions of $(t,r)$ only. 
Alternatively one can write the metric in terms
of the variables of a 3+1 decomposition as
\[ g = (-\alpha^2 +h^2\beta^2)\,dt^2 + 
2 h^2\beta\, dt\,dr + h^2 \,dr^2 + b^2 \,d\sigma^2, \]
where the lapse $\alpha$, the shift $\beta$, and the spatial metric functions 
$h$ and $b$ are functions of the coordinates $(t,r)$ only. They are given 
in terms of the metric components by
\[\alpha = \frac{1}{\sqrt{-g^{tt}}} = \sqrt{\frac{g_{tr}^2}{g_{rr}}-g_{tt}},
\qquad \beta = \frac{g_{tr}}{g_{rr}}, \qquad h = \sqrt{g_{rr}}.\]

The evolution variables can be chosen such that the resulting system of partial
differential equations is first order in time and second order in space. 
This can be done in various ways with different stability
properties. A simple choice is given by $g_{\mu\nu}$ and 
the auxiliary variables $\pi_{\mu\nu}=n^\lambda\partial_\lambda g_{\mu\nu}$ where 
$n^\lambda$ is the unit normal to the surfaces $t=\textrm{const.}$ given by
$n=\frac{1}{\alpha}\,(\partial_t-\beta\,\partial_r)$. We write the
system of evolution equations for the variables $(g_{\mu\nu}, \pi_{\mu\nu})$ as
\begin{eqnarray*}
\partial_t g_{\mu\nu} &=& \beta\,\partial_r g_{\mu\nu} + \alpha\,\pi_{\mu\nu}, \\
\partial_t \pi_{\mu\nu} &=& \beta\,\partial_r \pi_{\mu\nu} + 
\frac{\alpha}{g_{rr}}\,\partial^2_r g_{\mu\nu} + H_{\mu\nu}(g,\partial_r g,\pi,r),
\end{eqnarray*}
where the $H_{\mu\nu}$ are lower order terms. The initial data at $\{t=0\}$ is 
read from the explicit solution (\ref{cmc_ss}) and no boundary data is needed. 

The numerical code is based on finite differencing for the spatial
derivatives and method of lines for time integration.  We discretize
the computational domain by introducing the homogeneous grid
$r_j=(r_0+j\delta r)$.  The radial coordinate location of the
spacelike inner boundary inside the event horizon is denoted by $r_0$,
the grid spacing is denoted by $\delta r$ and $j=1,2,\dots,n$ where
$n+1$ is the total number of nominal grid points.  We define
difference operators $D_{\pm}$ by their action on a grid function
$\phi_j$ via $D_{+}\phi_j= (\phi_{j+1} - \phi_{j})/(\delta r)$ and
$D_{-}\phi_j= (\phi_j - \phi_{j-1})/(\delta r)$. We also define
$D_0=(D_{+}+D_{-})/2$. The spatial derivatives are replaced by fourth
order accurate discrete derivatives \be\label{discrete_der} \partial_r
\phi \to D_0\left(1-\frac{(\delta r)^2}{6}\,D_{+}D_{-}\right) \phi_j,
\qquad \partial^2_r\phi \to D_+ D_- \left(1-\frac{(\delta
  r)^2}{12}\,D_{+}D_{-}\right) \phi_j. \ee The calculation of
derivatives at the boundaries is described in \ref{sec:boundaries}.
To each evolution equation for a variable $\phi$ we add Kreiss-Oliger
type dissipation terms \cite{Kreiss73} given by
\[\phi^{\textrm{diss}}_i=
\epsilon\,\frac{(\delta r)^5}{2^6}\,(D_{+}D_{-})^3\phi_i,\]
where $\epsilon$ is a small dissipation coefficient. 
No dissipation is applied in a neighborhood of the boundaries. The time 
integration is done by a fourth order accurate Runge-Kutta scheme.
\subsection{Choice of the gauge source functions}
For the evolution of Einstein equations starting from some initial surface,
we prescribe the covariant components of the gauge source functions,
$F_\mu$. The form of the metric (\ref{ss_metric}) and the Einstein equations 
(\ref{ss_eq}) imply for the angular components 
\[ F_\vartheta = - \cot \vartheta, \qquad F_\varphi = 0. \]
The gauge source functions $F_t$ and $F_r$ are free. In spherical symmetry 
they can be prescribed in a way that distinguishes the radial coordinate $r$. 
For any metric adapted to spherical symmetry, 
one can find a radial coordinate such that $g_{\vartheta\vartheta}(t,r)=r^2$ 
or $b(t,r)=r$. Such coordinates are called \textit{areal} as in these 
coordinates any $r=\textrm{const.}$ 2-surface on a time slice has the area 
$4\pi r^2$. 

In the ADM formalism, the condition that $r$ is an areal coordinate is 
called the "area locking" condition. It determines the shift by an algebraic 
relation \cite{Kelly01}.
In the general wave gauge, the condition that $r$ is an areal 
coordinate implies a relation between the gauge source functions $F_t$ and 
$F_r$. This relation can be derived from the Einstein equations by setting 
$g_{\vartheta\vartheta}(t,r)=r^2$ or $b(t,r)=r$, 
into the angular component of the equations. 
Without compactification ($T_{\mu\nu}=0$ in (\ref{ss_eq})) and 
constraint adjustment ($A_{\mu\nu}=0$), we get the following relation 
between the gauge source functions $F_t$ and $F_r$
\[ g_{tt}-g_{tr}^2+g_{tt}g_{rr}+ r\, (F_r\, g_{tt} - F_t\, g_{tr}) = 0.\]
This relation can be used to determine for example $F_t(g_{tt},g_{tr},g_{rr},F_r,
r)$ for a given $F_r$. In terms of variables of the 3+1 decomposition we get 
\be \label{f1f0} F_t(\alpha,\beta,h,F_r,r) = 
-\frac{\alpha^2}{\beta}\left(\frac{1}{h^2}-\frac{\beta^2}{\alpha^2}\right)F_r 
-\frac{\alpha^2}{r\beta}\left(\frac{1}{h^2}-\frac{\beta^2}{\alpha^2}+1\right),
\ee
for non-vanishing shift. 
For the conformally transformed Einstein equations one gets the relation
\be \label{areal} F_t = \frac{F_r g_{tt}}{g_{tr}} + \frac{1}{g_{tr}r}
\left( (-g_{tr}^2+g_{tt}g_{rr})\Omega + g_{tt}\right)
-\frac{3 g_{tt}}{g_{tr}\Omega}.\ee
Unfortunately, the areal condition in this form is not suitable for our 
calculations in general, because the shift (or equivalently $g_{tr}$) vanishes 
in the interior. This is because we require 
the shift to point into the computational domain at both boundaries.
For the interior calculation around the excision region of the black hole, 
the shift is required to be positive so that the inner boundary is a spacelike
surface inside the event horizon. At the outer boundary, the shift becomes
negative since $\scri^+$ is an ingoing null surface. The change of sign in the 
shift seems to be a general feature of hyperboloidal foliations that can be
used for excision.

For the numerical tests, we do not use (\ref{areal}). We read the gauge source 
functions from the explicit solution (\ref{cmc_ss}). 
In spherical symmetry, the condition that 
$F^r|_{\scri^+}=0$ corresponds to $F_t|_{\scri^+}=0$. 
A simple calculation shows that this condition is satisfied in our case.
\subsection{Numerical treatment of grid boundaries}\label{sec:boundaries}
We introduce ghost points for the discussion of grid boundaries. To build 
fourth order finite differences via (\ref{discrete_der}) at the boundaries
we need two ghost points. Our method for the numerical treatment of boundaries 
for a first order in time second order in space system relies on 
\cite{Calabrese:2005fp}. Below, we give prescriptions for the grid functions
at the inner ghost points $j=-2,-1$ and at the outer ghost points $j=n+1,n+2$.
For the simulations, we choose $m=1$ and set the boundaries at
\[r_0=\frac{2}{3}-\frac{2}{n}, \qquad r_n = 1 - \frac{1}{n}, \]
which implies a grid spacing of $\delta r=(r_n-r_0)/n=1/(3n)-1/n^2$. 
The event horizon is at $2/3$ and $\scri^+$ is at $r=1$. 
For large $n$, the distance of the inner boundary to the horizon is about 
$6\,(\delta r)$ and the distance of the outer boundary to $\scri^+$ is about 
$3\,(\delta r)$. Note that the coordinate distance decreases with increasing 
resolution.
\subsubsection{Inner Boundary}
The inner boundary is a spacelike surface inside the event horizon of the
black hole as depicted in the Penrose diagram Fig.~\ref{fig:cmc_ss_conformal} 
or in the diagram of the causal structure on the grid 
Fig.~\ref{fig:cmc_ss_causal}.
No boundary data is needed on a spacelike hypersurface so that we can excise
the singularity from the computational domain. The computation outside the 
event horizon should not be disturbed by the excision inside the black hole.

The numerical outflow boundary conditions consist of a fifth order 
extrapolation for $g$ and fourth order extrapolation for $\pi$ 
\cite{Calabrese:2005fp}. We set
\begin{eqnarray*}
(\delta r)^5 D_{+}^5 g_{-1} = 0   &\Rightarrow& 
g_{-1}=5\,g_{0}-10\,g_{1}+10\,g_{2}-5\,g_{3}+g_{4}, \\
(\delta r)^5 D_{+}^5 g_{-2} = 0   &\Rightarrow& 
g_{-2}=5\,g_{-1}-10\,g_{0}+10\,g_{1}-5\,g_{2}+g_{3}, \\
(\delta r)^4 D_{+}^4 \pi_{-1} = 0 &\Rightarrow&  
\pi_{-1}=4\,\pi_{0}-6\,\pi_{1}+4\,\pi_{2}-\pi_{3}, \\
(\delta r)^4 D_{+}^4 \pi_{-2} = 0 &\Rightarrow&  
\pi_{-2}=4\,\pi_{-1}-6\,\pi_{0}+4\,\pi_{1}-\pi_{2}. 
\end{eqnarray*}

\begin{figure}
  \centering
  \psfrag{scr+}{$\scri^+=\{r=1\}$}
  \psfrag{t}{$t$}
  \psfrag{r}{$r$}
  \psfrag{pn-6}{$g_{n-6}$}
  \psfrag{pn-5}{$g_{n-5}$}
  \psfrag{pn-4}{$g_{n-4}$}
  \psfrag{pn-3}{$g_{n-3}$}
  \psfrag{pn-2}{$g_{n-2}$}
  \psfrag{pn-1}{$g_{n-1}$}
  \psfrag{pn}{$g_{n}$}
  \psfrag{pn+1}{$g_{n+1}$}
  \psfrag{pn+2}{$g_{n+2}$}
  \includegraphics[width=0.6\textwidth]{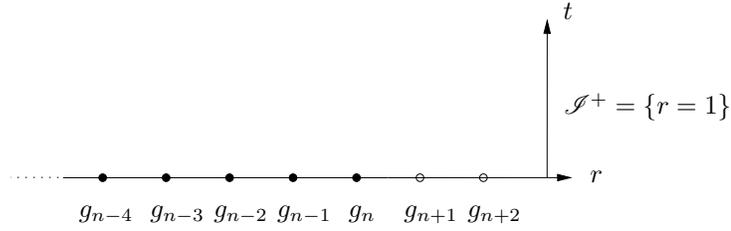}
  \caption{Stencil for extrapolation at the outer boundary.
  \label{fig:extrapol}}
\end{figure}

\subsubsection{Outer Boundary}
There are no characteristics entering the computational domain from the outer
boundary therefore no analytic boundary
data are needed. A numerical treatment, however, is necessary 
to deal with the appearance of divisions 
by factors of $\Omega$ on the right hand side (\ref{pres_tmunu}). 
Two possibilities are:
\begin{itemize}
\item Calculate the limit for $T_{\mu\nu}$ at $\scri^+$ by numerical techniques 
(for example by using a discrete version of the l'Hospital rule).
\item Extrapolate the solution $(g_{\mu\nu}, \pi_{\mu\nu})$ at the outer boundary.
\end{itemize}
I chose the second option, not because it is a clean treatment, but 
because it is a simple method to implement. In the simulations presented below,
$\scri^+$ is about three grid points away from the outer grid boundary 
(Fig.~\ref{fig:extrapol}). The ghost points are filled by the outflow 
conditions as for the inner boundary
\begin{eqnarray*}
(\delta r)^5 D_{-}^5 g_{n+1} = 0   &\Rightarrow& 
g_{n+1}=5\,g_{n}-10\,g_{n-1}+10\,g_{n-2}-5\,g_{n-3}+g_{n-4}, \\
(\delta r)^5 D_{-}^5 g_{n+2} = 0   &\Rightarrow& 
g_{n+2}=5\,g_{n+1}-10\,g_{n}+10\,g_{n-1}-5\,g_{n-2}+g_{n-3}, \\
(\delta r)^4 D_{-}^4 \pi_{n+1} = 0 &\Rightarrow&  
\pi_{n+1}=4\,\pi_{n}-6\,\pi_{n-1}+4\,\pi_{n-2}-\pi_{n-3}, \\
(\delta r)^4 D_{-}^4 \pi_{n+2} = 0 &\Rightarrow&  
\pi_{n+2}=4\,\pi_{n+1}-6\,\pi_{n}+4\,\pi_{n-1}-\pi_{n-2}. 
\end{eqnarray*}
Note that in this method, the outer grid points move along a timelike 
curve and $\scri^+$ is not on a grid point. Clearly, this method is not 
optimal. Extrapolation near $\scri^+$ is not very accurate. An accurate
calculation in a neighborhood of $\scri^+$ is however important for a general 
calculation of gravitational radiation that depends via the rescaled
Weyl tensor on third derivatives of the metric. 
To establish the idea presented in \ref{sec:hyp_evol} and \ref{sec:gauge}, 
a more sophisticated numerical treatment of the formally singular terms 
will be required. 
As mentioned before, our aim in this section is not to construct a general 
numerical method, but to test the suggested idea on a special example 
using simple techniques.
\subsection{Test results}
\begin{figure}[t]
  \begin{minipage}[ht]{0.32\textwidth}
    \centering
    \psfrag{i0}{$i^0$}
    \psfrag{h}{$\mathcal{H}$}
    \psfrag{h-}{$\mathcal{H}^-$}
    \psfrag{scr-}{$\scri^-$}
    \psfrag{scr+}{$\scri^+$}
    \psfrag{r=0}{$\tilde{r}=0$}
    \includegraphics[width=\textwidth]{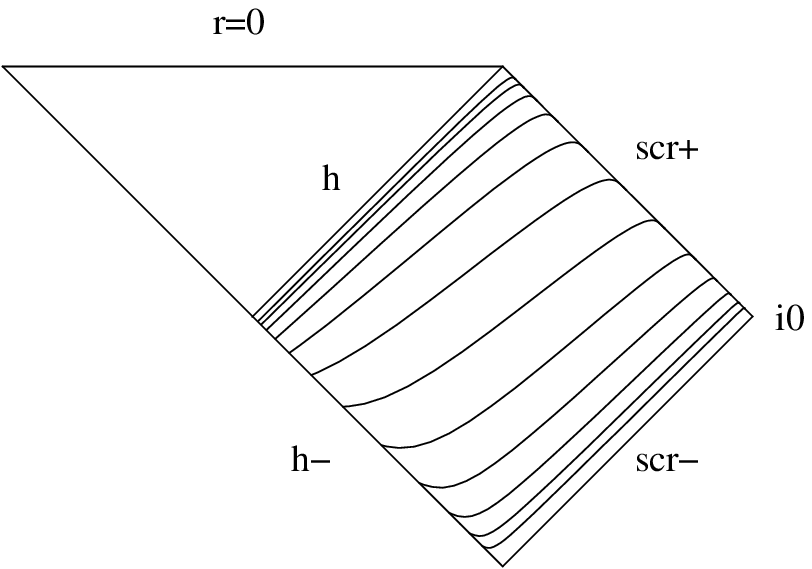}
  \end{minipage}%
  \hspace{0.01\linewidth}%
  \begin{minipage}[ht]{0.32\textwidth}
    \centering
    \psfrag{i0}{$i^0$}
    \psfrag{h}{$\mathcal{H}$}
    \psfrag{scr-}{$\scri^-$}
    \psfrag{scr+}{$\scri^+$}
    \psfrag{r=0}{$\tilde{r}=0$}
    \includegraphics[width=\textwidth]{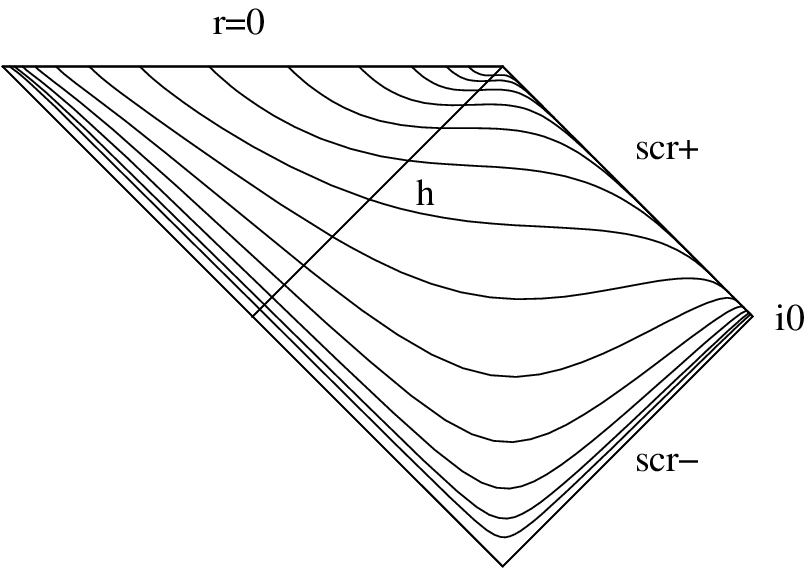}
  \end{minipage}
  \hspace{0.01\linewidth}%
  \begin{minipage}[ht]{0.32\textwidth}
    \centering
    \psfrag{i0}{$i^0$}
    \psfrag{h}{$\mathcal{H}$}
    \psfrag{scr-}{$\scri^-$}
    \psfrag{scr+}{$\scri^+$}
    \psfrag{r=0}{$\tilde{r}=0$}
    \includegraphics[width=\textwidth]{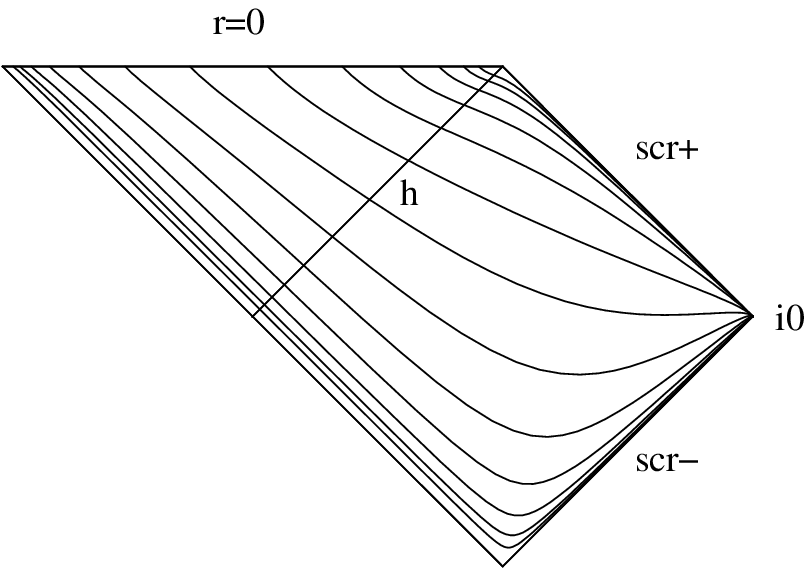}
  \end{minipage}
    \caption{Penrose diagrams of CMC foliations of the Schwarzschild 
      spacetime with  $C=2$, $m=1$ and $\tilde{K}=\{1,0.3,0.07\}$.
      See also Fig.~\ref{fig:cmc_ss_causal}. \label{fig:cmc_ss_conformal}}
\end{figure}

We test our method on the example of the extended Schwarzschild
spacetime in a CMC foliation presented in \ref{sec:cmc_ss}.  Malec and
Murchadha discuss in \cite{MalecMurch03} the asymptotic behavior of
the embedded CMC slices depending on the parameters $\tilde{K}$ and
$C$.  The surfaces $\tilde{K}>0$ reach future null infinity.  Their
behavior in the interior depends on the choice of $C$. Below the
critical value $C=8 \tilde{K} m^3/3$ the surfaces pass the
Schwarzschild throat below the bifurcation sphere.
Fig.~\ref{fig:cmc_ss_conformal} shows foliations of the extended
Schwarzschild spacetime with the same value of $C=2, m=1$ and three
different values of $\tilde{K}=\{1,0.3,0.07\}$. In the case
$\tilde{K}=1$ the surfaces go through the past horizon into the white
hole.  Then the shift vector given in (\ref{eq:cmc_ss_3+1}) by \mbox{$\beta =
-\frac{J(r)-C(1-r)^3}{r^2}\,\alpha$} is negative over the computational
domain and excision can not be expected to work. In this case, no
numerically stable evolution was possible.  Note that in a black hole
spacetime formed by gravitational collapse, no bifurcation sphere and
no past horizon exists.

Below we discuss numerical test results for the cases $\tilde{K}=0.3$ and 
$\tilde{K}=0.07$ where the condition $C > 8 \tilde{K} m^3/3$ is satisfied with 
$C=2$ and $m=1$. As seen in Fig.~\ref{fig:cmc_ss_conformal} these surfaces 
go through the future horizon. The dissipation coefficient has been chosen to 
be $\epsilon = 0.05$ and no constraint adjustment terms have been added.
The solutions have been calculated on the domain $r\in[2/3-2/n,1-1/n]$ where
$n+1$ is the number of grid points. We use $501, 1001$ and $2001$ grid points 
on a homogeneous grid. In terms of the Schwarzschild radius $\tir$, the 
lowest resolution correspond to the domain $\tir\in[1.96m,499m]$ while the 
highest resolution corresponds to $\tir\in[1.99m,1999m]$.

\begin{figure}[!ht]
  \begin{minipage}[hbt]{0.49\textwidth}
    \centering
    \psfrag{K=0.3}{$\tilde{K}=0.3$}
    \includegraphics[width=0.9\textwidth]{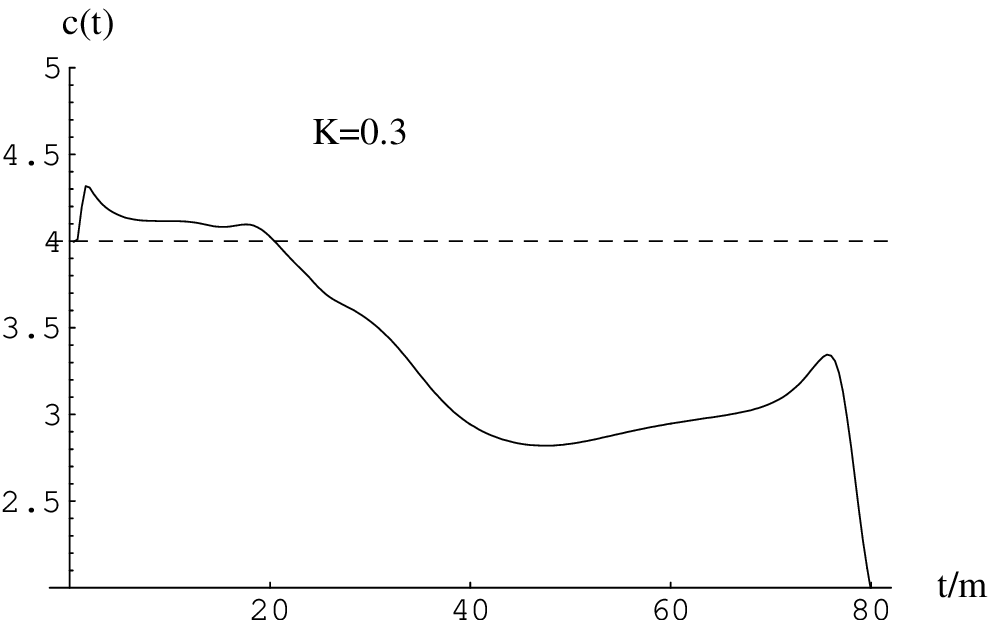}
  \end{minipage}
  \hspace{0.02\linewidth}%
  \begin{minipage}[hbt]{0.49\textwidth}
    \centering
    \psfrag{K=0.07}{$\tilde{K}=0.07$}
    \includegraphics[width=0.9\textwidth]{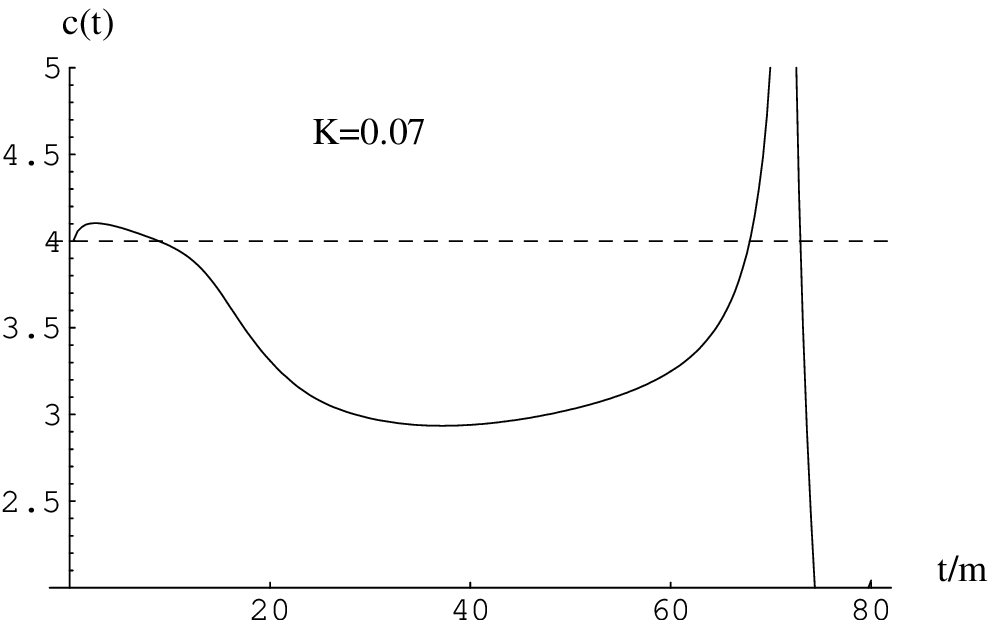}
  \end{minipage}%
  \caption{Convergence factors in $g_{tt}$ with $m=1$ and $C=2$.
    \label{fig:conv_fac}}
\end{figure}

Fig.~\ref{fig:conv_fac} shows the convergence factors $c(t)$ in the $L_2$-norm
for the metric component $g_{tt}$ as a function of time. 
The convergence factor $c(t)$ for a grid function $\phi$ is calculated by 
\[ c(t) = \log_2\,\frac{\norm{\phi^{med}-\phi^{ex}}_{L_2}}
{\norm{\phi^{high}-\phi^{ex}}_{L_2}} \]
where $\phi^{med}$ is the numerical solution in medium resolution, 
$\phi^{high}$ in high resolution and $\phi^{ex}$ is the explicit solution. 
The $L_2$ norm of a real function $\phi\in L^2(\mathcal{S}_t,\mathbb{R})$ 
over the one dimensional computational domain $\mathcal{S}_t$ reads
\be \norm{\phi}_{L_2} = \left(\int_{\mathcal{S}_t}\phi^2\,dr 
\right)^{\frac{1}{2}}.\ee
The discretized version of the $L_2$-norm reads
\be\norm{\phi}_{L_2}=\left( \frac{1}{n+1}\sum_{j=0}^{n} 
\phi_{j}^2\right)^{\frac{1}{2}},\ee
To compare different resolutions for the convergence tests, 
the sum over the grid points in the $L_2$
norm is sampled according to the resolution.

\begin{figure}[!ht]
  \begin{minipage}{0.49\textwidth}
    \centering
    \includegraphics[width=0.97\textwidth]{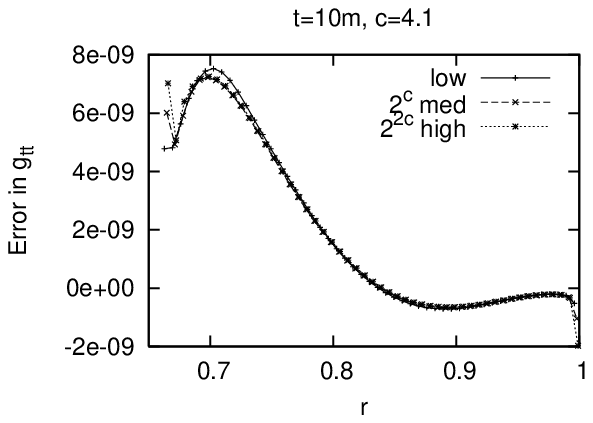}
  \end{minipage}%
  \hspace{0.02\linewidth}%
  \begin{minipage}{0.49\textwidth}
    \centering
    \includegraphics[width=0.97\textwidth]{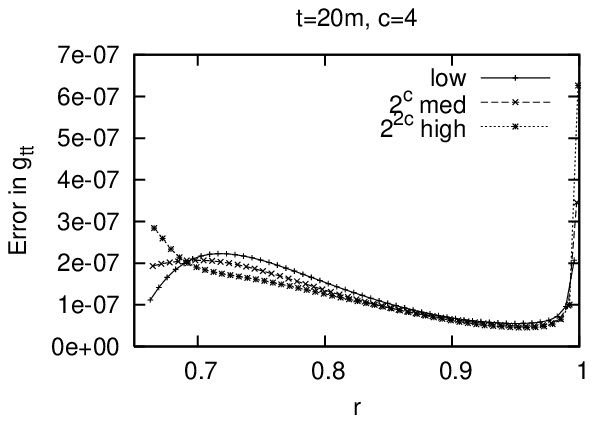}
  \end{minipage}\\[3pt]
  \begin{minipage}{0.49\textwidth}
    \centering
    \includegraphics[width=0.97\textwidth]{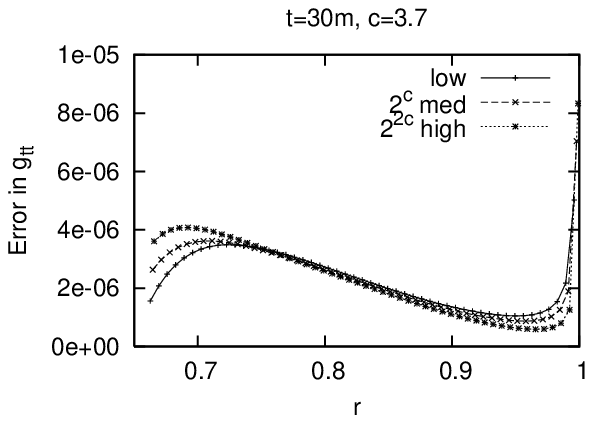}
  \end{minipage}%
  \hspace{0.02\linewidth}%
  \begin{minipage}{0.49\textwidth}
    \centering
    \includegraphics[width=0.97\textwidth]{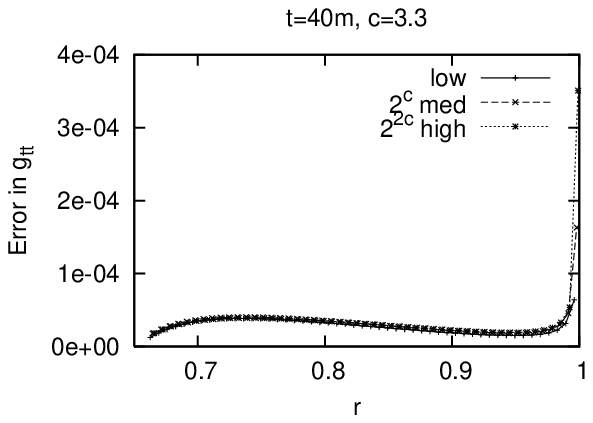}
  \end{minipage}
  \caption{Rescaled error plots for $\tilde{K}=0.3$ \label{fig:conv_k03}}
  \vspace{3mm}%
  \begin{minipage}{0.49\textwidth}
    \centering
    \includegraphics[width=0.97\textwidth]{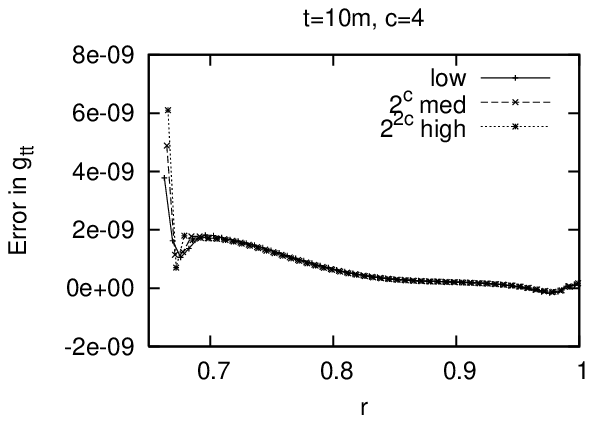}
  \end{minipage}%
  \hspace{0.02\linewidth}%
  \begin{minipage}{0.49\textwidth}
    \centering
    \includegraphics[width=0.97\textwidth]{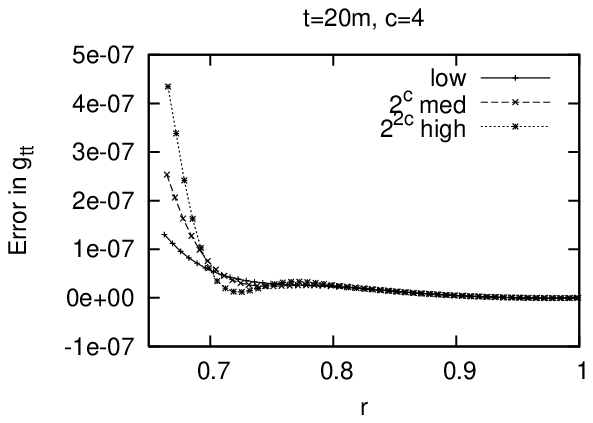}
  \end{minipage}\\[3pt]
  \begin{minipage}{0.49\textwidth}
    \centering
    \includegraphics[width=0.97\textwidth]{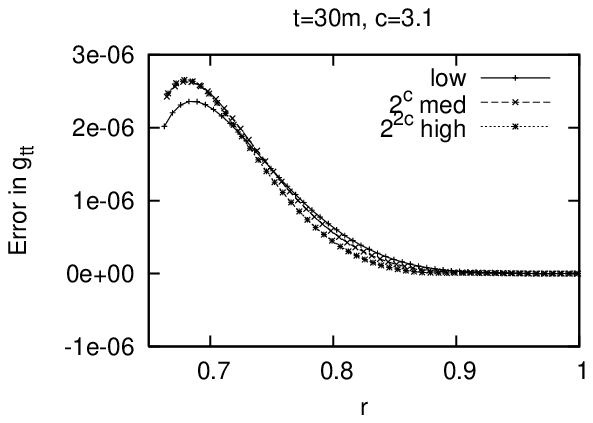}
  \end{minipage}%
  \hspace{0.02\linewidth}%
  \begin{minipage}{0.49\textwidth}
    \centering
    \includegraphics[width=0.97\textwidth]{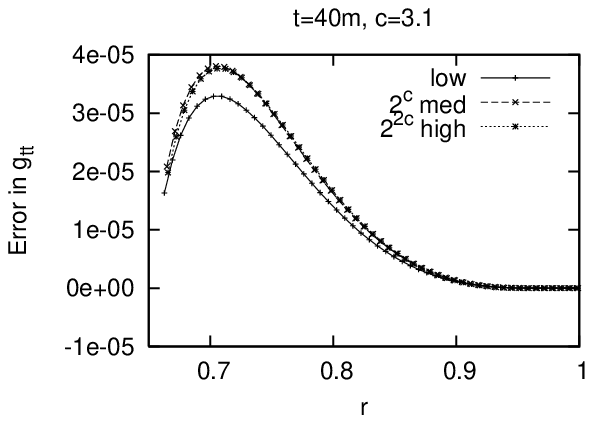}
  \end{minipage}
  \caption{Rescaled error plots for $\tilde{K}=0.07$ \label{fig:conv_k007}}
\end{figure}

We observe in Fig.~\ref{fig:conv_fac} for $\tilde{K}=0.3$ that the
convergence factor drops to roughly 3 after about $40m$.  For
$\tilde{K}=0.07$ the decrease is faster.  A global convergence factor
of $3$ might be expected due to our numerical outer boundary
treatment. After about $80m$ the codes crash. To localize the problem
of the crash, we plot in Fig.~\ref{fig:conv_k03} and
Fig.~\ref{fig:conv_k007} the radial dependence of the error for
$g_{tt}$ at times $t=\{10m,20m,30m,40m\}$.  The aim is to see on which
part of the computational domain the errors become large and where
convergence is lost.

For representing the errors of different resolutions in the same plot, 
we scale them depending on a convergence factor $c$.
For example when the convergence factor is expected to be
4, the error for the solution in medium resolution is scaled by $2^4=16$ and
the error for the solution in high resolution is scaled by $(2^2)^4=256$. 
When the curves are on top of each other, 
we can assume that the overall convergence factor is close to the chosen one
with which the errors have been scaled. This visual test gives us a possibility
to localize deviations from convergence along the grid.

In Fig.~\ref{fig:conv_k007} we see that already at $t=20m$ the convergence 
in the interior is disturbed. 
At $t=30m$ the convergence factor has dropped to about $3.1$ while the lowest 
resolution does not align with the medium and the high resolution. 
This deviation of the low resolution can be seen more clearly at $40m$. 
For $\tilde{K}=0.07$ the largest errors appear in the interior where the code
eventually crashes.

In Fig.~\ref{fig:conv_k03} with $\tilde{K}=0.3$ we see that the decrease in the
convergence factor is slower in accordance with Fig.~\ref{fig:conv_fac} but 
this time the largest errors are due to the outer boundary treatment. One 
observes deviations from convergence in the interior too but the error is 
dominated by the region close to $\scri^+=\{r=1\}$ and the code crashes in this
region.

For small $\tilde{K}$ the numerical errors produced in the interior propagate
outwards slower compared to large $\tilde{K}$ as can also be seen in the plots 
of causal structures Fig.~\ref{fig:cmc_ss_causal}. Therefore the numerical
errors in a neighborhood of $\scri^+$ stay small for small $\tilde{K}$
while the errors in the interior grow until the code crashes in contrast to 
large $\tilde{K}$ which crashes close to $\scri^+$.
We observe that different choices of $\tilde{K}$ have a strong effect on 
numerical errors. A related effect of the foliation has been studied in 
\cite{Frauendiener:2004bj} where it was shown for a certain system 
that the choice of the time slicing influences propagation properties 
of constraint errors. 

Summarizing, we can say that although our numerical setup does not
allow us to do long time evolutions of the extended Schwarzschild
spacetime, a piece of null infinity can be calculated with the method
suggested in \ref{sec:hyp_evol} and \ref{sec:gauge} even with simple
choices of variables and numerical boundary treatment.
\clearpage

\section{Discussion}\label{sec:outlook}
The main content of this chapter has been the development of a
numerical method that allows us to construct solutions to the Einstein
equations including null infinity in the computational domain based on a
general wave gauge with $\scri^+$-fixing coordinates.  The method
employs a suitable coupling of the conformal and the coordinate gauge
to establish expansion freeness of $\scri^+$ via an
appropriate choice of the gauge source functions for the coordinates.
Under our assumptions on initial data and our gauge conditions, 
each formally singular term arising from the conformal
compactification attains a regular limit at $\scri^+$ which needs to
be calculated by numerical techniques.

I presented a numerical test of this method in spherical symmetry in which 
the outer boundary treatment is based on extrapolation near $\scri^+$. 
The code is afflicted by an unbounded growth of numerical errors.
One could see, however, that the treatment of the outer boundary does not lead
to immediate problems so that a piece of null infinity can be calculated.
It is an outstanding question whether one can calculate the formally
singular terms arising from the conformal compactification numerically
in a stable manner for dynamical spacetimes. To achieve this one might
need to use intrinsic properties of the characteristic surface
$\scri^+$ to a larger extent.

As a next step, one should try the method described in \ref{sec:hyp_evol} and 
\ref{sec:gauge} in a setting including radiation using a more sophisticated 
choice of variables and numerical boundary treatment. 
Techniques developed in various groups over many years 
working with the general wave gauge reduction might be useful in that
context \cite{Babiuc05,Lindblom05,Pazos06,Pretorius06,Scheel06,Szilagyi06}. 
One should also study the freedom in prescribing gauge source functions 
further to derive useful gauge conditions which lead to a convenient 
representation of the solution metric along $\scri^+$. 
A numerical finite differencing scheme might be adapted to certain expected 
properties of the solution metric as in the characteristic approach.

As $\scri^+$ has topology $\mathbb{R}\times S^2$, a general numerical
calculation in which $\scri^+$ is fixed to a spatial coordinate
location which corresponds to the outer grid boundary should be able
to handle spherical grid topology. To my knowledge, there are
currently two approaches in numerical relativity that avoid coordinate
singularities on a sphere and use the general wave gauge
\cite{Pazos06,Scheel06}.  Both of these methods seem promising for
trying the suggested idea in a general setting.

In our studies we used a general wave gauge reduction. However, the
idea to solve a hyperboloidal initial value problem for conformally
compactified Einstein equations with a prescribed representation of
the conformal factor is independent of the reduction.  The open
question for other reductions of Einstein equations is whether the
conditions (\ref{expansion}, \ref{gauge}) can be satisfied during time
evolution in a well-posed Cauchy problem.  This question needs to be
studied for each reduction separately.  

The maximal development of hyperboloidal initial data does not yield
the global spacetime.  We do not get access to spatial infinity in a
hyperboloidal initial value problem and in $\scri^+$-fixing
coordinates, also timelike infinity can not be reached.  While one can
argue that these points are not of physical interest and possibly
irrelevant for the comparison of observational data with numerical
calculations, it would be desirable to have access to the global
structure for various reasons.  In the hyperboloidal approach it is
not clear whether the cut of the initial hyperboloidal surface at null
infinity is close to timelike infinity or to spatial infinity. One
would also like to be able to relate asymptotic quantities as mass or
momentum defined at null infinity to corresponding quantities at
spatial infinity.

%% file: i0.tex
\chapter{Spatial Infinity}\label{chapter:i0}
In this chapter we want to solve numerically a Cauchy problem for the
Einstein equations starting from a Cauchy hypersurface including
spatial and null infinity in the numerical domain. This gives us, in
principal, access to the global spacetime solution. We will see that
in spherical symmetry the maximal development of Schwarzschild-Kruskal
initial data given on a Cauchy hypersurface can be calculated. For
more general situations including gravitational radiation we will
focus our interest on the calculation of the detailed structure of
gravitational fields in a neighborhood of spatial infinity including a
piece of null infinity.

We make certain assumptions on initial data that allow us to discuss
the main features of our problem while simplifying the calculations
involved.  We consider asymptotically flat, time symmetric, vacuum
initial data, i.e.~we are given
$(\tilde{\mathcal{S}},\tilde{h}_{\alpha\beta})$, where
$\tilde{\mathcal{S}}$ is a three dimensional manifold with an
asymptotically flat end and $\tilde{h}_{\alpha\beta}$ is a positive
definite, asymptotically flat, Riemannian metric on
$\tilde{\mathcal{S}}$. The data satisfies the vacuum Einstein
constraint equations with $\tilde{K}_{\alpha\beta}=0$.  We mean by
asymptotic flatness for
$(\tilde{\mathcal{S}},\tilde{h}_{\alpha\beta})$ that the complement of
a compact set in $\tilde{\mathcal{S}}$ is diffeomorphic to the
complement of a closed ball in $\mathbb{R}^3$ and in the chart
$\{\tilde{x}^\alpha\}$ given by this diffeomorphism the following
fall-off conditions are required \be\label{eq:as_flat_h}
\tilde{h}_{\alpha\beta}=\left(1+\frac{2m}{\tir}\right)
\delta_{\alpha\beta}+ O_k(\tir^{-(1+\epsilon)}), \quad
\mathrm{as}\quad \tir^2=\delta_{\alpha\beta}\tilde{x}^\alpha
\tilde{x}^\beta\to\infty, \qquad \alpha,\beta=1,2,3, \ee with
$\epsilon>0$, $k\geq 2$ and $m$ is the ADM-mass of the initial data
set. The fall-off conditions above are written with respect to a
coordinate system. To discuss such conditions without relying on
specific coordinate systems one can use conformal techniques along the
lines of Penrose's considerations.  In \cite{Geroch70} Geroch uses
conformal compactification techniques to represent spatial infinity as
a single point $i$ so that asymptotic properties of fields on
$\tilde{\mathcal{S}}$ may be treated as local geometric properties at
$i$.  The construction is similar to the conformal mapping of
$\mathbb{R}^3$ onto the 3-sphere by adding a single point at infinity.
We require that there exists a manifold
$\mathcal{S}=\tilde{\mathcal{S}}\cup\{i\}$ and a conformal factor
$\phi$ with $\phi>0$ on $\tilde{\mathcal{S}}$ satisfying
\begin{itemize}
\item $\bar{h} = \phi^2 \tilde{h}$ is a smooth metric on $\mathcal{S}$,
\item $\phi=0,\ \bar{D}_\alpha \phi = 0,
\ \bar{D}_\alpha \bar{D}_\beta \phi=2\, \bar{h}_{\alpha\beta}$ at $i$,
\end{itemize}
where $\bar{D}$ is the covariant derivative operator defined by $\bar{h}$. 
The described notion of asymptotic flatness is a property of the conformal 
structure in the sense that the conditions are invariant under conformal 
rescalings $\bar{h}\to \omega^2 \bar{h},\ \phi\to\omega \phi$ with positive 
smooth functions $\omega$ satisfying $\omega(i)=1$. In the examples 
that we will study, $\bar{h}$ will not just be smooth but analytic at $i$.

To calculate numerically an entire, asymptotically flat spacetime
including spatial, null and timelike infinity, we might wish to solve
Friedrich's conformally regular field equations presented in
\cite{Friedrich81a}.  The reason why we can not do this is that the
initial data for the conformal field equations blow up at spatial
infinity when represented as a point $i$ for non-vanishing ADM-mass.
Specifically, denote by $r(p)$ the distance of the point
$p\in\mathcal{S}$ from the point $i$ in terms of the metric
$\bar{h}_{\alpha\beta}$ on the Cauchy surface $\mathcal{S}$.  It turns
out as a consequence of the constraint equations implied by the
conformal field equations on $\mathcal{S}$ that the rescaled Weyl
tensor blows up like \cite{Friedrich98b} \be\label{eq:singular}
W^{i}_{\ jkl} = O\left(\frac{1}{r^3}\right) \quad \mathrm{as} \quad
r\to 0 \quad \mathrm{with} \quad m\ne 0. \ee

\begin{figure}[ht]
  \centering
  \begin{minipage}{0.5\textwidth}
    \centering
    \psfrag{t}{{$t$}}
    \psfrag{r}{{$r$}}
    \psfrag{scri+}{{$\scri^+$}}
    \psfrag{i0}{{$i^0\sim i$}}
    \psfrag{S}{{$\mathcal{S}$}}
    \includegraphics[width=\textwidth]{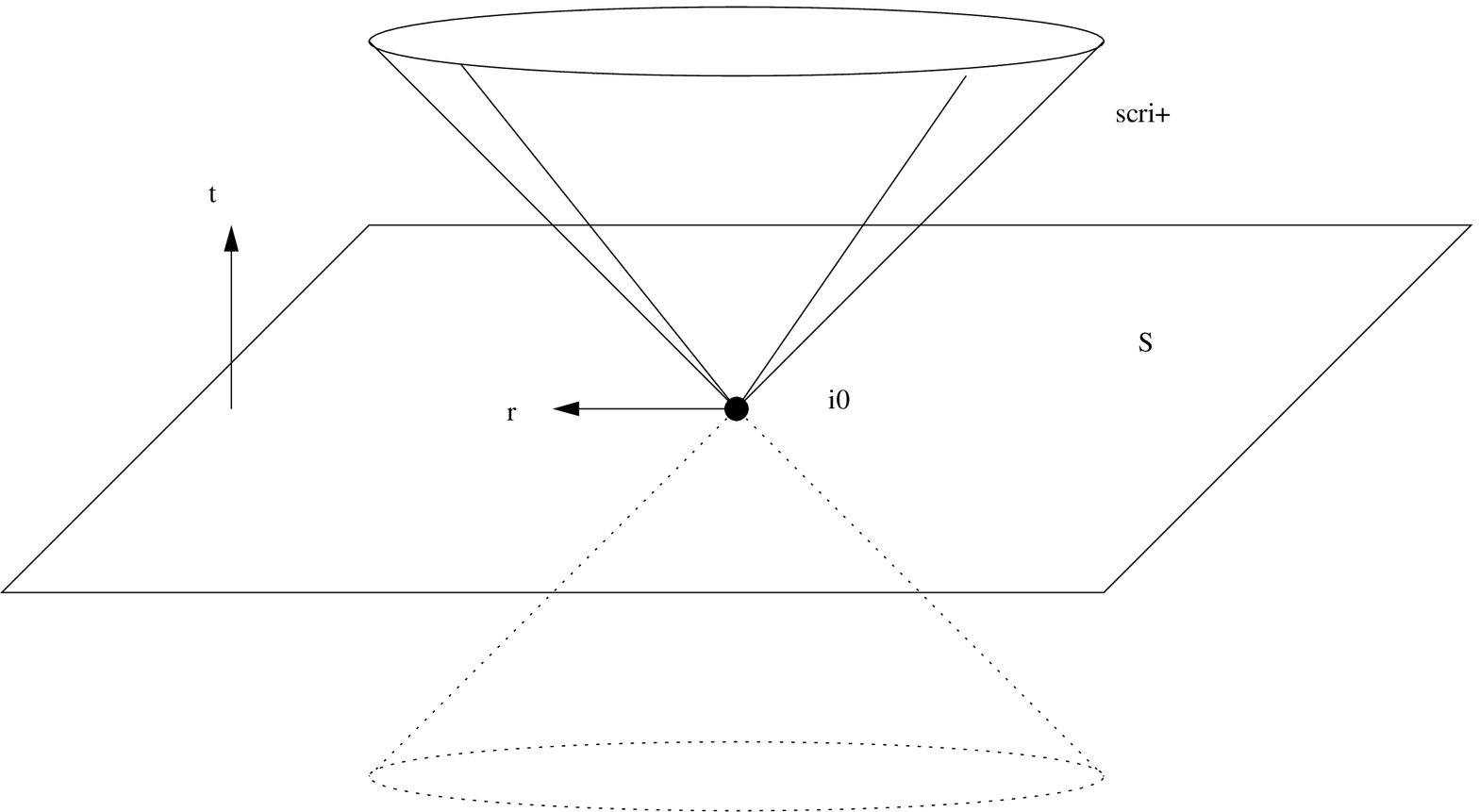}
  \end{minipage}\hfill
  \begin{minipage}{0.4\textwidth}
    \centering
    \psfrag{t}{{$t$}}
    \psfrag{r}{{$r$}}
    \psfrag{scri+}{{$\scri^+$}}
    \psfrag{I0}{{$\mathcal{I}^0$}}
    \psfrag{I+}{{$\mathcal{I}^+$}}
    \psfrag{S}{{$\mathcal{S}$}}
    \includegraphics[width=\textwidth]{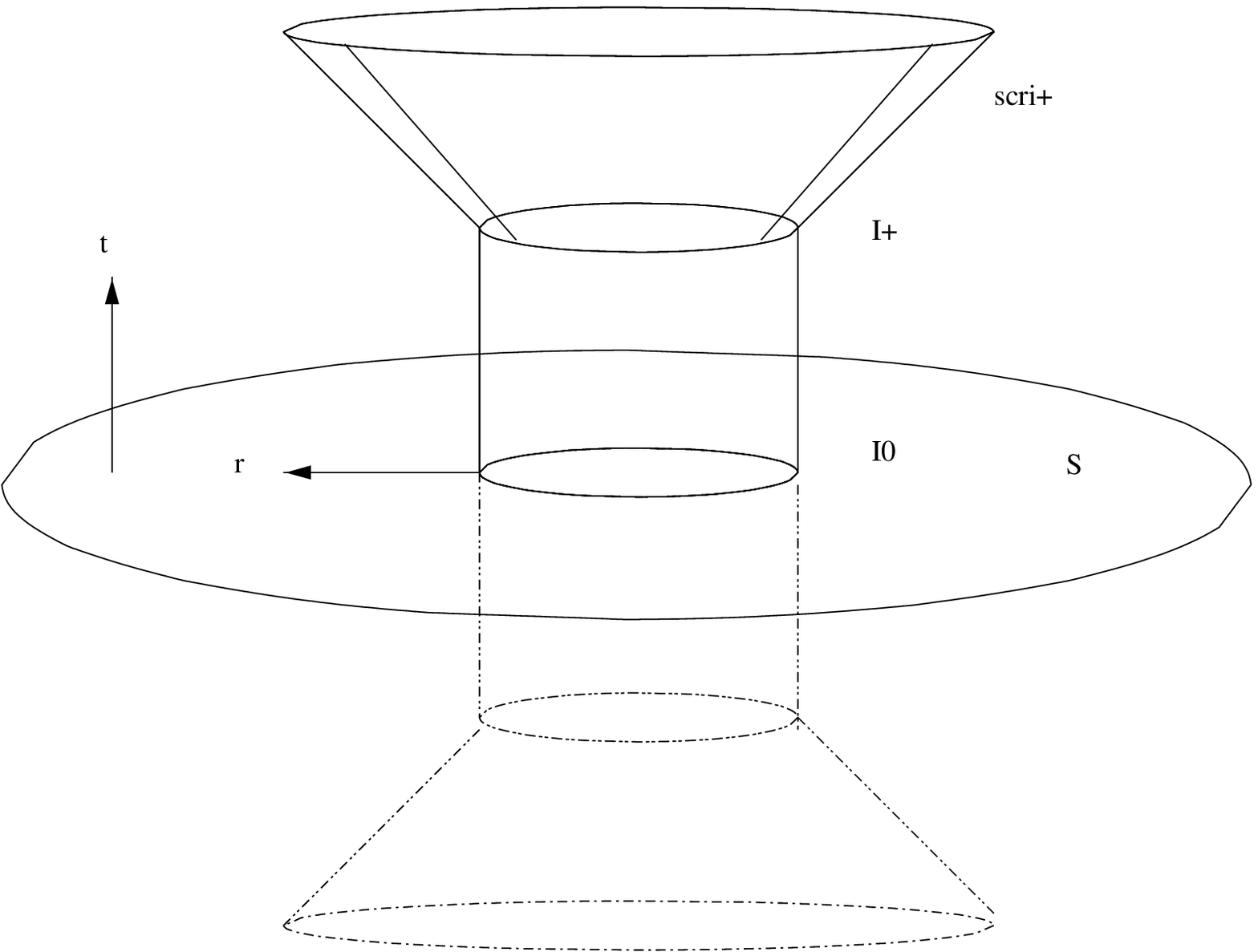}
  \end{minipage}
  \caption{Point compactification and the cylinder at spatial
    infinity.\label{fig:point-cylinder}}
\end{figure}

A regular, finite, initial value problem near spatial infinity could
be formulated by Friedrich based on an extended system of conformal field
equations and a certain conformal gauge near spatial infinity, called the 
conformal Gauss gauge. This gauge allows a representation of spatial infinity 
as a cylinder denoted by $\mathcal{I}$. In this representation, 
the point $i$ is blown up to a sphere $\mathcal{I}^0$ 
(Fig.~\ref{fig:point-cylinder}). The blow-up procedure is
such that a suitable rescaling of fields on $\mathcal{S}$ results in smooth
conformal data up to and beyond $\mathcal{I}^0$. The development of this data 
can be studied further by the evolution equations. A basic difficulty in this 
new representation is the degeneracy of the equations at the sets 
$\mathcal{I}^\pm$ where null infinity meets spatial infinity.

The construction of the cylinder at spatial infinity is a delicate
procedure which relies on a deliberate choice of the coordinate, the
conformal and the frame gauge in accordance with a suitable notion of
asymptotic flatness. For further details of this construction that we
do not discuss in this chapter, the reader is referred to the original
paper \cite{Friedrich98} and the review articles
\cite{Friedrich02,Friedrich04}. Further applications of the cylinder
at spatial infinity in analytic work can be found for example in
\cite{ValienteKroon:2003ux,ValienteKroon06,ValienteKroon07}.

We are concerned in this chapter with a numerical implementation of the 
extended system of conformal field equations with the cylinder at spatial 
infinity. First we discuss the numerical construction of a conformal 
Gauss gauge in the simple examples of Schwarzschild and Kerr spacetimes. 
After discussing some aspects of the reduced general conformal field equations,
we integrate the equations in spherical symmetry for the Schwarzschild-Kruskal 
spacetime. To include radiation into our discussion in a simple setting 
in which the problem (\ref{eq:singular}) does not appear, we calculate 
initial data with vanishing ADM-mass but a non-vanishing radiation 
field along $\scri^+$. The numerical development of this data is discussed
in a neighborhood of spatial infinity represented both as a point and as a 
cylinder. The numerical implementation of the equations with the cylinder is 
such that the code can also be used to study physically more interesting 
spacetimes with non-vanishing ADM-mass. Such studies are left for future work.

In the following, we will use a frame formalism. 
Latin letters are used for frame indices with $i,j,k,\ldots=0,1,2,3$ 
and $a,b,c,\ldots=1,2,3$. Greek letters are used for coordinate indices with 
$\mu,\nu,\lambda,\ldots=0,1,2,3$ and $\alpha,\beta,\gamma,\ldots=1,2,3$. 
\section{The conformal Gauss gauge}\label{cgg}
The main reference for the properties of the conformal Gauss gauge 
that we discuss below is \cite{Friedrich03}. This gauge is based on conformal 
geodesics.
\subsection{Conformal geodesics}
Null geodesic congruences, when they are smooth, provide a valuable
tool to study the asymptotic and causal structure of spacetimes.  As
null geodesics are invariants of the conformal structure, one might
presume that spacelike or timelike curves that are conformal
invariants might also be useful in such studies. In general, geodesics
with respect to a metric are not geodesic with respect to a
conformally rescaled metric. Conformal geodesics, however, are
conformally invariant in the sense that, as point sets, they are
independent of the metric chosen in the conformal class. They are
autoparallel curves with respect to a Weyl connection defined in
\ref{sec:con_geo}.

A solution to the conformal geodesic equations given below does not only 
provide a spacetime curve, but along the curve also a Weyl 
connection, a conformal factor and a frame 
which is orthonormal for a metric in the conformal class. While the
equations are independent of coordinates, we will have to write them in some 
coordinate system, as we will study them in numerical applications. 

Given a metric $\tilde{g}$, the equations that define a conformal geodesic 
$x^\mu(\tau)$ are written for its tangent vector $\dot{x}^{\mu}(\tau)$ and a 
covector $\tilde{f}_{\mu}(\tau)$ as
\begin{eqnarray}\label{eq:congeo} 
(\tilde{\nabla}_{\dot{x}}\dot{x})^{\mu} + S(\tilde{f})^{\ \mu}_{\lambda \ \rho}
\dot{x}^{\lambda}\dot{x}^{\rho} &=& 0, \nonumber \\
(\tilde{\nabla}_{\dot{x}}\tilde{f})_{\mu}- \frac{1}{2}\tilde{f}_{\lambda} 
S(\tilde{f})^{\ \lambda}_{\rho \ \mu}\dot{x}^{\rho}&=& 
\tilde{L}_{\lambda\mu}\dot{x}^{\lambda},
\end{eqnarray}
where $S(\tilde{f})$ has been given in (\ref{intro:sf}) and $\tilde{L}_{\mu\nu}=
\frac{1}{2}\,\tilde{R}_{\mu\nu}-\frac{1}{12}\,\tilde{R}\,\tilde{g}_{\mu\nu}$ 
is called the Schouten tensor. It is related to the Schouten tensor 
$\hat{L}_{\mu\nu}$ of a Weyl connection 
$\hat{\nabla}=\tilde{\nabla}+S(\tilde{f})$ by
$\hat{L}_{\mu\nu}=\tilde{L}_{\mu\nu}-\tilde{\nabla}_\mu \tilde{f}_\nu + 
\tilde{f}_\mu \tilde{f}_\nu - \frac{1}{2}\tilde{g}_{\mu\nu} 
\tilde{g}^{\lambda\rho} \tilde{f}_\lambda \tilde{f}_\rho$. 
By setting $\tilde{f}_\mu=0$ in (\ref{eq:congeo}) we see that every physical 
geodesic of a vacuum spacetime is also a conformal geodesic.

We can write the equations (\ref{eq:congeo}) using the Weyl connection 
$\hat{\nabla}$ that is defined by the 1-form $\tilde{f}$ along the curve 
$x^\mu(\tau)$ by $\hat{\nabla} = \tilde{\nabla} + S(\tilde{f})$. The equation
for $\dot{x}^\mu$ becomes $\hat{\nabla}_{\dot{x}}\dot{x}^\mu=0$, so that 
the curve $x^\mu(\tau)$ is autoparallel with respect to the Weyl connection 
$\hat{\nabla}$. The equation for $\tilde{f}$ becomes 
$\dot{x}^\mu \hat{L}_{\mu\nu}=0$.
We will also use a frame that is $\hat{\nabla}$-parallelly transported along the
conformal geodesics. The equation for the components of the frame vector fields
$e_k^{\ \mu}(\tau)$ reads $\hat{\nabla}_{\dot{x}}e_k^{\ \mu}=0$. 
We will take $\dot{x}$ to be the timelike frame vector field $e_0$. 

We see by (\ref{weyl-conn}) that the metric $g=\Omega^2 \tilde{g}$ is
$\hat{\nabla}$-parallelly transported along a given curve $x(\tau)$ in
$\mathcal{M}$ if the function $\Omega>0$ satisfies on $x(\tau)$ the
equation \be\label{eq:transport} \hat{\nabla}_{\dot{x}} \Omega =
\Omega \, \tilde{f}_\mu\dot{x}^\mu. \ee In a spacetime solution to the
vacuum Einstein equations one can derive for the conformal factor
$\Omega(\tau)$ satisfying (\ref{eq:transport}) the equation
$\tilde{\nabla}^3_{\dot{x}}\Omega=0$ (see \cite{Friedrich95} or
\cite{Friedrich03}). This equation can be solved explicitly so that
the conformal factor $\Omega(\tau)$ is known a priori in terms of
initial data. This is a remarkable property of conformal geodesics
which will play an important role in this chapter. For given initial data on
$\mathcal{S}=\{\tau=0\}$
\[ \Omega(0)=\Omega_\ast, \quad x^\mu(0) = x^\mu_\ast,\quad 
\dot{x}^\mu(0)=\dot{x}^\mu_\ast,\quad \tilde{f}_\mu(0)=(\tilde{f}_\ast)_\mu,\]
the conformal factor $\Omega(\tau)$ along the conformal geodesic can be 
written explicitly as
\[ \Omega(\tau)=\Omega_\ast \left(1+\tau\,(\tilde{f}_{\mu}\dot{x}^{\mu})|_{\tau=0}+
\frac{1}{4}\tau^2\, (\tilde{g}_{\mu\nu}\dot{x}^\mu \dot{x}^\nu)|_{\tau=0}\,
(\tilde{g}^{\lambda\rho}\tilde{f}_{\lambda}\tilde{f}_{\rho})|_{\tau=0} \right).\]

It is useful to distinguish between the conformal compactification 
of the induced metric $\tilde{h}$ on the initial Cauchy hypersurface 
$S=\{\tau=0\}$ and the conformal compactification 
of the spacetime metric $\tilde{g}$. 
We write for the compactification of the induced spatial metric 
$\bar{h}=\phi^2\tilde{h}$, with $\phi$ satisfying the relations given in the 
beginning of the chapter. 
The choice of $\Omega_\ast$ can be made such that spatial infinity is 
represented either by a point or by a cylinder. We introduce a free function
$\kappa$ by setting $\Omega_\ast=\frac{\phi}{\kappa}$. When we choose 
$\kappa|_i=1$, $d\kappa|_i=0$, then spatial infinity $i$ corresponds to the 
point $i^0$ of the spacetime. Choosing 
$\kappa|_i=0$, $d\kappa|_i\ne 0$, results in the representation of spatial 
infinity as a cylinder $\mathcal{I}$ (Fig.~\ref{fig:point-cylinder}). 
We will make use of this freedom in later sections of this chapter. 

We have $h=\Omega_\ast^2 \tilde{h}=\frac{\phi^2}{\kappa^2} \tilde{h}=
\frac{1}{\kappa^2}\bar{h}$. 
We choose initial data for the conformal geodesics such that 
$(\tilde{f}_\ast)_{\mu}x_\ast^{\mu}=0$. On the initial hypersurface $\tilde{S}$ 
we set $(\tilde{f}_\ast)_{\mu}=\phi^{-1}\partial_\mu\phi$. 
By requiring $g(\dot{x},\dot{x})|_{\tau=0}=-1$, we have 
$\tilde{g}(\dot{x},\dot{x})_{\tau=0} = -\frac{\kappa^2}{\phi^2}$. 
Then the conformal factor becomes 
\be \label{eq:conf_fac} \Omega(\tau) = 
\frac{\phi}{\kappa}\left(1-\tau^2\frac{\kappa^2}{\omega^2}\right) \qquad 
\mathrm{with} \qquad \omega =  \frac{2 \ \phi}{\sqrt{\, \bar{h}^{\mu\nu}\, 
\partial_{\mu}\phi \,\partial_\nu \phi}}. \ee
Another important role of the free function $\kappa$ is seen by this formula. 
We can control the value $\tau_{\scri^{\pm}}=\pm\frac{\omega}{\kappa}$ 
of the time coordinate at which the conformal geodesic cuts 
$\scri^{\pm}=\{\Omega(\tau_{\scri^\pm})=0\}$ 
by the choice of $\kappa$. 

Beside the conformal factor, the field $d_k:=\Omega\,\tilde{f}_\mu e^{\ \mu}_{k}$
can also be determined explicitly. It turns out that 
\be \label{eq:dk} d_k =(\dot{\Omega}(\tau),d_a(0)) = \left(-2\tau\frac{\kappa\,
\phi}{\omega^2},\frac{1}{\kappa}(e_\ast)_a^{\ \mu}\partial_\mu\phi\right),\ee
where $(e_\ast)_a^{\ \mu} := e_a^{\ \mu}(0)$. For a derivation of these results, 
see \cite{Friedrich95}. The knowledge of the conformal factor by 
(\ref{eq:conf_fac}) will be very useful in our numerical studies in later 
sections.
\subsection{Construction of the conformal Gauss gauge}\label{sec:cgg}
To construct a conformal Gauss gauge one uses conformal geodesics in a similar 
way as one uses metric geodesics to construct the Gauss gauge. One specifies a 
congruence of timelike vectors, a conformal factor and a 1-form on an initial 
hypersurface.  The timelike conformal geodesics starting from this surface 
provide the conformal Gauss coordinates. Spatial coordinates are dragged along.

Assume a solution $(\widetilde{\mathcal{M}},\tilde{g})$ to the vacuum Einstein 
equations has been given. We construct a conformal Gauss gauge 
with timelike conformal geodesics and an orthonormal frame along them, 
$(\,x^\mu(\tau),\tilde{f}_\mu(\tau),e_a^{\ \mu}(\tau)\,)$, as follows:
\begin{enumerate}
\item We find a conformal extension $(\mathcal{M},\breve{g},\theta)$ of 
the solution $(\widetilde{\mathcal{M}},\tilde{g})$. In general, we will need 
different coordinates in different asymptotic regions which are used for the
calculation of conformal geodesics via (\ref{congeo-component}).
\item On a spacelike slice with a spatial metric $\tilde{h}$, we introduce 
compactifying coordinates and rescale $\tilde{h}$ with a suitable conformal 
factor $\phi$, such that in these coordinates we have 
$\bar{h}=\phi^2 \tilde{h}$. The metric $\bar{h}$ is used in the calculation of 
the conformal factor (\ref{eq:conf_fac}).
\item We set initial data 
$(x_\ast^\mu,\dot{x}_\ast^\mu,(\tilde{f}_\ast)_\mu,(e_\ast)_a^{\,\mu})\equiv
(x^\mu,\dot{x}^\mu,\tilde{f}_\mu,e_a^{\,\mu})|_S$ according to 
\[ (\tilde{f}_\ast)_\mu = \phi^{-1} \partial_\mu \phi, \quad 
(\tilde{f}_\mu\,\dot{x}^\mu)|_{S} = 0, \quad \dot{x}_\ast\perp S, \quad 
g(\dot{x},\dot{x})|_{S}= \frac{\phi^2}{\kappa^2}\tilde{g}
(\dot{x},\dot{x})|_{S}=-1,\] \[g(e_a, \dot{x})|_{S} = 0, \qquad 
g(e_a,e_b)|_{S} = \delta_{ab}, \qquad \mathrm{where} \qquad a,b=1,2,3. \] 
The timelike frame vector is given by $\dot{x}$ itself. The frame is not 
unique, the freedom corresponds to the freedom of spatial rotations. 
One also needs to choose the free function $\kappa$ that determines the 
representation of spatial infinity and the value of the time coordinate on 
$\scri$ in the conformal Gauss gauge.
\item We solve the following system of ordinary differential equations
\begin{eqnarray}\label{congeo-component}
(\partial_\tau x)^\mu &=& \dot{x}^\mu, \nonumber \\
(\partial_\tau\dot{x})^{\mu} &=& - \breve{\Gamma}_{\lambda\ \rho}^{\ \mu} 
\dot{x}^\lambda\dot{x}^\rho -2(\breve{f}_{\nu}\dot{x}^{\nu})\,\dot{x}^{\mu} + 
(\breve{g}_{\lambda\rho}\dot{x}^{\lambda}\dot{x}^{\rho})\,\breve{g}^{\mu\nu}
\breve{f}_{\nu},\nonumber \\
(\partial_\tau \breve{f})_{\mu} &=& \breve{\Gamma}_{\mu\ \lambda}^{\ \rho} 
\dot{x}^\lambda \breve{f}_\rho + (\breve{f}_{\nu}\dot{x}^{\nu})\, \breve{f}_{\mu} -
\frac{1}{2}(\breve{g}^{\lambda\rho}\breve{f}_{\lambda}\breve{f}_{\rho})\, 
\dot{x}^{\mu} + \breve{L}_{\mu\nu}\dot{x}^{\nu}, \nonumber \\
(\partial_\tau e_a)^{\mu} &=& - \breve{\Gamma}_{\lambda\ \rho}^{\ \mu} 
\dot{x}^\lambda e_a^{\,\rho} - (\breve{f}_{\nu} e_a^{\,\nu})\,\dot{x}^\mu - 
(\breve{f}_{\nu}\dot{x}^{\nu})e_{a}^{\,\mu} + 
(\breve{g}_{\lambda\rho}e_a^{\,\lambda}\dot{x}^{\rho})\, 
\breve{g}^{\mu\nu}\breve{f}_{\nu}.
\end{eqnarray}
The one-form $\breve{f}$ for which we solve the equations is related to the 
Weyl connection $\hat{\nabla}=\tilde{\nabla}+S(\tilde{f})$ by 
$\hat{\nabla}=\breve{\nabla}+S(\breve{f})$ where 
$\breve{\nabla}=\tilde{\nabla}+S(\theta^{-1}d\theta)$, so $\breve{f}=\tilde{f}+
\theta^{-1}d\theta$. The initial data for $\breve{f}$ is chosen accordingly as 
$\breve{f}_\ast=(\phi/\theta)^{-1}d(\phi/\theta)$.
\end{enumerate}
In a numerical calculation, we check the quality of the solution using 
(\ref{eq:conf_fac}) and (\ref{eq:dk}).

We introduced various conformal metrics to do the calculation. 
The physical metric is denoted by $\tilde{g}$, it induces a spatial metric 
$\tilde{h}$ on the initial hypersurface $\tilde{S}$ which we compactify on
$S=\tilde{S}\cup \{i\}$ by rescaling 
$\bar{h}=\phi^2\tilde{h}$. As the conformal geodesics cover asymptotic 
regions where the physical metric $\tilde{g}$ becomes singular, 
we also introduced a compactified spacetime metric $\breve{g}=\theta^2\tilde{g}$
to calculate the right hand side of (\ref{congeo-component}). 
Another conformal spacetime metric is the one that is $\hat{\nabla}$-parallelly
transported along the conformal geodesics with the conformal factor satisfying 
(\ref{eq:transport}). It is acquired by $g=\Omega^2 \tilde{g}$ 
with $\Omega$ given explicitly by (\ref{eq:conf_fac}).

\subsection{Numerical experiments}\label{sec:cgg_background}
In \cite{Friedrich03} Friedrich analytically constructs conformal Gauss 
coordinates in the Schwarz-schild-Kruskal spacetime and shows that they cover 
the conformal extension in a smooth way. 
In the following, we will construct such coordinates numerically. 
Going beyond the analytical studies, we will also see in our numerical studies
that one can construct conformal Gauss coordinates on the Kerr spacetime.

For the numerical experiments presented below, I calculated the right hand side
of (\ref{congeo-component}) using the computer algebra package 
\texttt{MathTensor}. For the integration of the system of ordinary
differential equations (ODE's), I used a 4th order 
Runge-Kutta integration algorithm. 
\subsubsection{The Schwarzschild-Kruskal spacetime}\label{cgg_ss}
\begin{figure}[t]
\centering
\psfrag{0}{$2$}
\psfrag{2}{$4$}
\psfrag{10}{$12$}
\psfrag{300}{$300$}
\psfrag{theta=0}{$\{ \Omega = 0 \} $}
\psfrag{Scri}{${\scri^{+}}$}
\psfrag{I}{${\mathcal{I}}$}
\psfrag{i+}{$i^{+}$}
\psfrag{congeo}{conformal geodesics}
\psfrag{tau}{$\tau$}
\psfrag{v=const}{$v=$const}
\psfrag{w=const}{$u =$const}
\psfrag{r=2m}{event horizon}
\psfrag{r=0}{singularity}
\psfrag{r}{$\tilde{r}_s/m$}
\includegraphics[width=0.9\textwidth,height=0.3\textheight]
{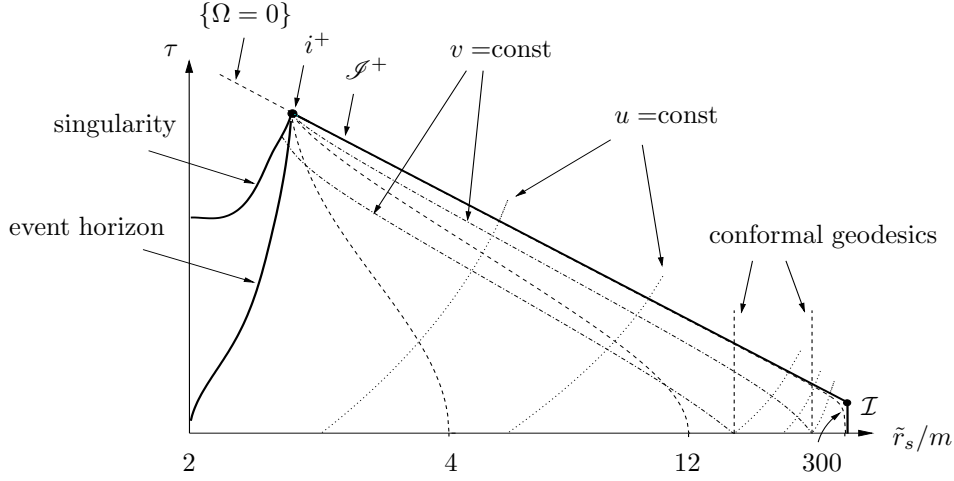}
\caption{Schwarzschild-Kruskal spacetime in a conformal Gauss gauge.
\label{fig:cgg}}
\end{figure}

\begin{figure}[tbp]
\flushleft
\psfrag{pi/2}{$\frac{\pi}{2}$}
\psfrag{theta=0}{$\{ \Omega = 0 \} $}
\psfrag{Scri}{${\scri^{+}}$}
\psfrag{chi}{$\chi$}
\psfrag{I0}{${\mathcal{I}}^{0}$}
\psfrag{I}{${\mathcal{I}}$}
\psfrag{i+}{$i^{+}$}
\psfrag{I+}{${\mathcal{I}}^{+}$}
\psfrag{pi}{$\pi$}
\psfrag{congeo}{conformal geodesics}
\psfrag{tau}{$\tau$}
\psfrag{v=const}{$v=$const}
\psfrag{w=const}{$u =$const}
\psfrag{r=2m}{event horizon}
\psfrag{r=0}{singularity}
\includegraphics[width=0.93\textwidth]{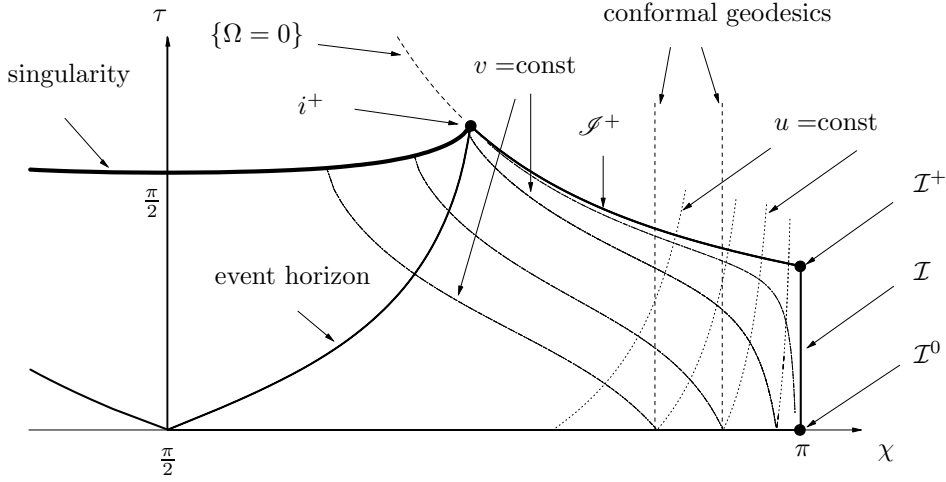}
\caption{Schwarzschild-Kruskal spacetime in a conformal Gauss gauge.
\label{fig:cgg_isotropic}}
\end{figure}

We calculate initial data for the conformal Gauss gauge in the 
Schwarzschild-Kruskal spacetime using different coordinates and 
conformal compactifications. Resulting gauges are illustrated in the 
numerically generated conformal diagrams Fig.~\ref{fig:cgg} and 
Fig.~\ref{fig:cgg_isotropic}. 

The physical Schwarzschild metric reads
\[ \tilde{g}_s = - \left( 1 - \frac{2m}{\tilde{r}_s}\right)\, d\tilde{t}^2+  
\left( 1 - \frac{2m}{\tilde{r}_s}\right)^{-1} \, d\tilde{r}_s^2 + 
\tilde{r}_s^2\,d\sigma^2, \]
where $\tilde{r}_{s}$ is the Schwarzschild radial coordinate and 
$\tilde{r}_{s}>2m$. 
We transform the metric using retarded and advanced null coordinates 
\be \label{wv} u=\tilde{t}-\left(\tilde{r}_s+
2m\,\ln\,\left(\tilde{r}_s-2m\right)\right), \quad 
v=\tilde{t}+\tilde{r}_s+ 2m\,\ln\,\left(\tilde{r}_s-2m\right),\ee
After the inversion $r(\tilde{r}_s) = \frac{1}{\tilde{r}_S}$ and the rescaling 
with $\theta=r$ following $\breve{g} = \theta^{2} \tilde{g}$ we get 
\[ \breve{g} = - r^{2} ( 1-2mr)\,du^{2} + 2\, du\, dr + d\sigma^{2}, 
\qquad \breve{g} = - r^{2} ( 1-2mr)\,dv^{2} - 2\, dv \, dr + d\sigma^{2} \]
We use the metric $\breve{g}$ to calculate the right hand side of 
(\ref{congeo-component}).
The coordinates $u,v$ extend analytically into regions where 
$\tilde{r}_s \leq 2 m$. 
For a simulation through $\scri^+$ we use the retarded null coordinate $u$, 
for a simulation through the future horizon we use the advanced null 
coordinate $v$. 

We give initial data on the $\tit=0$ slice. From the transformation (\ref{wv}) 
and the inversion $r=\frac{1}{\tilde{r}_s}$, we have on this slice
\[ du = \frac{1}{r^2(1-2mr)}dr, \qquad dv = - \frac{1}{r^2(1-2mr)}dr, \]
which leads to the induced spatial metric
\[ \tilde{h} = \frac{1}{r^4(1-2mr)}\,dr^2 + \frac{1}{r^2}\,d\sigma^2. \]
From now on, we choose $m=2$ and present the calculation only for the retarded
null coordinate $u$. The domain from the event horizon 
to spatial infinity corresponds to $r\in[\frac{1}{4},0]$. 
We compactify $\tilde{h}$ using 
the conformal factor $\phi=\frac{2r^2}{1-2r}$, so that 
$\bar{h}=\phi^2 \tilde{h}$ is diffeomorphic
to the standard metric on the three sphere  
$\bar{h}=d\chi^2 + \sin^2\chi\, d\sigma^2$, as can be shown by the 
transformation $r(\chi)=\frac{\sin\chi}{2(1+\sin\chi)}$. 
The starting point of the conformal geodesic on the $\tit=0$ hypersurface 
depending on $r_\ast$ is given by 
\[x_\ast=(u_\ast,r_\ast)=\left(-\frac{1}{r_\ast}-
4 \ln\left(\frac{1}{r_\ast}-4\right),r_\ast\right),\]
where we suppressed the angular coordinates. The initial data for the 1-form 
$\tilde{f}$ is given as $\tilde{f}_\ast=\frac{2(1-r_\ast)}{r_\ast(1-2r_\ast)}dr$.
The orthogonality condition $(\tilde{f}_\mu \dot{x}^\mu)_{\ast}=0$ and the 
normalization requirement delivers 
$\dot{x}_\ast= \frac{1-2r_\ast}{2 r_\ast^2\sqrt{1-4 r_\ast}}\kappa\,\partial_u$. 
The timelike frame vector is given by $e_0 = \dot{x}$. Initial data for the 
spatial frame vector reads
\[ (e_\ast)_1=\kappa\,\frac{1-2 r_\ast}{2 r_\ast^2\sqrt{1-4 r_\ast}}\partial_u + 
\kappa \, \frac{(1-2r_\ast)\sqrt{1- 4 r_\ast}}{2}\partial_r. \]
The time coordinate at which the conformal geodesics cross $\scri$ is given by 
$\tau_{\scri^\pm} =\pm\frac{\omega}{\kappa}$. We have 
$\omega = \frac{2 r}{\sqrt{1-4r}(1-r)}$. Note that $\omega$ vanishes at spatial
infinity and becomes unbounded at the intersection of the event horizon with
the initial hypersurface.

The resulting conformal Gauss gauge can be visualized in a conformal
diagram which depends on the choice of the spatial coordinate and the
free function $\kappa$. Fig.~\ref{fig:cgg} is the result of a
numerical calculation where $\kappa=\omega/(k r+1)$ has been chosen in
a neighborhood of spatial infinity such that $\scri^+$ is a straight
line in the corresponding conformal Gauss gauge with the slope $k>0$.
At the intersection of the event horizon with the initial hypersurface
where $\omega\to\infty$, this choice would lead to
$\kappa\to\infty$. To avoid this, we choose a different $\kappa$ in
the interior of the event horizon which reads $\kappa = 2r/(1-2r)$.
There is a smooth transition region between the domain inside the
event horizon and the exterior region.

Illustrated in Fig.~\ref{fig:cgg} is the "upper right part" of the Penrose 
diagram for the Schwarzschild-Kruskal spacetime (Fig.~\ref{fig:schwarzschild}).
The lower horizontal line corresponds to the 
hypersurface ${\{\tit=0\}}$ in the standard Schwarzschild coordinates, 
where also $\{\tau=0\}$. We see that the conformal geodesics 
cover in a smooth way spacelike, null and timelike infinity and the 
domain close to the singularity. This result is robust in the sense that it 
allows variations of initial data which satisfy the orthogonality condition
$(\tilde{f}_\mu \dot{x}^\mu)_{\ast}=0$, or the use of different compactifications
$\breve{g}$ based on different coordinates.

Another representation of the Schwarzschild-Kruskal spacetime is given in 
Fig.~\ref{fig:cgg_isotropic}. For this representation a spatial coordinate 
$\chi$ has been chosen via $r(\chi)=\frac{\sin\chi}{2(1+\sin\chi)}$.
The domain from the event horizon to spatial infinity corresponds to
$\chi\in[\pi/2,\pi]$. We have in this coordinate  
\[\phi = \frac{\sin^2\chi}{2(1+\sin\chi)},\qquad 
\bar{h}=d\chi^2+\sin^2\chi\,d\sigma^2,\qquad 
\omega = \frac{2\sin\chi(1+\sin\chi)}{\abs{\cos\chi}(2+\sin\chi)} \]
We choose $\kappa=\sin\chi$. The causal structure near spatial infinity 
can be clearly seen in Fig.~\ref{fig:cgg_isotropic}. Denoted are the sets 
$\mathcal{I}^0,\mathcal{I}$ and $\mathcal{I}^+$ all represented by $i^0$ 
in the one-point compactification. We will refer to these sets in the 
following sections in the context of the regular finite initial value problem 
at spatial infinity. Note that the presented conformal diagrams are not Penrose
diagrams as the light rays are not represented by straight segments with 45 
degrees to the horizontal. In a Penrose diagram, spatial infinity is 
necessarily represented by a point.

We can do the calculation of the conformal Gauss gauge using other 
compactifications of the Schwarzschild spacetime as well. As an example, 
consider compactifying isotropic coordinates given by the radial transformation
\[\rho(r)=2\,\left(\tir_s-m+\sqrt{\,\tir_s\,(\tir_s-2m)\,}\right)^{-1}.\] 
The Schwarzschild metric becomes
\[ \tilde{g} = -\left( \frac{1-m\rho/2}{1+m\rho/2} \right)^2 d\tit^2+ 
\left[\frac{(1+m\rho/2)^2}{\rho^2} \right]^{2}(d \rho^2 + \rho^2 d\sigma^2)\]
Choosing as our initial hypersurface the $\tit=0$ hypersurface and setting the 
spatial conformal factor to $\phi=\rho^2/\left(1+m\rho/2 \right)^2$ we get 
$\bar{h}= \phi^2\tilde{h}=d\rho^2+\rho^2d\sigma^2$. \\
The initial point has $(\tilde{t}_\ast,\rho_\ast)=(0,\rho_\ast)$. We set further
\[ \tilde{f}_\ast=\phi^{-1}d\phi=\frac{2}{\rho_\ast(1+m\rho_\ast/2)}\,d\rho_\ast,
\qquad \dot{x}_\ast = \kappa\,\frac{(1+m\rho_\ast/2)^3}{\rho_\ast^2
(1-m\rho_\ast/2)}\,\partial_{\tilde{t}} \]
The spatial frame vector on the initial hypersurface reads 
$e_1=\kappa\partial_\rho$. 
For the choice of $\kappa$ we calculate $\omega = \rho (1+m \rho/2)$. The 
choice $\kappa=\rho$ leads to a similar conformal diagram as 
Fig.~\ref{fig:cgg}.
The use of different coordinate systems or different compactifications 
for calculating the right hand side of (\ref{congeo-component}) 
does not affect the quality of the calculation.

A nice feature of the conformal geodesics is that we have some global
control on their behavior in a given spacetime. This analytic
knowledge has been used to test the quality of numerical
calculations. One test is given by the double role that the conformal
factor plays. While $\Omega$ is some known function given by
(\ref{eq:conf_fac}) in terms of coordinates and initial data, the set
where it vanishes corresponds to the conformal boundary of the
spacetime and has therefore a special meaning. The spatial coordinate
value on the conformal geodesics must correspond to
$\tilde{r}_s\to\infty$ at the set $\{\Omega=0\}$.

A stronger test has been made using the behavior of the frame. The 1-form 
$d_k=\Omega\,\tilde{f}_\mu e_k^{\ \mu}$ is known explicitly in terms of the
initial data and the coordinates via the relation (\ref{eq:dk}). 
Comparing the evolution of $d_k$ 
with its a priori known form (\ref{eq:dk}) delivers a strong test 
for the numerical calculation.
\subsubsection{The Kerr spacetime}
\begin{figure}[tbp]
\flushleft
\psfrag{theta=0}{$\{ \Omega = 0 \} $}
\psfrag{Scri}{${\scri^{+}}$}
\psfrag{chi}{$\tilde{r}=1/r$}
\psfrag{I0}{${\mathcal{I}}^{0}$}
\psfrag{I}{${\mathcal{I}}$}
\psfrag{i+}{$i^{+}$}
\psfrag{Iplus}{${\mathcal{I}}^{+}$}
\psfrag{congeo}{conformal geodesics}
\psfrag{tau}{$\tau$}
\psfrag{r=2m}{Event horizon}
\psfrag{v=const}{$v=$const}
\psfrag{w=const}{$u =$const}
\psfrag{r=0}{Cauchy horizon}
\includegraphics[width=0.9\textwidth]{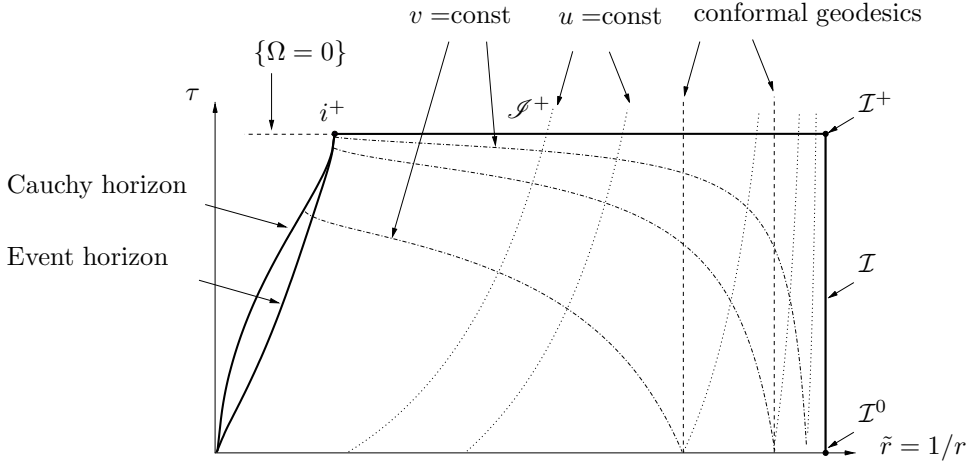}
\caption{Kerr spacetime in a conformal Gauss gauge.
\label{fig:cgg_kerr}}
\end{figure}

Numerical techniques allow us to go beyond the spherically symmetric
case in the study of the conformal Gauss gauge.  As shown in this
subsection and seen by the numerically generated conformal diagram
Fig.~\ref{fig:cgg_kerr}, we can solve the conformal geodesic equations
in the Kerr spacetime covering null infinity, timelike infinity and
the Cauchy horizon.

We take the Kerr metric in Boyer-Lindquist coordinates with $m>a$ where $m$ is
the mass and $a$ is the angular momentum of the Kerr spacetime. We make 
an Eddington-Finkelstein-like transformation using ingoing and outgoing 
light rays as coordinate lines (see \cite{Hawking73} for the transformations). 
We introduce a compactifying coordinate $r$ which is related 
to the physical coordinate $\tilde{r}$ by the inversion $r(\tir)=1/\tilde{r}$. 
For the outgoing case with the retarded null coordinate $u$ we get for the Kerr
metric
\begin{eqnarray*}
\tilde{g} &=& -\left(1-\frac{2m}{r\Sigma}\right) du^2 - \frac{2}{r^2} du\,dr - 
\frac{4 a m}{r\Sigma}\sin^2\!\vartheta \,du\,d\varphi \\
 & + & \frac{2a}{r^2}\sin^2\!\vartheta\,dr\,d\varphi+\Sigma\,d\vartheta^2 +
\frac{1}{\Sigma}\left(\left(\frac{1}{r^2}+a^2\right)^2 - 
\triangle a^2 \sin^2\!\vartheta\right) \sin^2\!\vartheta\, d\varphi^2,
\end{eqnarray*}
where
\[ \Sigma = \frac{1}{r^2} + a^2 \cos^2\vartheta, \qquad \ 
\triangle = \frac{1}{r^2} - \frac{2m}{r} + a^2. \]
The metric $\breve{g}$ with respect to which we calculate the right hand side 
of (\ref{congeo-component}) is obtained by $\breve{g}=\theta^2\tilde{g}$ with
$\theta=r$ as before.

We give initial data on a $\tit=0$ hypersurface in the Kerr spacetime where 
$\tit$ is the timelike Boyer-Lindquist coordinate. Following 
\cite{Dain01a}, we do the coordinate transformation 
$\tir(\chi)=\frac{\sin \chi}{m\,\sin\chi+\sqrt{m^2-a^2}}$. 
We rescale $\tilde{h}$ with the conformal factor 
\[ \phi = \frac{\sin\chi}{\sqrt{\Sigma}}=\frac{r^2\,\sqrt{m^2-a^2}}
{(1-mr)\sqrt{1+a^2r^2cos^2\vartheta}}.\]
The compactified spatial metric reads
\[ \bar{h} = \phi^2 \tilde{h} = d\chi^2 + \sin^2\chi d\vartheta^2 + 
\sin^2\chi \sin^2\vartheta \left( 1+a^2 \frac{1+2m/(\Sigma\,r(\chi))}
{\Sigma \sin^2\chi} \sin^2\chi \sin^2\vartheta \right) d\varphi^2. \]
We write the initial data using the coordinate $r$. We get 
\[ \tilde{f}_\ast=\phi^{-1}\,d\phi = \frac{2-mr+a^2r^2\cos^2\!\vartheta}
{r(1-mr)(1+a^2r^2\cos^2\!\vartheta)}\,dr + \frac{a^2r^2\cos\vartheta 
sin\vartheta}{1+a^2r^2\cos^2\!\vartheta}\,d\vartheta,\]
and 
\[ \dot{x}_\ast = \frac{(1-mr)(1+a^2r^2\cos^2\!\vartheta)}{\sqrt{m^2-a^2} r^2 
\sqrt{1+a^2r^2\cos^2\!\vartheta-2mr}}\, \kappa \, \partial_u. \]
Note that for $a=0$ we get the formula for the Schwarzschild spacetime 
calculated in the previous subsection.

The conformal diagram in Fig.~\ref{fig:cgg_kerr} has been plotted with
the parameters chosen as $\vartheta=\pi/2$, $m=2$ and $a=1$.  For the
"flat" $\scri^+$ depicted in the figure, the free function $\kappa$
has been chosen according to $\kappa=\omega$. We made this choice only
to illustrate the resulting conformal diagram. For a numerical
solution to the Cauchy problem, this choice is bad as the grid speed
of ingoing characteristics becomes unbounded near $\scri^+$. It is bad
also in the interior of the event horizon as the Cauchy and the event
horizons meet on the initial hypersurface in such coordinates. 
It turns out, however, that the choice is useful in analytic studies
\cite{ValienteKroon04a,ValienteKroon07}.  

In both the Schwarzschild-Kruskal and the Kerr 
spacetimes the conformal geodesics are seen to be regular in the interior of 
event horizons. In the Kerr spacetime they pass regularly through the Cauchy 
horizon where one needs to change the coordinate representation of the metric 
to integrate the equations further. This property of the conformal geodesics 
suggests that the conformal Gauss gauge might be suitable to study the inner 
structure of black holes and questions on stability of Cauchy horizons 
\cite{Dafermos05,Ori99}. The behavior of conformal geodesics in the interior
of black holes needs to be analyzed in further detail before such studies
can be made. As our main interest in this thesis lies in the treatment 
of the asymptotic region we do not follow these questions further. 

\section{The reduced general conformal field equations}\label{sec:rgcfe}
To use the conformal geodesics as the underlying gauge, Friedrich extended the 
conformal field equations to admit Weyl connections. 
This extended system is called the general conformal field 
equations and is written for a conformally rescaled frame field 
with respect to a Weyl connection. 

The introduction of a rescaling and the transition to Weyl 
connections allows a more general conformal gauge freedom than the one studied 
in chapter \ref{chapter:null}. The conformal gauge now includes 
in addition to the conformal factor $\Omega$, a 1-form $f$.
We have seen that in vacuum, conformal geodesics determine 
a conformal factor and a 1-form defining a Weyl connection. 
Friedrich used the conformal Gauss gauge based on the conformal geodesics 
to fix the extended conformal gauge freedom. He derived a reduced system that
is equivalent to the vacuum Einstein equations and implies a symmetric 
hyperbolic system which preserves the constraints on a conformal extension. 
We call this system the reduced general conformal field equations.

Based on the reduced general conformal field equations, Friedrich formulated 
a regular finite initial value problem near spatial infinity 
in which the location of spatial and null infinity are known a priori
\cite{Friedrich98}. The construction allows one to study in detail 
properties of solutions to Einstein equations near spatial infinity 
(see \cite{Friedrich02,Friedrich04} for reviews). 

The reduced general conformal field equations are written for the following 
variables: a frame field $e^{\ \mu}_{k}$, 
a Weyl connection $\hat{\Gamma}^{\ k}_{i \ j}$, the Schouten 
tensor $\hat{L}_{ij} = \frac{1}{2}\hat{R}_{(ij)}-\frac{1}{4}\hat{R}_{[ij]} - 
\frac{1}{12}\hat{R}\,g_{ij}$, and the rescaled conformal Weyl tensor 
$W^{i}_{\ jkl} = \frac{1}{\Omega}\,C^{i}_{\ jkl}$. Note that the rescaling is 
such that if the conformal Weyl tensor satisfies the Sachs peeling behavior, 
the rescaled conformal Weyl tensor attains a regular limit at $\scri$. 
We set $\dot{x}^\mu =  e_0^{\ \mu}$ as in \ref{sec:cgg} to construct a conformal
Gauss gauge. We have as a consequence
\be\label{eq:con_gauge} e_0^{\ \mu} = \delta_0^{\ \mu},\quad 
f_0= f_\mu e_0^{\ \mu}=0,\quad \hat{\Gamma}_{0\ j}^{\ i} = 0, 
\quad \hat{L}_{0j} = 0. \ee
The reduced general conformal field equations read
\begin{eqnarray}\label{eq:reduced_conformal}
\partial_0 e^{\ \mu}_{a} &=& - \hat{\Gamma}^{\ l}_{a \ 0}e^{\ \mu}_{l},\nonumber \\
\partial_0 \hat{\Gamma}^{\ i}_{a \ j} &=& 
  -\hat{\Gamma}^{\ i}_{e \ j}\hat{\Gamma}^{\ e}_{a \ 0} + 
  \eta^{i}_{\ 0} \hat{L}_{aj} + \eta^{i}_{\ j} \hat{L}_{a0} - 
  \eta_{j0} \hat{L}^{\ i}_{a} + \Omega\, W^{i}_{\ j0a}, \nonumber \\
\partial_0 \hat{L}_{aj} &=& - \hat{\Gamma}^{\ e}_{a \ 0} \hat{L}_{ej}
  +d_{l}\, W^{l}_{\ j0a}, \nonumber \\
\nabla_{l} W^{l}_{\ ijk} &=& 0.
\end{eqnarray}
The functions $\Omega$ and $d_k$ are known a priori in terms of initial data 
and are given by (\ref{eq:conf_fac}) and (\ref{eq:dk}). 

The numerical and analytical advantages of the system 
(\ref{eq:reduced_conformal}) are similar. We take a numerical point of view:
\begin{itemize}
\item The system is regular for all values of the conformal factor so that 
no calculation of formally singular terms is needed that might introduce 
numerical instabilities.
\item The system consists mainly of ordinary differential equations (ODE's) 
except the Bianchi equation, which admits symmetric hyperbolic reductions. 
Besides enhancing the accuracy of the code and lowering the computational cost,
this property is especially advantageous when one is dealing with complicated 
geometries.
\item The location of the conformal boundary is known explicitly so that the 
code can be adapted to calculate the physical part of the conformal extension 
only.
\item The Weyl tensor is one of the variables. Combined with the knowledge
of the conformal factor, this property simplifies the numerical extraction of 
radiation.
\item There is a wealth of analytic knowledge on the solutions of the system 
near spatial infinity which can be used to test or improve the numerical code
\cite{Friedrich98,ValienteKroon04}.
\end{itemize}

Some difficulties with the system are the following
\begin{itemize}
\item The equations become degenerate at the sets $\mathcal{I}^\pm$ where null 
infinity meets spatial infinity.
\item For non-vanishing ADM-mass, the regular finite initial value problem at 
spatial infinity imposes a spherical grid topology so that simple codes based 
on Cartesian grids cannot be used.
\item The number of variables and equations is higher than in common reductions
of Einstein equations.
\item Numerical experience with systems based on frames is sparse.
\end{itemize}
Note that while the degeneracy of the equations at $\mathcal{I}^+$ is an 
intrinsic difficulty, the other items depend on the practical experience
gathered in numerical calculations. In any case, the advantages of the system 
seem to counteract its difficulties.

We will deal with the degeneracy of the equations at the set $\mathcal{I}^+$ 
by freezing the evolution in the unphysical domain given by $\Omega<0$ and by
choosing a suitable time stepping. To achieve this, a priori knowledge of the 
conformal factor along the conformal geodesics is very helpful.

As to the geometry imposed by the cylinder at spatial infinity, numerical codes
are available that can handle coordinate singularities on a spherical grid 
\cite{Lehner2005a,Scheel06,Thornburg00}. Indeed, we will use an infrastructure 
based on the Cactus framework \cite{Cactusweb} called 
\texttt{GZPatchSystem} that can handle different coordinate patches to cover 
the sphere in a smooth way \cite{Thornburg00,Thornburg04}. 
To implement a frame-based evolution system on a sphere, we need to deal not 
just with different coordinate systems but also with different frames 
as the sphere can not be covered by a single frame field. 
We will extend \texttt{GZPatchSystem} such that frame-based evolution systems 
can be solved on a spherical grid. The fact that the main part 
of the reduced general conformal field equations determining the geometry 
consists of ODE's will simplify the numerical implementation significantly. 

The high number of variables is a common feature in conformally regular field
equations. One may think of this as the price for having a system 
which is regular for all values of the conformal factor. 
In the reduction based on conformal geodesics, 
the high number of variables is mainly just a memory issue 
as the main part of the resulting equations consists of ODE's 
that are cheap in terms of computational time.
These issues will be discussed further in later sections. First, we rewrite the
equations in a form suitable for numerical implementation.
\pagebreak
\subsection{Rewriting the equations}
We want to write the equations for independent evolution variables
such that only spatial indices appear. The components of a Weyl
connection $\hat{\Gamma}_{i\ k}^{\ j}$ do not have the symmetries of
the components of a metric connection $\Gamma^{\ j}_{i\ k}$. The
components of a metric connection satisfies $\Gamma_{ijk} =
-\Gamma_{ikj}$ which is convenient in calculations. We therefore write
the Weyl connection in terms of components of a metric connection and
a 1-form. The relation of the Weyl connection components to the metric
connection components of $g=\Omega^2\tilde{g}$ reads
\be \label{eq:comp_conn} \hat{\Gamma}^{\ j}_{i\ k} =
\Gamma^{\ j}_{i\ k} + \delta_i^{\,j}f_k + \delta_k^{\,
  j}f_i-\eta_{ik}\eta^{jl}f_l. \ee We count 21 independent variables
$(f_a,\Gamma^{\ b}_{a\ 0},\Gamma^{\ b}_{a\ c})$.  We will rewrite the
equations in terms of these variables.  By (\ref{eq:con_gauge}) and
(\ref{eq:comp_conn}) we have
\[ \hat{\Gamma}_{a\ 0}^{\ 0} = f_a,\quad 
\hat{\Gamma}_{a\ b}^{\ 0} = \Gamma_{a\ 0}^{\ b}. \]
The equation for the frame components $e_a^{\ \mu}$ becomes by summation
\[\partial_0e^{\ \mu}_{a}=-f_a\delta_0^{\,\mu}-\Gamma^{\ b}_{a\ 0}e^{\ \mu}_{b}.\]
The equation for $\hat{\Gamma}^{\ i}_{a \ j}$ consists of three parts
\begin{eqnarray*}
\partial_0 f_a &=& -f_e\Gamma_{a\ 0}^{\ e}+\hat{L}_{a0},\\
\partial_0 \Gamma^{\ b}_{a \ 0} &=& - \Gamma^{\ b}_{e \ 0}\Gamma^{\ e}_{a \ 0} 
  - \eta_{00} \hat{L}^{\ b}_{a} + \Omega\,W^{b}_{\ 00a}, \\
\partial_0 \Gamma^{\ b}_{a \ c} &=& - \Gamma^{\ b}_{e \ c} \Gamma^{\ e}_{a \ 0} 
  + \Omega\,W^{b}_{\ c0a} + \delta^b_a f_e \Gamma^{\ e}_{c\ 0} - 
  \delta^b_a\hat{L}_{c0}-\eta_{ac}\eta^{be}f_d \Gamma^{\ d}_{e\ 0} +\\
  && + \eta_{ec}\eta^{bd}f_d \Gamma^{\ e}_{a\ 0}+
  \eta_{ac}\eta^{be}\hat{L}_{e0}-f_c\Gamma^{\ b}_{a\ 0}.
\end{eqnarray*}
For the Schouten tensor we get
\begin{eqnarray*}
\partial_0 \hat{L}_{a0} &=& -\Gamma^{\ e}_{a\ 0}\hat{L}_{e0}+d_e\,W^{e}_{\ 00a}, \\
\partial_0 \hat{L}_{ab} &=& -\Gamma^{\ e}_{a\ 0}\hat{L}_{eb}+d_0\,W^{0}_{\ b0a}+
d_e\,W^{e}_{\ b0a}.
\end{eqnarray*}
All together we get 45 ODE's for the variables $(e_a^{\ \mu}, f_a,
\Gamma_{a\ 0}^{\ b}, \Gamma_{a\ c}^{\ b}, \hat{L}_{aj})$.  The rest of
the system is given by the Bianchi equation that plays a fundamental
role. We set $n=e_0$ as in \cite{Friedrich99} and decompose the
Bianchi equation with respect to $n$ and its orthogonal component with
the induced metric $h_{ij}=g_{ij}+n_i n_j=
g_{ij}+\eta_{0i}\eta_{0j}$. We denote by $\epsilon_{lijk}$ the totally
antisymmetric tensor with $\epsilon_{0123}=1$ and set
$\epsilon_{ijk}=n^l \epsilon_{lijk}=\epsilon_{0ijk}$. The electric and
the magnetic parts of the rescaled Weyl tensor are defined by $E_{ij}=
h_i^m h_j^n W_{mknl} n^k n^l$ and $B_{ij}=h_i^m h_j^n W^\ast_{mknl}
n^k n^l$ where $W^\ast_{ijkl}=\frac{1}{2}
W_{ijmn}\epsilon^{mn}_{\,\quad kl}$, so we have $E_{ij}n^j=0,
E_{i}^{\ i}=0$ and $B_{ij}n^j=0, B_{i}^{\ i}=0$. Setting
$l_{ij}=h_{ij}+n_i n_j$ we write the splitting of the rescaled Weyl
tensor as
\[W_{ijkl}=-2(l_{j[k}E_{l]i}-l_{i[k}E_{l]j}-n_{[k}B_{l]m}\epsilon^m_{\ kl}-
n_{[i}B_{j]m}\epsilon^m_{\ kl}).\]
In our case where $n^i=\delta_0^i$ we get for the electric and magnetic parts 
$E_{ab}=W_{0a0b}=E_{ba}$ and 
$B_{ab}=W^{\ast}_{a0b0} = -\frac{1}{2}W_{0acd}\,\epsilon^{cd}_{\ \ b}=B_{ba}$. Using 
the relation $\epsilon_{acd}\epsilon^{bcd}=-2\delta_a^b$ we get 
$W_{0abc}= \epsilon_{bc}^{\ \ d}B_{da}$. 
The splitting of the rescaled Weyl tensor becomes
\begin{eqnarray} \label{eq:splitting}
W_{ijkl}= -\eta_{0i}E_{jk}\eta_{0l} + \eta_{0j}E_{ik}\eta_{0l} + 
\eta_{0i} E_{jl}\eta_{0k} - \eta_{0j}E_{il}\eta_{0k} - \nonumber \\
\eta_{0i}B_{j}^m \epsilon_{mkl} + \eta_{0j} B_{i}^m \epsilon_{mkl}+ 
\epsilon_{ij}^{\ \ m} B_{mk}\eta_{0l} - \epsilon_{ij}^{\ \ m} B_{ml}\eta_{0k} -
\epsilon_{ijm} E^{mn} \epsilon_{nkl}.
\end{eqnarray}
We use these relations to replace the Weyl tensor by its electric 
and magnetic parts. It is convenient for numerical implementation 
to write all evolution variables with lower frame indices. We get
\begin{eqnarray} \label{eq:odes}
\partial_0 e_{a}^{\ 0} &=& -f_a - \Gamma_{ac0}e_{d}^{\ 0}\delta^{cd}, \nonumber\\
\partial_0 e_{a}^{\ \alpha} &=& -\Gamma_{ac0}e_{d}^{\ \alpha}\delta^{cd}, \nonumber\\
\partial_0 f_a &=& -f_c\Gamma_{ad0} \delta^{cd}+\hat{L}_{a0}, \nonumber \\
\partial_0 \Gamma_{ab0} &=& - \Gamma_{cb0}\Gamma_{ad0}\delta^{cd}+\hat{L}_{ab}
- \Omega \,E_{ab}, \nonumber \\
\partial_0 \Gamma_{abc} &=& \left(- \Gamma_{dbc} \Gamma_{ae0} + 
\delta_{ab}f_d\Gamma_{ce0}-\delta_{ac}f_d\Gamma_{be0}\right)\delta^{de},\nonumber\\
&& - \delta_{ab}\hat{L}_{c0}  + \delta_{ac}\hat{L}_{b0} - f_c\Gamma_{ab0} + 
f_b \Gamma_{ac0} + \Omega\, \epsilon_{bcd} B_{ae}\delta^{de},\nonumber \\
\partial_0 \hat{L}_{a0} &=& (-\Gamma_{ac0}\hat{L}_{e0}-
d_c\,E_{ea})\delta^{ce}, \nonumber \\
\partial_0 \hat{L}_{ab} &=& (-\Gamma_{ac0}\hat{L}_{eb}+
d_c\,\epsilon_{ebd}B_{af} \delta^{df})\delta^{ce} - d_0\,E_{ab}.
\end{eqnarray}
The rescaled Weyl tensor has 10 independent components. We use the 
tracefreeness property of its electric and magnetic parts $(E_{ab},B_{ab})$ 
for the replacement $E_{33}=-E_{11}-E_{22}$ and $B_{33}=-B_{11}-B_{22}$. 

The splitting of the Bianchi equation, $\nabla_l W^{l}_{\ ijk}=0$, 
has been given in \cite{Friedrich96,Friedrich99}. We set 
$K_{ij}=h_i^k \nabla_k n_j,\ K=h^{ij}K_{ij},\ a^i=n^j\nabla_j n^i,\ 
\mathcal{D}_k E_{ij}=h_k^l h_i^m h_j^n \nabla_l E_{mn}$. In our gauge we have
\[ K_{ab}=-\Gamma_{a\ b}^{\ 0}, \quad K=-\delta^{ab}\Gamma_{a\ b}^{\ 0},\quad 
a^i = \Gamma^{\ i}_{0\ 0} = 0. \]
We get the following evolution equations for the electric and the magnetic 
parts of the rescaled Weyl tensor
\begin{eqnarray}\label{Weyl_evol} 
\mathcal{L}_n E_{ab} +\mathcal{D}_c B_{d(a}\epsilon_{b)}^{\ cd}-
3 E^c_{\ (a} K_{b)d}
+ K E_{ab} - \epsilon_a^{\ cd} \epsilon_b^{\ ef} E_{ce}K_{df} &=& 0, \nonumber \\
\mathcal{L}_n B_{ab} -\mathcal{D}_c E_{d(a}\epsilon_{b)}^{\ cd}-
3 B^c_{\ (a} K_{b)d}
+ K B_{ab} - \epsilon_a^{\ cd} \epsilon_b^{\ ef} B_{ce}K_{df} &=& 0,
\end{eqnarray}
where $\mathcal{L}_n$ denotes the Lie derivative along $n=e_0$. The system 
(\ref{Weyl_evol}) is not yet ready for implementation in a code based on 
method of lines. The derivatives along the spatial frame vector fields include 
time derivatives because the time components of the spatial frame do not vanish
in general, that is $e_a^{\ 0}\ne 0$. 
The system (\ref{Weyl_evol}) is a homogeneous, first order, symmetric 
hyperbolic system for the unknown $u=(E_{ab}, B_{ab})$, ignoring the 
tracefreeness of $E_{ab}$ and $B_{ab}$. We can write this system in the form
\be \label{eq:sym_hyp}
A^\mu \partial_\mu u + F u = 0, 
\ee
where $A^\mu$ and $F$ are matrix valued functions of the unknown $u$. 
For the time integration using method of lines we need to build 
$\partial_0 u=-(A^0)^{-1}(A^\alpha\partial_\alpha u+F u)$.
Instead of a direct calculation, it is more practical 
for the numerical implementation to calculate and store 
$(A^\alpha\partial_\alpha u+F u)$ using finite differencing and then to build 
$-(A^0)^{-1}(A^\alpha\partial_\alpha u+ F u)$. This leads to less terms in the
equations which reduces the time for computation as well as for the search of 
eventual errors in the code. We use the tracefreeness property of $E_{ab}$ and
$B_{ab}$ to replace $E_{33}$ and $B_{33}$.

The splitting of the Bianchi equation implies the following constraints
\begin{eqnarray}
\mathcal{D}^c E_{ca}+ 2 K^{bc} \epsilon^{d}_{\ c(a}B_{b)d} & = & 0\nonumber\\
\mathcal{D}^c B_{ca}+\epsilon^{\ cd}_{a}(2 K^b_{\ c}- K^{\ b}_{c}) E_{bd} & = & 0
\end{eqnarray}
The calculation of the constraints includes time derivatives if 
$e_a^{\ o}\ne 0$.

\subsection{Initial data for the reduced conformal field equations}
\label{sec:id_rcfe}
The construction of initial data for the reduced general conformal field 
equations from a given asymptotically flat solution to the vacuum constraint 
equations can be found in \cite{Friedrich04}. We will focus on the special case
of time reflection symmetric data for which the second fundamental form 
vanishes.

Assume we are given on a three dimensional manifold 
$\tilde{\mathcal{S}}$ with an asymptotically flat end an asymptotically flat 
solution metric $\tilde{h}$ to the vacuum Einstein constraint equations. 
The data we consider is such that $(\tilde{\mathcal{S}},\tilde{h})$ 
admits an analytic conformal compactification at spatial infinity.
The analyticity of the conformal compactification is not essential for our
studies. This assumption is made for convenience. 
To set initial data for the reduced general conformal field 
equations on $\mathcal{S}=\tilde{\mathcal{S}}\cup\{i\}$ we need to calculate 
from $\tilde{h}$ the following 55 variables: 
$(e_a^{\ \alpha},e_a^{\ 0}, \Gamma_{a\ c}^{\ b}, \Gamma_{a\ 0}^{\ b}, f_a,
\hat{L}_{ab}, \hat{L}_{a0}, E_{ab}, B_{ab})$. 

We introduce compactifying coordinates on $\tilde{S}$ and choose a 
conformal factor $\phi$ such that the rescaled metric $\bar{h}=\phi^2\tilde{h}$
is analytic on $\mathcal{S}$ and 
$\phi|_{i}=0$, $\bar{D}_\alpha\phi|_{i}=0$ and 
$\bar{D}_\alpha \bar{D}_\beta\phi|_{i}=2\,\bar{h}_{\alpha\beta}$, where 
$\bar{D}_\alpha$ is the Levi-Civita derivative on $\mathcal{S}$ defined by 
$\bar{h}$. In these coordinates we find an adapted spatial frame 
$\bar{e}_a^{\ \alpha}$ to $\bar{h}$ such that $\bar{h}(\bar{e}_a,\bar{e}_b)=
\delta_{ab}$. Note that this frame is not unique. We can use any other frame 
$\bar{e}'_a=R_a^{\ b} \bar{e}_b$ where $R_a^{\ b}$ is a rotation matrix such that
the orthonormality requirement $\bar{h}(\bar{e}'_a,\bar{e}'_b)=\delta_{ab}$ 
is fulfilled. 
We can also use different frames on different domains and patch the solution 
together. Having chosen a frame, we calculate the spatial connection 
coefficients $\bar{\Gamma}_{a\ b}^{\ c}$, the spatial Ricci tensor $\bar{r}_{ab}$ 
and the Ricci scalar $\bar{r}$ on $\mathcal{S}$ with respect to $\bar{e}_a$.

For the calculation of initial data we need the following intermediate 
quantities
\be \label{eq:intermed} 
\omega=\frac{2\ \phi}{\sqrt{\,\delta^{ab}\,\bar{D}_{a}\phi\,\bar{D}_b \phi}},
\qquad t_{ab}= \bar{D}_a \bar{D}_b \phi - \frac{1}{3}\,\delta_{ab}\,\delta^{cd}
\bar{D}_c \bar{D}_d \phi, 
\qquad s_{ab} = \bar{r}_{ab} - \frac{1}{3}\,\delta_{ab}\,\bar{r}. \ee
The derivative $\bar{D}_a$ is taken with respect to $\bar{e}_a$, so that for
example 
$\bar{D}_a \phi = \bar{e}_a(\phi) = \bar{e}_a^{\ \alpha}\partial_\alpha \phi$.

We choose a free function $\kappa$ which determines the value of the
time coordinate at $\scri$ and 
the representation of spatial infinity in conformal Gauss coordinates. 
Then we set the initial data for our 55 evolution variables on 
$\mathcal{S}$. The frame components $e^{\ \alpha}_{a}$ are given by the rescaling
$e^{\ \alpha}_{a} = \kappa\,\bar{e}^{\ \alpha}_{a}$ and we set $e^{\ 0}_{a}=0$. 
The spatial connection coefficients $\Gamma^{\ b}_{a\ c}$ are calculated
from the rescaled frame $e_{a}$. In the conformal Gauss gauge, $e_0$ is chosen
orthogonal to the initial hypersurface so the connection components 
$\Gamma^{\ b}_{a\ 0}$ correspond to the second fundamental form on the initial 
surface. It holds $\Gamma^{\ b}_{a\ 0}=0$ due to time reflection symmetry. 
The 1-form $f$ on $\mathcal{S}$ is given such that 
$\tilde{f}|_{\mathcal{S}}=\phi^{-1}d\phi$ which implies 
$f|_{\mathcal{S}}=\kappa^{-1}d\kappa$. Initial data for the Schouten tensor is 
symmetric due to time reflection symmetry and is given by
\be\label{eq:id_L} \hat{L}_{ab} = \kappa^2\, \left( - \frac{1}{\phi} \,t_{ab} +
\frac{1}{12}\,\bar{r}\,\delta_{ab}\right)=\hat{L}_{ba},\qquad\hat{L}_{a0}=0.\ee
The electric and the magnetic parts of the rescaled Weyl tensor read
\be \label{eq:id_eb} E_{ab}=-\kappa^3\,\left(\frac{1}{\phi^2}\,t_{ab} +
\frac{1}{\phi}\,s_{ab}\right), \qquad B_{ab}=0.\ee
\pagebreak

The a priori known variables for (\ref{eq:odes}) are as given in 
(\ref{eq:conf_fac}) and (\ref{eq:dk}) 
\[ \Omega(\tau)=\frac{\phi}{\kappa}\left(1-\tau^2\frac{\kappa^2}{\omega^2}
\right), \qquad d_0(\tau) = -2\,\tau\,\frac{\kappa\,\phi}{\omega^2}, \qquad 
d_a = \bar{e}_a^{\ \mu}\partial_\mu\phi.\]

The choice $\kappa=1$ corresponds to the point compactification at spatial 
infinity. We have seen in (\ref{eq:singular}) that the Weyl tensor blows up 
near spatial infinity with $1/r^3$ if spatial infinity is represented as a 
point. As seen in (\ref{eq:id_eb}), choosing $\kappa\sim r$ compensates for
this singular behavior. This choice leads to a representation of spatial 
infinity as a cylinder depicted in Fig.~\ref{fig:point-cylinder}.

\section{The Cauchy problem in spherical symmetry}\label{sec:ss}
A first step in numerical work with the reduced general conformal field 
equations is the spherically symmetric case. From the analytic work of 
Friedrich \cite{Friedrich03} we know that the conformal Gauss gauge covers 
the complete Schwarzschild-Kruskal solution. Our studies in \ref{cgg_ss}
show that the gauge is robust enough for numerical calculations. 
Therefore, we can expect that the Cauchy problem for the reduced general 
conformal field equations in spherical symmetry with initial data from the 
Schwarzschild-Kruskal spacetime can be solved numerically without major
difficulties.

We set the frame such that $e_1$ shows in the radial direction. 
Due to spherical symmetry we can assume that the frame matrix has the form
\[ e^{\ \mu}_{k} = \left( \begin{array}{cccc}
1 & 0 & 0 & 0 \\
e^{\ 0}_{1} & e^{\ 1}_{1} & 0 & 0 \\
0 & 0 &  e_2^{\ 2} & 0 \\ 
0 & 0 & 0 &  e_2^{\ 2}/\sin \vartheta 
\end{array} \right),\]
where the coefficients depend on $(t,r)$ only. In spherical symmetry 
the reduced general conformal field equations simplify considerably. 
If we are interested only in the single non-vanishing component 
of the Weyl tensor $E_{11}$, we just need to solve the following 3 coupled 
ordinary differential equations for $\Gamma^{\ 2}_{2 \ 0},\hat{L}_{22}$ and 
$E_{11}$ which decouple from the rest of the system
\begin{eqnarray*}
\partial_0 \Gamma^{\ 2}_{2 \ 0} &=& - (\Gamma^{\ 2}_{2 \ 0})^2 + \hat{L}_{22} + 
\frac{1}{2} \Omega \, E_{11},\\
\partial_0 \hat{L}_{22} &=&  - \Gamma^{\ 2}_{2 \ 0}\, \hat{L}_{22} + 
\frac{1}{2} d_0\, E_{11}, \\
\partial_0 E_{11} &=& - 3 \, \Gamma^{\ 2}_{2 \ 0} \, E_{11}. 
\end{eqnarray*}
The degeneracy of the equations at the set $\mathcal{I}^+$ is not present in 
this simple case.

We calculate the initial data using isotropic coordinates as in section 
\ref{cgg_ss} and in \cite{Friedrich98}. We set $\Gamma^{\ 2}_{2\ 0}|_{\tau=0}=0$
and choose $\kappa=\rho$. The initial data for the Schouten tensor reads
\[ \hat{L}_{22}|_{\tau=0} = - \frac{ m \rho}{(1+m \rho /2)^2}. \]
For the electric part of the Weyl tensor we get $E_{11}|_{\tau=0} = - 2 m$.
The conformal factor reads 
\[ \Omega = \frac{\rho}{(1+m \rho/2)^2} 
\left( 1- \frac{\tau^2}{(1+m \rho /2)^2} \right).\]
The one form $d_k$ becomes
\[ d_0 = - 2 \tau \frac{\rho}{(1+m \rho /2)^4}, \quad 
d_1 = \frac{2 \rho}{(1+m \rho /2)^3}.\]
The solution to the Cauchy problem generates the Schwarzschild-Kruskal 
spacetime in the conformal Gauss gauge. The resulting conformal diagram is
similar to the ones presented in \ref{cgg_ss}. 

To check the solution, we compare the Weyl tensor 
in the frame representation adapted to the conformal geodesics 
constructed in \ref{cgg_ss} in the given background 
of the Schwarzschild-Kruskal spacetime with the Weyl tensor 
that we get from the solution to the Cauchy problem 
described above. For this comparison, we need to calculate 
the timelike coframe $\sigma^i_{\ \mu}$ for the conformal 
geodesics satisfying the orthonormality relations
\be \label{eq:onm}e_i^{\ \mu}\sigma^j_{\ \mu}= \delta_i^{\,j} \quad 
\mathrm{and} \quad \sigma^i_{\ \mu}\sigma^j_{\ \nu}\,\eta_{ij}=g_{\mu\nu}. \ee
In a frame adapted to spherical symmetry we get
\[\sigma^0_{\ 0}=\frac{e_1^{\ 1}}{e_1^{\ 1}e_0^{\ 0}-e_0^{\ 1}e_1^{\ 0}}, \qquad 
\sigma^0_{\ 1} = - \frac{e_1^{\ 0}}{e_1^{\ 1}e_0^{\ 0} - e_0^{\ 1}e_1^{\ 0}}.  \]
The rescaled Weyl tensor in a conformal Gauss gauge on the 
Schwarzschild-Kruskal background is then calculated by
\[ - E_{11}= W^0_{\ 101} = \frac{1}{\Omega}\,C^0_{\ 101} = 
\frac{1}{\Omega} \ \sigma^0_{\ \mu} e^{\ \nu}_{1} e^{\ \lambda}_{0}
e^{\ \rho}_{1}\, C^\mu_{\ \nu\lambda\rho}.\] On the background of the
Schwarzschild-Kruskal spacetime in the conformal Gauss gauge using
isotropic coordinates we calculate
\begin{eqnarray*} 
C^0_{\ 101} &=& \frac{2 m \rho^3 (1-m \rho/2)^2}{(1+m
  \rho/2)^8}\,\sigma^0_{\ 1} \, ( - e_0^{\ 1}( e_1^{\ 0})^2 +
e_0^{\ 0}e_1^{\ 0}e_1^{\ 1}) + \\ &+& \frac{2 m }{\rho (1+m
  \rho/2)^2}\, \sigma^0_{\ 0}\, ( - e_0^{\ 1} e_1^{\ 0}e_1^{\ 1} +
e_0^{\ 0}(e_1^{\ 1})^2). 
\end{eqnarray*} 
This background calculation agrees with the result of the numerical
solution to the Cauchy problem up to converging numerical errors.

The calculation carried out above is the first numerical calculation
of an entire, asymptotically flat black hole spacetime including
spacelike, null and timelike infinity and the region close to the
singularity \cite{Zeng06a}.  Difficulties related to the
mathematical representation of spatial infinity and numerical
resolution loss in the neighborhood of timelike infinity prevented
earlier attempts of other authors to cover the entire
Schwarzschild-Kruskal solution (see \cite{Frauendiener04,Schmidt02}).

While numerical tests of tetrad formulations in spherical symmetry can be 
instructive \cite{Buchman05}, the simplicity of the reduced general
conformal field equations in spherical symmetry does not allow us to draw 
representative conclusions for the general case. Still, the study shows that
global numerical calculations are possible with this system.
\pagebreak

\section{Regular data at spatial infinity as a point}
\label{sec:point_data}
We want to include gravitational radiation into our discussion in a simple
setting. We would like to study numerically a spacetime that allows us to 
represent spatial infinity both as a point and as a cylinder. 
The point representation is regular only for vanishing ADM-mass. 

To any asymptotically flat, 
conformally non-flat, static vacuum solution one can construct conformally 
related initial data for a spacetime with vanishing ADM-mass 
but a non-vanishing radiation field along $\scri$ \cite{Friedrich88}. 
In this section we will see how such data can be calculated.
\subsection{Radiative solutions with vanishing ADM-mass}
An asymptotically flat, static solution $\tilde{g}$ to the Einstein vacuum 
field equations can be written in coordinates 
adapted to a hypersurface-orthogonal, timelike Killing field as
\be\label{eq:as_flat} \tilde{g} = - v^2 dt^2 + \tilde{h}, \qquad v=v(x^\alpha),
\qquad \tilde{h}= \tilde{h}_{\alpha\beta}dx^\alpha dx^\beta,\ee
where $v$ is the norm of the Killing field that in these coordinates takes the 
form $\partial_t$. The induced positive definite metric on the hypersurfaces 
$\tilde{\mathcal{S}}_t=\{t=\textrm{const.}\}$ is denoted by $\tilde{h}$. 
The Einstein vacuum field equations reduce to the static vacuum field equations
on $\tilde{\mathcal{S}}\equiv\tilde{\mathcal{S}}_t$
\be \label{eq:static} \tilde{r}_{\alpha\beta}[\tilde{h}] = 
\frac{1}{v}\,\tilde{D}_\alpha\tilde{D}_\beta v, \qquad \tilde{h}^{\alpha\beta}
\tilde{D}_\alpha\tilde{D}_\beta v = 0. \ee
In addition to the asymptotic flatness condition (\ref{eq:as_flat_h}) for
$\tilde{h}$ we require $v\to 1$ as $\tir\to\infty$.

Beig and Simon show in \cite{Beig80} that if the ADM-mass of an 
asymptotically flat, static spacetime does not vanish, the conformally rescaled
metric $\bar{h}$ in a suitable conformal and coordinate gauge is not just 
smooth, but analytic at spatial infinity. A suitable conformal gauge is given
by 
\be\label{eq:flat_gauge}\bar{h}=\phi^2\tilde{h}\qquad\mathrm{with}\qquad
\phi=\left(\frac{1-v}{m}\right)^2.\ee
In this conformal gauge, we have $\bar{r}[\bar{h}]=0$ as a consequence of the 
Einstein equations \cite{Friedrich06}. The rescaled metric $\bar{h}$ extends as
an analytic metric to $i$, but the conformal data for the evolution equations 
is still singular at $i$. Following \cite{Friedrich88} we construct from 
$\bar{h}$ conformally related initial data $\tilde{h}_{rad}$ for a 
spacetime with vanishing ADM-mass but a non-vanishing radiation field via
\be\label{radiative}\tilde{h}_{rad}=\sigma^{-2} \bar{h} \qquad \mathrm{with} 
\qquad \sigma=\left(\frac{2}{m}\frac{1-v}{1+v}\right)^2.\ee
The conformal data constructed from $(\bar{h}, \sigma)$ is regular at spatial 
infinity represented as a point $i$ which implies that the ADM-mass of 
$\tilde{h}_{rad}$ vanishes. 
If $\tilde{h}_{rad}$ is not conformally flat, the radiation field along null 
infinity does not vanish as shown in \cite{Friedrich88} and therefore, 
by the positive mass theorem, we conclude that
the initial hypersurface $\mathcal{S}$ can not be complete. Spacetimes 
constructed from such data will probably include naked singularities and do not
represent physically reasonable solutions. They serve, however, 
as a good testbed to study the neighborhood of spatial infinity 
in the presence of radiation. In this thesis we are not interested in the 
singularity and will confine our evolutions in a neighborhood of spatial 
infinity.

The static vacuum field equations (\ref{eq:static}) in the conformal gauge 
(\ref{eq:flat_gauge}) imply with the notation of (\ref{eq:intermed}) and 
$\phi=\frac{\sigma}{(1+\sqrt{\mu \sigma})^2}$ where $\mu=m^2/4$ the relation 
\cite{Friedrich04}
\[ \bar{D}_a \bar{D}_b \sigma - \frac{1}{3}\bar{\triangle}\sigma\,\delta_{ab}+
\sigma \left(1-\mu\sigma\right) \bar{r}_{ab} = 
t_{ab}+\sigma (1-\mu\sigma)s_{ab} = 0.\]
Using the above relation the initial data for the fields 
$\hat{L}_{ab}$ and $E_{ab}$ (\ref{eq:id_L}, \ref{eq:id_eb}) simplifies to
\be \label{eq:id} \hat{L}_{ab} = -\frac{\kappa^2}{\sigma}t_{ab},\qquad E_{ab} =
\frac{\kappa \mu}{1-\mu\sigma} \,\hat{L}_{ab}. \ee
\subsection{Weyl solutions}
In this section we calculate initial data of the type described above
from Weyl solutions. These are asymptotically flat, static,
axisymmetric, vacuum solutions to the Einstein equations
\cite{Weyl17}.  A metric from the class of Weyl solutions can be written in
standard spherical coordinates as
\[ \tilde{g} = - e^{2U}dt^2 + e^{2(K-U)} (d\tilde{r}^2 + 
\tilde{r}^2\,d\vartheta^2+e^{-2K}\tilde{r}^2\sin\vartheta^2 d\varphi^2),\]
where the functions $U$ and $K$ are $\varphi-$independent, 
$U$ satisfies the flat-space Laplace equation, $\triangle U = 0 =
 \left(\frac{1}{r^2} \partial_r(r^2\partial_r) + 
\frac{1}{r^2\sin^2\varphi}\partial^2_\vartheta\right)U$, 
and $K$ is determined from $U$ by quadrature up to an additive constant 
\cite{Bicak00,Stephani03}. Due to the requirement of asymptotic flatness, we 
have $U\to 0$ and $K\to 0$ as $\tir\to\infty$. Weyl solutions are uniquely 
parametrized by the asymptotically flat solutions $U$ to the Laplace equation 
in flat space which can be written in the form
\[ U = \sum_{n=0}^{\infty} a_n \tir^{-(n+1)}P_n(\cos\vartheta), \] 
where the $a_n$ are constants and the $P_n$ are the Legendre polynomials.
We introduce the compactifying coordinate 
$r=1/\tilde{r}$ which maps spatial infinity to the origin. 
The metric $\tilde{h}$ induced by $\tilde{g}$ on a $\{t=\mathrm{const.}\}$ 
hypersurface reads
\[ \tilde{h}=\frac{e^{2(K-U)}}{r^4}\,\left(dr^2+r^2\,d\vartheta^2 +
e^{-2K}r^2 \sin^2\vartheta\,d\varphi^2\right).\]
For Weyl solutions a simple choice for the conformal factor is given by
$\phi' = r^2\,e^{U-K}$ which leads to the analytic metric
$h'=\phi'^2\tilde{h}=dr^2+r^2\,d\vartheta^2+e^{-2K}r^2\sin^2\vartheta\,
d\varphi^2$. By comparing with (\ref{eq:as_flat}) we see that $v=e^U$, 
so that the conformal factor $\phi=\left(\frac{1-e^U}{m}\right)^2$ 
also leads to an analytic metric. To calculate initial data for the reduced 
general conformal field equations, we can use the simple metric $h'$ 
or the metric $\bar{h}=\phi^2 \tilde{h}$ which leads to the simplification 
(\ref{eq:id}). The conformal relations for these metrics are as follows
\begin{eqnarray}\label{eq:curzon} 
&h'=\phi'^{2}\tilde{h},\quad \phi'=r^2 e^{U-K},&\quad
h'=dr^2+r^2\,d\vartheta^2+e^{-2K}r^2\sin^2\vartheta\, d\varphi^2, \nonumber \\
& \bar{h} =\phi^2 \tilde{h},\quad \phi = \left(\frac{1-e^U}{m}\right)^2,&\quad
\bar{h}=(\phi \phi'^{-1})^2 h'=\left(\frac{4\,\sinh^2
\left(\frac{U}{2}\right)\, e^K}{m^2 r^2}\right)^2 h'. \qquad \end{eqnarray}
The physical initial metric induced by the radiative spacetime with vanishing 
ADM-mass is given by
\[ \tilde{h}_{rad} = \sigma^{-2} \bar{h}, \quad \mathrm{with} \quad 
\sigma = \left(\frac{2}{m}\,\tanh\frac{U}{2} \right)^2. \]
We can write it also as
\[ \tilde{h}_{rad}=\sigma^{-2}(\phi\phi'^{-1})^2 h'=\sigma'^{-2}
h',\quad \mathrm{with}\quad \sigma'=\frac{\phi' \sigma}{\phi}=
e^{-K} r^2 \cosh^{-2}\left(\frac{U}{2}\right). \]
The advantage of using $(h',\sigma')$ for calculating initial data
is the simplicity of $h'$. The pair $(\bar{h},\sigma)$ is advantageous because 
in this conformal gauge we have $\bar{r}[\bar{h}]=0$ and the related 
simplification (\ref{eq:id}) in the construction of initial data. 
\subsubsection{The Curzon solution}
The mathematically simplest Weyl solution is the Curzon solution \cite{Curzon} 
which is given by 
\[ U=-\frac{m}{\tilde{r}} ,\qquad K=-\frac{m^2 \sin^2\vartheta}{2\tilde{r}^2}.\]
In the compactifying coordinate $r=1/\tilde{r}$ we have
\[U=-m r,\qquad K=-\frac{1}{2}m^2 r^2\sin^2\vartheta.   \]
For an interpretation of this solution see \cite{Bicak00} and the 
references therein. 

\section{A Cartesian implementation}
\label{sec:cart_rgcfe}
We solve numerically the reduced general conformal field equations 
for a massless, axi-symmetric, asymptotically flat, radiative 
spacetime with initial data from the previous section. 
The code is based on an equidistant grid in Cartesian coordinates
$\{x,y,z\}$ and does not make use of the axisymmetry of the spacetime. 

We expect a singularity in the interior. 
No attempt has been made to study questions on the nature of this singularity, 
such as the (non-)existence of a horizon or  
the precise blow up behavior of fields. 
Our interest lies in a neighborhood of spatial infinity including a 
piece of $\scri^+$. Our aim is to calculate the radiation field along $\scri^+$
and to show that it does not vanish in accordance with the theorem presented in
\cite{Friedrich88}. The spacetime is axisymmetric and therefore 
we may expect that the radiation field in the direction of the axis will 
vanish.

\subsection{The initial data}
As we solve a frame-based system using Cartesian coordinates, we need to choose
a suitable frame in which the initial data for a numerical implementation 
of the reduced general conformal field equations can be calculated as 
described in \ref{sec:id_rcfe}.
For the presentation of an adapted frame I will use the metric $h'$ as the 
formula are shorter. I used the frame $\bar{e}_a$ in the actual calculations 
involving the simplification (\ref{eq:id}) which is related to $e'_a$ by the 
rescaling $\bar{e}_a=(\phi'\phi^{-1})e'_a$. 

Spatial infinity is mapped to the origin of our coordinate system.
The numerical code based on this one-point compactification uses Cartesian 
coordinates which are regular at the origin. A frame adapted 
to the axisymmetry of $h'$ can not be regular at the origin. 
To find a regular frame, we write the metric $h'$ in Cartesian coordinates
$(x,y,z)$
\begin{eqnarray*}
h' &=& dr^2+r^2\,d\vartheta^2+e^{-2K}\,r^2 d\varphi^2 = \\
&=& \frac{x^2+e^{-2 K} y^2}{x^2+y^2}\,dx^2 + \frac{(1-e^{-2 K}) x\,y}
{x^2+y^2}\,2\,dx\,dy + \frac{e^{-2 K} x^2+y^2}{x^2+y^2}\,dy^2 + dz^2.
\end{eqnarray*}
We choose one of the frame vector fields to be $\partial_z$. By choosing 
$e'_1$ to be proportional to $\partial_x$ we get
\begin{eqnarray*} 
e'_1 &=& \sqrt{\frac{x^2+y^2}{x^2+e^{-2 K}y^2}}\,\partial_x, \qquad 
e'_3 = \partial_z, \\
e'_2 &=& \frac{(1-e^{2 K}) x\,y}{\sqrt{(x^2+y^2)(x^2+e^{-2K}y^2)}}\,
\partial_x + \sqrt{\frac{e^{2 K} x^2+y^2}{x^2+y^2}}\,\partial_y.
\end{eqnarray*}
This frame is regular at the origin (remember that we require $K|_{i}=0$). 
Using cylindrical coordinates  $(\rho, z, \varphi)$ 
instead of Cartesian coordinates simplifies certain calculations. 
In cylindrical coordinates the metric reads 
$h'=d\rho^2+dz^2+e^{-2K}\,\rho^2 d\varphi^2$ and the frame above becomes
\begin{eqnarray*} 
e'_1 &=& \frac{1}{\sqrt{\cos^2\varphi+e^{-2K}\sin^2\varphi}}\left(
\cos\varphi\,\partial_\rho-\frac{\sin\varphi}{\rho}\,\partial_\varphi\right),\\ 
e'_2 &=& \frac{1}{\sqrt{\cos^2\varphi+e^{-2K}\sin^2\varphi}}\left(
e^{-K}\sin\varphi\,\partial_\rho+\frac{e^k \cos\varphi}{\rho}\,
\partial_\varphi\right),\quad e'_3 = \partial_z.\\
\end{eqnarray*}
We can do the calculations in cylindrical coordinates and then transform the 
data to Cartesian coordinates by a simple point transformation.

The rest of the calculation is tedious but straightforward and will 
not be presented. I just write the conformal factor because of its importance 
for the analysis and the conformal boundary. For $\kappa=1$ which results 
in the one-point compactification $i\sim i^0$, the conformal factor 
given by (\ref{eq:conf_fac}) becomes for data from the Curzon solution
\be\label{point_conf_fac} \Omega=\sigma\,\left(1-\frac{t^2}{\omega^2}\right)=
\frac{4}{m^2}\,\tanh^2\left(\frac{U}{2}\right)
\left(1-\frac{4 e^{m \left(m\rho ^2+4 r\right)} m^6 r^4 t^2}{\left(1-e^{m r}
\right)^6 \left(1+e^{m r}\right)^2}\right). \ee
Here, $m$ is the ADM-mass of the static Curzon solution. Remember that
the radiative solution that we construct has vanishing ADM-mass.

\pagebreak
\subsection{Form of $\scri^+$ and the computational domain}
\begin{figure}[ht]
  \centering
  \begin{minipage}{0.49\textwidth}
    \centering
    \psfrag{s}{$\mathcal{S}$}
    \psfrag{scri}{$\scri^+$}
    \includegraphics[width=\textwidth,height=0.23\textheight]
		    {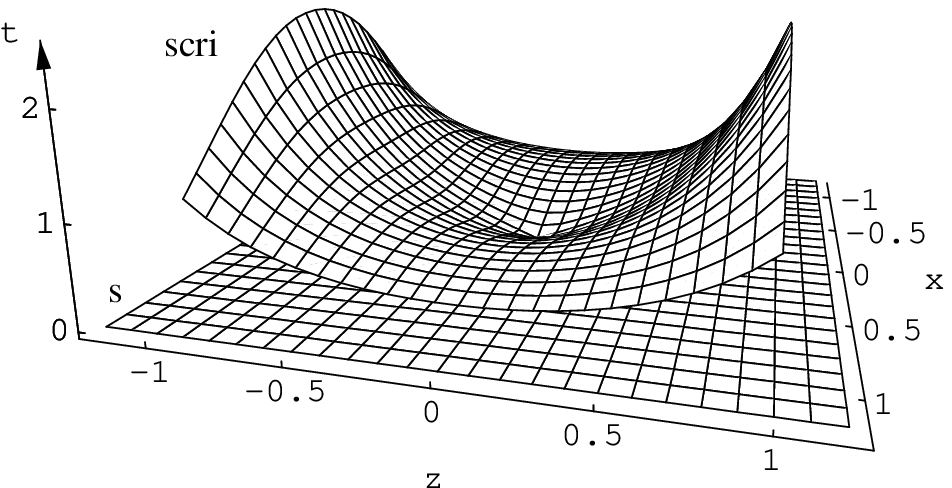}
    \caption{$\scri^+$ plotted on the $xz$-plane.\label{fig:scrixz}}
  \end{minipage}
  \hspace{0.01\linewidth}%
  \begin{minipage}{0.49\textwidth}
    \centering
    \psfrag{s}{$\mathcal{S}$}
    \psfrag{scri}{$\scri^+$}
    \includegraphics[width=\textwidth,height=0.23\textheight]
		    {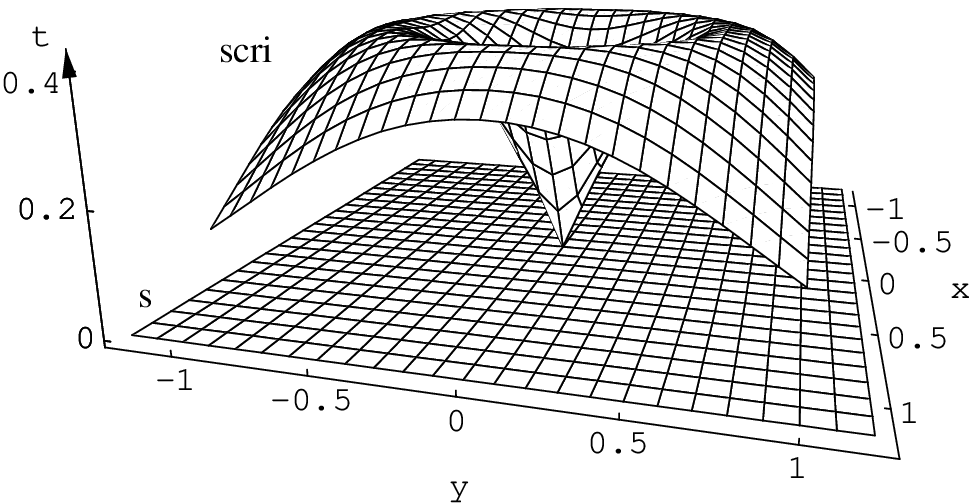}
    \caption{$\scri^+$ plotted on the $xy$-plane.\label{fig:scrixy}}
  \end{minipage}
\end{figure}

The conformal boundary of the spacetime is given by the zero set of
the conformal factor which is explicitly known (\ref{point_conf_fac}).
From this formula we read the value of the time coordinate at
$\scri^+$ depending on the space coordinates as \be\label{point_scri}
t_{\scri^+} = \frac{(1-e^{m r})^3 (1+e^{m r})}{2 m^3 r^2
  e^{\frac{m}{2}\left(m \rho^2+4 r \right)}}. \ee We refer to the
spatial variation of the value of the time coordinate along $\scri^+$
as the "form" of $\scri^+$. In the present compactification, the form
depends on the ADM-mass of the static Curzon metric, $m$. In the code,
we choose $m=2$. The resulting form of $\scri^+$ on the spatial
coordinate domain $[-1,1]$ is shown in Fig.~\ref{fig:scrixz} on the
$xz$-plane with $y=0$ and in Fig.~\ref{fig:scrixy} on the $xy$-plane
with $z=0$. We see that the form is symmetric on the $xy$-plane but
not on the $xz$-plane. The diagrams show an awkward behaviour for a
numerical calculation as the time coordinate of $\scri^+$ decreases
while we move to the interior in the $x$ and $y$ directions.

\begin{figure}[ht]
  \centering
    \psfrag{s}{$\mathcal{S}$}
    \psfrag{scri}{$\scri^+$}
    \includegraphics[width=0.66\textwidth]{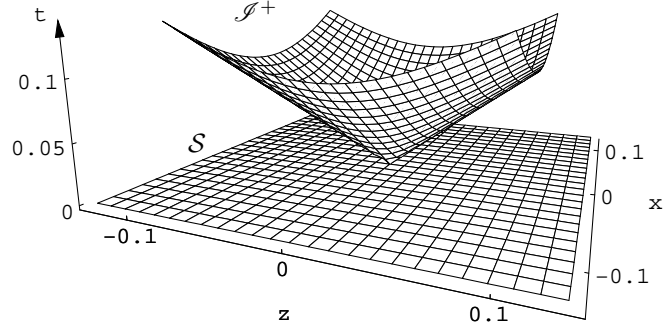}
    \caption{The domain of analysis in a neighborhood of spatial infinity. 
      \label{fig:curzon_scri}}
\end{figure}

This behavior can be changed by a suitable choice of the free function $\kappa$
away from $i^0$. As we are interested only in a neighborhood of spatial 
infinity, however, we will simply confine our analysis to a region 
depicted in Fig.~\ref{fig:curzon_scri}. 

Our computational domain is given by $\{x,y,z\}\in[-0.2,0.2]$ and 
$t\in[0,0.2]$. We restrict the analysis of the numerically generated
spacetime to the domain  $\{x,y,z\}\in[-0.1,0.1]$ and 
$t\in[0,0.1]$. In terms of the physical coordinate $\tilde{r}$, 
this domain corresponds to $\tilde{r}\in[10,\infty]$ on
the initial slice $\mathcal{S}$. Note that we are also able to calculate 
a piece of $\scri^+$ which is infinitely far away from the singularity 
in null directions. This allows us to calculate the Weyl component $\psi_4$ 
in terms of a suitably adapted frame at null infinity as described below.

In our calculations, spatial infinity is included in the numerical domain, 
but it is not on a grid point. The reason is that the numerical evaluation of 
initial data at $i$ causes difficulties  due to formally singular divisions by 
the conformal factor. This can be remedied by taking the limit analytically 
at $i$. For our purposes, however, using a staggered grid where $i$ is between
two grid points works sufficiently well. 
\subsection{The code}
The code for the numerical solution of the reduced general conformal field 
equations with the point compactification at spatial infinity
using Cartesian coordinates is based on the Cactus computational infrastructure
\cite{Cactusweb}. I used MathTensor and Kranc \cite{Husa:2004ip} to generate 
the formula for the initial data and the right hand side calculations. 
For the time integration I used a 4th order Runge-Kutta algorithm.
Spatial derivatives are calculated using second order accurate finite 
differences. For a grid function $\phi(x,y,z)\to\phi_{i,j,k}$, 
second order accurate finite differencing along a direction, 
say along the $x$-direction, is given by  
\be\label{second_der} \partial_x \phi(x,y,z) \ \to \quad D^x_0 \phi_{i,j,k}= 
\frac{1}{(2 \delta x)}\,(\phi_{i+1,j,k}-\phi_{i-1,j,k}).\ee

The reason for using second order accurate derivatives instead of fourth order 
derivatives lies in the boundary treatment. To avoid the presumably naked
singularity in the interior of the spacetime, we use an artificial timelike
inner boundary. In our case, the numerical \emph{outer} boundary, i.e.~the 
outer boundary of the computational domain, lies in the \emph{interior} 
domain of the spacetime. Therefore in the discussion, we refer to the 
computational outer boundary as the inner boundary. 

As the inner boundary of our evolution is not a generic aspect of our
approach, no special numerical inner boundary treatment has been
devised for our system.  The inner boundary is present in this test case
due to the naked singularity in the spacetime under study. In cases
with regular data given on a complete hypersurface
$\mathcal{S}$, we would not have such a timelike boundary. 
Therefore we are not concerned with the problem of 
boundary treatment and set the derivatives simply to zero at the
inner boundary.  This results in large numerical errors in the
interior domain which propagate outwards faster than the grid speed of
the physical characteristics because the treatment is not
well-posed. Second order accurate finite differencing results in a
slower propagation of errors than a fourth order accurate
differencing.  We will see that in the restricted
domain of our analysis the code is second order convergent.

In Fig.~\ref{fig:curzon_scri}, the domain of analysis has been plotted
for the medium resolution. The analysis has been done for two
resolutions: medium and high. The medium resolution has 40 points, the
high resolution has 80 points in each spatial direction on the
computational domain $[-0.2,0.2]$ which implies 20 and 40 points on
the domain of analysis in each spatial direction.
\subsection{The radiation field}\label{sec:plot}
We calculate gravitational radiation represented by the 
absolute value of the Weyl tensor component $\psi_4$ at null infinity
with respect to a suitably adapted Newman-Penrose tetrad in the $x$ and $z$ 
directions.

In an adapted complex Newman-Penrose tetrad $(l,k,m,\bar{m})$ along $\scri^+$
fulfilling the relations (\ref{eq:np_tetrad}), the complex rescaled Weyl tensor
component $\psi_4$ representing the radiation field is given by
\cite{Chandrasekhar83}
\be\label{eq:psi4} \psi_4 = W_{ijkl} l^i m^j l^k m^l. \ee
The advantage of knowing not only the location of $\scri^+$ but also the 
conformal factor in explicit form simplifies the numerical radiation extraction
considerably. 

The vector field $l$ along $\scri^+$ is given by 
$l_i=\nabla_i \Omega = e_i^{\ \mu}\partial_\mu \Omega$. We can calculate
$\partial_\mu \Omega|_{\scri^+}$ a priori from (\ref{point_conf_fac}) or we can
use the known field $d_i=\Omega f_i + \nabla_i \Omega$ from (\ref{eq:dk}).
I chose to use the explicit form of the conformal factor. 
By our gauge conditions we have $e_0^{\ \mu} = \delta_0^{\ \mu}$ 
and therefore also $l_0$ is known a priori. 
We only need to read the values $e_a^{\ \mu}$ from the 
numerically generated solution to calculate the adapted frame vector field $l$,
the rest is known explicitly. 

There is an arbitrary choice involved in the calculation of $\psi_4$ that we fix
in the following way. Given $l_i=(l_0,l_1,l_2,l_3)$ with respect to some
frame $e_i$, we can always find a new frame $e'_{i'}$, so that
$l$ takes the form $l_{i'}=(l_{0'},0,0,l_{0'})$ 
with respect to the new frame $e'_{i'}$. We have
$l_0=l_{0'}=\pm\sqrt{l_1^2+l_2^2+l_3^2}$ as $l$ is null and the rotation is 
purely spatial. The new frame $e'_{i'}$ is related to the original frame 
$e_i$ by $e_0=e'_{0}$ and $e'_{a'}=R_{a'}^{\ b}e_b$. The rotation may be thought
as consisting of two subsequent two-dimensional rotations resulting in 
\begin{displaymath} R_{a'}^{\ b} = \left( \begin{array}{ccc}
\cos\alpha_{12} & -\sin\alpha_{12} & 0 \\
\cos\alpha_{23}\,\sin\alpha_{12} & \cos\alpha_{23}\,\cos\alpha_{12} & 
-\sin\alpha_{23} \\ 
\sin\alpha_{23}\,\sin\alpha_{12} & \cos\alpha_{12}\,\sin\alpha_{23} & 
\cos\alpha_{23} \end{array} \right) 
\end{displaymath}

Assume without loss of generality that $l_2\ne 0$ and $l_3\ne 0$. 
The first rotation is done with the angle 
$\alpha_{12}=\arctan\left(\frac{l_1}{l_2}\right)$ 
on the $12$-plane to achieve $l_{1'}=0$ and the second one with the angle 
$\alpha_{23}=\arctan\left(\frac{\sqrt{l_1^2+l_2^2}}{l_3}\right)$ on the 
$23$-plane to achieve $l_{2'}=0$. The rotation matrix becomes
\begin{displaymath} R_{a'}^{\ b} = \left( \begin{array}{ccc}
\frac{l_2}{\sqrt{l_1^2+l_2^2}} & -\frac{l_1}{\sqrt{l_1^2+l_2^2}} & 0 \\
\frac{l_1\,l_3}{l_0\sqrt{l_1^2+l_2^2}} & \frac{l_2\,l_3}{l_0\sqrt{l_1^2+l_2^2}}
& -\frac{\sqrt{l_1^2+l_2^2}}{l_0} \\ 
l_1/l_0 & l_2/l_0 & l_3/l_0\end{array} \right)\end{displaymath}
We have used $\sqrt{l_1^2+l_2^2+l_3^2}=l_0$.
In this new frame, $m_{i'}$ can be written in accordance 
with the relations (\ref{eq:np_tetrad}) as
$m_{i'}=(0,0,a+i b, i a-b)$ where $i^2=-1$ and $a,b\in\mathbb{R}$ satisfy 
$a^2+b^2=\frac{1}{2}$. In our calculations, we choose $a=0$ and 
$b=\frac{1}{\sqrt{2}}$. The rotation matrix is used to calculate the electric 
and the magnetic parts of the rescaled Weyl tensor in the new frame via
$E_{a'b'} = R_{a'}^{\ a} R_{b'}^{\ b} E_{ab}$ and 
$B_{a'b'} = R_{a'}^{\ a} R_{b'}^{\ b} B_{ab}$.
Using the splitting (\ref{eq:splitting}), the tracefreeness of $E$ and 
$B$ and the definition of $\psi_4$ given in (\ref{eq:psi4}) we see that 
$\psi_4$ is given in the new frame by 
\be\label{psi4_adapted} \psi_4 = (l_0)^2 ( (-E_{1'1'}+E_{2'2'}-2\,B_{2'1'})+
i\,(B_{1'1'}-B_{2'2'}-2\,E_{2'1'})).\ee

We discuss the absolute value of $\psi_4$ calculated from (\ref{psi4_adapted}) 
along $\scri^+$ in the $x$ and $z$ directions in a neighborhood of spatial 
infinity on the domain illustrated by Fig.~\ref{fig:curzon_scri}. 
\begin{figure}[t]
  \begin{minipage}[ht]{0.49\textwidth}
    \centering
    \includegraphics[width=\textwidth]{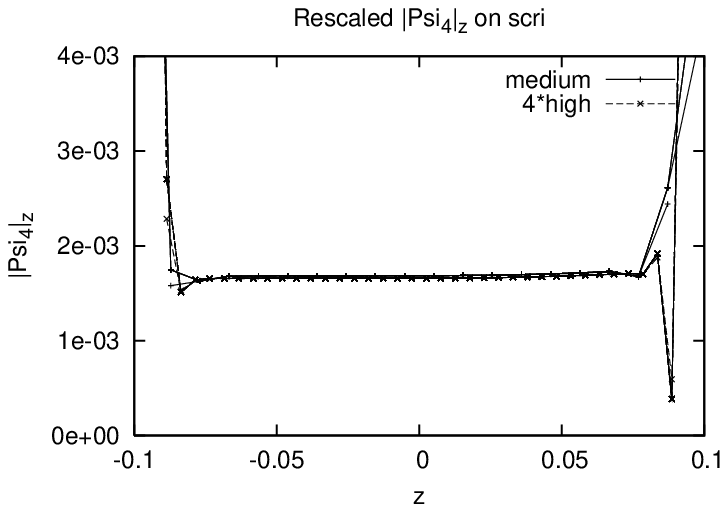}
    \caption{Absolute value of $\psi_4$ along $\scri^+$ in the z-direction 
      rescaled according to resolution. \label{fig:psi4z}}
  \end{minipage}%
  \hspace{0.02\linewidth}%
  \begin{minipage}[ht]{0.49\textwidth}
    \centering
    \includegraphics[width=\textwidth]{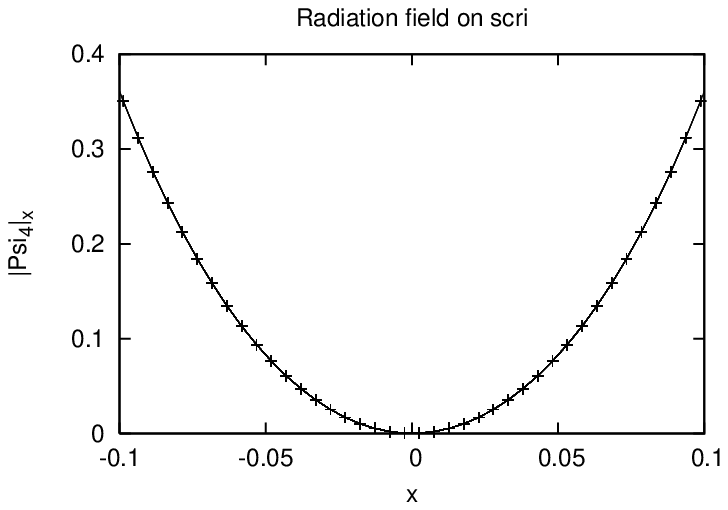}
    \caption{Absolute value of $\psi_4$ along $\scri^+$ in the x-direction 
      representing radiation. \label{fig:psi4x}}
  \end{minipage}
\end{figure}
While we did not make explicit use of the axisymmetry of the spacetime in the 
code, we expect that it makes itself manifest in the radiation field. 
The radiation field in the direction of the axis along $\scri^+$ should vanish.
This can be seen nicely in Fig.~\ref{fig:psi4z} which shows the absolute value 
of $\psi_4$ in the direction of the axis, the $z$-direction in our coordinates,
along $\scri^+$ in two different resolutions. We see that the absolute value 
of $\psi_4$ vanishes up to numerical errors with the expected
order of accuracy. The curve from the high resolution run has been multiplied 
with $2^2=4$ and lines up quite accurately with the curve from the medium 
resolution run which implies that the code is second order accurate 
in a neighborhood of spatial infinity. We also see that errors coming from the 
inner boundary destroy the convergence at a $z$-coordinate distance 
of about $0.08$ to spatial infinity along $\scri^+$. This problem 
can be dealt with by devising a better numerical boundary treatment, 
but we are interested in the radiation field only in a neighborhood
of spatial infinity and for this purpose the code is good enough.

The radiation field along the $x$-direction has been plotted in 
Fig.~\ref{fig:psi4x} for the high resolution run. 
To see the effect of the inner boundary on this plot, 
we need to go further in the direction of the source.
We see that the radiation field grows monotonously as we move away 
from spatial infinity along $\scri^+$. The source of radiation is 
presumably the naked singularity in the interior of the spacetime.

We note that the simplification of radiation extraction over earlier forms 
of conformally regular field equations is two-fold. Firstly, we know the 
location of $\scri^+$ a priori by (\ref{point_scri}) and therefore do not need 
to find numerically the zero set of the conformal factor (note though that the 
zero set is also known in a $\scri^+$-fixing gauge in the context of a 
hyperboloidal initial value problem \cite{Frauendiener98b}). 
Secondly, we know the explicit form of the conformal factor 
(\ref{point_conf_fac}) and not only its zero set.
This allows us to write down analytic expressions for the derivatives 
$\partial_\mu\Omega|_{\scri^+}$, which would have to be calculated numerically 
on $\scri^+$ if the conformal factor would be one of the evolution variables. 
A disadvantage over a $\scri^+$-fixing gauge is that the location of
the extraction surface is not fixed on the grid.
\section{Numerical implementation of frames on $S^2$}\label{sec:gzps}
The numerical simulation presented in the previous section is based on
the assumption of a regular point compactification at spatial infinity
and cannot be generalized to include the more interesting cases with
non-vanishing ADM-mass.  For a generalizable numerical code, we need
to implement the equations in a gauge in which spatial infinity is
represented by a cylinder.  The cylinder implies a spherical geometry,
because its construction and the resulting regularity of the conformal
data distinguishes the direction towards the cylinder in a substantial
way as will be discussed in the next section. The main difficulty with
implementing a spherical grid geometry is that the sphere can not be
covered by a single coordinate chart in a non-singular manner.

One possibility to deal with this problem is to use pseudo-spectral methods 
which are promising due to their high accuracy and low memory requirements 
\cite{Boyd00,Fornberg98}. These techniques apply spectral methods for spatial 
derivatives and method of lines for time integration. 
In pseudo-spectral methods, the expected singular behavior of variables 
at coordinate singularities can be dealt with by an appropriate choice of 
function spaces which respect certain parity properties
\cite{Bartnik99,Boyd00,Fornberg98,Kidder04}. 

The pseudo-spectral approach does not seem to be appropriate for our problem 
for two reasons. The first reason is the nature of our initial data. 
We would like to avoid effects from the interior where presumably a naked
singularity resides and want to study only a small neighborhood of spatial 
infinity. Pseudo-spectral methods, however, are not local methods
in contrast to finite differencing methods. One might introduce some 
localization by using spectral elements instead of global basis 
functions. This roughly corresponds in the language of finite differences 
to high order accurate stencils \cite{Boyd00,Fornberg98}. 
Remember, however, that in the previous section 
we chose a second order accurate finite differencing stencil to 
separate, as much as possible, the study of a neighborhood of spatial infinity 
from the influence of the inner boundary.
It seems desirable for the studies we are interested 
in to keep the localization that finite difference methods offer. The second
reason for not working with pseudo-spectral methods is that we do not only need
to deal with the coordinate singularity at the poles but also with the 
singularity of the frame because $S^2$ can not be covered by a single frame in a
regular way (see \cite{Beyer-phd} for a treatment of $S^3$).

A natural solution to the problem described above
is to cover the sphere in a non-singular manner using multiple charts
on which different frames can be defined. 
It is not a coincidence that this idea which underlies the concept of a 
manifold has found its way into numerical relativity. 

To my knowledge, there are currently two approaches in numerical relativity 
in the context of an evolution problem based on finite differences that use 
multiple coordinate patches on a sphere.
They differ by the way neighboring patches are organized. 
The "penalty method" uses touching patch boundaries over which 
characteristic information for a first order hyperbolic system is exchanged.  
Penalty terms drive ingoing modes of one patch to the outgoing modes of a 
neighboring patch \cite{Diener05,Lehner2005a,Schnetter06a}. 
The "ghost-zones method" uses overlapping grids with redundant 
points at the boundaries of the patches, called ghost points, where  
information from the interior of a neighboring patch is interpolated 
to \cite{Reisswig06,Thornburg00,Thornburg04}. As the equations we are 
interested in are naturally written in first order symmetric hyperbolic form
we can use both methods. In this thesis, the second approach will be followed. 
The reasons for this choice are rather circumstantial than fundamental. 
\subsection{Coordinates of \texttt{GZPatchSystem}}
The ghost-zones method has been implemented by Jonathan Thornburg in a code 
called the \texttt{GZPatchSystem}. The acronym \texttt{GZ} stands for ghost 
zone. I used a version of \texttt{GZPatchSystem} based on the Cactus 
computational infrastructure \cite{Cactusweb} and the Carpet driver 
\cite{carpet}. 
In the following we will discuss certain properties of \texttt{GZPatchSystem} 
that are relevant for the implementation of a frame-based evolution system.
Information on further details can be found in the references 
\cite{Thornburg00,Thornburg04}.

From different ways to cover a sphere with non-singular coordinate
charts,\linebreak 
\texttt{GZPatchSystem} uses the ``inflated cube'' coordinates with 6 
patches as described below. Another notable possibility is given by two 
stereographic coordinate charts \cite{Bishop96CCM}. Due to large 
coordinate distortions in stereographic coordinates along patch boundaries, 
however, these seem to result in less accurate codes \cite{Reisswig06}. 

In the following discussion, it will be convenient to think of the
spheres \mbox{$r=\mathrm{const}.$} as embedded into $\mathbb{R}^3$
with its standard global Cartesian coordinates $\{x,y,z\}$. \linebreak
\texttt{GZPatchSystem} uses 3 angular coordinates denoted by $\mu,\nu$
and $\varphi$ to cover $S^2$ regularly which correspond to the
rotation angles around the Cartesian coordinate axes.  Their relation
to the Cartesian coordinates is given by
\[ \mu(x,y,z) = \arctan(y/z), \quad \nu(x,y,z) = \arctan(x/z), \quad 
\varphi(x,y,z) = \arctan(y/x). \]
We enumerate the patches as: $0,1,2,3,4,5\to +z,+x,+y,-x,-y,-z$. 
We use the pair $(\nu,\varphi)$ on the $\pm x$- respectively $(1,3)$-patches, 
$(\mu,\varphi)$ on the $\pm y$- respectively $(2,4)$-patches and $(\nu,\mu)$ on
the $\pm z$- respectively $(0,5)$-patches such that the local coordinates on 
each patch are regular covering a neighborhood of the Cartesian axes. 
The internal coordinates of \texttt{GZPatchSystem} denoted by $(\rho,\sigma)$ 
are defined on the interval $[-\pi/4,\pi/4]$. 
This implies the following translations for the coordinates on the patches 
\begin{eqnarray*}
&P_{\pm x}\equiv P_{(1,3)}: (\nu,\varphi) = (\rho\mp \pi/2,\sigma), \quad 
& P_{+z}\equiv P_{(0)}: (\nu,\mu) = (\rho,\sigma), \\
&P_{\pm y}\equiv P_{(2,4)}: (\mu,\varphi) = (\rho\mp \pi/2,\sigma\mp \pi/2),\quad
& P_{-z}\equiv P_{(5)}: (\nu,\mu) = (\rho - \pi,\sigma - \pi).
\end{eqnarray*}
The standard polar coordinates $(\vartheta, \varphi)$ on the unit sphere 
are related to $(\nu,\mu)$ via
\[ \tan\vartheta(\nu,\mu) = \sqrt{\tan^2\mu+\tan^2\nu}, \quad 
\tan\varphi(\nu,\mu) = \tan\mu/\tan\nu. \]
The standard metric on $S^2$ in polar coordinates reads 
$ds^2=d\vartheta^2+\sin\vartheta\,d\varphi^2$. In local patch 
coordinates $(\rho,\sigma)$ this metric takes the form
\be\label{eq:rhosigma} ds^2= \frac{1}{(1-\sin^2\rho\,\sin^2\sigma)^2}\,
\left(\cos^2\sigma\,d\rho^2-2\sin\sigma\cos\sigma\sin\rho\cos\rho\,d\rho\,
d\sigma+\cos^2\rho\,d\sigma^2\right).\ee

The main task of \texttt{GZPatchSystem} is the interpolation of a specified set
of functions from the interior of neighboring patches into each patch's 
interpatch boundary ghost zones. We refer to this process as "synchronizing" 
the interpatch boundary ghost zones in accordance with \cite{Thornburg04}. 
On each patch the local coordinates are used to define 
the tensor basis. To communicate the information between different patches, one
needs to do a a point transformation and a tensor transformation of the field
variables. As we use a frame basis, we do not need the tensorial coordinate 
transformation except for the vector fields that constitute the frame. Instead
we need to implement the rotation of frame-based variables. Later, we will 
refer to the rotation of the variables as being part of the synchronizing 
process. First we discuss the choice of frames on the unit sphere.

\subsection{Choice of frames on $S^2$}\label{sec:s2frames}
The sphere can neither be regularly covered by a single coordinate chart nor
by a single frame field. The coordinate singularity is independent of the 
singularity of the frame. Any frame field on the unit sphere in standard 
coordinates $(\vartheta,\varphi)$ can be written as
\[e_1=\cos\alpha\,\partial_\vartheta-\frac{\sin\alpha}{\sin\vartheta}\, 
\partial_\varphi, \qquad e_2=\sin\alpha\,\partial_\vartheta-
\frac{\cos\alpha}{\sin\vartheta}\,\partial_\varphi, \]
where $\alpha=\alpha(\vartheta,\varphi)$ is an angle of rotation. 
We see that at least one of the frame vector fields becomes singular at 
$\vartheta=0,\pi$ independent from
$\alpha$ and therefore independent from the choice of the frame. For 
$\alpha=\pi/2$ we recover the standard frame on the sphere given by 
$e_\vartheta=\partial_\vartheta$ and  
$e_\varphi = \frac{1}{\sin\vartheta}\,\partial_\varphi$. In this choice, the 
singularity of the frame coincides with the singularity of the coordinates 
so that the frame vector field $e_\varphi$ becomes singular at the center of the
patches $(0,5)$. 

Given coordinates $\{x^1,x^2\}$ and a metric $g$ on $S^2$, 
we can use the one-parameter rotation freedom to adapt a frame to a 
coordinate direction so that one of the frame components vanishes. 
Choosing arbitrarily $e_1^{\ 2}=0$ we get by orthonormality relations 
up to reflections
\be\label{adapted} e_1^{\ 1}=\frac{1}{\sqrt{g_{11}}},\quad e_1^{\ 2}=0,\qquad
e_2^{\ 1} = \sqrt{\frac{g_{11}}{\det g}},\quad 
e_2^{\ 2} = -\frac{g_{12}}{\sqrt{g_{11}\det g}}. \ee
Using the above relations we can set the frame according to the metric given
in local patch coordinates by (\ref{eq:rhosigma}). It turns out that the
standard frame $(e_\vartheta,e_\varphi)$ in the inflated cube coordinates is 
adapted so that one of the frame components vanishes on the patches (1,2,3,4).
In terms of local patch coordinates the standard frame on $S^2$ is given by
\begin{eqnarray*}
e_\vartheta &\overset{(1,3)}{=}& 
\frac{1-\cos^2\nu\sin^2\varphi}{\cos\varphi}\,\partial_\nu \overset{(2,4)}{=} 
\frac{1-\cos^2\mu\cos^2\varphi}{\sin\varphi}\,\partial_\mu, \\
e_\varphi &\overset{(1,3)}{=}& \sqrt{1-\cos^2\nu\sin^2\varphi}\,\left(
-\tan\varphi\cos\nu\,\partial_\nu+\frac{1}{\sin\nu}\,\partial_\varphi\right)
\overset{(2,4)}{=} \\&\overset{(2,4)}{=}& \sqrt{1-\cos^2\mu\cos^2\varphi}\,
\left(\frac{\cos\mu}{\tan\varphi}\,\partial_\mu+
\frac{1}{\sin\mu}\,\partial_\varphi\right).
\end{eqnarray*}
On the patches $(0,5)$ the standard frame is not adapted and becomes singular. 
There we choose a different frame given by 
\begin{eqnarray*}
e_1 &\overset{(0,5)}{=}&\frac{1}{\cos\nu}(1-\sin^2\mu\sin^2\nu)\,\partial_\mu,\\
e_2 &\overset{(0,5)}{=}& -\sqrt{1-\sin^2\mu\sin^2\nu}\,
\left(\frac{1}{\cos\mu}\,\partial_\nu+\sin\mu\tan\nu\,\partial_\mu\right).
\end{eqnarray*}
It is convenient to have the frames written in global Cartesian coordinates
\[
e_\vartheta=\frac{1}{r\,\sqrt{x^2+y^2} r}\left( xz \,\partial_x+yz\,\partial_y-
(x^2+y^2)\,\partial_z\right), \quad e_\varphi = \frac{1}{\sqrt{x^2+y^2}}
\left(-y \,\partial_x + x\,\partial_y\right), \]
\be\label{frame_cart} 
e_1 = \frac{1}{r\,\sqrt{x^2+z^2}}\left( xy\,\partial_x+(x^2+z^2)\,\partial_y-
yz\,\partial_z\right), \qquad e_2 = \frac{1}{\sqrt{x^2+z^2}}
\left(-z \,\partial_x + x\,\partial_z\right).\ee
Note that we have chosen only two frames on $S^2$, namely the standard frame
$(e_\vartheta,e_\varphi)$ on the (1,2,3,4) patches and an adapted frame 
$(e_1,e_2)$ on the (0,5) patches. This means that we need the rotation
implied by the different frames only between the 2 polar patches and the 4 
equatorial patches in contrast to the coordinate transformations which need
to be done between each pair of neighboring patches. 
The number of frame systems to cover $S^2$ should not be relevant for the 
accuracy of finite differencing in contrast to the number of coordinate charts 
because finite difference operators act along the coordinate lines and not 
along the integral curves of the frame vector fields.
\subsubsection{Rotation of the frames}
To communicate information on evolution variables between different patches, 
we need to use the rotation between the frames 
$e_{\bar{A}} = R_{\bar{A}}^{\ B} e_{B}$. The rotation matrix can be written as
\[ R_{\bar{A}}^{\ B}=\left( \begin{array}{cc} 
\cos\alpha & -\sin\alpha \\ \sin\alpha & \cos\alpha \end{array}\right). \]
Using the orthonormality relation in (\ref{eq:onm}), 
the rotation matrix can be calculated by
\be\label{rotation} R_{\bar{A}}^{\ B}=
R_{\bar{A}}^{\ C}e_{C}^{\ \mu}\sigma_{\ \mu}^{B}= 
e_{\bar{A}}^{\ \mu} \sigma_{\ \mu}^{B}.\ee
There are different possibilities to implement this calculation and the 
transformation of frame-based grid functions between neighboring patches.
We shall discuss this question on a simple example in the next subsection.
\subsection{The eigenvalue equation for the Laplace-operator on $S^2$}
\label{sec:test_lapl}
It is instructive to discuss the numerical implementation of frames using
\linebreak \texttt{GZPatchSystem} on a simple example that does not involve time
evolution. We check the code by verifying the following eigenvalue equation for
the Laplace-operator on an embedded sphere of radius $r=\sqrt{x^2+y^2+z^2}$ 
\[\triangle Y_{lm} = -\frac{l(l+1)}{r^2}\,Y_{lm}, \]
We use this relation on a three dimensional domain to test the numerical 
implementation of frames on a spatial slice. 
We calculate the Laplacian of $Y_{lm}$ in first order and second order form. 
We write $\phi:=Y_{lm}$ for notational simplicity. 
The Laplacian of $\phi$ can be calculated by
\be\label{second_order} \triangle \phi= 
\delta^{ab}\left(e_a^{\ \alpha}e_b^{\ \beta}\,\partial_\alpha\partial_\beta\phi
+e_a^{\ \alpha}\,(\partial_\alpha e_b^{\ \beta}) \partial_\beta 
\phi-\Gamma_{a\ b}^{\ c}e_c^{\ \beta}\partial_\beta \phi\right). \ee
This formula includes second order derivatives of $\phi$. 
Therefore we refer to the above calculation as the second order form. 
Another option is to define auxiliary variables $\phi_a:=e_a(\phi)=e_a^\alpha
\partial_\alpha \phi$. Then the Laplacian can be written in first order form as
\be\label{first_order} \triangle \phi= \delta^{ab}\left(e_a^{\ \alpha}
\partial_\alpha \phi_b-\Gamma_{a\ b}^{\ c} \phi_c\right). \ee
The main steps in the calculation of the Laplacian of the spherical harmonics 
using \texttt{GZPatchSystem} in first order form (\ref{first_order}) are as 
follows: 
\begin{itemize}
\item Fix the geometry.\\ 
We fix the geometry by setting the frame fields
$(e_{\vartheta},e_\varphi)$, $(e_1,e_2)$, the radial frame vector
field $e_r$ and the related connection coefficients
$\Gamma_{a\ b}^{\ c}$. It is convenient to use Cartesian coordinates
for the input (\ref{frame_cart}). The radial frame vector field is the same on
each patch.
\[ e_r = e_3 = \frac{1}{r}(x\,\partial_x + y\,\partial_y+z\,\partial_z).\]
The connection coefficients $\Gamma_{\vartheta\ \vartheta}^{\ r}=
\Gamma_{\varphi\ \varphi}^{\ r}=\Gamma_{1\ 1}^{\ 3}=\Gamma_{2\ 2}^{\ 3}
= \frac{1}{r}$ are the same for both frame fields. The only non-vanishing 
connection coefficient that depends on the choice of our frame is
\[ \Gamma_{\varphi\ \vartheta}^{\ \varphi}=\frac{z}{r\,\sqrt{x^2+y^2}}, \qquad 
\Gamma_{2\ 1}^{\ 2} = -\frac{y}{r\,\sqrt{x^2+z^2}}.\]
\item Set the function $\phi=Y_{lm}$.\\
We set the function for some $l$ and $m=0,1,\dots,l$. 
We give the real part of the spherical harmonics as initial data 
in global Cartesian coordinates. We set, up to the normalization factor, 
\[ Y_{10}=\frac{z}{r}, \qquad  \Re(Y_{11})=\frac{x}{r}, \qquad 
Y_{20}=-1+\frac{3z^2}{r^2},\ \dots\]
\item Transform the data to local patch coordinates.\\
We let \texttt{GZPatchSystem} transform the grid functions to each patch's 
local coordinates and coordinate basis. 
Only a point transformation is needed for the spherical harmonics and the 
connection coefficients. For the frame fields a point transformation 
as well as a transformation of the coordinate basis needs to be done. 
These transformations are implemented in \texttt{GZPatchSystem} and are
steered by the user via interface and parameter files. 

\item Calculate the auxiliary variables $\phi_a$.\\
We calculate $\phi_a=e_a^{\ \alpha}\partial_\alpha \phi$ 
in local coordinates on each patch using fourth order accurate discrete 
derivatives as in (\ref{discrete_der}). For the numerical approximation of 
$\partial_\alpha \phi$ using fourth order accurate derivatives on each nominal 
grid point, the values of the function on two neighboring points in each 
direction along the coordinate line of $x^\alpha$ is needed.
We do not calculate the derivatives on the ghost zones. 

Fig.~\ref{fig:gzps_grid} illustrates a two-dimensional non-trivial stencil 
geometry for overlapping grids with two ghost points.
The thick line is the common boundary of adjacent 
patches denoted by $P$ and $\bar{P}$. The solid lines are the coordinate lines 
of the patch $P$, the dashed lines are the coordinate lines of the patch 
$\bar{P}$. Note that the patches share the coordinate perpendicular to their 
common boundary. The ghost zone of $P$ is in the interior of $\bar{P}$.
For clearness the ghost zone of $\bar{P}$ has not been drawn. 
The filled small circles correspond to grid points of $P$, 
the empty ones correspond to grid points of $\bar{P}$. 
No distinction has been made in the circles between nominal and ghost points.
Say, we want to calculate the derivative of $\phi$ in patch $P$ on the boundary
line at the grid point denoted by $n$ in the figure. 
Beside the nominal points $n-2,n-1,n$, we need the values of 
$\phi$ at $n+1$ and $n+2$.
When initial data has been set correctly on the whole grid including the ghost 
points, the derivatives $\partial_\alpha \phi$ can be calculated on the nominal
points.
\begin{figure}[ht]
  \centering
  \psfrag{n-2}{$n-2$}
  \psfrag{n-1}{$n-1$}
  \psfrag{n}{$n$}
  \psfrag{n+1}{$n+1$}
  \psfrag{n+2}{$n+2$}
  \psfrag{p}{$P$}
  \psfrag{bp}{$\bar{P}$}
  \includegraphics[width=0.75\textwidth,height=0.25\textheight]
  {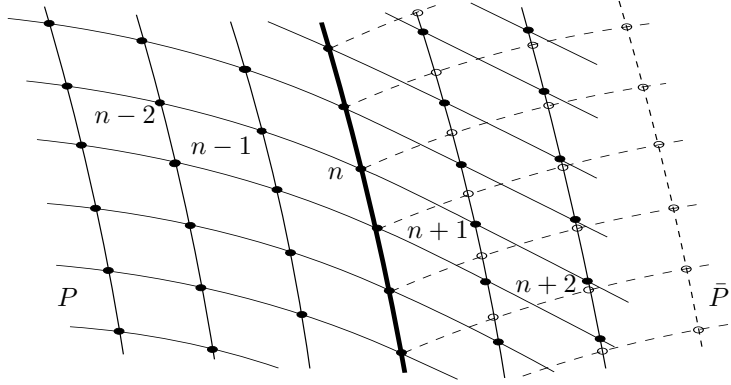}
  \caption{Stencil geometry with overlapping grids.\label{fig:gzps_grid}}
\end{figure} 
\item Synchronize the auxiliary variables.\\ As we will derive the
  auxiliary variables for the calculation of the Laplacian in first
  order form (\ref{first_order}), we need their correct values also on
  the ghost zones. These are not known, neither in patch $P$ nor in
  patch $\bar{P}$, but they can be calculated by interpolation from
  $\bar{P}$.  This calculation is done in \texttt{GZPatchSystem} using
  one-dimensional interpolation in the direction parallel to the
  boundary.  In Fig.~\ref{fig:gzps_grid} \texttt{GZPatchSystem} uses
  the empty circles in the nominal domain of $\bar{P}$ along the two
  coordinate lines parallel to the thick boundary line to calculate
  the values of the auxiliary variables on the full circles. 
\item Rotate the auxiliary variables.\\
For the calculation of the auxiliary variables on the ghost zone of patch $P$, 
we need a transformation in accordance with the rotation of the frame. We set
$\phi_A = R_A^{\bar{B}} \phi_{\bar{B}}$ where the rotation matrix $R_A^{\bar{B}}$ 
is calculated as in (\ref{rotation}). For this calculation we have two options:
\begin{itemize}
\item Analytic calculation: We calculate the rotation matrix  analytically 
using the explicitly known frames $(e_\vartheta,e_\varphi)$ and $(e_1,e_2)$
in local coordinates.
\item Numeric calculation: We use the algebraic relation (\ref{rotation})
numerically.
\end{itemize}
The code for the analytic calculation includes switches by patch number 
which determines the coordinates and by grid side 
which determines the rotation matrix (each of the 6 patches has 4 sides).
The code for the numeric calculation on the other hand is very short. 
The solution of (\ref{rotation}) can be written independent of local 
coordinates, so the second option results in a code which is more 
efficient than the input of the analytic rotation matrix. 
This method can also be generalized to the case in which the frame 
components are evolution variables and the rotation matrix is not known 
a priori. We choose the numeric calculation.
\item Build the Laplacian (\ref{first_order}).
\item Calculate the error by
\[\mathrm{lapl\_error}=
\triangle\phi-\left(-\frac{l(l+1)}{r^2}\phi\right).\]
\end{itemize}
The calculation in low resolution is made with 20 points, the medium one with
40 points and the high one with 80 points in both angular directions on each
patch. The radial domain of calculation is $r\in[1,2]$.

\begin{table}[ht]\centering
  \begin{tabular}{|cc|ccc|ccc|}\hline
    & & & low-med & & & med-high  & \\\hline
    $l$ & $m$ & (0,5) & (1,3) & (2,4) & (0,5) & (1,3) & (2,4) \\
    \hline
    
    \hline
    
    \hline
    1 & 0 & 3.96 & 3.96 & 3.96 & 3.99 & 3.99 & 3.99 \\\hline
    1 & 1 & 3.96 & 3.96 & 3.96 & 3.99 & 3.99 & 3.99 \\\hline
    2 & 0 & 3.94 & 3.95 & 3.95 & 3.99 & 3.99 & 3.99 \\\hline
    2 & 1 & 3.95 & 3.95 & 3.95 & 3.99 & 3.99 & 3.99 \\\hline
    2 & 2 & 3.95 & 3.94 & 3.94 & 3.99 & 3.99 & 3.99 \\\hline
    3 & 0 & 3.92 & 3.94 & 3.94 & 3.98 & 3.98 & 3.98 \\\hline
    3 & 1 & 3.93 & 3.93 & 3.94 & 3.98 & 3.98 & 3.98 \\\hline
    3 & 2 & 3.93 & 3.93 & 3.93 & 3.98 & 3.98 & 3.98 \\\hline
    3 & 3 & 3.94 & 3.92 & 3.93 & 3.98 & 3.98 & 3.98 \\\hline
  \end{tabular}
  \caption{Convergence factors for the calculation of the Laplacian in second 
    order form. \label{tab:conv_ylm_second}}
\end{table}

In Table \ref{tab:conv_ylm_second} convergence factors are listed
for the second order form of the Laplacian. They are calculated in the 
$L_2$-norm for $l$ up to $3$ on a sphere with radius $r=1.5$.
The convergence factors are independent of the sphere on 
which they have been calculated. For their calculation I used the code 
\texttt{reduce} by Christian Reisswig available from 
\cite{Cactusweb} including documentation. The factors have been 
given for pairs of patches that use the same coordinate system.
We see a clean fourth order convergence. 
Note that for the calculation of the Laplacian in second order form, 
we do not need any synchronization.

\begin{table}[ht]\centering
  \begin{tabular}{|cc|ccc|ccc|}\hline
    & & & low-med & & & med-high  & \\\hline
    $l$ & $m$ & (0,5) & (1,3) & (2,4) & (0,5) & (1,3) & (2,4) \\
    \hline
    
    \hline
    
    \hline
    1 & 0 & 3.33 & 3.02 & 3.02 & 3.17 & 2.96 & 2.96 \\\hline
    1 & 1 & 3.00 & 3.33 & 3.00 & 2.96 & 3.17 & 2.96 \\\hline
    2 & 0 & 3.75 & 4.06 & 4.06 & 3.51 & 3.61 & 3.61 \\\hline
    2 & 1 & 2.82 & 2.84 & 2.88 & 2.90 & 2.91 & 2.92 \\\hline
    2 & 2 & 4.09 & 3.79 & 3.79 & 3.58 & 3.46 & 3.46 \\\hline
    3 & 0 & 3.31 & 3.21 & 3.21 & 3.18 & 3.12 & 3.12 \\\hline
    3 & 1 & 3.23 & 3.21 & 3.42 & 3.14 & 3.11 & 3.23 \\\hline
    3 & 2 & 3.04 & 3.30 & 3.30 & 3.02 & 3.17 & 3.17 \\\hline
    3 & 3 & 3.27 & 3.23 & 3.20 & 3.14 & 3.13 & 3.12 \\\hline
  \end{tabular}
  \caption{Convergence factors for the calculation of the Laplacian in first 
    order form. \label{tab:conv_ylm_first}}
\end{table}

Table \ref{tab:conv_ylm_first} lists convergence factors for the first order 
form of the Laplacian. We see that the calculation is roughly third order 
accurate. At first sight it seems that a higher number of grid points results 
in a drop of convergence which would be problematic. 
A closer look, however, reveals that the convergence factor approaches 
consistently 3 when we increase the resolution as seen for example in the case 
with $l=2$ and $m=1$. The study of point-wise convergence factors using gnuplot
or \texttt{reduce} shows that the convergence on the ghost points is 
third order which dominates in the $L_2$-norm. 

The reason seems to be the interpolation of spatially derived functions 
and not the rotation. This claim can be tested by synchronizing the derivatives
$\partial_\alpha \phi$ as covariant tensors instead of synchronizing the 
frame-based auxiliary variables $\phi_a$ as functions. In that case no rotation
is needed, but the spatial derivation $\partial_\alpha \phi$ is interpolated
and derived again for the calculation of the Laplacian. One observes a similar
behavior of the convergence factors as in Table \ref{tab:conv_ylm_first}.

In an evolution scheme, a grid function that is derived on a time step is not 
interpolated because synchronization takes place immediately 
after each evolution step in the \texttt{MoL\_PostStep} schedule bin of 
Cactus (see \cite{Cactusweb} for a documentation on schedule bins of Cactus). 
The next subsection discusses a test case with a simple evolution 
system where we see fourth order convergence as expected. The issue with the
interpolation of numerical derivatives will not play any role in our later 
discussions. 
\subsection{Wave equation with source terms}
As a simple test for an evolution system, we solve the wave equation on 
a Minkowski background for a scalar function $\phi$ with source $f$
\[ \Box \phi = f. \]
Writing out the equation in terms of frames we get
\[ \Box \phi = \eta^{ij} \nabla_i \nabla_j \phi = \eta^{ij} \nabla_i e_j(\phi) = 
\eta^{ij} [ e_i\, (e_j(\phi)) - \Gamma^{\ k}_{i \ j} e_k(\phi) ] = f \]
The above system can be brought into first order form by using 
auxiliary variables $\phi_k:=e_k(\phi)$, which obey an integrability 
condition $\nabla_j\phi_k-\nabla_k\phi_j=0$. Take $e_0$ to be a timelike 
frame field and $e_a$ to be spacelike frame fields. 
The evolution system reads in terms of the frame and 
the connection coefficients
\begin{eqnarray*}
e_0(\phi_0)-\delta^{ab}e_a(\phi_b)&=& - \delta^{ab}\,\Gamma^{\ 0}_{a\ b}\, \phi_0 
+ (\Gamma^{\ c}_{0\ 0} - \delta^{ab}\,\Gamma^{\ c}_{a\ b})\, \phi_c - f, \\
e_0(\phi_a)-e_a(\phi_0) &=&  \Gamma^{\ 0}_{0\ a} \phi_0 + 
(\Gamma^{\ c}_{0\ a} - \Gamma^{\ c}_{a\ 0})\, \phi_c, \\
e_0(\phi)&=&\phi_0.
\end{eqnarray*}
For the tests we use the Minkowski background in standard spherical coordinates
with $e_3$ taken to be the radial direction. The wave equation takes the form
\begin{eqnarray}\label{test_wave}
\partial_t\,\phi_0 &=&  e_1^{\ 1}\partial_1\phi_1  + e_1^{\ 2}\partial_2\phi_1 + 
e_2^{\ 1}\partial_1\phi_2 + e_2^{\ 2}\partial_2\phi_2 + \partial_3\phi_3 + 
\Gamma^{\ 2}_{2\ 1}\phi_1 + \frac{2}{r}\phi_3 - f, \nonumber\\
\partial_t\,\phi_1 &=&  e_1^{\ 1}\partial_1\phi_0 + e_1^{\ 2}\partial_2\phi_0, 
\nonumber \\
\partial_t\,\phi_2 &=&  e_2^{\ 1}\partial_1\phi_0 + e_2^{\ 2}\partial_2\phi_0, 
\nonumber \\
\partial_t\,\phi_3 &=&\partial_3\phi_0, \nonumber \\
\partial_t \phi &=& \phi_0.
\end{eqnarray}
An explicit solution to this symmetric hyperbolic system is given with the
source term $f= -\frac{l(l+1)}{r^2}\phi$, by $\phi(t,r,\vartheta,\varphi) = 
\phi(t,r)\, Y_{lm}(\vartheta,\varphi)$ with
\be\label{wave_sol}  
\phi(t,r) = \frac{1}{2 r}((r + t)e^{-\frac{(r + t)^2}{2 \sigma^2}} +
(r - t) e^{-\frac{(r - t)^2}{2 \sigma^2}}), \ee
where $\sigma$ is some constant. The evolution consists of the following steps:
\begin{itemize}
\item Fix the geometry.\\ We set the frame fields and the connection
  coefficients as in \ref{sec:test_lapl}.
\item Set initial data.\\
The initial data consists of $\phi(0,x,y,z)$ and 
$\phi_i(0,x,y,z)=(e_i^{\ \mu}\partial_\mu\phi(t,x,y,z))|_{t=0}$ calculated from 
the explicit solution (\ref{wave_sol}) in Cartesian coordinates. Note that 
the input for the auxiliary variables $\phi_A$ depends on the frame and 
therefore also on the patch number.
\item Transform the data to each patch's local coordinate system. 
\item Calculate the right hand side of the system (\ref{test_wave}) and apply 
boundary conditions for integration using method of lines.\\
No boundary conditions are needed in the angular directions. 
The boundary conditions we apply in the radial direction are the same as for 
the reduced general conformal field equations. We set the derivatives at the 
outer boundary to zero. 
\item Do an evolution step.
\item Analyze the numerical solution.\\
Our analysis consists of building the difference between the numerical 
solution and the analytic solution given by (\ref{wave_sol}).
\item Synchronize the auxiliary variables $\phi_i$. \\
This step is discussed in some detail below. 
\item Calculate the right hand sides and apply boundary conditions.
\item Do an evolution step \ldots
\end{itemize}
Due to its importance in the code, we go through the steps included in
the synchronization. We use the frame components to illustrate the
steps involved.  We write a two-dimensional frame field $e_A$ in
angular coordinates $x^\Lambda$ on a patch $P$ with grid points $n$ as
$^{P}\!e_A^{\ \Lambda}(x_n)$. The frame, the coordinates, the
coordinate basis, and the grid points of a neighboring patch $\bar{P}$
will be denoted by barred indices. A synchronization step transforms
the information from a neighbouring patch into the current patch
$^{P}\!e_A^{\ \Lambda}(x_n)\leftarrow\,^{\bar{P}}\negthinspace
e_{\bar{A}}^{\ \bar{\Lambda}}(\bar{x}_{\bar{n}})$.  The steps of the
synchronization are (compare Fig.~\ref{fig:gzps_grid})

\begin{centering}
\vspace{1mm}
\begin{tabular}{l r @{$\leftarrow$} l}
\vspace{1mm}
Copying into current patch: &
$^{P}\!e\ $ & $\ ^{\bar{P}}\!e$\\\vspace{1mm}
Point transformation: &
$e(x)\ $ & $\ e(\bar{x})$ \\\vspace{1mm}
Transformation of coordinate basis: &
$e^{\ \Lambda}\ $ & $\ e^{\ \bar{\Lambda}}$\\\vspace{1mm}
Interpolation: &
$e(\bar{x}_n)\ $ & $\ e(\bar{x}_{\bar{n}})$\\\vspace{1mm}
Frame rotation: & 
$e_A\ $ & $\ e_{\bar{A}}$
\end{tabular}\\
\end{centering}
The steps except the rotation are implemented by Thornburg in
\texttt{GZPatchSystem}.

There are three main sources of numerical errors in our evolution system 
that can be controlled to some degree: 
the round-off, the interpolation and the finite differencing. 
The convergence factors presented below show that round-off and 
interpolation errors are negligible compared to finite differencing errors. 
Of course, this can only be expected in a high enough resolution and a higher 
interpolation order than the order of finite differencing. 
For the tests presented below a 6th order interpolation has been chosen.

Table \ref{tab:conv_wave} shows convergence factors for the numerical 
solution of the symmetric hyperbolic wave equation (\ref{test_wave}). 
The source terms have $l=2$, $m=1$ and $l=3, m=2$. 
The convergence factors in Table \ref{tab:conv_wave} have been calculated 
on a sphere with $r=1.5$.

\begin{table}[ht]\centering
  \begin{tabular}{|c|cc|cc|}\hline
    & $l=2$, & $m=1$ & $l=3$, & $m=2$ \\ \hline
    time   &  low-med & med-high & low-med & med-high \\ \hline
    0.0125 &          & 4.000    &         & 4.001    \\ \hline
    0.025  &  3.995   & 3.999    & 4.003   & 3.999    \\ \hline
    0.0375 &          & 3.998    &         & 3.998    \\ \hline
    0.05   &  3.994   & 3.997    & 3.999   & 3.994    \\ \hline
    0.0625 &          & 3.995    &         & 3.990    \\ \hline
    0.075  &  3.992   & 3.994    & 3.994   & 3.987    \\ \hline
    0.0875 &          & 3.992    &         & 3.984    \\ \hline
    0.1    &  3.974   & 3.990    & 3.981   & 3.981    \\ \hline
  \end{tabular}
  \caption{Convergence factors for the wave equation averaged over 6 patches.
    \label{tab:conv_wave}}
\end{table}
We see fourth order convergence. Tests with other choices of
parameters result in the same qualitative convergence behavior with
slightly different numerical factors.

\section{Implementation of the cylinder at infinity}\label{sec:cylinder}
In the previous section we discussed the principles of a frame-based evolution
code using overlapping grids. In this section we will apply this technique 
in a local study of spatial infinity represented as a cylinder. 
The spacetime under study is the same as in \ref{sec:cart_rgcfe}. 
We use a spherical grid topology instead of a Cartesian one. 
The code is designed to be usable also in studies of spacetimes with 
non-vanishing ADM-mass.

We make some remarks on the structure of the reduced general conformal
field equations with respect to the cylinder at infinity.  The system
consist of ordinary differential equations and the Bianchi equation
for the rescaled Weyl tensor that implies a symmetric hyperbolic
system as in (\ref{eq:sym_hyp}). In coordinates $x^0=t$, $x^\Lambda,
\Lambda=1,2$, and $x^3=r$ we can write the Bianchi equation in the
form
\[ (A^t \partial_t + A^r \partial_r + A^\Lambda \partial_\Lambda) u + F u = 0.\]
A rescaling with $\kappa|_{i}=0$, $d\kappa|_{\scri^+}\ne 0$ leads to the 
blow-up of spatial infinity to a cylinder $\mathcal{I}$ 
which acts as a boundary surface to our evolution equations. 
This does not imply an initial boundary value problem 
as the boundary $\mathcal{I}$ is totally characteristic 
in the sense that $A^r=0$ on $\mathcal{I}$ \cite{Friedrich98,Friedrich02} .
We get interior symmetric hyperbolic equations on $\mathcal{I}$  
and no prescription of boundary data is required or allowed.
The solution is determined uniquely by Cauchy data. 

The blow-up of spatial infinity takes care 
of the singular behavior of conformal initial data for non-vanishing ADM-mass
(\ref{eq:singular}), such that we get a regular finite initial value
problem near spatial infinity.
This delicate interplay between the regularization of the conformal data for 
the field equations and the blow-up procedure of spatial infinity is the main
reason for insisting on a spherical grid topology for our numerical 
calculations. The procedure distinguishes radial and angular directions in a 
substantial way. The smoothness of conformal data and the property that 
$\mathcal{I}$ is a totally characteristic surface is related to the vanishing 
of the radial frame vector field on the cylinder. 
We can implement radial and angular frame vector fields 
also in Cartesian coordinates, so that the radial and angular directions 
are distinguished geometrically near the totally characteristic surface 
$\mathcal{I}$. The points of the cylinder, however, would not correspond to 
points of our grid and this might require a more complicated numerical boundary
treatment. The massless case studied in this section might still
be numerically stable, but the code might not be applicable in studies of 
physically interesting spacetimes.

Another remark concerns the time evolution. As described in \cite{Friedrich02},
certain entries of the matrix valued function $A^t$ vanish on the set 
$\mathcal{I}^+$ where null infinity meets spatial infinity. This degeneracy
of the evolution equations is the main difficulty in clarifying the open 
problems regarding the regular finite initial value problem near spatial 
infinity. It causes also numerical difficulties. 
Remember that we build $(A^t)^{-1}$ in the calculation of the right hand side.
This calculation is singular at $\mathcal{I}^+$ where the matrix $A^t$ is 
degenerate. The main feature that allows us to deal with this problem is that
the location of $\mathcal{I}^+$ is known a priori. A suitable
choice of time stepping and freezing the evolution in the unphysical 
domain takes care of this issue, at least in the massless case that
we study. Whether this treatment of the problem is sufficient in cases
with non-vanishing mass remains to be seen.
\subsection{The initial data}
For numerical calculations with \texttt{GZPatchSystem} we do not map spatial 
infinity to $r=0$ as we did in the Cartesian 
case. Instead we map spatial infinity to a finite coordinate radius. 
The reason is that spherical coordinates are not well-defined at the origin.
Therefore \texttt{GZPatchSystem} assumes that the computational domain is a 
shell bounded by two spheres with non-vanishing radial coordinate values.

We compactify the physical Weyl solution such that the interval 
$\tilde{r}\in [0,\infty)$ is mapped to $r\in (1,\infty)$ 
via the coordinate transformation
\[\tilde{r}(r)=\frac{1}{r-1},\qquad r(\tilde{r})=\frac{1}{\tilde{r}}+1. \]
It might better reflect the physical relations to map spatial infinity to some 
large radius and change the sign in the coordinate transformation 
so that large values of the compactifying radial coordinate 
correspond to the far field zone away from the source of radiation. 
In our case, however, we are only interested in a
neighborhood of spatial infinity and not in the source. Therefore we have made 
the above choice of a compactifying radial coordinate.

The conformally rescaled metric $\bar{h}$ in these compactifying coordinates 
takes the form (compare (\ref{eq:curzon}))
\[ \bar{h} = \phi^2 \tilde{h} = \left( \frac{4\,\sinh^2\frac{U}{2}\ 
e^K}{m^2 (r-1)^2}\right)^2 (d r^2+ (r-1)^2\,d\vartheta^2+
e^{-2K} (r-1)^2\sin^2\vartheta\, d\varphi^2).\]
The Curzon solution from which we calculate conformal data is given by 
$U=-m(r-1)$ and ${K=-\frac{1}{2}m^2 (r-1)^2 \sin^2\vartheta}$. 
The metric $\bar{h}$ is analytic at spatial infinity so it can be extended 
beyond  $r=1$. We can use the conformal factor 
$\sigma=\left(\frac{2}{m}\,\tanh\frac{U}{2} \right)^2$ as before to generate
initial data from $(\bar{h},\sigma)$.
Another option is to use the frame adapted to $h'$ and the conformal factor 
$\sigma'=(r-1)^2\,e^{-K}/\cosh(\frac{U}{2})$. I have done the calculation 
presented below using the pair $(h',\sigma')$.

A frame adapted to $h'=(\phi'\phi^{-1})^{-2}\bar{h}$ reads in spherical and 
Cartesian coordinates as
\begin{eqnarray*} 
e'_\vartheta&=&\frac{1}{r-1}\,\partial_\vartheta = \frac{1}{(r-1)\sqrt{x^2+y^2}}\,
\left( xz\,\partial_x + yz \partial_y - (x^2+y^2)\,\partial_z\right), \\
e'_\varphi&=&\frac{e^{K}}{(r-1)\sin\vartheta}\,\partial_\varphi = \frac{r\,e^K}
{(r-1)\sqrt{x^2+y^2}}\,\left( -y\,\partial_x + x \partial_y\right),\\
e'_r &=& \partial_r = \frac{1}{r}\,(x\,\partial_x+y\,\partial_y+z\,\partial_z).
\end{eqnarray*}
The frame $(e_\vartheta,e_\varphi)$ is used on the equatorial patches 
$(1,2,3,4)$. On the polar patches $(0,5)$ around the $\pm z$-axis, we choose a 
different frame $(e'_1,e'_2)$ which is adapted the to local coordinates of the
$(0,5)$-patches in the sense of section \ref{sec:s2frames}. The frame
$(e'_1,e'_2)$ is given in Cartesian coordinates by
\begin{eqnarray*}
e'_1 &=& \frac{e^{K} \sqrt{x^2+y^2}}{(r-1) f}\left((y^2+z^2)\, \partial_x +
 x y\, \partial_y + x z \,\partial_z \right),\\
e'_2 &=& \frac{r}{(r-1)f\sqrt{x^2+y^2}}((1-e^{2K})x y z\,\partial_x + 
(e^{2K}x^2+y^2)\,\partial_y + (x^2+y^2) y\, \partial_z),
\end{eqnarray*}
where $f=\sqrt{y^2(y^2+z^2)+x^2(y^2+e^{2K}z^2)}$. This frame is by construction
regular at $x=0,y=0$. While the input of the frame into \texttt{GZPatchSystem} 
is made in Cartesian coordinates, the calculation of initial data using this 
representation is difficult because none of the components of the frame vector
fields vanish and all coordinates $\{x,y,z\}$ appear many times in the 
expressions. Instead, for the calculation of the conformal initial data 
we use the frame written in spherical polar coordinates as
\begin{eqnarray*} 
e'_1 &=&\frac{r^2\,e^K}{(r-1) f}\left(\cos\varphi\sin\vartheta\cos\vartheta\,
\partial_\vartheta - \sin\varphi\,\partial_\varphi\right), \\
e'_2 &=& \frac{r^2}{(r-1) f}\left(\sin\varphi\sin\vartheta\,
\partial_\vartheta - e^{2K}\cos\varphi\cos\vartheta\,\partial_\varphi
\right),\end{eqnarray*} with 
$f=r^2\sin\vartheta \sqrt{e^{2K}\cos^2\varphi\cos^2\vartheta+\sin^2\varphi}$.
Components of the frame $(e'_1,e'_2)$ in this coordinate representation 
become singular at the poles given by $\vartheta=0,\pi$ and these 
points are part of the patches $(0,5)$, but this is a coordinate singularity of
the coordinates $(\vartheta,\varphi)$ and not a singularity of the frame 
as discussed in section \ref{sec:s2frames} on a simple example.
We do not evaluate the above expressions on the patches $(0,5)$. 
We use it for the analytic calculation of the initial data as described in 
\ref{sec:id_rcfe}. After the calculation has been done, 
we transform the data to regular Cartesian coordinates for the input. 
Because the frame is geometrically regular on the patches $(0,5)$, the 
calculated data will also be regular.
One should reorganize the result of the calculation such that no formal
divisions by the coordinates $x$ and $y$ appear (with formal division I mean
expressions of essentially the form $\frac{1}{1/x}$). Divisions by $z$ 
will appear, but they cause no problems because on the patches $(0,5)$ we 
have $z\ne 0$. That such a reformulation can be done is due to the regularity 
of the data. It is a lengthy but a useful test to calculate 
the conformal data using the frames $(e'_\vartheta,e'_\varphi)$ and $(e'_1,e'_2)$
and to check that the rotation of frame-based tensors agrees with the rotation 
of the frames on the initial hypersurface.

Another point concerning the input of initial data into the evolution code is
the divisions by $\phi$ appearing in the construction of initial data 
(\ref{eq:id_L}, \ref{eq:id_eb}). Though the data extends analytically through 
$\mathcal{I}^0$, a straightforward evaluation of the input causes problems. 
Therefore, we evaluate the formula analytically at $\mathcal{I}^0$. 

The calculation of the full data set described in \ref{sec:id_rcfe} results 
in long expressions that will not be given as they are not relevant for our 
discussion. 
\subsection{Form of $\scri^+$ and the computational domain}
Looking at the initial data for the frame $e'_a$ we observe that the components
of the angular frame field $e'_A$ become singular at $r=1$ which 
corresponds to spatial infinity $i$. The rescaling $e_a=\kappa e'_a$ with 
$\kappa\sim (r-1)$ results in the blow-up of the point $i$ to a sphere 
$\mathcal{I}^0$. It leads to regular angular frame vector fields while the 
frame vector field in the radial direction vanishes on $\mathcal{I}^0$.

\begin{figure}[ht]
  \centering
  \psfrag{t}{$t$}
  \psfrag{theta}{$\vartheta$}
  \psfrag{scri}{$\scri^+$}
  \psfrag{I}{$\mathcal{I}$}
  \psfrag{r}{$r$}
  \psfrag{s}{$\mathcal{S}$}
  \includegraphics[width=0.7\textwidth,height=0.17\textheight]
  {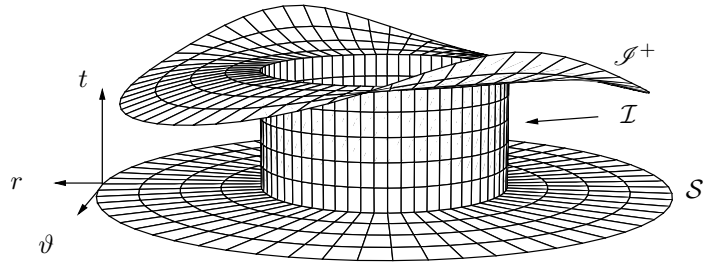}
  \caption{The cylinder at infinity and the form of $\scri^+$ for 
    $\kappa=(r-1)$. \label{fig:badscri}}
\end{figure}

A simple choice for $\kappa$ motivated by the regularization of conformal data
would be $\kappa = (r-1)$. Fig.~\ref{fig:badscri} shows the resulting 
coordinate representation of the cylinder, $\mathcal{I}$, 
and null infinity, $\scri^+$. 
The value of the time coordinate on $\scri^+$ is seen to depend on the angular 
coordinate $\vartheta$ in a way that seems bad for numerical calculations. 
It decreases as we move away from the cylinder in certain directions so that 
future null infinity is running backwards in grid time.
This behavior leads to singularities in the numerical solution 
once the conformal geodesics approach $\scri^+$.

\begin{figure}[t]
  \centering
  \psfrag{t}{$t$}
  \psfrag{theta}{$\vartheta$}
  \psfrag{scri}{$\scri^+$}
  \psfrag{I}{$\mathcal{I}$}
  \psfrag{r}{$r$}
  \psfrag{s}{$\mathcal{S}$}
    \includegraphics[width=0.7\textwidth,height=0.17\textheight]
    {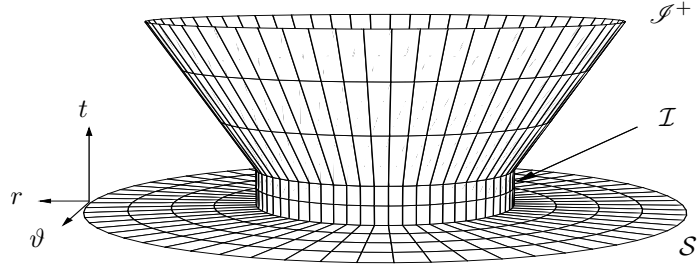}
    \caption{The cylinder at infinity and the form of $\scri^+$ for 
      $\kappa=\frac{\omega}{k (r-1)+d}$. \label{fig:goodscri}}
\end{figure}

We have complete control over the form of $\scri^+$. A good choice in
the asymptotic region that we studied also in \ref{sec:cgg_background}
seems to be \be\label{eq:goodscri} \kappa=\frac{\omega}{k\,(r-1)+d}
\qquad \Rightarrow \qquad t_{\scri^+}=k\,(r-1)+d, \ee with positive
real parameters $k$ and $d$. The parameter $d$ determines the height
of the cylinder at $r=1$, the parameter $k$ determines the slope of
$\scri^+$ (see Fig.~\ref{fig:goodscri}).

\pagebreak
\subsection{The code}
The code is based on the Cactus infrastructure \cite{Cactusweb} and uses the 
thorn \texttt{GZPatchSystem} \cite{Thornburg04}. We use second order 
finite differencing and fourth order Runge-Kutta time integration. No 
dissipation terms have been added to the evolution equations. 

We solve the reduced general conformal field equations numerically on the 
radial computational domain $r\in[1,1.2]$. The time domain is given by 
$t \in [0,t_{\scri^+}+ \triangle t]$. We stop the evolution 
after $\scri^+$ has been reached and one time step into the unphysical 
spacetime has been calculated. 
A reason for stopping the evolution is that the numerical solution 
in the unphysical region becomes singular near the set $\mathcal{I}^+$ 
where spatial infinity meets null infinity. Another reason is 
computational efficiency. We are not interested in the domain beyond $\scri^+$.

The fact that a large part of our field equations consist of 
ordinary differential equations plays a simplifying role in the implementation 
of the spherical grid geometry. 
We need to implement the rotation of frame-based variables, but only for those
for which numerical derivatives need to be calculated. 
These are the 10 components of the rescaled Weyl tensor. 
We do not need to implement the rotation for the remaining 45 variables. 
This leads to a simplification in the synchronization step. 
If the evolution of the frame components and the
connection coefficients is governed by partial differential equations (such
as for the Friedrich-Nagy system \cite{Friedrich99}), the rotation might cause 
difficulties with \texttt{GZPatchSystem} because for the 
derivative of the transformation of the connection coefficients we would need 
to derive synchronized frame variables. 

As the geometry is not given a priori, the rotation matrix is
calculated after each evolution step from the evolved frame
components. We proceed in this calculation as follows:
\begin{itemize}
\item After an evolution step and before the synchronization we store the 
components of the evolved frame field of the local patch in a temporary grid 
function.
\item We synchronize, without rotation, the 10 components of the rescaled Weyl 
tensor, and also the frame field although no numerical derivatives of the frame
field need to be calculated. 
\item After the synchronization step, the temporary grid function has the 
information on the local frame field while the synchronized frame field 
corresponds to the frame field of the neighboring patch. By using this 
information we calculate the rotation matrix on the ghost zone via 
(\ref{rotation}).
\item We rotate the components of the rescaled Weyl tensor as covariant 
two-tensors. After this step we can build derivatives and proceed with the 
evolution.
\end{itemize}

We have two disjoint numerical boundaries in contrast to the evolution
with point compactification at spatial
infinity.  The inner boundary in the physical domain (the outer
boundary of the computational domain) is as in the Cartesian case an
artificial timelike surface.  Our treatment consists of freezing the
evolution at this artificial boundary which is not a well-posed
treatment.  We emphasize again that in the physically interesting
cases no artificial boundary will be present. As this inner boundary
is not a general feature of the method we develop, we are not
concerned with this issue.

The boundary which was not present in the Cartesian case is given by the 
cylinder at infinity. The cylinder is the natural boundary to our evolution 
system. It is a characteristic surface so that no boundary data are to be 
prescribed. Further, as it is totally characteristic, no radial derivatives 
need to be calculated. Therefore no one-sided differencing needs to
be implemented on the cylinder. For second order accurate finite differencing 
(\ref{second_der}) also the next point to the cylinder does not require any 
special treatment.
\subsection{The radiation field}
The form of $\scri^+$ in our choice of $\kappa$ 
does not depend on angular coordinates as seen in Fig.~\ref{fig:goodscri}. 
Therefore, on $\scri^+$, the angular derivatives of the conformal factor 
vanish.

\begin{figure}[ht]
  \centering
  \includegraphics[width=0.5\textwidth]{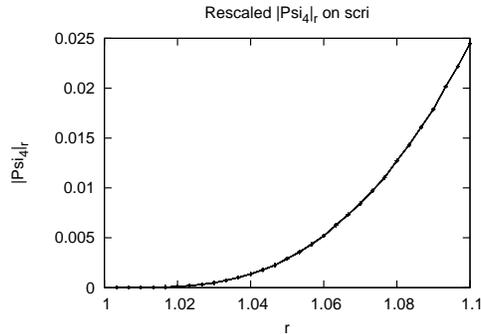}
  \caption{Absolute value of $\psi_4$ along $\scri^+$ in the r-direction 
    representing radiation. \label{fig:psi4r}}
\end{figure}
Fig.~\ref{fig:psi4r} shows $|\psi_4|$ calculated along $\scri^+$ in
the direction of the Cartesian $x$-axis. We observe the same
qualitative behavior as in Fig.~\ref{fig:psi4x}. The radiation field
grows as we move to the interior of the spacetime. The numerical
values depend on the choice of the free function $\kappa$. The
parameter $d$ that determines the height of the cylinder at $r=1$
according to (\ref{eq:goodscri}) is chosen to be small so that we can
calculate a piece of $\scri^+$ before the ill-posed inner boundary
treatment can destroy the solution. For the evolution leading to
Fig.~\ref{fig:psi4r} we have chosen $d=0.02$ and $k=0.7$.

No quantitative comparison has been made between the calculation with the 
one-point compactification and with the cylinder. 
Certainly, more tests and experiments can be done in the massless axisymmetric 
case. It seems, however, more interesting to move on to studies of spacetimes 
with non-vanishing ADM-mass using further developed numerical methods. 
This will be part of future work. 

\section{Discussion}
The main content of this chapter has been the development of a
numerical method that allows us to construct solutions to the Einstein
equations including spatial and null infinity by solving a
regular finite initial value problem near spatial infinity based on the
reduced general conformal field equations.  We have implemented this
system numerically with spatial infinity represented as a cylinder.

The method presented in this chapter is based on solving a frame-based
evolution system with smooth inner and outer boundaries.  Certain
difficulties we had to deal with are specific to numerical
calculations, such as the coordinate and the frame singularity on
spheres.  Other difficulties, such as the degeneracy of the equations
at the set $\mathcal{I}^+$ where spatial infinity meets null infinity
are shared by the numerical and analytic studies. It would be
interesting to see whether our numerical treatment can deal with some
mild singularity of the conformal structure at null infinity.

We have studied a neighborhood of spatial infinity of a radiative massless 
spacetime. The study of spacetimes with non-vanishing ADM-mass may require 
technical improvements of the code. We may use higher order information on the
solution along the cylinder from analytic studies 
\cite{Friedrich98,ValienteKroon04}. Parallelization would allow more extensive 
and accurate studies of the spacetimes in question. A possibility would be to
use parallel multi-block methods to solve evolution systems using multiple 
patches \cite{Lehner2005a,Schnetter06a}. 

As shown in conformal diagrams of section \ref{cgg}, the speed of outgoing 
characteristics approaches zero near spatial infinity. 
This property may be used to construct adapted finite differencing stencils.
One should also consider adapting the stencil to the a priori known location of
$\scri^+$ such that boundary grid points lie on $\scri^+$. 
For long time evolutions aiming at timelike infinity, one may 
add additional grid points dynamically into the computational domain 
such that the resolution loss due to the form of $\scri^+$ is compensated 
numerically. 

A major advantage of the conformal Gauss gauge is that the location of
the conformal boundary can be prescribed a priori in terms of
coordinates and initial data by the choice of a free function
$\kappa$, assuming the underlying conformal geodesics are
well-behaved. A convenient choice for numerical calculations seems to
be (\ref{eq:goodscri}). Given a time scale and an accuracy
requirement, we may choose a very large $k$ that would allow an
efficient and accurate calculation of gravitational radiation in the
conformal Gauss gauge without much resolution loss in the physical
part of the conformal extension.  It is an open issue which choices of
$\kappa$ are good in the interior of asymptotically flat
spacetimes. We have seen in numerical solutions of conformal geodesics
on given backgrounds, that certain choices of $\kappa$ are bad in the
interior of black holes. These problems seem, however, of secondary
importance. The main question that one should study concerns the
feasibility of the conformal Gauss gauge in general studies of the
asymptotic region.  If it turns out that the conformal geodesics are
not well-behaved in the interior region, one can still devise matching
methods or choose another gauge in a domain where the conformal factor
has been set initially to unity. One can only hope to answer the
question on the feasibility of the conformal Gauss gauge
gradually. Possible steps for further research are outlined in section
\ref{sec4:outlook}.

%% file: conclusion.tex
\chapter{Summary and Outlook} \label{sec:conclusion}
In this thesis I took a conformal approach to the numerical calculation
of asymptotically flat spacetimes. To avoid certain conceptual and technical 
deficiencies of current numerical methods discussed in the introduction, I 
devised and studied new numerical methods which have, as expected, 
difficulties of their own. 

In the summary I discuss the achievements of the thesis and point out some 
open problems of the methods introduced. In the outlook I try to give 
a glimpse of the possibilities opened up by the methods assuming their 
problems can be solved. The thesis ends with remarks on the idea of 
conformal infinity.
\section{Summary}
The question that has been raised in the preface concerns the 
numerical methods for calculating asymptotically flat spacetimes, not the 
statements that can be made on such spacetimes. Accordingly the main content of
this thesis does not lie in numerical studies of the solution space to the 
Einstein equations but in the development of new methods for future studies of 
the solution space. I tried to motivate the search for new methods in the 
introduction by stating some problems of currently available numerical methods.
The results of this thesis have been presented in two chapters, each 
concentrating on a different asymptotic domain: 
null infinity and spatial infinity. 
\subsubsection{Chapter 2. Null Infinity}
\begin{itemize}
\item Construction of a $\scri$-fixing gauge in spherical symmetry, 
\ref{sec:scr_fix}.

This study refutes claims made at various places (for example in 
\cite{Calabrese:2005rs}) that the conformal approach necessarily leads to a 
resolution loss in the physical spacetime. The explicit 
examples written in a $\scri$-fixing gauge in spherical symmetry show clearly 
that the allegedly necessary resolution loss is a property of a bad coordinate 
choice and not of the conformal approach itself, a fact sufficiently 
supported by existence results. 
The examples also serve as a testbed for new ideas in the conformal approach.
\pagebreak

\item The development of a numerical method that includes null infinity
  in the computational domain to solve a hyperboloidal
  initial value problem for the Einstein equations in a general wave
  gauge, \ref{sec:hyp_evol}, \ref{sec:gauge}.

A basic difficulty with the equations used in the conformal approach has been
that they are very different from common formulations of the Einstein 
equations. Experience gained within the standard approach could so far not be 
applied in the conformal approach, and the clean framework that the idea of 
conformal infinity yields could not be used in conventional  
numerical relativistic calculations. 
Motivated by this difficulty, I constructed a method to 
include null infinity in the computational domain using a common reduction 
of the Einstein equations.

Assuming hyperboloidal initial data whose maximal development admit a smooth 
conformal boundary,
I showed that a $\scri$-fixing gauge can be constructed during the solution
of a hyperboloidal initial value problem using a suitable coupling
of the conformal and the coordinate gauge freedom in a general wave gauge. 
This result opens up the possibility to attach smoothly a conformally 
compactified asymptotic region, where conformal techniques are applied, 
to successful numerical calculations of the interior domain in a direct way, 
using the same set of variables and the same set of equations.

\item A numerical test of the introduced method in spherical symmetry, 
\ref{ss_num}.

The main difficulty with the aforementioned method is the appearance
of formally singular terms in the evolution equations. Within the
class of spacetimes that admit a smooth conformal boundary and in a
$\scri$-fixing gauge, each of these terms attains at null infinity a
regular limit. This limit can, in principle, be calculated by
numerical methods.  One can not assert without extensive numerical
studies, however, that such a calculation will, in practice, result in
a numerically stable code for highly dynamical spacetimes.  As a first
test I studied the spherically symmetric case in a stationary gauge
using simple numerical methods. One could observe, in this special
case, that a numerical treatment of the outer boundary based on
extrapolation of the variables allows the calculation of a piece of
null infinity.
\end{itemize}

\subsubsection{Chapter 3. Spatial Infinity}

The hyperboloidal approach does not allow us access to spacetimes in their 
entirety. To make statements on global properties of spacetimes 
one would like to get access to the conformal boundary including
spatial infinity. The main problem in this context is 
the singular behaviour of conformal initial data at spatial 
infinity for non-vanishing ADM-mass. This problem was treated by Friedrich who
formulated a regular finite initial value problem near spatial infinity using 
the reduced general conformal field equations. The construction allows a 
smooth extension of conformal data with non-vanishing ADM-mass through spatial 
infinity which is blown up to a sphere on the initial hypersurface. 
The development of the data results in a representation of spatial infinity 
as a cylinder. Currently the only system that allows a numerical study of 
spatial infinity including a piece of null infinity is given by the reduced 
general conformal field equations. The following steps have been taken in the 
course of a numerical implementation of this system.
\pagebreak
\begin{itemize}
\item Numerical construction of a conformal Gauss gauge on the 
Schwarzschild-Kruskal and the Kerr spacetimes, \ref{cgg}.

The reduced general conformal field equations are based on the
conformal Gauss gauge.  Though it has been shown analytically that the
conformal Gauss gauge is well-behaved in strong field regions in
spherical symmetry \cite{Friedrich03}, the general behavior of the
underlying conformal geodesics is unknown.  Besides, it can not be
guaranteed that analytical methods work well in numerical
calculations.  To test the numerical feasibility of the conformal
Gauss gauge I reproduced the analytic construction of a conformal
Gauss gauge in the Schwarzschild-Kruskal spacetime covering the entire
spacetime in a smooth way. Going beyond available analytical studies I
found out numerically that one can also cover the Kerr spacetime using
conformal geodesics including null infinity, timelike infinity and the
Cauchy horizon.

\item Numerical simulation of an entire, asymptotically flat, black hole
spacetime, \ref{sec:ss}.

As a first test of the numerical feasibility of the reduced general
conformal field equations, I solved a Cauchy problem based on this
system in spherical symmetry with data from the Schwarzschild-Kruskal
spacetime.  The equations in spherical symmetry become a system of
ordinary differential equations. This is a major simplification so
that no statements on the applicability of the system in the general
case can be deduced from this study. It should still be noted that
this is the first numerical calculation of an entire,
asymptotically flat black hole spacetime including spacelike, null and
timelike infinity and the region close to the singularity.

\item Numerical calculation of a radiative spacetime including spatial 
infinity and a piece of null infinity, \ref{sec:cart_rgcfe}.

In order to study cases with non-vanishing gravitational radiation, 
I wrote an evolution code for the reduced general conformal field equations 
based on a three dimensional Cartesian grid using the software 
infrastructures Cactus \cite{Cactusweb} and Kranc \cite{Husa:2004ip}.
This code can only deal with the one point compactification at spatial 
infinity which results in singular conformal 
data for non-vanishing ADM-mass. Therefore I calculated asymptotically 
flat, axisymmetric initial data based on studies by Friedrich which have 
vanishing ADM-mass but whose development has a non-vanishing radiation field. 
I analyzed the numerical development of this data and showed that the radiation
content of the spacetime does not vanish but grows towards the interior where 
presumably a naked singularity resides. 

\item The development of a three dimensional code to solve Cauchy problems for
frame-based evolution systems with smooth inner and outer boundaries, 
\ref{sec:gzps}.

I wrote a numerical code within the Cactus infrastructure
\cite{Cactusweb} using a Cactus thorn written by Thornburg 
\cite{Thornburg04} which implements spherical coordinates using multiple 
patches. I extended this infrastructure to include frame rotations so that 
Cauchy problems for frame-based evolution systems can be solved numerically.

\item Numerical studies of spatial infinity represented as a cylinder, 
\ref{sec:cylinder}.

In the representation of spatial infinity as a cylinder, the equations 
degenerate at the set where the cylinder meets null infinity. This difficulty 
has been dealt with by freezing the evolution in the unphysical
domain and choosing a time stepping that avoided an evaluation of the 
equations at the critical set. An essential property of the conformal 
Gauss gauge which allows us such a treatment of the problem is the a priori 
knowledge of the conformal factor in terms of initial data and grid 
coordinates. I suggested a convenient choice for the remaining gauge freedom 
in the conformal Gauss gauge for numerical calculations. I did numerical 
studies of the cylinder in the massless, axisymmetric case 
which is special in the sense that the spacetime under study 
allows the one-point compactification.

The main open problem with this approach is related to the degeneracy of the 
equations at the critical set. Extensive studies and sophisticated numerical 
techniques are required before this approach can be applied to study 
interesting mathematical and physical questions concerning the asymptotic 
behavior of solutions to Einstein equations.
\end{itemize}

The studies show that the a priori knowledge of the conformal factor is a very 
convenient feature in numerical studies. It seems essential that the conformal
boundary and also the conformal factor should be controlled to some degree in a
numerical calculation. One can expect that this feature will play an
important role in the future development of the conformal approach.

\section{Outlook}\label{sec4:outlook}
The discussed methods suggest a wide range of prospects to study the solution 
space to the Einstein equations in a way that has not been possible before. 
Some suggestions are listed below.

\begin{itemize}
\item A simple but illustrative experiment would be to solve a wave equation 
on the Schwarzschild and the Kerr spacetimes written in a $\scri$-fixing gauge 
using a three dimensional code that can handle spherical grid
topology. This might allow an accurate study of quasi-normal modes
and tail behaviour.

\item The method of solving an hyperboloidal initial value problem for the 
conformally transformed Einstein equations in a general wave gauge 
by prescribing the coordinate representation of a conformal factor should be 
tested in three dimensions for dynamical spacetimes. The basic infrastructure 
required for this test is already available \cite{Pazos06} so that one can 
concentrate on the main intrinsic problem of the method, namely the numerical 
calculation of the formally singular terms in the course of the evolution.

\item The idea of using compactifying coordinates without transforming the 
metric seems to work on Cauchy hypersurfaces \cite{Pretorius05b}, 
although this is very awkward to study radiation as discussed 
in \ref{sec:coord_comp}. It would be interesting to try the coordinate 
compactification technique using a hyperboloidal foliation.

\item One should construct hyperboloidal initial data 
for a radiative spacetime admitting a smooth conformal boundary.
There are many interesting and open questions related to the interpretation of 
hyperboloidal data that may be studied by evolving them.

\item A straightforward test of the numerical techniques presented in
  chapter 3 would be to evolve conformal data corresponding to static
  or stationary spacetimes with non-vanishing ADM-mass in a
  neighbourhood of spatial infinity and to check that their radiation
  field vanishes.

\item One should construct solutions with a complete initial
  hypersurface so that no artificial inner boundary needs to be
  introduced. The question how to construct numerically initial data
  for spacetimes which admit a smooth conformal boundary is open. Data
  whose development leads to some mildly singular behavior at null
  infinity can be constructed by solving numerically a Lichnerowicz
  equation for the initial conformal factor \cite{Dain01a}.  To
  numerically calculate conformal initial data for the reduced general
  conformal field equations one needs to build numerical
  divisions of the data by the conformal factor \ref{sec:id_rcfe}.
  This might not to be an easy task, but it
  seems worth the effort. A numerical study on smoothness properties
  of solutions near null infinity has not yet been done.  A strong
  interaction between mathematical and numerical studies in this
  question might give new impulses in both directions.

\item Another interesting study related to the asymptotic behavior of 
solutions to the Einstein equations would be to evolve data which might
results in different degrees of smoothness at future and past null infinity as 
suggested in \cite{ValienteKroon06}.

\item The setting with the cylinder at spatial infinity would allow us to 
construct solutions in which pure gravitational radiation collapses to form 
a black hole as suggested in \cite{Beig91, Beig93}
(see also the discussion in \cite{Friedrich:2005kk}).

\item One might try to use use the conformal Gauss gauge in a metric 
formulation Einstein equations. This might lead
to a simplification in the equations and require less variables.

\item The implementation of a frame-based evolution system should allow us to 
implement the Friedrich-Nagy system \cite{Friedrich99}. An interesting study
with a successful implementation of the Friedrich-Nagy system and the reduced
general conformal field equations would be to evolve a simple but radiative 
spacetime both with and without an artificial, timelike outer boundary. This 
would allow us to compare the approximation given by the timelike outer 
boundary at a finite distance away from the source with the idealization given 
by conformal infinity such that systematic errors can be estimated for 
different kinds of boundary data. Studies along these lines have been 
made without numerical access to null infinity using solutions for 
which accurate analytic expectations can be calculated \cite{Pazos06,Rinne07}.
\end{itemize}
One of the motivations behind the work of this thesis has been to develop 
techniques to calculate numerically entire, asymptotically flat, radiative 
spacetimes. This still remains an important goal for future work.
A combination of numerical calculations near spatial infinity and null 
infinity might allow us to accomplish this task. An achievement of this
goal would deliver a starting point for extensive studies of the solution space
to the Einstein equations using numerical methods based on rigorous analysis.
\section{Concluding remarks}
A general critique concerning the concept of null infinity and its role in 
physical questions holds that null infinity is "too far away".
After all, the Earth where we are doing 
our measurements is not infinitely far away from the sources and we do 
not move along null geodesics in contrast to an observer along null infinity. 

In the Schwarzschild spacetime, we have a length scale $m$ at our disposal, 
so we can discuss astrophysical distances in numerical calculations. 
To give a notion for the scales in the conformal diagram Fig.~\ref{fig:cgg}, 
three curves with constant Schwarzschild radius  
$\tilde{r}_s=4m, 12m, 300 m$ have been plotted. 
The curve at Schwarzschild radius $300m$ corresponds to about $1800$ km 
for a black hole with four solar masses 
but it can hardly be distinguished from $\scri^+$ in the conformal 
diagram. Even if we consider super-massive black holes, 
the distance corresponding to $300 m$ is incomparably small 
contrasted to the thousands or millions of light years 
that separates us from the astrophysical sources. 
In this representation, the numerical effort for simulating the region from 
$300 m$ to $\scri^+$ is very small whereas in the standard approach putting the
outer boundary from $300 m$ to $1000 m$ costs considerable effort in terms of 
numerical techniques and computational sources. We note that this behavior 
depends on the conformal gauge. One has the freedom to set $\Omega=1$ 
initially on a given domain if more resolution is desired in that domain. 
Further, we have seen in chapter 2 that a positive smooth function for the 
coordinate representation of the conformal factor may be prescribed 
in the context of the hyperboloidal initial value problem such that 
certain spacetime domains can be emphasized for the calculation while still 
compactifying the asymptotic region. We may conclude that the 
conformal compactification technique is not only promising for an unambiguous 
outer boundary treatment and radiation extraction, but also for a 
computationally efficient numerical code to deal with the asymptotic region. 

If some astrophysically motivated length scale can be determined, as 
suggested for example in \cite{Cox07}, such that an observer "nearly at rest 
in the frame of the isotropic cosmic microwave background radiation" at that 
distance from the source can regard the source as isolated, 
this length scale will naturally be given in astronomical units. 
The example above shows that a timelike curve at such a distance away from 
the source is almost indistinguishable from $\scri^+$ 
in the conformally compactified picture 
while the standard approach does not even allow the discussion of 
such length scales. Therefore, the arguments presented in \cite{Cox07}
can as well be considered as supporting the notion of conformal infinity.
After having calculated the entire spacetime, we can still discuss the 
calculation of a field which represents radiation (having a proper limit 
with respect to an adapted tetrad at null infinity to which we would have 
direct access) on a timelike surface arbitrarily far away from the source 
of radiation. Within the standard approach, however, it seems hardly possible 
to calculate an astronomical domain for an isolated system.

In this context, I would like to emphasize that the idea of conformal infinity
is an idealization, in contrast to an approximation. The difference can be
illustrated on the idea of a number, say a real number. 
We do not have physical access to
real numbers. Results of measurements are in a sense fuzzy. They are given in 
terms of rational numbers with error bars regarded as approximations to real
numbers.
Physics without real numbers would be, however, very inconvenient. 
Most physicists would agree that the notion 
of a real number is a useful one. 
The idealization of a real number allows a clean modeling 
in various fields of science. Similarly, the concept of 
conformal infinity allows a clean modeling in general relativity. 

In \cite{Cox07} it is stated that the conformal approach implies "introducing 
an arbitrary amount of empty space not reflecting astronomical reality". 
A similar statement can be made concerning real numbers by claiming that they 
imply an arbitrary amount of accuracy beyond the Planck scale not 
reflecting physical reality. The question whether real numbers are real 
in an ontological sense belongs, however, to the realm of philosophy. 
If an idealization allows a clean modeling of the 'actual reality' we can 
observe and if it is of practical use, than it is a viable physical viewpoint 
to use the concept while being aware of the limits of its application. 

Another argument claims that there is not much motivation for using
the idea of conformal infinity, especially regarding the impressive
milestones achieved recently within the standard approach.  While a
case can be made that the deficiencies of the standard approach are
not relevant for the detection of gravitational waves as the accuracy
of the waveforms seems to be enough for detection, there is also the
viewpoint holding that interesting physics will be hidden behind the
first couple of digits in the observational data.  Having predictions
which are accurate only up to a couple of percents due to systematic
errors will most probably not be enough to answer many interesting
questions in the prospective field of gravitational wave astronomy.
One should also not forget that beside the effort of calculating
waveforms there remains the task of exploring the solution space.
Here it is unclear how useful the conventional approaches are if
global questions are concerned.

If an objection is to be made against the conformal approach, I
believe, it should be made not on conceptual but rather on practical
grounds.  Whatever level of conceptual clarity, geometric appeal or
computational efficiency the conformal approach promises, in the end,
its application in numerical relativity must be judged after its
practicability in numerical studies of the solution space to the
Einstein equations.

It seems to me that the most important task for future work in the
context of our discussion is to devise from the conformal approach a
\emph{practical tool} for numerical calculations of asymptotically
flat spacetimes.  This could not be achieved yet for the calculation
of highly dynamical, asymptotically flat spacetimes.  Unfortunately,
the results presented in this thesis also do not allow the conclusion
that the suggested methods can be used in general calculations.  I
believe that this task can only be managed if insights that have been
gained in the standard approach can be conveniently combined with the
calculation of the asymptotic region.  The ideas presented in this
thesis, especially the a priori knowledge of the conformal factor in
numerical calculations, seem promising to accomplish this task. One
may hope that future work will allow us to construct from the
beautiful idea of conformal infinity a practical tool for numerical
calculations.

%% file: appendix.tex
\begin{appendix}
\chapter{Calculation of causal diagrams}\label{app:A}
In this appendix, we present how Penrose diagrams like 
Fig.~\ref{fig:schwarzschild} or Fig.~\ref{fig:cmc_ss_conformal} and
radial light rays on the grid like in Fig.~\ref{fig:ief_grid} or in
Fig.~\ref{fig:cmc_ss_causal} for a given Schwarzschild-Kruskal background 
are calculated. We omit angular dimensions.
\section{Penrose diagrams}\label{app:sk}
The two-dimensional Schwarzschild metric in standard coordinates is given by
\[ \tilde{g}_s = - \left( 1 - \frac{2m}{\tilde{r}_s}\right)\, d\tilde{t}^2+  
\left( 1 - \frac{2m}{\tilde{r}_s}\right)^{-1} \, d\tilde{r}_s^2. \]
The coordinate singularity at the event horizon $\tilde{r}_s=2m$ 
can be removed by introducing advanced and retarded null coordinates 
\be \label{app:vu} \tilde{v}=\tilde{t}+ \tilde{r}_\ast, \qquad 
\tilde{u}=\tilde{t}- \tilde{r}_\ast \qquad \mathrm{where} \qquad  
\tilde{r}_\ast = \tilde{r}_s+2m\,\ln\,\left(\frac{\tilde{r}_s}{2m}-1\right).\ee
The Schwarzschild metric in these coordinates takes the forms
\be\label{app:ef} \tilde{g}_s=-\left( 1-\frac{2m}{\tilde{r}_s}\right)\, 
d\tilde{v}^2 + 2\,d\tilde{v}\,d\tilde{r}_s, \qquad 
\tilde{g}_s=-\left( 1-\frac{2m}{\tilde{r}_s}\right)\, d\tilde{u}^2 - 
2\,d\tilde{u}\,d\tilde{r}_s.\ee
These line elements can be analytically extended beyond the event
horizon to $r>0$.

We introduce the Kruskal-Szekeres coordinates for $\tilde{r}_s>2m$ by
$\tilde{V}=e^{\tilde{v}/4m}>0$ and  $\tilde{U}=-e^{-\tilde{u}/4m}<0$.
The metric written in double-null coordinates becomes
\[ \tilde{g}_s = -\left(1-\frac{2m}{\tilde{r}_s}\right)\,
d\tilde{u} d\tilde{v} = -\frac{32 m^3}{\tilde{r}_s}\,e^{-\tilde{r}_S/2m}\,
d\tilde{U} d\tilde{V}, \]
where $\tilde{r}_s=\tilde{r}_s(\tilde{u},\tilde{v})=
\tilde{r}_s(\tilde{U},\tilde{V})$ respectively. The Kruskal-Szekeres 
coordinates can be analytically extended to $\tilde{U}>0$ und $\tilde{V}<0$.
The Schwarzschild coordinates are given by the implicit relations
\[ \tilde{U}\tilde{V} = -e^{\tilde{r}_s/2m}\left(\frac{\tilde{r}_s}{2m}-1\right), 
\quad  \frac{\tilde{V}}{\tilde{U}}=\mp e^{\tilde{t}/2m}.\]
Compactifying coordinates can be introduced by $U = \arctan\tilde{U}$ and
$V=\arctan\tilde{V}$. These are defined in the range 
$(-\pi/2\leq U,V \leq \pi/2)$. We define new time and space 
coordinates by setting
\be\label{app:comptr} 
t = \frac{1}{2}\left(V+U\right),\qquad r = \frac{1}{2}\left(V-U\right).\ee
In the Penrose diagrams, the $t$-axis runs vertically and the $r$-axis
runs horizontally. Summarizing the coordinate relations, we have 
$\tan V=e^{\tilde{v}/4m},\quad \tan U=\mp e^{-\tilde{u}/4m}$, and 
\be\label{app:cts} 
\tan U \tan V = -e^{\tilde{r}_s/2m}\left(\frac{\tilde{r}_s}{2m}-1\right), \qquad
\frac{\tan V}{\tan U}=\mp e^{\tilde{t}/2m}.\ee The compactifying
coordinates $(V,U)$ have been used to plot the Penrose diagrams of the
Schwarzschild-Kruskal spacetime in Fig.~\ref{fig:schwarzschild} or the
extended Schwarzschild spacetime in Fig.~\ref{fig:ief}.
\subsection{Ingoing Eddington-Finkelstein coordinates}
We introduce the coordinate $\tilde{\tau} = \tilde{v} - \tilde{r}_s$. 
The metric becomes
\be \label{app:ief}
\tilde{g}_s=-\left(1-\frac{2m}{\tilde{r}_s}\right)\,d\tilde{\tau}^2 + 
\frac{4m}{\tilde{r}_s} \, d\tilde{\tau} d\tilde{r}_s+\left( 1 + \frac{2m}
{\tilde{r}_s}\right) \, d\tilde{r}_s^2. \ee
These coordinates are sometimes referred to as ingoing Eddington-Finkelstein 
coordinates in the numerical literature. 
The spatial surfaces that are used in a simulation based on ingoing 
Eddington-Finkelstein coordinates are given by the level sets of the function
\[\tilde{\tau}(\tilde{t},\tilde{r}_s)= 
\tilde{v}-\tilde{r}_s = \tilde{t} + \tilde{r}_\ast - \tilde{r}_s = 
\tilde{t}+2m\,\ln\,\left(\frac{\tilde{r}_s}{2m}-1\right). \]
To plot the surfaces $\tilde{\tau}=\mathrm{const.}$ in the Penrose diagram of 
the extended Schwarzschild spacetime in Fig.~\ref{fig:ief}, we use 
(\ref{app:cts}) and write (see also \cite{Martel00})
\[ V(\tilde{r}_s) = \arctan\left(e^{\tilde{\tau}/4m} 
e^{\tilde{r}_s/4m} \right),\qquad
U(\tilde{r}_s)=\arctan\left(-e^{-\tilde{\tau}/4m} e^{\tilde{r}_s/4m}
\left(\frac{\tilde{r}_s}{2m}-1\right)\right).\] 
The Schwarzschild coordinate $\tilde{r}_s$ is regarded as a parameter
along the curves \linebreak $(V(\tilde{r}_s), U(\tilde{r}_s))$ with
constant values of $\tilde{\tau}$. We transform the above expressions using the
compactifying time and space coordinates $(t,r)$ from
(\ref{app:comptr}). The resulting curves are plotted on the
$(t,r)$-plane with \texttt{Mathematica} using the function
\texttt{ParametricPlot}.

For plotting the hypersurfaces $\tilde{r}_s=\textrm{const.}$ we can write 
a similar parametric representation.
Using the relations (\ref{app:vu}) and (\ref{app:cts}), we write
\[ \tan V = e^{\tilde{t}/4m} 
e^{\tilde{r}_s/4m} \sqrt{\frac{\tilde{r}_s}{2m}-1}, \qquad 
\tan U=-e^{-\tilde{t}/4m} e^{\tilde{r}_s/4m}
\sqrt{\frac{\tilde{r}_s}{2m}-1}. \]
We regard $\tilde{t}$ to be the parameter along the curves 
$(V(\tilde{t}),U(\tilde{t}))$ with constant values of $\tilde{r}_s$.

\subsection{Constant mean curvature foliation}\label{app:cmc}
In Fig.~\ref{fig:cmc_causal} and Fig.~\ref{fig:cmc_ss_conformal} we plotted 
constant mean curvature surfaces in the extended Schwarzschild spacetime
using compactifying coordinates. The properties of these surfaces 
have been discussed in \cite{MalecMurch03}. 

Surfaces in the Schwarzschild spacetime that are dragged along the
timelike Killing vector field $\partial_{\tilde{t}}$ can be written as
level sets of a function
$\Phi(\tilde{t},\tilde{r}_s)=\tilde{t}-h(\tilde{r}_s)$. Here, 
$h(\tilde{r}_s)$ is referred to as the height function. To find a
CMC-slicing of the Schwarzschild spacetime one requires that the
mean extrinsic curvature $\tilde{K}$ of the surfaces
$\Phi=\mathrm{const}.$ is constant. We get
\[ \tilde{K} =\tilde{\nabla}_\mu \tilde{n}^\mu = \frac{1}{\tilde{r}_s^2} 
\partial_{\tilde{r}_s}\left(\frac{\tilde{r}_s^2 h' \left(1-\frac{2m}{\tilde{r}_s}
\right)}{\sqrt{\left(1-\frac{2m}{\tilde{r}_s}\right)^{-1}-h'^2\left(1-\frac{2m}
{\tilde{r}_s}\right)}}\right).\]
For $\tilde{K}=\mathrm{const.}$ this can be integrated once with an integration
constant $C$. A following algebraic manipulation results in the differential 
equation
\be\label{app:height} h'=\frac{\frac{\tilde{K}\tilde{r}_s^3}{3} - C}
{\left(1-\frac{2m}{\tilde{r}_s}\right)\tilde{P}(\tilde{r}_s)},\ee
where $\tilde{P}(\tilde{r}_s)$ is as defined in \ref{sec:cmc_ss}. 
We can not integrate this equation explicitly to get $h(\tilde{r}_s)$
in closed form but we can integrate it numerically. 
A difficulty is the numerical integration for the function $h$
at the horizon for $\tilde{r}_s\to 2m$.
The integral can be calculated in the Cauchy principal value sense
\cite{MalecMurch03}. 
We build the integral in the regions 
$\tilde{r}_s>2m$ and $\tilde{r}_s<2m$ respectively via
\[ h(\tilde{r}_s)=\int_{2m+\epsilon}^{\tilde{r}_s}h'(x)\,dx, \qquad
\mathrm{and} \qquad h(\tilde{r}_s)=\int_{\tilde{r}_s}^{2m-\epsilon}h'(x)\,dx,\]
where $\epsilon$ is a positive small constant. 
We subsequently match the resulting 
curves together at the event horizon. 
The embedding of the foliation is calculated as follows. We define the embedded 
surfaces by level sets of a function which can be written as
\[ \Phi= \tilde{t}-h(\tilde{r}_s) =  \tilde{t}+\tilde{r}_\ast - \tilde{r}_\ast -
h(\tilde{r}_s) = \tilde{v} -\tilde{r}_\ast - h(\tilde{r}_s).\]
so that $\tilde{v} = \Phi +(\tilde{r}_\ast+h(\tilde{r}_s))$.
We use the relations (\ref{app:cts}) and write 
\begin{eqnarray*}
\tan V &=& e^{\Phi/4m} e^{\tilde{r}_s/4m}\sqrt{\frac{\tilde{r}_s}{2m}-1}\ 
e^{h(\tilde{r}_s)/4m},\\ \tan U &=& - e^{-\Phi/4m} e^{\tilde{r}_s/4m}
\sqrt{\frac{\tilde{r}_s}{2m}-1}\ e^{-h(\tilde{r}_s)/4m}.\end{eqnarray*}
The numerical integration gives us a list of pairs of 
$(\tilde{r}_s,h(\tilde{r}_s))$. The surfaces corresponding to different
values of $\Phi$ can be plotted with \texttt{Mathematica} using the 
function \texttt{ListPlot}.
\pagebreak
\section{Null rays on the grid}
To visualize the causal structure on a numerical grid for a spherically 
symmetric spacetime in a stationary gauge we calculate the ingoing and outgoing
radial null rays with respect to grid coordinates $(t,r)$. 
The metric can be written as
\[g=g_{tt}(r) \,dt^2 + 2g_{tr}(r) \,dt\,dr + g_{rr}(r) \,dr^2.\]
A parametrized null curve $t(r)$ fulfills the relation
\[ g_{tt}\, \left(\frac{dt}{dr}\right)^2 + 2 g_{tr}\,\frac{dt}{dr}+g_{rr} = 0. \]
For $g_{tt}\ne 0$, we can calculate the curve $t(r)$ by integrating
\be \label{app:null_cone}
t(r) = \int\frac{-g_{tr} \pm \sqrt{g_{tr}^2 - g_{tt}g_{rr}}}{g_{tt}} \ dr'.\ee
The positive sign corresponds to ingoing null rays. In the following we 
discuss the visualisation of the causal structure on the grid for the 
Schwarzschild spacetime based on different coordinates.
\subsection{Ingoing Eddington-Finkelstein coordinates}
For Fig.~\ref{fig:ief_grid} we use the form of the metric in ingoing 
Eddington-Finkelstein coordinates (\ref{app:ief}) and integrate 
(\ref{app:null_cone}) to get
\[ \tilde{\tau}_{in}(\tilde{r}_s)=-\tilde{r}_s+c_{in}, \qquad \tilde{\tau}_{out}
(\tilde{r}_s)=\tilde{r}_s+4m\ln\abs{\tilde{r}_s-2m} + c_{out}, \] 
for ingoing and outgoing null rays. We plot different light rays by varying the 
constants $c_{in}$ and $c_{out}$ To include the excision region in 
Fig.~\ref{fig:ief_grid} we plot the null rays inside and outside 
the event horizon separately and match them together.
\subsection{Coordinate compactification at spatial infinity}
To generate the plots in Fig.~\ref{fig:i0_grid1} and Fig.~\ref{fig:i0_grid2} 
we introduce a compactifying radial coordinate in (\ref{app:ief}) by 
\[\tilde{r}_s(r)=\frac{r}{1-r}, \quad d\tilde{r}_s = \frac{1}{(1-r)^2}\,dr^2,\]
The domain $\tilde{r}_s\in [3/2,\infty)$ corresponds to $r\in[3/5,1]$
and the event horizon is at $r=2/3$. The metric (\ref{app:ief}) becomes
\[ \tilde{g} = - \left(1-\frac{2m}{r}(1-r)\right)\,d\tilde{\tau}^2 + 
\frac{4m}{r(1-r)}\,d\tilde{\tau}\,dr + \left(1+\frac{2m}{r}(1-r)\right)\,
\frac{1}{(1-r)^2}\,dr^2.\]
This metric is singular at $r=1$ as expected. Integration of 
(\ref{app:null_cone}) results in long expressions that can be written in 
explicit form. Note that the ingoing null rays in Fig.~\ref{fig:i0_grid2} that 
seem to come in from spatial infinity start at a small vicinity of
spatial infinity.  
\subsection{Cauchy-Characteristic Matching}
The left part of Fig.~\ref{fig:ccm_grid} has been plotted using the timelike 
ingoing Eddington-Finkelstein coordinates as described above. For the right
part we write the Schwarzschild metric in the advanced null coordinate 
$\tilde{u}$ as in (\ref{app:ef}) and subsequently introduce a compactifying 
radial coordinate $r$ via $\tilde{r}_s = r/(1-r)$ to get
\[ \tilde{g}_s = -\left(1-\frac{2m}{r}(1-r)\right)\,d\tilde{u}^2 -\frac{2}{1-r}
\,d\tilde{u}\,dr. \]
The light cones in the parametrization $\tilde{u}(r)$ are then given by 
solutions to the differential equation
\[\left(1-\frac{2m}{r}(1-r)\right)\,\left(\frac{d\tilde{u}}{dr}\right)^2 + 
\frac{2}{1-r}\,\frac{d\tilde{u}}{dr}=0. \]
The solutions are given by
\[ \tilde{u}_{in}(r)=2\,\ln\abs{1-r}-\frac{4m \ln\abs{r-2m(1-r)}}{1+2m}+ 
c_{in} \qquad \tilde{u}_{out}(r)= c_{out}.\]
Note that we need these light rays only outside the event horizon in the 
matching region.
\subsection{Constant mean curvature slicing}
Fig.~\ref{fig:cmc_grid} and Fig.~\ref{fig:cmc_ss_causal} have been plotted 
using the form of the Schwarzschild metric given in (\ref{cmc_ss}). For this
metric we can not integrate (\ref{app:null_cone}) to get a closed form for the 
ingoing and outgoing light rays. One needs to calculate the integrals
numerically. 

\end{appendix}